\documentclass[a4paper,fleqn,usenatbib,useAMS]{mnras}

\usepackage[T1]{fontenc}
\usepackage{ae,aecompl}

\usepackage{graphicx}   
\usepackage{amsmath}    
\usepackage{amssymb}    
\usepackage{pdflscape}  
\usepackage{gensymb}    
\usepackage{array}      
\usepackage{enumerate}  

\title[The SONS survey of debris discs]{SONS: The JCMT legacy survey of debris discs in the submillimetre}
\author[W. S. Holland et al.]
     {Wayne S. Holland,$^{1,2}$\thanks{E-mail:~wayne.holland@stfc.ac.uk}
     Brenda C. Matthews,$^{3,4}$
     Grant M. Kennedy,$^{5}$
\newauthor
     Jane S. Greaves,$^{6}$\thanks{Present address: School of Physics \& Astronomy, Cardiff University, 5 The Parade, Cardiff, CF24 3AA, UK}
     Mark C. Wyatt,$^{5}$
     Mark Booth,$^{7,8}$
     Pierre Bastien,$^{9}$
\newauthor
     Geoff Bryden,$^{10}$
     Harold Butner,$^{11}$
     Christine H. Chen,$^{12}$
     Antonio Chrysostomou,$^{13}$\thanks{Present address: SKA Organisation, Jodrell Bank Observatory, Lower Withington, Macclesfield, Chesire, SK11 9DL, UK}
\newauthor
     Claire L. Davies,$^{6}$\thanks{Present address: School of Physics, University of Exeter, Physics Building, Stocker Road, Exeter, EX4 4QL, UK}
     William R. F. Dent,$^{14}$
     James Di Francesco,$^{3,4}$
\newauthor
     Gaspard Duch{\^e}ne,$^{15,16}$
     Andy G. Gibb,$^{17}$
     Per Friberg$,^{18}$\thanks{Present address: East Asian Observatory, 660 N. A`oh\={o}k\={u} Place, University Park, Hilo, HI 96720, USA}
     Rob J. Ivison$,^{2,19}$
\newauthor
     Tim Jenness$,^{18}$\thanks{Present address: LSST Project Office, 950 N. Cherry Avenue, Tucson, AZ 85719, USA}
     JJ Kavelaars,$^{3,4}$
     Samantha Lawler,$^{3,4}$
     Jean-Fran{\c c}ois Lestrade$,^{20}$
\newauthor
     Jonathan P. Marshall,$^{21,22,23}$
     Amaya Moro-Martin$,^{12,24}$
     Olja Pani{\'c}$,^{5}$\thanks{Present address: School of Physics and Astronomy, E C Stoner Building, University of Leeds, Leeds, LS2 9JT, UK}
\newauthor
     Neil Phillips,$^{14}$
     Stephen Serjeant,$^{25}$
     Gerald H. Schieven$,^{3,4}$
     Bruce Sibthorpe$,^{26}$\thanks{Present address: Astrium Airbus Defence and Space, Gunnels Wood Road, Stevenage, SG1 2AS, UK}
\newauthor
     Laura Vican,$^{27}$
     Derek Ward-Thompson,$^{28}$
     Paul van der Werf,$^{29}$
\newauthor
     Glenn J. White$,^{25,30}$
     David Wilner,$^{31}$
     Ben Zuckerman$^{27}$
\newauthor
\small\it{Affiliations are listed at the end of the paper.}
}
\date{Accepted XXX. Received YYY; in original form ZZZ}

\pubyear{2016}

\begin{document}
\label{firstpage}
\pagerange{\pageref{firstpage}--\pageref{lastpage}}
\maketitle


\begin{abstract}

Debris discs are evidence of the ongoing destructive collisions between planetesimals, and their presence around stars also suggests that planets exist in these systems. In 
this paper, we present submillimetre images of the thermal emission from debris discs that formed the SCUBA-2 Observations of Nearby Stars (SONS) survey, one of seven legacy 
surveys undertaken on the James Clerk Maxwell telescope between 2012 and 2015. The overall results of the survey are presented in the form of 850\,$\umu$m (and 450\,$\umu$m, 
where possible) images and fluxes for the observed fields. Excess thermal emission, over that expected from the stellar photosphere, is detected around 49 stars out of the 
100 observed fields. The discs are characterised in terms of their flux density, size (radial distribution of the dust) and derived dust properties from their spectral 
energy distributions. The results show discs over a range of sizes, typically 1--10 times the diameter of the Edgeworth-Kuiper Belt in our Solar System. The mass of a disc, 
for particles up to a few millimetres in size, is uniquely obtainable with submillimetre observations and this quantity is presented as a function of the host stars' age, 
showing a tentative decline in mass with age. Having doubled the number of imaged discs at submillimetre wavelengths from ground-based, single dish telescope observations, 
one of the key legacy products from the SONS survey is to provide a comprehensive target list to observe at high angular resolution using submillimetre/millimetre 
interferometers (e.g., ALMA, SMA).

\end{abstract}




\section{Introduction}
\label{sec:introduction}

Debris discs represent the longest-lived phase in the lifetime of circumstellar discs. Following the decline of the gas-rich protoplanetary phase when agglomeration processes 
prevail, the remnant mass of circumstellar discs is dominated by planetesimals, which undergo collisional grinding down to smaller and smaller bodies, until particles reach the 
blow-out size determined by the radiation pressure from the host star \citep[e.g.][]{Wyatt2008, Krivov2010}. The presence of these unseen planetesimals can be inferred through 
scattered light or thermal emission from micron to millimetre-sized dust grains. The dust must be continuously replenished, by ongoing collisions between the aforementioned 
planetesimals, since the timescales for dust grains to be removed from the system are significantly shorter than the ages of the stars around which they are observed 
\citep{Backman+Paresce1993}. It appears to be the case that debris discs can persist over \emph{all} stages following the pre-main-sequence phase of stellar evolution 
\citep[e.g.][]{Bonsor2013}, even including white dwarfs \citep[e.g.][]{Farihi2016}.

\vskip 1mm

Observations at submillimetre/millimetre wavelengths are immensely valuable to the study of debris discs in that they trace the Rayleigh-Jeans tail of the outer cold dust in a 
system \citep{Matthews2014}. For example, they probe substantially different (thermal) emission mechanisms than scattered light observations, and lower characteristic 
temperatures for the material than far-infrared (far-IR) data. These long wavelengths also provide an important anchor to the flux energy distribution (loosely referred to in 
this paper as the spectral energy distribution, or SED) in an otherwise poorly constrained wavelength range, and can indicate the presence of any (cold) disc components not 
detectable at shorter wavelengths. By probing the Rayleigh-Jeans tail of the spectrum, the effect of any possible bias introduced by modelling the dust temperatures from the 
observed data is minimised, thus allowing information to be derived on the radial distribution of the disc and the size distribution of the emitting grains \citep{Ertel2012, 
Marshall2014b}. The slope of the spectrum constrains the dust size distribution, providing a test of whether or not the solids in the disc are undergoing a steady-state 
collisional cascade. Critically, since the emission is optically thin, the dust mass for grain sizes up to $\sim$1\,mm is uniquely determined from submillimetre data.

\vskip 1mm

The disc component (in millimetre-sized grains) probed in the submillimetre is also unique from the perspective of understanding disc dynamics. These relatively large dust 
grains are less affected by the radiation or stellar wind pressure \citep{Burns1979} and therefore trace the location of their parent planetesimal belts more reliably than 
smaller grains at shorter wavelengths. Debris discs act as important pointers to planetary systems \citep{Kospal2009} with features in the discs having the potential to 
highlight the presence of planets, even in cases where the planet is as yet undetected, or would be difficult to detect by any other method, including direct imaging 
\citep[e.g.][]{Wyatt2003, Wyatt2006}. For example, the planet around $\beta$ Pictoris was predicted due to structure in the debris disc through scattered light imaging before 
the planet was found \citep{Mouillet1997, Heap2000}. Whilst scattered light observations are sensitive to the small grains around a given star, the bulk of the mass resides in 
the largest grains most detectable at submillimetre to centimetre wavelengths. These grains are most likely to be located in or near the planetesimal belts, and hence may show 
evidence of perturbed geometries due to resonances with long-period planets \citep{Wyatt2006}.

\vskip 1mm

The James Clerk Maxwell telescope (JCMT) has a long history of debris disc studies \citep[e.g.,][]{Zuckerman&Becklin1993}, including some of the earliest imaging using the SCUBA 
camera \citep{Holland1999}. At the time of its decommissioning in 2005, half of the resolved images of debris discs (about a dozen in total) were due to submillimetre imaging 
with SCUBA \citep[e.g.,][]{Holland1998,Greaves2005}. Subsequent surveys in the mid-far IR (e.g. using \emph{Spitzer}, \emph{AKARI} and \emph{Herschel}) identified a large sample 
of discs in the solar vicinity (to a distance of $\sim$100\,pc). For example, the \emph{Herschel} DEBRIS (Disk Emission via a Bias-Free Reconnaissance in the 
Infrared/Submillimetre) survey observed the nearest $\sim$90 stars in each of the spectral type groups A, F, G, K and M, obtaining a disc detection rate of 17 per cent based on 
100\,$\umu$m and 160\,$\umu$m results, corresponding to 77 out of a total of 446 targets detected \citep{Matthews2014}. Similarly, the \emph{Herschel} DUNES (DUst around NEarby 
Stars) survey detected an incidence of 20 per cent for nearby Sun-like stars, probing to the photospheric level \citep{Eiroa2013, Montesinos2016}. In terms of limits to the 
detectable flux, \emph{Spitzer} and \emph{Herschel} achieved average sensitivities, expressed as fractional dust luminosities (see Section~\ref{sec:fractional_luminosities}) of 
$\sim$10$^{-5}$ and $\sim$10$^{-6}$, respectively. These levels compare to $\sim$10$^{-7}$ for the Edgeworth-Kuiper belt in our Solar System. Crucially, the surveys by 
\emph{Herschel} spatially resolved half of the detected discs, many for the first time. Other surveys, in the near-mid IR with \emph{AKARI}, have probed ``warmer'' debris discs 
($T \geq 150\,\rm{K}$), i.e. material to be found closer to the central star, with incidence rates typically 3 per cent, much lower than for the ``cooler'' discs detected at 
longer wavelengths \citep{Fujiwara2013}.

\vskip 1mm

The SCUBA-2 Observations of Nearby Stars (SONS) survey was one of the seven original legacy surveys undertaken on the James Clerk Maxwell telescope between 2012 and 2015 
\citep{Chrysostomou2010}. The survey set out to target 115 known disc host stars (within 100\,pc of the Sun) searching for debris signatures in the form of dust emission at 
850\,$\umu$m. The aim of the SONS survey was to characterise these discs to the fullest extent possible by: (1) providing direct dust masses that cannot be obtained from 
shorter wavelengths alone; (2) adding to the far-IR/submillimetre spectrum to constrain the dust size distribution; (3) using the power of a 15\,m telescope to resolve disc 
structures around the nearest systems, and (4) looking for evidence of resonant clumps and other features in resolved structures that could be indicative of unseen perturbers, 
such as planets. This paper presents the first results from the full survey. Future work will concentrate on detailed modelling of the disc structures, further investigations 
into dust grain properties and size distributions, and their interpretation in terms of the relationship to possible planetary systems.

\section[]{Survey history and target selection}
\label{sec:history}

\begin{figure}
\includegraphics[width=83mm]{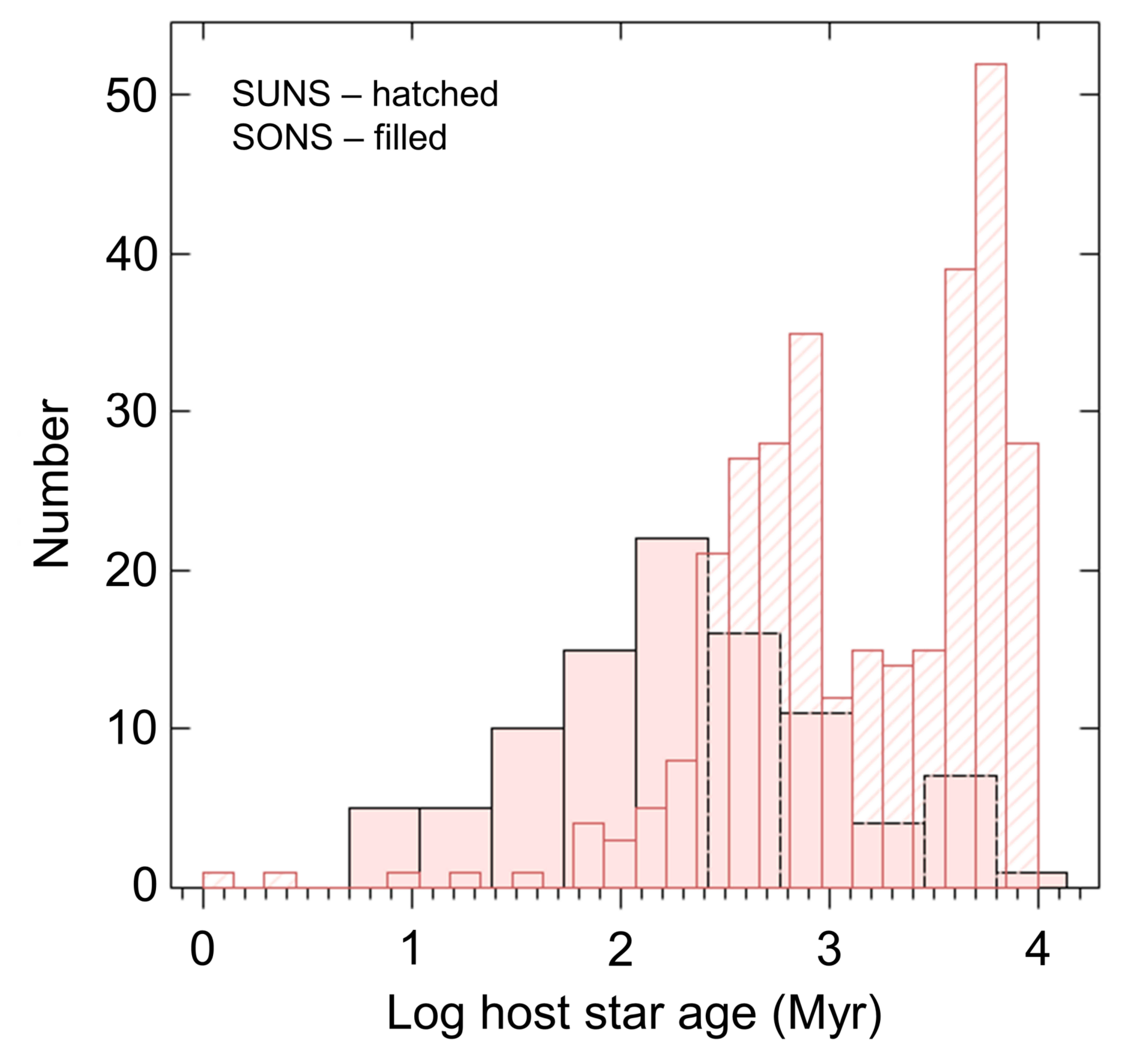}

\caption{Histogram showing the difference in age distribution between the targets in the original SUNS (SCUBA-2 Unbiased Nearby Stars) survey of 500 stars, and the re-scoped 
SONS survey of 115 targets.}

\label{fig:age_distribution}
\end{figure}

The original concept was for a volume-limited, unbiased survey of 500 stars, the 100 nearest in each of the spectral type groups A, F, G, K, and M \citep{Matthews2007}. First 
formulated in 2004, this was called the SCUBA-2 Unbiased Nearby Stars (SUNS) survey. The aforementioned extensive surveys by \emph{Spitzer} and \emph{Herschel} during the period 
2004 -- 2012, together with a shortfall in instrument sensitivity of approximately a factor of 2, however, meant that the greatest potential legacy lay in a revamped JCMT/SCUBA-2 
survey to target a more modest number of known debris disc hosts. Discs would have to be very cold to be detectable with SCUBA-2 but below the detection threshold of, for example, 
the \emph{Herschel} DEBRIS survey. Hence the SONS survey became targeted towards younger stars, and stars with known infrared excesses, with a higher expectation of detection at a 
wavelength of 850\,$\umu$m over the original volume-limited survey. Fig.~\ref{fig:age_distribution} shows the distribution of stars by age, emphasising the survey bias towards 
younger targets.

\vskip 1mm

The revised target list was assembled in 2011 from \emph{IRAS} and \emph{Spitzer} published data \citep{Low2005, Su2006, Beichman2006, Rhee2007, Trilling2007, Trilling2008, 
Bryden2009, Plavchan2009, Koerner2010, Zuckerman2011, Morales2011}, unpublished data from \emph{Spitzer}, \emph{Herschel} DEBRIS and DUNES, the \emph{Herschel} Guaranteed Time discs 
programme, \emph{Herschel} GASPS (GAS in Protoplanetary Systems; \citealt{Dent2013}), and several smaller programmes on planet/disc hosts. Flux densities at 850\,$\umu$m were 
therefore predicted based on existing photometric data and a fit to the IR excess from the target. Simply assuming a standard blackbody spectrum would result in an over-estimate of 
the 850\,$\umu$m flux by about a factor of four \citep{Wyatt2007}. Hence, particularly in the cases where few photometric points existed, predictions were based on 
modified blackbody spectra, $B_\nu(\lambda/\lambda_0)^{-\beta}$, assuming a critical wavelength, $\lambda_0$\,=\,200\,$\umu$m and a dust emissivity index, $\beta$\,=\,1.0 
\citep{Wyatt2008,Phillips2011}. Targets were then classified according to the likelihood of a 3$\sigma$ detection being achievable at 850\,$\umu$m with a flux density of at least 
3\,mJy.

\vskip 1mm

\begin{figure*}
\includegraphics[width=140mm]{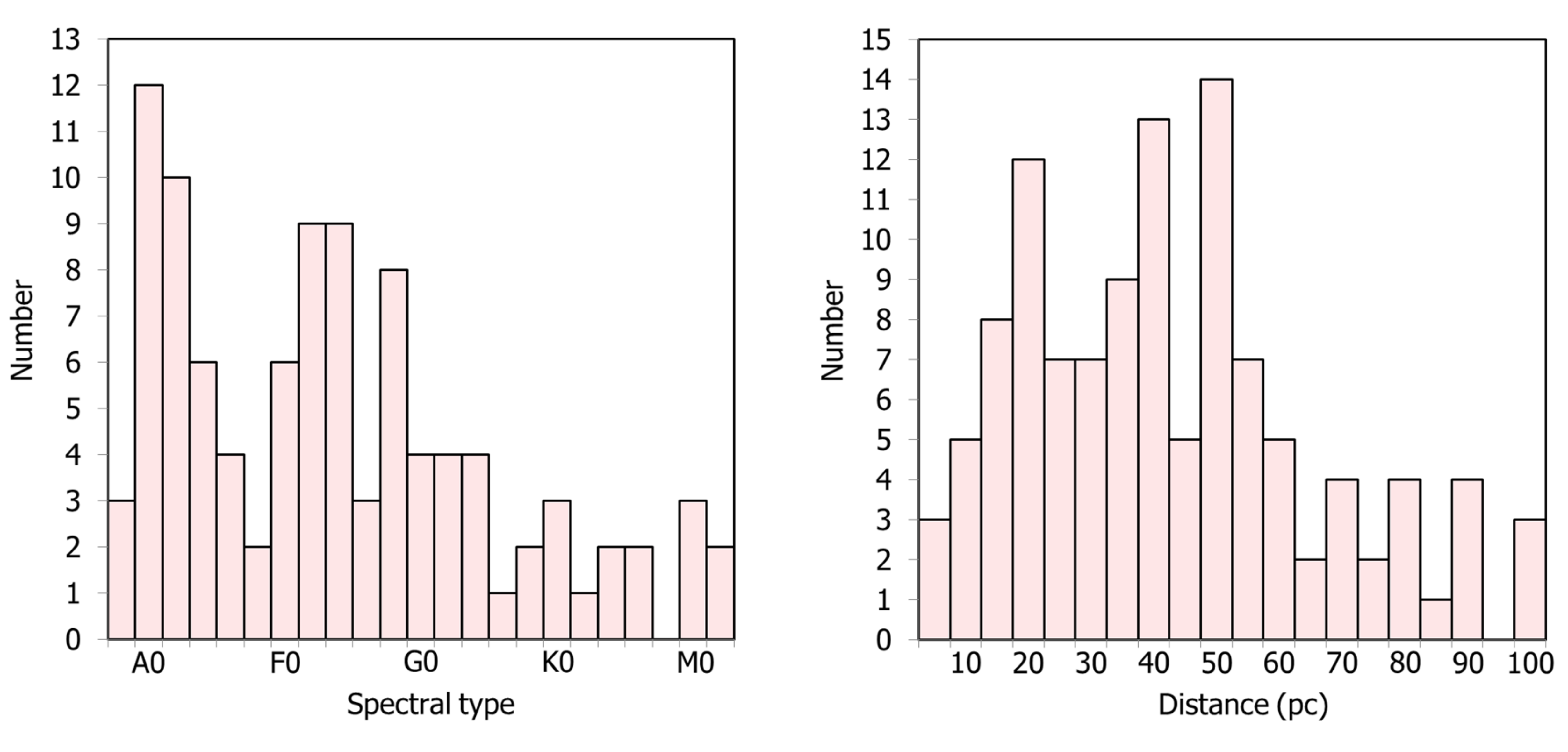}

\caption{The distribution of SONS survey targets as a function of host star spectral type (left) and distance (right).}

\label{fig:distributions}
\end{figure*}

Those sources classified as having guaranteed, likely or hard to quantify fluxes (i.e., with an unconstrained dust temperature) were retained if they were within 100\,pc, had 
declinations between $-$40 to +80\degree\ and predicted 850\,$\umu$m fluxes of $>$1\,mJy (or above $-$60\degree\ dec with predicted fluxes exceeding 15\,mJy, to include several 
southern bright targets). It was accepted that there was still up a factor 3 uncertainty in the 850\,$\umu$m flux predictions for some targets. This uncertainty arises because 
the grain properties and size distribution are generally unknown; characterising these was one of the key science goals of this survey. This method produced a candidate list of 
115 targets (see Table~\ref{tab:table1}) with 37 (i.e. one-third) of these in the guaranteed or likely detection categories. The selection criteria led to the expectation of a 
high detection rate, given the evidence of discs at multiple wavelengths for many targets. Fig.~\ref{fig:distributions} shows the distribution of targets by host star spectral 
type and distance.

\vskip 1mm

The survey was formally allocated 270\,hrs of observing time on the JCMT, equally split between weather bands 2 and 3, equivalent to 225\,GHz zenith optical depths in the range 
0.05 -- 0.08 for band 2 and 0.08 -- 0.12 for band 3\footnote{Zenith optical depths at 225\,GHz in the range 0.05 -- 0.12 correspond to line-of-sight precipitable water vapour 
levels of approximately 1 -- 2.5\,mm.}. The time allocated was sufficient to reach a 1\,$\sigma$ sensitivity limit of 1.4\,mJy at 850\,$\umu$m for each of the 115 fields. 
Furthermore, to maximise the chances of a disc detection a ``quick-look'' approach to the observing methodology was adopted, in which each star was initially targeted for a 
minimum 1\,hr observation block, typically achieving a 1$\sigma$ noise level of 1.5 $-$ 2\,mJy. Although SCUBA-2 operates simultaneously at wavelengths of 450\,$\umu$m and 
850\,$\umu$m, the allocation of band 2 and 3 weather meant that it was unlikely any significant number of discs would be detected at 450\,$\umu$m. Such detections, however, were 
never a goal of the survey as it was planned to follow up possible 450\,$\umu$m detections with future observations, most likely requiring the best weather conditions (``band 
1'').

\vskip 1mm

\begin{figure*}
\includegraphics[width=140mm]{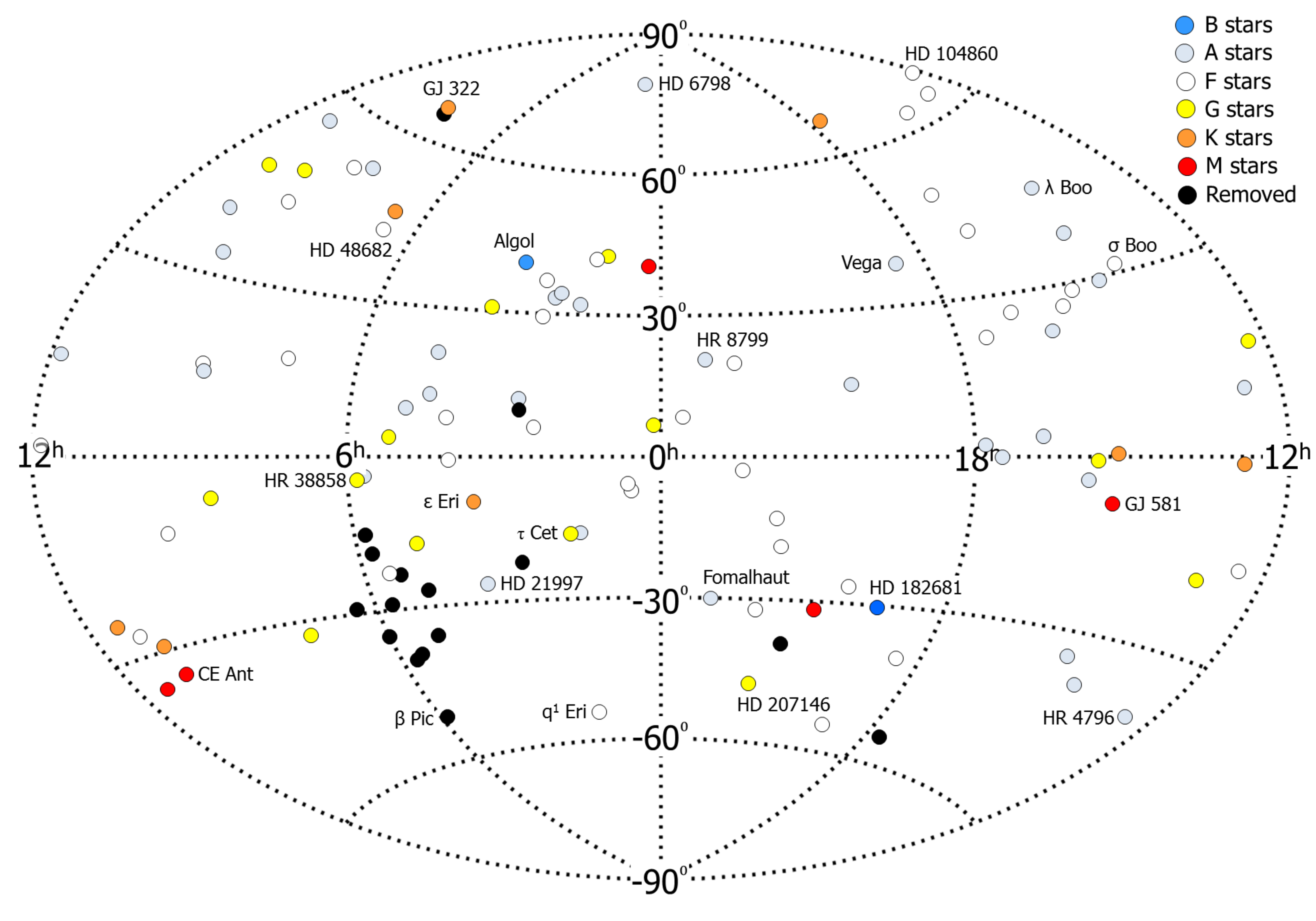}

\caption{The distribution of SONS survey targets on the sky by spectral type. The spectral type for the stars is indicated by the colour table. Filled black circles are stars
that were removed from the original 115 target list. For information, names for some of the stars are also labelled.}

\label{fig:target_positions}
\end{figure*}

Any stars with detectable flux were immediately prioritised for more observing time (as needed, to boost the significance of the detection), with the possibility of returning 
to the others if and when time allowed. Although the time was scheduled as 3 -- 4 night blocks on the telescope, roughly spaced every few months during the 3\,yr survey 
period, the fact that the survey had targets all over the sky (as shown in Fig.~\ref{fig:target_positions}) meant that SONS observations benefited from gaps in the schedules 
for the other legacy surveys. The broad sky distribution was the main reason that SONS became the first survey to be completed in terms of time in 2014 August. In addition to 
the 270\,hr formal allocation the data presented in this paper also includes observations from the initial survey verification phase (2012 Jan) amounting to 26\,hrs, 4.5\,hrs 
from the SCUBA-2 Guaranteed Time allocation (PI: Holland) for observations of $\epsilon$ Eridani, a further 21\,hrs from the survey extension programme in late 2014/early 
2015, and 2\,hrs each from the PI programmes M12AC17 and M13AC19 (PIs: Brenda Matthews and Christine Chen, respectively). The total observing time for the survey data was 
therefore 325.5\,hrs.

\vskip 1mm

During the observing campaign, some observations were prioritised to confirm (or rule out) a previous marginal disc detection. Hence, more observing time was expended on a 
handful of disc candidates than was originally planned. Together with an over-allocation of sources and fields around 05\,hrs RA from the entire Legacy Survey programme, it 
became necessary to remove 15 targets from the list of 115. These were mainly around 05\,hrs RA (see Table~\ref{tab:table1}), and included $\beta$ Pictoris as it has been 
well-characterised in the past at 850\,$\umu$m \citep[e.g.][]{Holland1998, Dent2014}, and 14 further targets least likely to yield a disc detection based on the criteria 
outlined above.

\vskip 1mm

\begin{table*}
  \centering
  \caption{The target list for the SONS survey. RA/Dec positions and spectral types are from the SIMBAD database \citep{Wenger2000}, and stellar distances from
the \emph{Hipparcos} catalogue \citep{Perryman1997, van Leeuwen2007}. Stellar ages are referenced individually in Section~\ref{sec:target_discussion}.}
  \vskip 0.5mm
  \label{tab:table1}

  \begin{tabular}{lllllllll}
\hline

HD             &   Other           &   RA (J2000)   &   Dec (J2000)   & Spectral      &  Distance  &   Age         &  Association     &    Notes                       \\
number         &   names           &                &                 & type          &    (pc)    &  (Myr)        &                  &                                \\

\hline

377            &                   &  00 08 25.75   &  +06 37 00.49   &   G2-V        &   39.1     &   170         &  Field           &                                \\

               &   HIP 1368        &  00 17 06.38   &  +40 56 53.87   &   M0.5-V      &   14.7     &   500         &  Field           &                                \\

3126           &                   &  00 34 27.17   & $-$06 30 14.05  &   F2-V        &   41.5     &  1500         &  Field           &                                \\

3296           &   LTT 317         &  00 36 01.85   & $-$05 34 14.59  &   F5-D        &   47.2     &  1700         &  Field           &                                \\

6798           &                   &  01 12 16.82   &  +79 40 26.27   &   A3-V        &   82.8     &   320         &  Field           &                                \\

7590           &   V445 And        &  01 16 29.25   &  +42 56 21.90   &   G0-V        &   23.6     &  1820         &  Field           &                                \\

8907           &                   &  01 28 34.36   &  +42 16 03.68   &   F8-D        &   34.2     &   320         &  Field           &                                \\

9672           &   49 Cet          &  01 34 37.78   & $-$15 40 34.90  &   A1-V        &   59.4     &    40         &  Argus           &                                \\

10647          &   q$^1$ Eri       &  01 42 29.32   & $-$53 44 27.00  &   F9-V        &   17.4     &  1600         &  Field           & Known planet host              \\

10700          &   $\tau$ Cet      &  01 44 04.08   & $-$15 56 14.93  &   G8.5-V      &   3.7      &  7650         &  Field           & Possible planet host           \\

10638          &                   &  01 44 22.81   &  +32 30 57.16   &   A3-E        &   69.3     &    50         &  Field           &                                \\

13161          &  $\beta$ Tri      &  02 09 32.63   &  +34 59 14.27   &   A5-III      &   38.9     &   730         &  Field           & Binary                         \\

\hline

14055          &  $\gamma$ Tri     &  02 17 18.87   &  +33 50 49.90   &   A1-Vnn      &   34.4     &   230         &  Field           &                                \\

15115          &                   &  02 26 16.25   &  +06 17 33.19   &   F2-D        &   45.2     &    23         &  $\beta$ Pic MG  &                                \\

15257          &  12 Tri           &  02 28 09.98   &  +29 40 09.59   &   F0-III      &   49.8     &  1000         &  Field           &                                \\

15745          &                   &  02 32 55.81   &  +37 20 01.04   &   F2-V        &   63.5     &    23         &  $\beta$ Pic MG  &                                \\

17094          &  87 Cet           &  02 44 56.54   &  +10 06 50.91   &   F0-IV       &   25.8     &  1500         &  Field           & Not observed                   \\

17093          &  38 Ari           &  02 44 57.58   &  +12 26 44.73   &   A7-III      &   36.3     &   580         &  Field           &                                \\

17390          &                   &  02 46 45.11   & $-$21 38 22.28  &   F3-IV/V     &   48.0     &   600         &  Field           & Not observed                   \\

19356          &  $\beta$ Per; Algol   &  03 08 10.13   &  +40 57 20.33   &   B8-V    &   27.6     &   450         &  Field           & Triple star system             \\

21997          &                   &  03 31 53.65   & $-$25 36 50.94  &   A3-IV/V     &   71.9     &    30         &  Columba         &                                \\

22049          & $\epsilon$ Eri    &  03 32 55.85   & $-$09 27 29.73  &   K2-V        &   3.22     &   850         &  Field           & Possible planet host           \\

22179          &  V* 898 Per       &  03 35 29.90   &  +31 13 37.44   &   G5-IV       &   16.0     &    16         &  Field           &                                \\

25457          &                   &  04 02 36.75   & $-$00 16 08.12  &   F6-V        &   18.8     &   130         &  AB Doradus      &                                \\

\hline

25570          &                   &  04 03 56.60   &  +08 11 50.16   &   F2-V        &   34.9     &   600         &  Hyades          &                                \\

28226          &                   &  04 28 00.78   &  +21 37 11.66   &   A5-C        &   47.1     &   600         &  Hyades          &                                \\

28355          &  79 Tau           &  04 28 50.16   &  +13 02 51.37   &   A7-V        &   48.9     &   600         &  Hyades          &                                \\

30447          &                   &  04 46 49.53   & $-$26 18 08.85  &   F3-V        &   80.3     &    30         &  Columba         & Not observed                   \\

30495          &  58 Eri           &  04 47 36.29   & $-$16 56 04.04  &   G1.5-V      &   13.3     &   650         &  Field           &                                \\

31392          &                   &  04 54 04.21   & $-$35 24 16.27  &   G9-V        &   25.7     &  3700         &  Field           & Not observed                   \\

31295          &  7 Ori            &  04 54 53.73   &  +10 09 03.00   &   A0-V        &   37.0     &   125         &  Field           &                                \\

33636          &                   &  05 11 46.45   &  +04 24 12.73   &   G0-V        &   28.4     &  2500         &  Field           &                                \\

34324          &                   &  05 15 43.90   & $-$22 53 39.70  &   A3-V        &   85.8     &   450         &  Field           & Not observed                   \\

35650          &                   &  05 24 30.17   & $-$38 58 10.77  &   K6-V        &   18.0     &     ?         &  Field           & Not observed                   \\

35841          &                   &  05 26 36.59   & $-$22 29 23.72  &   F3-V        &   96.0     &    30         &  Columba         &                                \\

36968          &                   &  05 33 24.07   & $-$39 27 04.64  &   F2-V        &  140.0     &    20         &  Octans          & Not observed                   \\

\hline

37484          &                   &  05 37 39.63   & $-$28 37 34.66  &   F3-V        &   59.5     &    30         &  Columba         & Not observed                   \\

37594          &                   &  05 39 31.15   & $-$03 33 52.93  &   A8-V        &   42.6     &   650         &  Field           &                                \\

38206          &                   &  05 43 21.67   & $-$18 33 26.92  &   A0-V        &   69.2     &    30         &  Columba         & Not observed                   \\

38678          & $\zeta$ Lep       &  05 46 57.34   & $-$14 49 19.02  &   A2-IV/Vn    &   21.6     &    23         &  $\beta$ Pic MG  & Not observed                   \\

39060          &  $\beta$ Pic      &  05 47 17.09   & $-$51 47 17.09  &   A6-V        &   19.3     &    23         &  $\beta$ Pic MG  & Not observed;                  \\

38858          &                   &  05 48 34.94   & $-$04 05 40.72  &   G4-V        &   15.2     &  4700         &  Field           & Known planet host              \\

40540          &                   &  05 57 52.60   & $-$34 28 34.01  &   A8-IVm      &   89.9     &   170         &  Field           & Not observed                   \\

45184          &                   &  06 24 43.88   & $-$28 46 48.41  &   G1.5-V      &   21.9     &  4400         &  Field           & Not observed                   \\

48682          &  56 Aur           &  06 46 44.34   &  +43 34 38.73   &   F9-V        &   16.7     &  6000         &  Field           &                                \\

49601          &  GJ 249           &  06 51 32.39   &  +47 22 04.14   &   K6-V        &   18.6     &     ?         &  Field           &                                \\

57703          &                   &  07 23 04.61   &  +18 16 24.27   &   F2 D        &   41.4     &   600         &  Field           &                                \\

61005          &  ``The Moth''     &  07 35 47.46   & $-$32 12 14.04  &   G8-V        &   35.3     &    40         &  Argus?          &                                \\

\hline

70313          &                   &  08 23 48.50   &  +53 13 10.96   &   A3-V        &   50.4     &   200         &  Field           &                                \\

73350          &  V401 Hya         &  08 37 50.29   & $-$06 48 24.78  &   G5-V        &   24.0     &   300         &  Field           &                                \\

72905          &                   &  08 39 11.70   &  +65 01 15.27   &   G1.5-Vb     &   14.4     &   490         &  Ursa Major      & Not observed                   \\

               &  GJ 322; HIP 43534    &  08 52 00.34   &  +66 07 53.37   &   K5 D    &   16.5     &   490         &  Ursa Major      &                                \\

75616          &                   &  08 53 06.10   &  +52 23 24.83   &   F5 D        &   35.4     &  1400         &  Field           &                                \\

76543          &  62 Cnc           &  08 57 14.95   &  +15 19 21.95   &   A5-III      &   45.7     &   400         &  Field           &                                \\

76582          &  63 Cnc           &  08 57 35.20   &  +15 34 52.61   &   F0-IV       &   46.1     &   540         &  Field           &                                \\

82943          &                   &  09 34 50.74   & $-$12 07 46.37  &   F9-VFe      &   27.5     &   430         &  Field           & Multiple planet host           \\

84870          &                   &  09 49 02.85   &  +34 05 07.40   &   A3          &   88.0     &   100         &  Field           &                                \\

85301          &                   &  09 52 16.77   &  +49 11 26.85   &   G5-V        &   32.8     &   600         &  Hyades          &                                \\

91312          &                   &  10 33 13.89   &  +40 25 32.02   &   A7-IV       &   34.6     &   410         &  Field           &                                \\

91782          &                   &  10 36 47.84   &  +47 43 12.47   &   G0+M9V      &   61.4     &  1580         &  Field           & Binary?                        \\

\hline
  \end{tabular}
\end{table*}

\begin{table*}
  \centering
  \contcaption{ }
  \vskip 0.5mm
  \begin{tabular}{lllllllll}
\hline
HD             &   Other           &  RA (J2000)    &  Dec (J2000)    &  Spectral     &  Distance  &  Age          &  Association      &    Notes                       \\
number         &   names           &                &                 &  type         &   (pc)     &  (Myr)        &                   &                                \\
\hline

               &  TWA7; CE Ant     &  10 42 30.11   & $-$33 40 16.21  &   M2-Ve       &   50       &     9         &  TW Hydrae        &  Distance is uncertain         \\

92945           &                  &  10 43 28.27   & $-$29 03 51.43  &   K1-V        &   21.4     &   200         &  Field            &                                \\

95418           &  $\beta$ UMa     &  11 01 50.48   &  +56 22 56.73   &   A1-V        &   24.4     &   490         &  Ursa Major       &                                \\

95698           &                  &  11 02 24.45   & $-$26 49 53.42  &   F1-V        &   56.1     &  1520         &  Field            &                                \\

                &  TWA13           &  11 21 17.24   & $-$34 46 45.5   &   M1-Ve       &   50.0     &     9         &  TW Hydrae        &                                \\

98800           &                  &  11 22 05.29   & $-$24 46 39.76  &   K5-V        &   50.0     &     9         &  TW Hydrae        & Pre-MS dwarf                   \\

102647          &  $\beta$ Leo     &  11 49 03.58   &  +14 34 19.41   &   A3-V        &   11.0     &    45         &  IC 2391          & Binary                         \\

102870          &  $\beta$ Vir     &  11 50 41.72   &  +01 45 52.99   &   F9-V        &   10.9     &  2900         &  Field            &                                \\

104860          &                  &  12 04 33.73   &  +66 20 11.72   &   F8          &   45.5     &   200         &  Field            &                                \\

107146          &                  &  12 19 06.50   &  +16 32 53.86   &   G2-V        &   27.5     &   100         &  Field            &                                \\

109085          &  $\eta$ Crv      &  12 32 04.23   & $-$16 11 45.62  &   F2-V        &   18.3     &  1380         &  Field            &                                \\

109573          &  HR 4796; TWA 11   &  12 36 01.03   & $-$39 52 10.23  &   A0-V      &   72.8     &     9         &  TW Hydrae        & Binary                         \\

\hline

110411          &  $\rho$ Vir      &  12 41 53.06   &  +10 14 08.25   &   A3-V        &   36.3     &    90         &  Field            &                                \\

111631          &                  &  12 50 43.57   & $-$00 46 05.26  &   K7          &   10.6     &   600         &  Field            &                                \\

113337          &                  &  13 01 46.93   &  +63 36 36.81   &   F6-V        &   36.9     &  1450         &  Field            & Binary; Planet host            \\

115617          &  61 Vir          &  13 18 24.31   & $-$18 18 40.30  &   G7-V        &   8.6      &  6300         &  Field            & Multiple planet host           \\

122652          &                  &  14 02 31.64   &  +31 39 39.08   &   F8          &   39.3     &   300         &  Field            &                                \\

125162          & $\lambda$ Boo    &  14 16 23.02   &  +46 05 17.90   &   A0p         &   30.4     &  2800         &  Field            &                                \\

125473          & $\psi$ Cen       &  14 20 33.43   & $-$37 53 07.06  &   A0-IV       &   79.4     &   300         &  Field            & Binary                         \\

127821          &                  &  14 30 46.07   &  +63 11 08.83   &   F4-IV       &   31.8     &  1020         &  Field            &                                \\

127762          & $\gamma$ Boo     &  14 32 04.67   &  +38 18 29.70   &   A7-III      &   26.6     &   950         &  Field            & Binary                         \\

128167          & $\sigma$ Boo     &  14 34 40.82   &  +29 44 42.46   &   F2-V        &   15.8     &  1000         &  Field            &                                \\

131625          &                  &  14 55 44.71   & $-$33 51 20.82  &   A0-V        &   77.8     &   200         &  Field            &                                \\

135502          & $\chi$ Boo       &  15 14 29.16   &  +29 09 51.46   &   A2-V        &   77.4     &   200         &  Field            &                                \\

\hline

135599          &                  &  15 15 59.17   &  +00 47 46.89   &   KO-V        &   15.8     &  1300         &  Field            &                                \\

                & GJ 581; HIP 74995   &  15 19 26.82  & $-$07 43 20.21  &   M3       &   6.2      &  5700         &  Field            & Multiple planet host           \\

139006          & $\alpha$ CrB     &  15 34 41.27   &  +26 42 52.89   &   A1-IV       &   23.0     &   490         &  Ursa Major       & Binary                         \\

139590          &                  &  15 39 01.06   & $-$00 18 41.38  &   G0-V        &   55.8     &  5000         &  Field            &                                \\

141378          &                  &  15 48 56.80   & $-$03 49 06.64  &   A5-IV       &   54.0     &   150         &  Field            &                                \\

143894          & 44 Ser           &  16 02 17.69   &  +22 48 16.03   &   A3-V        &   54.9     &   300         &  Field            &                                \\

149630          & $\sigma$ Her     &  16 34 06.18   &  +42 26 13.35   &   B9-V        &   96.5     &   700         &  Field            & Binary                         \\

150378          & 37 Her           &  16 40 38.69   &  +04 13 11.23   &   A1-V        &   90.0     &   200         &  Field            & Binary                         \\

150682          & 39 Her           &  16 41 36.70   &  +26 55 00.77   &   F3-V        &   43.7     &  1700         &  Field            &                                \\

151044          &                  &  16 42 27.81   &  +49 56 11.19   &   F8-V        &   29.3     &  3900         &  Field            &                                \\

157728          & 73 Tau           &  17 24 06.59   &  +22 57 37.01   &   F0-IV       &   42.7     &   530         &  Field            &                                \\

158633          &                  &  17 25 00.10   &  +67 18 24.15   &   K0-V        &   12.8     &  5900         &  Field            &                                \\

\hline

158352          &                  &  17 28 49.66   &  +00 19 50.25   &   A8-V        &   59.6     &   750         &  Field            &                                \\

161868          & $\gamma$ Oph     &  17 47 53.56   &  +02 42 26.20   &   A0-V        &   31.5     &   185         &  Field            &                                \\

170773          &                  &  18 33 00.92   & $-$39 53 31.28  &   F5-V        &   37.0     &   200         &  Field            &                                \\

172167          & $\alpha$ Lyr; Vega  &  18 36 56.34  &  +38 47 01.28  &   A0-V       &   7.7      &   700         &  Field            &                                \\

181327          &                  &  19 22 58.94   & $-$54 32 16.97  &   F5-V        &   51.8     &    23         & $\beta$ Pic MG    &                                \\

182681          &                  &  19 26 56.48   & $-$29 44 35.62  &   B8-V        &   69.9     &   145         &  Field            &                                \\

191089          &                  &  20 09 05.22   & $-$26 13 26.53  &   F5-V        &   52.2     &    23         & $\beta$ Pic MG    &                                \\

192425          & $\rho$ Aql       &  20 14 16.62   &  +15 11 51.39   &   A2-V        &   47.1     &   410         &  Field            &                                \\

197481          & AU Mic           &  20 45 09.53   & $-$31 20 27.24  &   M1-Ve       &   9.9      &    23         & $\beta$ Pic MG    &                                \\

202560          &                  &  21 15 15.27   & $-$38 52 02.50  &   M0-Ve       &   3.9      &    ?          &  Field            & Not observed                   \\

\hline

205674          &                  &  21 37 21.11   & $-$18 26 28.25  &   F4-IV       &   51.8     &   130         &  AB Doradus?      &                                \\

206893          &                  &  21 45 21.91   & $-$12 47 00.07  &   F5-V        &   38.3     &   860         &  Field            &                                \\

207129          &                  &  21 48 15.75   & $-$47 18 13.02  &   G2-V        &   16.0     &  3800         &  Field            & Binary                         \\

209253          &                  &  22 02 32.96   & $-$32 08 01.48  &   F6.5-V      &   30.1     &   950         &  Field            &                                \\

212695          &                  &  22 26 14.44   & $-$02 47 20.32  &   F5          &   46.5     &  2300         &  Field            &                                \\

213617          &  39 Peg          &  22 32 35.48   &  +20 13 48.06   &   F1-V        &   50.3     &  1200         &  Field            &                                \\

216956          &  $\alpha$ PsA; Fomalhaut  &  22 57 39.05  & $-$29 37 20.05  &   A4-V  &   7.7   &   440          &  Field            & Triple star; planet host       \\

218396          &  HR 8799         &  23 07 28.72   &  +21 08 03.31   &   A5-V        &   39.4     &    30         &  Columba          & Multiple planet host           \\

221853          &                  &  23 35 36.15   &  +08 22 57.43   &   F0          &   68.4     &   100         &  Local            &                                \\

\hline
  \end{tabular}
\end{table*}

\section{Observations and data reduction}
\label{sec:observations}

\subsection{Observations}
\label{sec:observations_intro}

The SCUBA-2 camera \citep{Holland2013} on the JCMT was used to take the survey data between 2012 Jan and 2015 Feb. The wavelengths of observation were 850\,$\umu$m and 
450\,$\umu$m, where the primary beam sizes are 13.0\arcsec\ and 7.9\arcsec\ (measured Full-Width at Half Maximum; FWHM) \citep{Dempsey2013}. The data were taken exclusively 
using the constant speed DAISY observing mode, which maximises the observing time in the central 3\,arcmin$^2$ region of a field \citep{Bintley2014}. This mode is appropriate 
for compact sources of less than a few arcminutes in diameter and so is well-suited to the observations of debris discs within the SONS survey. Each observation was taken as one 
continuous scan with a duration of approximately 30\,min. The data is saved as 30\,sec sub-scans, with each observation resulting in a total of 55 sub-scans, including a 
flat-field measurement at the start and end of each observation \citep{Holland2013}. The data were calibrated in flux density against the primary calibrators Uranus and Mars, 
and also secondary calibrators CRL\,618 and CRL\,2688 from the JCMT calibrator list \citep{Dempsey2013}, with estimated calibration uncertainties amounting to 20 and 7 per cent 
at 450\,$\umu$m and 850\,$\umu$m, respectively. Accurate telescope pointing was crucial to these observations and was regularly checked with reference to nearby bright sources 
(e.g. compact HII regions or blazars), with RMS pointing errors of less than 2\arcsec.

\subsection{Data reduction}
\label{sec:data_reduction}

\subsubsection{Original approach: ``Blank field''}
\label{sec:blank_field}

The data were reduced using the Dynamic Iterative Map-Maker within the Starlink SMURF package \citep{Chapin2013} called from the ORAC-DR automated pipeline \citep{Jenness2015}.
The original data reduction approach, used in the ``First Results'' paper \citep{Panic2013}, saw the data heavily high-pass filtered at 1\,Hz, corresponding to a gradual spatial
cut-off centred at $\sim$150\arcsec\, for a typical DAISY scanning speed of $\sim$150\arcsec\,s$^{-1}$. The filtering was necessary to remove low-frequency noise originating
from the detectors and readout electronics \citep{Holland2013}. To account for the attenuation of the signal, as a result of the time series filtering, the pipeline would re-make
each map with a fake 10\,Jy Gaussian added to the raw data, but offset from the nominal map centre by 30\arcsec\, to avoid contamination with any detected source. The amplitude
of the Gaussian in the output map gave the signal attenuation, and this correction was applied along with the flux conversion factor derived from the calibrator observations.
This method produced satisfactory results for unresolved compact sources but residual noise artefacts often remained in the images.

\vskip 1mm

\subsubsection{Revised approach: ``Zero masking''}
\label{sec:zero_masking}

The heavy high-pass filtering led not only to residual instrumental noise but also to an under-estimation of the flux density as a function of source-scale size. This effect is 
not surprising as when sources become larger, they contain more power at lower frequencies, meaning that a fixed-frequency, high-pass filter will remove more flux. Hence, SONS 
adopted a revised map-maker configuration optimised for known position, compact and moderately-extended sources. It used the technique of ``zero masking'' in which the map is 
constrained to a mean value of zero in all cases outside a radius of 60\arcsec\ from the centre of the field, for all but the final iteration of the map-maker 
\citep{Chapin2013}. The technique not only helped convergence in the iterative part of the map-making process but suppressed the large-scale ripples that can produce ringing 
artefacts. The results were more uniform, lower noise (by an average of $\sim$20 per cent) final images, largely devoid of gradients and artefacts \citep{Chapin2013}. Each 
output map was regridded with 1\arcsec\ pixels at both wavelengths, and then smoothed with a 7\arcsec\ Gaussian using the Starlink package KAPPA recipe {\sc gausmooth} 
\citep{Currie2013}. Flux conversion factors (FCFs) were derived from the calibrator observations taken on the same night as the observations, reduced in exactly the same way as 
the source data, and applied to calibrate each map in flux density. FCFs were calculated based on the diameter of the observed disc. For unresolved discs the FCF was measured in 
a beam-sized aperture from the calibrator observation (often referred to as ``per beam'' fluxes), whereas for resolved discs the conversion factor was based on an aperture 
diameter appropriate for the disc size (often referred to as ``integrated'' or ``aperture'' fluxes). The final images were made by coadding two or more maps using inverse-variance 
weighting implemented by the Starlink package PICARD recipe {\sc mosaic jcmt images} \citep{Gibb2013}.

\subsubsection{Noise analysis}
\label{sec:noise_analysis}

Output data files from the map-maker were written in Starlink $\emph{N}$-Dimensional Data Format \citep{Jenness2015b} and contain the rebinned image (``DATA'' array) in terms of 
signal per output map pixel, together with a variance array representing the spread in values falling in a map pixel (``VARIANCE'' array). Signal-to-noise maps were produced 
(Starlink package KAPPA recipe {\sc makesnr}) creating a new NDF file by dividing the DATA component by the square root of the VARIANCE array. This method, however, provides a 
noise estimate that is only representative of the true noise in an image if there are no residual features on a scale larger than approximately half a beam diameter. For 
unresolved discs the noise in an image was also obtained directly from the DATA array by taking integrated flux measurements within multiple beam-sized areas, spaced by 
4\arcsec, from the central few arcminutes of the image. Similarly, for resolved discs the noise was estimated from apertures appropriate for the diameter of the disc, spaced by 
the one-third of the aperture diameter. In both cases the resulting flux distribution was fitted by a Gaussian (IDL {\sc histogauss}) with the noise level corresponding to the 
standard deviation of the fit. For the vast majority of the SONS survey measurements, both methods gave very similar results. This similarity was expected since the ``zero 
masking'' technique is very effective at ensuring the final image is devoid of instrumental artefacts. The noise estimates reported in this paper are based on the measurements 
directly from the rebinned image (DATA array), and cases of residual instrumental artefacts in the images are discussed in Section~\ref{sec:target_discussion} for individual 
targets. The errors reported for integrated fluxes (aperture photometry) are similarly derived from overlapping apertures of the same diameter used to determine the source flux.

\begin{figure}
\includegraphics[width=82mm]{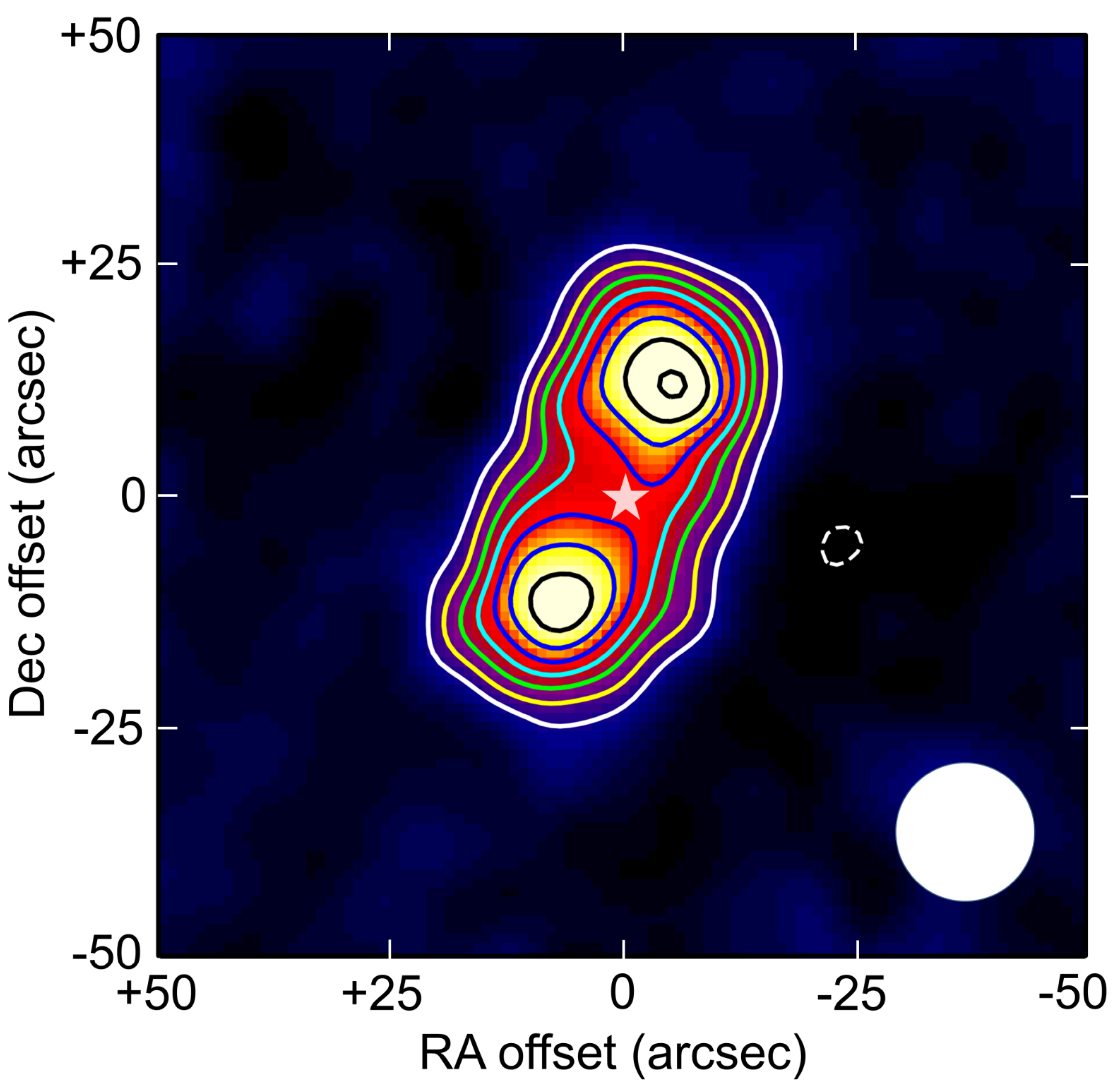}

\caption{The Fomalhaut debris disc imaged by SCUBA-2 at 850\,$\umu$m. This signal-to-noise image is colour scaled from $-$5$\sigma$ (black) to the maximum S/N in the image
at 26$\sigma$ (white). The dashed contour is at $-$5$\sigma$ and the solid contours start at 5$\sigma$ and increase in 3$\sigma$ steps. The star symbol shows the position of
Fomalhaut with respect to the disc (after proper motion corrections). The white circle represents the approximate FWHM beam size at 850\,$\umu$m after smoothing.}

\label{fig:fomalhaut}
\end{figure}

\subsection{The Fomalhaut debris disc: A test case}
\label{sec:fomalhaut}

As discussed in Section~\ref{sec:data_reduction}, due to the excess low-frequency noise, one of the major challenges for the SONS survey was the reduction and calibration of
moderately extended disc structures. The recovery of large-scale structure is a common issue for all observations undertaken with SCUBA-2. To test the robustness of the
``zero mask'' data reduction method an extended, well-characterised debris disc was adopted as a test case. HD 216956 (Fomalhaut; $\alpha$ PsA) has been well-studied at all
wavelengths from the optical to the millimetre, and the associated debris disc has a well-sampled spectral energy distribution. The disc is also one of the most extended so far
discovered. The SCUBA-2 signal-to-noise (S/N) image of the Fomalhaut debris disc at 850\,$\umu$m (Fig.~\ref{fig:fomalhaut}) shows an identical structure to previous
submillimetre observations \citep{Holland2003}. Furthermore, the measured integrated flux within a 60\arcsec\ diameter aperture is 91 $\pm$ 3 mJy, compared to the 97 $\pm$ 5 mJy
measured from the SCUBA observations \citep{Holland2003}, consistent within the measured errors. Hence, the ``zero mask'' data reduction method works well for discs extending to at
least 1\arcmin\ in diameter, although there remains some uncertainty for targets such as Vega (see Section~\ref{sec:vega}).

\section{Survey outputs and interpretation}
\label{sec:survey}

\subsection{Survey outputs}
\label{sec:survey_outputs}

\begin{figure}
\includegraphics[width=85mm]{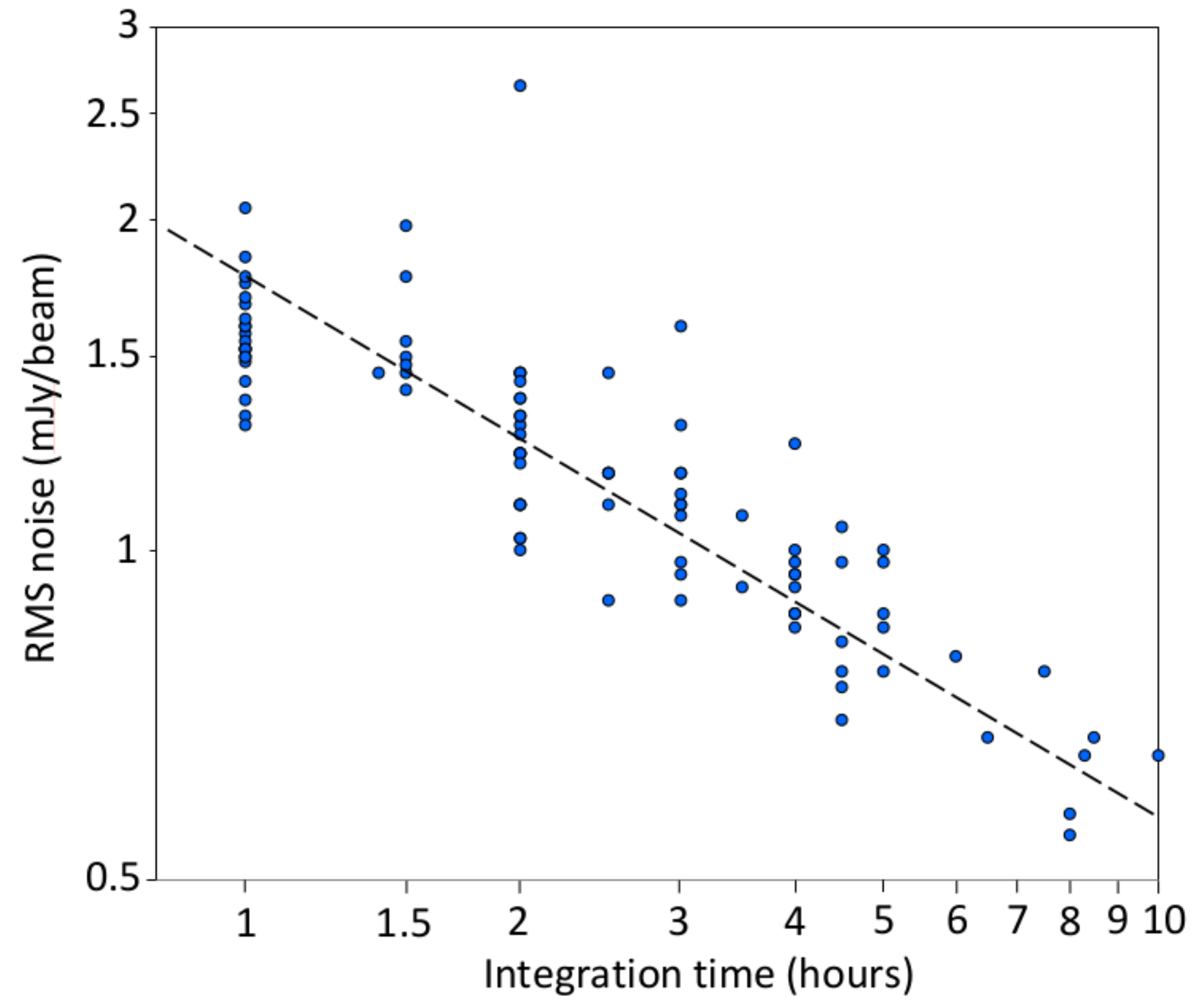}

\caption{The measured RMS noise (mJy/beam) at 850\,$\umu$m as a function of integration time ($t_{obs}$) for all 100 target fields in the SONS survey sample. The trend line 
indicates a noise decrease as $t_{obs}^{-0.5}$.}

\label{fig:RMSnoise}
\end{figure}

The outputs of the survey are a catalogue of images and fluxes (including errors) for the 100 observed fields at 850\,$\umu$m\footnote{The complete catalogue of images and 
spectral energy distributions for the entire sample of targets in the SONS survey, including the non-detections, is available online (doi:10.11570/17.0005).}. 
Table~\ref{tab:table2} lists the measured flux densities for all 100 stars and indicates whether the discs are resolved or unresolved by these observations. Signal-to-noise 
images for the detected fields are presented in Appendix~\ref{sec:appendixA}. Integration times ranged from a minimum of 1\,hr to a maximum of 10\,hrs. (The 5\,hrs or longer 
observations were largely undertaken during the survey verification programme time, prior to the formal start of the SONS survey.) The results for each star are described in 
Section~\ref{sec:target_discussion}. As shown in Fig.~\ref{fig:RMSnoise}, the RMS noise (determined from beam-sized apertures in all cases) within the inner 3\arcmin\ 
diameter region of each image decreases as $t_{obs}^{-0.5}$, where $t_{obs}$ is the integration time. The spread in RMS values is due to the relatively wide range of sky 
transmissions during the observations (i.e., the zenith sky opacity as defined by the allocated weather band combined with the airmass of the source at the time of the 
observation). Table~\ref{tab:table2} also lists five targets for which the peak is offset from the star by more than half a beam diameter ($\sim$7.5\arcsec\ for the smoothed 
850\,$\umu$m beam). In each of these cases the peak is interpreted as more likely to be a background object rather than a disc about the star (see 
Section~\ref{sec:target_discussion}).

\begin{table*}

\centering

  \caption{Integration times and measured fluxes for the SONS survey sample. The column ``Disc'' refers to whether the observed structure is unresolved (``P'' - point-like) or 
resolved (``E'' - extended compared to the beam diameter) at 850\,$\umu$m. Fluxes are presented at both wavelengths with 3\,$\sigma$ and 5\,$\sigma$ upper limits quoted at 
850\,$\umu$m and 450\,$\umu$m, respectively, in the case of a non-detected excess. Calibration uncertainties, as described in Section~\ref{sec:observations_intro}, are not 
included in the listed fluxes. For resolved (extended) structures the flux quoted is from aperture photometry (see Section~\ref{sec:zero_masking}) with the diameter of the 
aperture given in the individual source descriptions in Section~\ref{sec:target_discussion}.}

\vskip 0.5mm

  \label{tab:table2}

  \begin{tabular}{lllllll}
\hline
HD number  &  Other names  &  Time      &  Disc  &  850\,$\umu$m flux       &  450\,$\umu$m flux      & Notes                                                     \\
           &               &  (hrs)     &        &  (mJy)                   &  (mJy)                  &                                                           \\
\hline

377        &               &   4.0      &  P     &  5.3 $\pm$ 1.0           &  $<$90                  &                                                           \\

           &  HIP 1368     &   1.0      &  --    &  $<$4.2                  &  $<$105                 &                                                           \\

3126       &               &   1.0      &  --    &  $<$4.5                  &  $<$95                  &                                                           \\

3296       & LTT 317       &   2.0      &  --    &  $<$3.0                  &  $<$100                 &                                                           \\

6798       &               &   4.5      &  P     &  7.2 $\pm$ 1.0           &  $<$65                  &                                                           \\

7590       & V445 And      &   3.5      &  --    &  $<$3.3                  &  $<$110                 &                                                           \\

8907       &               &   2.0      &  P     &  7.8 $\pm$ 1.2           &  51 $\pm$ 10            &                                                           \\

9672       & 49 Cet        &   2.5      &  P     & 13.5 $\pm$ 1.5           & 125 $\pm$ 18            & Detected and resolved at 450\,$\umu$m                     \\

10647      & q$^1$ Eri     &   2.0      &  P     & 20.1 $\pm$ 2.7           &  $<$2800                & Background galaxy contamination to east of star           \\

10700      & $\tau$ Ceti   &   8.0      &  P     &  4.4 $\pm$ 0.6           &  25 $\pm$ 4.5           &                                                           \\

\hline

10638      &               &   4.0      &  P     &  5.1 $\pm$ 0.9           &  $<$105                 &                                                           \\

13161      & $\beta$ Tri   &   4.0      &  P     &  5.1 $\pm$ 0.9           &  $<$85                  &                                                           \\

14055      & $\gamma$ Tri  &   4.0      &  P     &  7.2 $\pm$ 1.0           &  $<$70                  &                                                           \\

15115      &               &   3.0      &  P     &  8.2 $\pm$ 1.1           &  $<$150                 & Peak is 5\arcsec\ offset from the star                    \\

15257      & 12 Tri        &   2.0      &  P     & 10.3 $\pm$ 1.2           &  56 $\pm$ 11            &                                                           \\

15745      &               &   3.0      &  E     & 12.0 $\pm$ 1.4           &  $<$110                 & Marginally resolved at 850\,$\umu$m                       \\

17093      & 38 Ari        &   5.0      &  --    &  (8.8 $\pm$ 0.9)         &  $<$110                 & Peak is 10\arcsec\ offset; likely a background object     \\

19356      & $\beta$ Per; Algol & 4.0   &  P     &  6.4 $\pm$ 0.9           &  $<$35                  & Is the emission from a debris disc?                       \\

21997      &               &   4.5      &  E     &  10.7 $\pm$ 1.5          &  $<$125                 & Marginally resolved at 850\,$\umu$m                       \\

22049      & $\epsilon$ Eri &  4.5      &  E     &  31.3 $\pm$ 1.9          &  181 $\pm$ 15           & Detected at 450\,$\umu$m, but poor S/N in clumps          \\

\hline

22179      & V* 898 Per    &   2.0      &  --    &  (7.0 $\pm$ 1.3)         &  $<$100                 & Peak is 18\arcsec\ offset; likely a background object     \\

25457      &               &   2.0      &  P     &  6.2 $\pm$ 1.4           &  $<$145                 &                                                           \\

25570      &               &   3.0      &  --    &  $<$3.6                  &  $<$120                 &                                                           \\

28226      &               &   1.0      &  --    &  $<$4.5                  &  $<$100                 &                                                           \\

28355      & 79 Tau        &   1.0      &  --    &  $<$3.9                  &  $<$150                 &                                                           \\

30495      & 58 Eri        &   1.0      &  --    &  $<$4.8                  &  $<$150                 &                                                           \\

31295      & 7 Ori         &   2.0      &  --    &  $<$4.5                  &  $<$165                 &                                                           \\

33636      &               &   3.0      &  --    &  $<$3.3                  &  $<$37                  &                                                           \\

35841      &               &   4.5      &  P     &  3.5 $\pm$ 0.8           &  $<$35                  & Peak is 5\arcsec\ offset from the star                    \\

37594      &               &   3.0      &  --    &  $<$3.6                  &  $<$200                 &                                                           \\

\hline

38858      &               &  10.0      &  E     &  7.5 $\pm$ 1.4           &  $<$55                  & Extended structure with background source?                \\

48682      &  56 Aur       &   8.0      &  E     &  3.9 $\pm$ 0.8           &  $<$25                  & Possibly resolved. Peak is 6.5\arcsec\ offset from star   \\

49601      &  GJ 249       &   1.0      &  --    &  $<$4.5                  &  $<$100                 &                                                           \\

57703      &               &   2.0      &  --    &  $<$3.9                  &  $<$150                 &                                                           \\

61005      & ``The Moth''  &   1.5      &  P     & 13.5 $\pm$ 2.0           &  $<$200                 &                                                           \\

70313      &               &   1.0      &  --    &  $<$4.5                  &  $<$135                 &                                                           \\

73350      &  V401 Hya     &   1.5      &  --    &  $<$4.5                  &  $<$115                 &                                                           \\

           &  GJ 322       &   2.0      &  P     &  7.3 $\pm$ 1.4           &  57 $\pm$ 11            &                                                           \\

75616      &               &   2.0      &  --    &  $<$3.3                  &  $<$45                  &                                                           \\

76543      &  62 Cnc       &   2.5      &  --    &  $<$3.6                  &  $<$100                 &                                                           \\

\hline

76582      &  63 Cnc       &   5.0      &  P     &  5.7 $\pm$ 1.0           &  89 $\pm$ 17            &                                                           \\

82943      &               &   1.0      &  --    &  $<$4.5                  &  $<$225                 &                                                           \\

84870      &               &   4.0      &  P     &  6.2 $\pm$ 1.0           &  $<$50                  &                                                           \\

85301      &               &   2.0      &  --    &  $<$4.5                  &  $<$75                  &                                                           \\

91312      &               &   3.0      &  --    &  $<$3.3                  &  $<$110                 &                                                           \\

91782      &               &   1.5      &  --    &  $<$4.8                  &  $<$350                 &                                                           \\

           &  TWA7; CE Ant    &   4.0   &  P     &  7.2 $\pm$ 1.3           &  $<$300                 &  Peak is 6\arcsec\ offset from the star                   \\

92945      &               &   3.0      &  P     &  8.6 $\pm$ 1.1           &  $<$90                  &  Background source contamination to south of star         \\

95418      & $\beta$ UMa   &   1.0      &  --    &  $<$4.8                  &  $<$100                 &                                                           \\

95698      &               &   1.0      &  --    &  $<$5.4                  &  $<$500                 &                                                           \\

\hline

           & TWA13         &   2.0      &  --    &  $<$3.0                  &  $<$70                  &                                                           \\

98800      & LTT 317       &   1.5      &  P     & 93.6 $\pm$ 1.5           &  242 $\pm$ 14           &                                                           \\

102647     & $\beta$ Leo   &   1.0      &  --    &  $<$4.5                  &  $<$75                  &                                                           \\

102870     & $\beta$ Vir   &   1.0      &  --    &  $<$4.2                  &  $<$145                 &                                                           \\

104860     &               &   5.0      &  P     &  6.5 $\pm$ 1.0           &  $<$135                 &                                                           \\

107146     &               &   1.0      &  P     & 20.6 $\pm$ 2.1           &  $<$375                 &  Peak is 4\arcsec\ offset from the star                   \\

109085     & $\eta$ Crv    &   8.3      &  E     & 15.4 $\pm$ 1.1           &  54 $\pm$ 12            &  Detected and resolved at 450\,$\umu$m, but poor S/N      \\

109573     & HR 4796; TWA 11 &   1.0    &  P     & 14.4 $\pm$ 1.9           &  117 $\pm$ 21           &                                                           \\

110411     & $\rho$ Vir    &   1.5      &  --    &  $<$4.5                  &  $<$165                 &                                                           \\

111631     &               &   3.5      &  --    &  $<$0.9                  &  $<$110                 &                                                           \\

\hline
  \end{tabular}

\end{table*}

\begin{table*}
  \centering
  \contcaption{ }
  
  \vskip 0.5mm
  
\begin{tabular}{lllllll}
\hline
HD number  &  Other names  &  Time      &  Disc  &  850\,$\umu$m flux       &  450\,$\umu$m flux      &  Notes                                                    \\
           &               &  (hrs)     &        &  (mJy)                   &  (mJy)                  &                                                           \\
\hline

113337     &               &   2.0      &  --    &  $<$3.6                  &  $<$75                  &                                                           \\

115617     & 61 Vir        &   7.5      &  E     &  5.8 $\pm$ 1.0           &  $<$70                  &  Marginally resolved at 850\,$\umu$m                      \\

122652     & 12 Tri        &   1.0      &  --    &  $<$5.1                  &  $<$200                 &                                                           \\

125162     & $\lambda$ Boo &   4.5      &  P     &  3.9 $\pm$ 0.8           &  $<$30                  &                                                           \\

125473     & $\psi$ Cen    &   2.0      &  --    &  $<$3.9                  &  $<$65                  &                                                           \\

127821     &               &   8.5      &  P     &  5.8 $\pm$ 0.7           &  $<$60                  &  Interpreted as a single dust peak                        \\

127762     & $\gamma$ Boo  &   2.0      &  --    &  $<$3.6                  &  $<$105                 &                                                           \\

128167     & $\sigma$ Boo  &   5.0      & --     &  (4.1 $\pm$ 0.9)         &  $<$80                  &  Peak is 11\arcsec\ offset; likely a background object    \\

131625     &               &   2.0      &  --    &  $<$4.5                  &  $<$180                 &                                                           \\

135502     & $\chi$ Boo    &   2.0      &  --    &  $<$4.2                  &  $<$115                 &                                                           \\

\hline

135599     &               &   2.0      &  --    &  $<$3.6                  &  $<$120                 &                                                           \\

           & GJ 581; HIP 74995  &  3.0  &  --    &  $<$4.8                  &  $<$115                 &                                                           \\

139006     & $\alpha$ CrB  &   1.0      &  --    &  $<$5.1                  &  $<$170                 &                                                           \\

139590     &               &   1.0      &  --    &  $<$3.9                  &  $<$60                  &                                                           \\

141378     &               &   1.0      &  --    &  (8.5 $\pm$ 1.8)         &  $<$165                 &  Peak is 13\arcsec\ offset; likely a background object    \\

143894     & 44 Ser        &   4.0      &  E     & 10.1 $\pm$ 1.2           &  $<$80                  &  Peak is 4\arcsec\ offset from the star                   \\

149630     & $\sigma$ Her  &   1.4      &  --    &  $<$4.5                  &  $<$75                  &                                                           \\

150378     & 37 Her        &   2.0      & --     & (10.2 $\pm$ 1.1)         &  $<$54                  & Peak is 17\arcsec\ offset; likely a background object     \\

150682     & 39 Her        &   4.0      &  P     &  5.5 $\pm$ 0.9           &  $<$60                  &                                                           \\

151044     &               &   3.0      &  --    &  $<$2.7                  &  $<$35                  &                                                           \\

\hline

157728     & 73 Tau        &   1.0      &  --    &  $<$4.5                  &  $<$155                 &                                                           \\

158633     &               &   1.0      &  --    &  $<$4.8                  &  $<$85                  &                                                           \\

158352     &               &   4.0      &  P     &  5.3 $\pm$ 1.0           &  $<$100                 &                                                           \\

161868     & $\gamma$ Oph  &   4.5      &  E     &  7.1 $\pm$ 1.0           &  $<$90                  &  Marginally resolved at 850\,$\umu$m                      \\

170773     &               &   1.5      &  E     & 26.0 $\pm$ 2.3           &  $<$225                 &                                                           \\

172167     & $\alpha$ Lyr; Vega  & 6.0  &  E     & 34.4 $\pm$ 1.4           & 229 $\pm$ 14            &  Also resolved at 450\,$\umu$m                            \\

181327     &               &   1.0      &  P     & 23.6 $\pm$ 3.4           &  $<$7500                &                                                           \\

182681     &               &   2.5      &  P     &  6.8 $\pm$ 1.2           &  $<$65                  &  Could be more than one clump?                            \\

191089     &               &   2.5      &  P     &  4.9 $\pm$ 0.9           &  $<$70                  &                                                           \\

192425     & $\rho$ Aql    &   1.5      &  --    &  $<$4.2                  &  $<$85                  &                                                           \\

\hline

197481     & AU Mic        &   5.0      &  P     & 12.6 $\pm$ 0.8           &  57 $\pm$ 9             &  Also resolved at 450\,$\umu$m                            \\

205674     &               &   5.0      &  P     &  4.0 $\pm$ 0.7           &  $<$40                  &  Peak is 6\arcsec\ offset. Two point sources?             \\

206893     &               &   2.0      &  P     &  4.5 $\pm$ 1.1           &  $<$75                  &  Peak is 4\arcsec\ offset from the star                   \\

207129     &               &   2.5      &  E     & 10.8 $\pm$ 1.8           &  $<$210                 &  Marginally resolved at 850\,$\umu$m                      \\

209253     &               &   1.0      &  --    &  $<$4.8                  &  $<$200                 &                                                           \\

212695     &               &   2.5      &  P     &  5.7 $\pm$ 1.1           &  $<$85                  &                                                           \\

213617     & 39 Peg        &   3.0      &  P     &  4.6 $\pm$ 1.3           &  $<$215                 &  Peak is 4\arcsec\ offset from the star                   \\

216956     & $\alpha$ PsA; Fomalhaut & 2.0 &  E  & 91 $\pm$ 2.5             & 475 $\pm$ 21            &  Also resolved at 450\,$\umu$m                            \\

218396     & HR 8799       &   3.0      &  E     &  17.4 $\pm$ 1.5          & 346 $\pm$ 34            &  Likely background cloud contamination at 450\,$\umu$m    \\

221853     &               &   2.0      &  --    &  $<$3.9                  &  $<$100                 &                                                           \\

\hline
  \end{tabular}
\end{table*}

\vskip 1mm

Although the primary wavelength of observation for the survey was 850\,$\umu$m, reflected by the allocation of weather bands 2 and 3 at the JCMT, a significant number of the 
targets showed flux excesses at 450\,$\umu$m. Experience has shown that in band 2 and 3 weather conditions, which is far from optimal for 450\,$\umu$m observations, the chances 
of a false detection are minimised by adopting a detection threshold of 5$\sigma$. Hence, whilst a level of 3$\sigma$ has been used when reporting 850\,$\umu$m results in this 
paper, a level of 5$\sigma$ has been adopted at 450\,$\umu$m. A total of 14 targets in the sample reached this threshold, all of which were also detected at 850\,$\umu$m. The 
450\,$\umu$m photometry has been used to constrain further the fitting of the SED (Section~\ref{sec:dust_temperature}) and offers improved angular resolution for extended 
structures (i.e., $\sim$10\arcsec\ when the image is smoothed with a 7\arcsec\ FWHM Gaussian). The results from the 450\,$\umu$m observations, where applicable, are noted in 
Table~\ref{tab:table2}, and discussed in the individual source descriptions in Section~\ref{sec:target_discussion}.

\subsection{Dust temperature and emissivity}
\label{sec:dust_temperature}

Photometry for the target stars has been compiled from the optical to the millimetre from a wide variety of sources, including all-sky surveys such as \emph{IRAS}
\citep{Moshir1990}, \emph{Hipparcos} \citep{Perryman1997}, 2MASS Point Source Catalog \citep{Cultri2003}, \emph{AKARI} \citep{Ishihara2010} and \emph{WISE} \citep{Wright2010},
as outlined in Section~\ref{sec:history}. Further data are also provided by the surveys undertaken by \emph{Spitzer} (including IRS data from the CASSIS database)
\citep[e.g.,][]{Lebouteiller2011} and \emph{Herschel} \citep[e.g.,][]{Booth2013}. Specific references to the photometric points provided for each target are given in the
individual source descriptions in Section~\ref{sec:target_discussion}. The photometric data allow a spectral energy distribution to be assembled for each of the target
stars, and these are shown in the figures presented in Appendix~\ref{sec:appendixA}, together with the 850\,$\umu$m S/N images.

\vskip 1mm

The SED modelling adopted in this paper \citep{Kennedy2012} has been successfully implemented for other surveys of debris discs, including the \emph{Herschel} DEBRIS survey
\citep[e.g.][]{Booth2013,Thureau2014}. Photometry shortward of about 10\,$\umu$m was first used to model the stellar photospheric emission, and the estimated contributions to
the remaining infrared and submillimetre/millimetre photometric fluxes were then subtracted. These photometric points were then fitted by one or two component Planck functions
in which a pure blackbody spectrum is modified beyond a critical wavelength, $\lambda_0$, as described in Section~\ref{sec:history}. The best fitting model was found by a
least-squares minimisation method. The model therefore accounts for inefficient emission by grains that are small relative to the wavelength of emission. The factor $\lambda_0$
is therefore representative of the grain size that dominates the emission spectrum, whilst the parameter $\beta$ is an index that describes the emissivity of the dust grains as
well as being indicative of the size distribution of the dust. 

\vskip 1mm

In many cases, even with the 850\,$\umu$m photometry provided by the SONS survey, the sparse data coverage at submillimetre and millimetre wavelengths means that both 
$\lambda_0$ and $\beta$ are poorly constrained by the modelling. Moreover, in cases of only a single measurement beyond 160\,$\umu$m, $\lambda_0$ and $\beta$ become strongly 
degenerate and no unique solution is possible. For some targets, a well-defined second component fit to the SED may possibly indicate the presence of multiple planetesimal belts 
\citep[e.g.,][]{Morales2011,Chen2014,Kennedy2014}. The main output parameters from the SED fitting (for one or more components), relevant to the interpretation of the IR/submm 
flux excess, are the dust temperature ($T_{\rm{d}}$), critical wavelength ($\lambda_0$) and the dust emissivity index ($\beta$). These values are listed in 
Table~\ref{tab:table3} for the SONS survey sample, in which the derived $\beta$'s are presented as a range of values (i.e., all values within the quoted range are possible).

\begin{table*}

  \centering

  \caption{The derived parameters from the SED fitting, measurements (and upper limits) from the radial profile fitting, and dust mass calculations for the SONS survey sample. 
Note that this list does not include the five ``extreme`` offset cases, in which the flux peak is observed to be equal to, or greater than 10\arcsec\ from the star (as 
indicated in Table~\ref{tab:table2}). In the cases where the disc is unresolved the parameter $R_{fit}$ represents the upper limit to disc radius, corresponding to the beam 
radius at the distance of the star (see also the scale bars in the figures of Appendix~\ref{sec:appendixA}). }
  
  \label{tab:table3}

  \vskip 0.5mm

  \begin{tabular}{llllllllll}

\hline
HD      &   Other     &  $L_{\star}^{(1)}$ & $L_{\rm{disc}}/L_{\star}$ &  $\lambda_0$  &  $\beta$  &  $T_{\rm{d}}$  &  $R_{\rm{BB}}^{(1)}$  &  $R_{\rm{fit}}$ & $M_{\rm{dust}}$    \\
number  &   names     &  ($L_{\sun}$)      & ($\times$10$^{-4}$)       &  ($\umu$m)    &           &  (K)           &  (au)             &  (au)   &  ($\times$0.01\,M$_{\earth}$)  \\
\hline

377     &                     &  1.2 $\pm$ 0.03  &  3.8 $\pm$ 1.4       &  --           &  0.0 -- 1.4  &  56 $\pm$ 6  &  27 $\pm$ 4   &  $<$290        &  3.6 $\pm$ 0.8       \\

6798    &                     &  35.2 $\pm$ 0.6  &  1.7 $\pm$ 1.1       &  $<$142       &  0.2 -- 0.9  &  48 $\pm$ 8  & 200 $\pm$ 48  &  $<$615        &  25.4 $\pm$ 5.5      \\

8907    &                     &  2.2 $\pm$ 0.03  &  2.5 $\pm$ 0.2       &  $<$655       &  0.4 -- 2.1  &  51 $\pm$ 5  &  44 $\pm$ 5   &  $<$175        &  4.4 $\pm$ 0.8       \\

9672    & 49 Cet              &  16.9 $\pm$ 0.3  &  8.5 $\pm$ 0.6       &  79 -- 129    &  0.9 -- 1.1  &  59 $\pm$ 2  &  92 $\pm$ 5   &  421 $\pm$ 16$^{(2)}$  & 20.0 $\pm$ 2.2   \\

10647   & q$^1$ Eri           &  1.6 $\pm$ 0.04  &  3.2 $\pm$ 0.8       &  $<$75        &  0.6 -- 0.8  &  44 $\pm$ 2  &  49 $\pm$ 6   &  $<$125        &  3.3 $\pm$ 0.5       \\

10700   & $\tau$ Cet          &  0.49 $\pm$ 0.02 &  0.08 $\pm$ 0.02     &  --           &  0.0 -- 1.1  &  71 $\pm$ 23 &  11 $\pm$ 5   &  $<$18         &  0.021 $\pm$ 0.007   \\

10638   &                     &  7.8 $\pm$ 0.1   &  5.0 $\pm$ 1.5       &  $<$357       &  $>$0.6      &  34 $\pm$ 15 & 187 $\pm$ 114 &  $<$515        &  17.9 $\pm$ 8.4      \\

13161   & $\beta$ Tri         &  73.5 $\pm$ 1.4  &  0.29 $\pm$ 0.02     &  113 -- 204   &  0.6 -- 1.7  &  84 $\pm$ 5  &  94 $\pm$ 8   &  $<$290        &  2.2 $\pm$ 0.4       \\

14055   & $\gamma$ Tri        &  25.3 $\pm$ 0.5  &  0.89 $\pm$ 0.03     &  155 -- 250   &  0.7 -- 1.5  &  77 $\pm$ 2  &  66 $\pm$ 3   &  $<$260        &  2.8 $\pm$ 0.4       \\

15115   &                     &  3.3 $\pm$ 0.1   &  5.5 $\pm$ 1.0       &  $<$149       &  0.5 -- 0.9  &  57 $\pm$ 4  &  44 $\pm$ 5   &  $<$340        &  7.3 $\pm$ 1.1       \\

\hline

15257   & 12 Tri              &  14.7 $\pm$ 0.3  &  1.1 $\pm$ 0.4       &  --           &  0.0 -- 1.9  &  53 $\pm$ 10 & 106 $\pm$ 27  &  $<$240        &  10.8 $\pm$ 2.3      \\

15745   &                     &  3.3 $\pm$ 0.1   &  20.1 $\pm$ 0.9      &  --           &  0.0 -- 0.9  &  89 $\pm$ 2  &  18 $\pm$ 1   &  514 $\pm$ 111 &  13.3 $\pm$ 1.6      \\

19356   & $\beta$ Per         &  101 $\pm$ 2.5   &  0.09 $\pm$ 2.34     &  --           &  0.0 -- 2.0  &  27 $\pm$ 17 & 1106 $\pm$ 976 & $<$210        &  4.5 $\pm$ 2.9       \\

21997   &                     &  11.7 $\pm$ 0.2  &  5.5 $\pm$ 0.2       &  86 -- 371    &  0.4 -- 1.4  &  64 $\pm$ 1  &  66 $\pm$ 3   &  813 $\pm$ 69  &  21.5 $\pm$ 3.0      \\

22049   & $\epsilon$ Eri      &  0.34 $\pm$ 0.01 &  1.2 $\pm$ 0.5       &  $<$200       &  0.6 -- 1.0  &  44 $\pm$ 8  &  23 $\pm$ 6   &  67 $\pm$ 2    &  0.16 $\pm$ 0.03     \\

25457   &                     &  2.1 $\pm$ 0.03  &  1.1 $\pm$ 1.6       &  $<$786       &  $>$0.1      &  50 $\pm$ 12 &  44 $\pm$ 15  &  $<$280        &  1.1 $\pm$ 0.4       \\

35841   &                     &  2.4 $\pm$ 0.3   &  14.3 $\pm$ 2.3      &  --           &  0.0 -- 2.7  &  71 $\pm$ 3  &  24 $\pm$ 5   &  $<$725        &  11.4 $\pm$ 2.7      \\

38858   &                     &  0.83 $\pm$ 0.02 &  0.82 $\pm$ 0.13     &  --           &  $\sim$0.0   &  50 $\pm$ 10 &  28 $\pm$ 8   &  192 $\pm$ 18  &  0.86 $\pm$ 0.24     \\

48682   &  56 Aur             &  1.9 $\pm$ 0.1   &  0.67 $\pm$ 0.10     &  $<$182       &  0.7 -- 1.8  &  43 $\pm$ 12 &  57 $\pm$ 22  &  184 $\pm$ 33  &  0.62 $\pm$ 0.21     \\

61005   & ``The Moth''        &  0.58 $\pm$ 0.01 &  27.1 $\pm$ 0.7      &  109 -- 217   &  0.4 -- 0.7  &  61 $\pm$ 1  &  16 $\pm$ 1   &  $<$265        &  6.9 $\pm$ 1.0       \\

\hline

        &  GJ 322             &  0.10 $\pm$ 0.01 &  3.2 $\pm$ 0.7       &  --           &  $>$0.1      &  24 $\pm$ 10 &  43 $\pm$ 27  &  $<$85         &  2.1 $\pm$ 1.0       \\

76582   &  63 Cnc             &  8.9 $\pm$ 0.2   &  2.3 $\pm$ 0.3       &  185 -- 547   &  $>$1.1      &  52 $\pm$ 2  &  85 $\pm$ 5   &  $<$230        &  5.7 $\pm$ 1.0       \\

84870   &                     &  7.6 $\pm$ 0.1   &  5.0 $\pm$ 2.1       &  $<$324       &  0.3 -- 1.2  &  50 $\pm$ 6  &  85 $\pm$ 14  &  $<$660        &  23.5 $\pm$ 4.5      \\

        & CE Ant              &  0.05 $\pm$ 0.01 &  16.9 $\pm$ 1.7      &  --           &  0.0 -- 2.5  &  19 $\pm$ 8  &  49 $\pm$ 30  &  $<$380        &  24.1 $\pm$ 11.8      \\

92945   &                     &  0.37 $\pm$ 0.01 &  6.5 $\pm$ 0.5       &  $<$270       &  0.4 -- 1.1  &  42 $\pm$ 7  &  27 $\pm$ 7   &  $<$160        &  2.4 $\pm$ 0.5       \\

98800   & LTT 317             &  0.98 $\pm$ 0.03 &  1090 $\pm$ 39       &  --           &  0.0 -- 0.1  & 156 $\pm$ 3  &  3.1 $\pm$ 0.2  & $<$255       &  36.9 $\pm$ 0.9      \\

104860  &                     &  1.2 $\pm$ 0.01  &  6.2 $\pm$ 0.4       &  138 -- 296   &  0.4 -- 1.2  &  47 $\pm$ 3  &  39 $\pm$ 4   & $<$340         &  7.1 $\pm$ 1.2       \\

107146  &                     &  0.99 $\pm$ 0.01 &  10.8 $\pm$ 3.2      &  276 -- 425   &  0.8 -- 1.0  &  41 $\pm$ 2  &  46 $\pm$ 5   & $<$210         &  9.3 $\pm$ 1.1       \\

109085  & $\eta$ Crv          &  5.2 $\pm$ 0.1   &  1.4 $\pm$ 0.3       &  $<$806       &  0.2 -- 1.0  &  41 $\pm$ 7  &  87 $\pm$ 18  & 190 $\pm$ 7    &  3.1 $\pm$ 0.6       \\

109573  & HR 4796             &  24.1 $\pm$ 0.8  &  56.5 $\pm$ 4.7      &  $<$72        &  0.7 -- 1.0  &  99 $\pm$ 3  &  38 $\pm$ 3   & $<$370         &  18.9 $\pm$ 2.5      \\

\hline

115617  & 61 Vir              &  0.83 $\pm$ 0.02 &  0.28 $\pm$ 0.04     &  --           &  0.0 -- 1.5  &  65 $\pm$ 11 &  17 $\pm$ 4   & 40 $\pm$ 13    &  0.16 $\pm$ 0.04     \\

125162  & $\lambda$ Boo       &  17.1 $\pm$ 0.3  &  0.30 $\pm$ 0.03     &  $<$216       &  0.3 -- 1.8  &  87 $\pm$ 6  &  43 $\pm$ 4   & $<$230         &  1.0 $\pm$ 0.2       \\

127821  &                     &  3.1 $\pm$ 0.1   &  1.9 $\pm$ 0.2       &  94 -- 199    &  0.9 -- 1.9  &  47 $\pm$ 3  &  61 $\pm$ 12  & $<$235         &  3.1 $\pm$ 0.6       \\

143894  & 44 Ser              &  29.1 $\pm$ 0.5  &  0.32 $\pm$ 0.04     &  --           &  0.0 -- 1.1  &  53 $\pm$ 9  & 148 $\pm$ 37  & 561 $\pm$ 69   &  14.1 $\pm$ 3.0      \\

150682  & 39 Her              &  6.8 $\pm$ 0.1   &  0.13 $\pm$ 0.03     &  --           &  0.0 -- 2.2  &  32 $\pm$ 14 & 201 $\pm$ 130 & $<$330         &  8.1 $\pm$ 3.8       \\

158352  &                     &  18.9 $\pm$ 0.3  &  0.78 $\pm$ 0.04     &  $<$797       &  0.1 -- 2.2  &  62 $\pm$ 7  &  88 $\pm$ 13  & $<$450         &  7.5 $\pm$ 1.6       \\

161868  & $\gamma$ Oph        &  26.2 $\pm$ 0.6  &  1.1 $\pm$ 0.2       &  109 -- 217   &  0.8 -- 1.4  &  68 $\pm$ 3  &  85 $\pm$ 6   & 246 $\pm$ 35   &  2.5 $\pm$ 0.4       \\

170773  &                     &  3.6 $\pm$ 0.1   &  5.2 $\pm$ 0.3       &  101 -- 343   &  0.5 -- 1.4  &  46 $\pm$ 2  &  70 $\pm$ 4   & 252 $\pm$ 26   &  19.1 $\pm$ 1.9      \\

172167  & $\alpha$ Lyr; Vega  &  48.4 $\pm$ 0.9  &  0.17 $\pm$ 0.12     &  $<$80        &  1.2 -- 1.6  &  46 $\pm$ 7  & 260 $\pm$ 58  & 73 $\pm$ 3$^{(2)}$ &  1.1 $\pm$ 0.2   \\

181327  &                     &  3.3 $\pm$ 0.1   &  26.5 $\pm$ 5.2      &  $<$90        &  0.5 -- 0.6  &  63 $\pm$ 3  &  47 $\pm$ 4   & $<$390         &  24.8 $\pm$ 3.7      \\

\hline

182681  &                     &  25.7 $\pm$ 0.3  &  2.7 $\pm$ 0.2       &  $<$331       &  0.2 -- 1.0  &  80 $\pm$ 7  &  62 $\pm$ 8   & $<$525         &  10.4 $\pm$ 1.0      \\

191089  &                     &  3.0 $\pm$ 0.1   &  14.2 $\pm$ 0.5      &  $<$672       &  0.3 -- 2.3  &  89 $\pm$ 5  &  17 $\pm$ 2   & $<$395         &  3.7 $\pm$ 0.7       \\

197481  & AU Mic              &  0.09 $\pm$ 0.01 &  3.9 $\pm$ 0.3       &  --           &  0.0 -- 0.3  &  50 $\pm$ 8  &   9 $\pm$ 2   & 70 $\pm$ 10$^{(2)}$  &  0.60 $\pm$ 0.10   \\

205674  &                     &  2.9 $\pm$ 0.1   &  3.8 $\pm$ 0.2       &  $<$752       &  0.1 -- 1.1  &  60 $\pm$ 2  &  37 $\pm$ 2   & $<$390         &  4.4 $\pm$ 0.7       \\

206893  &                     &  2.5 $\pm$ 0.1   &  2.4 $\pm$ 0.1       &  94 -- 423    &  0.5 -- 2.7  &  54 $\pm$ 2  &  43 $\pm$ 4   & $<$290         &  3.0 $\pm$ 0.8       \\

207129  &                     &  1.3 $\pm$ 0.03  &  1.1 $\pm$ 0.2       &  $<$156       &  0.4 -- 1.1  &  46 $\pm$ 8  &  41 $\pm$ 10  & 159 $\pm$ 23   &  1.5 $\pm$ 0.4       \\

212695  &                     &  3.0 $\pm$ 0.05  &  0.49 $\pm$ 0.13     &  --           &  0.0 -- 2.7  &  35 $\pm$ 19 &  108 $\pm$ 83 & $<$350         &  8.6 $\pm$ 5.0       \\

213617  &                     &  5.5 $\pm$ 0.3   &  1.0 $\pm$ 0.1       &  --           &  0.0 -- 2.1  &  59 $\pm$ 6  &  52 $\pm$ 8   & $<$380         &  4.9 $\pm$ 1.5       \\

216956  & Fomalhaut           & 16.5 $\pm$ 0.3   &  0.58 $\pm$ 0.16     &  $<$87        &  1.0 -- 1.3  &  41 $\pm$ 3  &  185 $\pm$ 22 & 151 $\pm$ 8    &  3.2 $\pm$ 0.3       \\

218396  & HR 8799             &  5.4 $\pm$ 0.1   &  3.1 $\pm$ 0.3       &  $<$449       &  0.6 -- 1.9  &  43 $\pm$ 5  &  98 $\pm$ 18  & 395 $\pm$ 42   &  15.6 $\pm$ 2.4      \\

\hline
  \end{tabular}

\begin{flushleft} $^{(1)}$The error quoted for $L_*$ and $R_{BB}$ includes the uncertainty in the distance of the star from the \emph{Hipparcos} catalogue \citep{Perryman1997, van 
Leeuwen2007}. This is typically less than 5 percent of the total error, which is dominated by errors in the fitting.

\vskip 0.5mm

$^{(2)}$Based on the fitted radius from the 450\,$\umu$m image.
\end{flushleft}

\end{table*}

\vskip 1mm

\subsection{Disc radius and orientation}
\label{sec:disc_radius}

The disc radius can also be estimated from derived parameters from the SED fit, assuming that the dust grains behave as a blackbody, and are uniformly distributed in a disc at a 
distance $R_{\rm{BB}}$ from the star \citep{Wyatt2008}. In the cases where the emission is optically thin, the dust temperature can be used as a proxy for the radial separation 
from the star, which is given by:

\begin{equation}
   R_{\rm{BB}} = \left(\frac{278.3}{T_{\rm{d}}}\right)^2 L_*^{0.5}
   \label{eq:radius}
\end{equation}

\vskip 1mm

\noindent where, if the dust temperature ($T_{\rm{d}}$) is measured in Kelvins, and the stellar luminosity ($L_*$) in Solar luminosities, then $R_{\rm{BB}}$ is in astronomical 
units. The disc radii estimated from this method are presented in Table~\ref{tab:table3}.

\vskip 1mm

As already discussed in Section~\ref{sec:dust_temperature}, the far-IR/submillimetre emission from discs is typically modelled using a modified blackbody spectrum. 
Therefore, the true radius of the disc is expected to be significantly larger in most cases \citep{Rodriguez2012,Booth2013,Pawellek2014}, as is further discussed in 
Section~\ref{sec:disc_morphology}. Hence, whilst modelling the SED provides a wealth of information about the disc properties, measuring the radius, directly from an image, 
allows better constraints to be placed on other physical properties such as the particle size distribution \citep{Wyatt+Dent2002}. For the cases in which discs are spatially 
resolved the 850\,$\umu$m (or 450\,$\umu$m, as available) images are fitted using a 2-D Gaussian function ({\sc idl} routine {\sc mpfit2dfun}; \citealt{Markwardt2009}) to 
estimate the radial extent of the disc. The fitted disc major and minor axes are deconvolved with the beam size (including the broadening effect of the smoothing factor), and 
the major axis multiplied by the distance of the star to give an estimate of the true disc radius according to:

\begin{equation}
   R_{\rm{fit}} = \frac{d}{2} \sqrt{{\rm FWHM_{fit}^2} - ({\rm FWHM_{beam}^2} + {\rm FWHM_{smo}^2)}}
   \label{eq:radius_fit}
\end{equation}

\vskip 1mm

\noindent where $d$ is the distance of the star, ${\rm FWHM_{fit}}$ is the measured major axis of the Gaussian fit, ${\rm FWHM_{beam}}$ is the beam diameter (assumed
circular and with a measurement variation of $\pm$ 0.2\arcsec), and ${\rm FWHM_{smo}}$ is the Gaussian smoothing component (by default 7\arcsec)\footnote{If the distance of the
star is measured in parsecs and the FWHM beam in arcseconds, then the disc radius as specified in Equation~\ref{eq:radius_fit} will be in astronomical units.}.

\vskip 1mm

\begin{figure*}
\includegraphics[width=150mm]{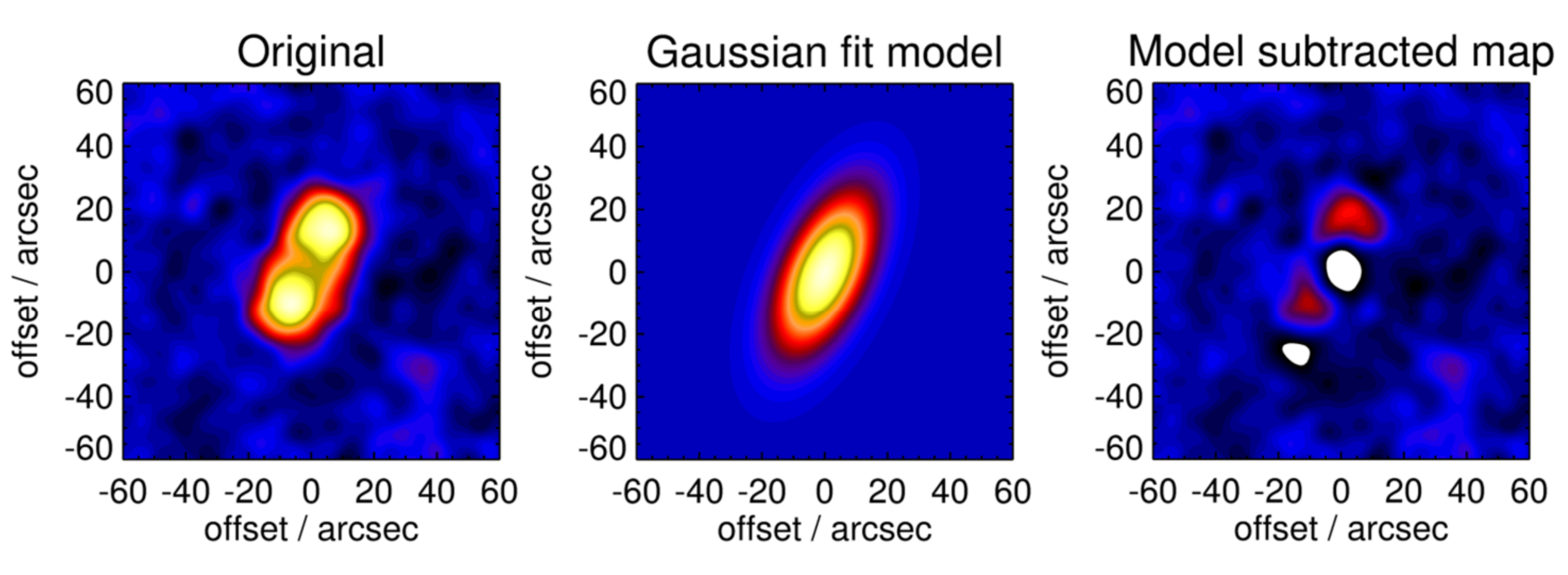}

\caption{(left) The original 850\,$\umu$m observed image for HD 216956 (Fomalhaut); (centre) The Gaussian model is based on the fit to the radial extent; (right) The result of 
subtracting the observed image from the model map. Note that the residual emission at the star is not due to the star, but is due to over-subtraction, as the Gaussian model used 
is centrally peaked but the real flux distribution is not.}

\label{fig:model_image}
\end{figure*}

There are a few cases where the emission is not well-approximated by a Gaussian profile. An example of this is Fomalhaut (HD 216956) where there are two equidistant lobes offset 
from the star position, and so the emission is not centrally concentrated. As discussed in Section~\ref{sec:fomalhaut_obs}, \emph{Herschel} and ALMA observations show that the 
emission is confined to a thin belt with the mid-point at a radius of $\sim$20\arcsec\ from the star. Fig.~\ref{fig:model_image} presents the results of the radial fit 
transposed into a model image and compared to the observed result. The model-subtracted map shows a residual peak at the star position (and to a lesser extent in the south-east 
lobe) but the fit, for the purposes of this paper, is a reasonable representation of the overall disc size. The Fomalhaut case is somewhat of an extreme example, and for the 
vast majority of stars in the sample the fitting is well-suited to the disc morphology.

\vskip 1mm

The inclination angle of the disc to the plane of the sky is derived from the axial ratio of the deconvolved major and minor axes fits (noting a 90\degree\ angle degeneracy). 
Finally, the position angle (PA) of the major axis of the disc is measured north through east (also noting an angle degeneracy of 180\degree). The estimated disc radius, 
including upper limits for the cases in which the disc is unresolved, are given in Table~\ref{tab:table3}. Full details of the measured major and minor axes radii, as well as 
the inclination and position angles, are presented in Table~\ref{tab:table4} for the 16 resolved discs in the SONS survey sample.

\begin{table*}

  \centering

  \caption{Derived parameters from the radial extent fitting for resolved sources in the SONS survey sample. The waveband indicates the wavelength from which
the measurements were made.}
  \label{tab:table4}

\begin{tabular}{llccccccc}
\hline

HD      &   Other                  & Waveband   & \multicolumn{2}{l}{Measured disc FWHM}    & Decconvolved            & Disc radius     & Inclination    &  Position angle        \\
number  &   names                  & ($\umu$m)  & major (\arcsec)     & minor (\arcsec)     & disc radius (\arcsec)   & $R_{\rm{fit}}$ (au) & (\degree)  &  (\degree\ E of N)     \\

\hline

9672    & 49 Cet                   &  450       & 17.5 $\pm$ 0.5      & 11.0 $\pm$ 0.7      & 7.1 $\pm$ 0.5           & 420 $\pm$ 16    &  74 $\pm$ 13   &  130 $\pm$ 10          \\

15745   &                          &  850       & 21.9 $\pm$ 3.5      & $<$15               & 8.1 $\pm$ 3.5           & 514 $\pm$ 111   &  $\geq$22      &  164 $\pm$ 21          \\

21997   &                          &  850       & 27.0 $\pm$ 1.9      & $<$15               & 11.3 $\pm$ 1.9          & 813 $\pm$ 69    &  $\geq$48      &  27 $\pm$ 12           \\

22049   & $\epsilon$ Eri           &  850       & 44.1 $\pm$ 1.2      & 40.2 $\pm$ 1.3      & 20.8 $\pm$ 1.2          & 67 $\pm$ 4      &  26 $\pm$ 2    &  61 $\pm$ 3            \\

38858   &                          &  850       & 29.3 $\pm$ 2.4      & 18.2 $\pm$ 2.9      & 12.7 $\pm$ 2.4          & 192 $\pm$ 18    &  65 $\pm$ 18   &  75 $\pm$ 11           \\

48682   &                          &  850       & 26.5 $\pm$ 3.9      & $<$15               & 11.0 $\pm$ 3.9          & 184 $\pm$ 33    &  $\geq$47      &  94 $\pm$ 19           \\

109085  & $\eta$ Crv               &  850       & 25.5 $\pm$ 1.2      & 21.8 $\pm$ 1.2      & 10.4 $\pm$ 1.2          & 190 $\pm$ 7     &  44 $\pm$ 4    &  132 $\pm$ 5           \\

115617  & 61 Vir                   &  850       & 17.5 $\pm$ 3.0      & $<$15               & 4.7 $\pm$ 3.0           & 40 $\pm$ 13     &  $\geq$0       &  66 $\pm$ 25           \\

143894  & 44 Ser                   &  850       & 25.2 $\pm$ 2.5      & 16.8 $\pm$ 2.7      & 10.2 $\pm$ 2.5          & 561 $\pm$ 69    &  67 $\pm$ 24   &  70 $\pm$ 15           \\

\hline

161868  & $\gamma$ Oph             &  850       & 21.5 $\pm$ 2.2      & $<$15               & 7.8 $\pm$ 2.2           & 246 $\pm$ 35    &  $\geq$16      &  75 $\pm$ 17           \\

170773  &                          &  850       & 20.1 $\pm$ 1.4      & 16.0 $\pm$ 1.6      & 6.8 $\pm$ 1.4           & 252 $\pm$ 26    &  63 $\pm$ 17   &  140 $\pm$ 13          \\

172167  & $\alpha$ Lyr; Vega       &  450       & 21.6 $\pm$ 0.7      & 18.1 $\pm$ 0.8      & 9.5 $\pm$ 0.7           & 73 $\pm$ 3      &  34 $\pm$ 2    &  40 $\pm$ 17           \\

        &                          &  850       & 38.1 $\pm$ 0.8      & 32.3 $\pm$ 0.9      & 17.5 $\pm$ 0.8          & 135 $\pm$ 3     &  35            &  45 $\pm$ 19           \\

197481  & AU Mic                   &  450       & 17.5 $\pm$ 2.1      & 13.5 $\pm$ 2.3      & 7.1 $\pm$ 2.1           & 70 $\pm$ 10     &  52 $\pm$ 16   &  127 $\pm$ 15          \\

207129  &                          &  850       & 24.8 $\pm$ 2.9      & 15.2 $\pm$ 3.1      & 10.0 $\pm$ 2.9          & 159 $\pm$ 23    &  80 $\pm$ 69   &  115 $\pm$ 13          \\

216956  & Fomalhaut                &  850       & 42.0 $\pm$ 1.8      & 21.1 $\pm$ 2.7      & 19.7 $\pm$ 0.5          & 151 $\pm$ 8     &  68 $\pm$ 5    &  156 $\pm$ 3           \\

218396  & HR 8799                  &  850       & 24.9 $\pm$ 2.1      & 18.8 $\pm$ 2.8      & 10.1 $\pm$ 2.1          & 395 $\pm$ 42    &  55 $\pm$ 6    &  71 $\pm$ 16           \\

\hline
  \end{tabular}
\end{table*}

\subsection{Fractional luminosities}
\label{sec:fractional_luminosities}

The amount of dust in debris discs is often quantified in terms of the fractional luminosity, and can be determined from the SED fits to both the stellar photospheric and excess 
thermal emission. This quantity is defined as the ratio of the IR luminosity from the dust to that of the star, $f = L_{\rm{IR}}/L_{\star}$ \citep[e.g.][]{Wyatt2008}, and can be 
estimated from the wavelength and flux of the maximum in the emission spectra of the star and the disc, according to:

\begin{equation}
   f = \frac{F_{\rm{d(max)}}}{F_{\rm{\star(max)}}}\frac{\lambda_{\rm{\star(max)}}}{\lambda_{\rm{d(max)}}}.
   \label{eq:luminosities}
\end{equation}

\vskip 1mm

As Equation~\ref{eq:luminosities} represents only an approximation to the fractional luminosity, $f$ is determined for the targets in this paper by the ratio of the integrated 
areas under the star and disc SED component fits. When there are two temperature components of the SED fit to the IR excess, then the fractional luminosity is derived from the 
sum of both. A defining property of a debris disc is that, in general, it has a fractional luminosity of $f < 10^{-2}$ \citep{Lagrange2000} in contrast to protoplanetary 
discs, which have higher fractional luminosities. This criterion is certainly met by cool Edgeworth-Kuiper belt analogues, but falls down for stars at an earlier evolutionary 
phase where planet formation is believed to be ongoing, and where the flux excess tends to peak at mid-IR wavelengths \citep[e.g.][]{Melis2010, Fujiwara2012, Vican2016}. The 
fractional luminosities for the SONS surveys discs are given in Table~\ref{tab:table3}, and all but one (the exception being HD 98800) of the discs detected fall into this 
``debris'' classification.

\subsection{Dust masses}
\label{sec:dust_masses}

Although the fractional luminosity can be converted into an estimate of dust mass by assuming all dust grains have the same diameter and density, dust masses are usually
derived directly from the 850\,$\umu$m flux density measurement. Since the emission from debris discs is optically thin at these wavelengths, their mass is directly
proportional to the emission, according to,

\begin{equation}
M_{\rm{d}} = \frac{F_{\nu} d^2}{\kappa_{\nu} B_{\nu}(T_{\rm{d}})}
   \label{eq:dust_mass}
\end{equation}

\vskip 1mm

\noindent where $F_{\nu}$ is the measured flux density, $d$ is the distance of the target, $\kappa_\nu$ is the dust opacity  which is assumed to be 1.7\,cm$^2$\,g$^{-1}$ at
850\,$\umu$m in accordance with similar studies \citep{Pollack1994,Dent2000}, and $T_{\rm{d}}$ is the dust temperature derived from the SED fit. In the Rayleigh-Jeans
limit, the mass becomes a linear function of temperature and so equation~\ref{eq:dust_mass} reduces to,

\begin{equation}
M_{\rm{d}} [M_{\earth}] = 5.8 \times 10^{-10} \frac{F_{\nu}[\rm{mJy}] d[\rmn{pc}]^2 \lambda[\rm{\umu}m]^2}{\kappa_{\nu}[\rmn{cm^2 g^{-1}}] T_{\rm{d}}[\rm{K}]}
\end{equation}

\vskip 1mm

The calculated dust masses are summarised in Table~\ref{tab:table3} for the SONS survey sample, with quoted uncertainties based on the errors in the 850\,$\umu$m flux and fitted
dust temperature only.

\vskip 1mm

\section{Individual targets discussion}
\label{sec:target_discussion}

This section of the paper provides a discussion of each of the targets for which emission was detected in the vicinity of the star. Each sub-section summarises the new SONS 
survey results at 850\,$\umu$m (and 450\,$\umu$m, if available) within the context of previous observations, as well as the results of the modelling of the SED and the 
estimation of the dust mass from the 850\,$\umu$m flux. Full details, including estimated errors on the modelled and calculated parameters such as $T_{\rm{d}}$, $R_{\rm{BB}}$, 
$R_{\rm{fit}}$, $\beta$ and $M_{\rm{d}}$, are given in Tables~\ref{tab:table3} and \ref{tab:table4}.

\subsection{HD 377}
\label{sec:hd377}

HD 377 is a Solar-type star (G2V) at a distance of 39.1\,pc with an estimated age of around 170\,Myr \citep{Chen2014}, but could be as young as 32\,Myr \citep{Hillenbrand2008} 
or as old as 250\,Myr \citep{Choquet2016}. The SONS survey image, as shown in Fig.~\ref{fig:figureA1}a, reveals emission centred on the stellar position with flux density 
of 5.3\,$\pm$\,1.0\,mJy at 850\,$\umu$m. (This value is revised slightly higher from the result reported in \citealt{Panic2013}.) Interpreting this peak as an unresolved disc 
about the star gives an upper limit to the radius of 290\,au.

\vskip 1mm

\emph{HST}/NICMOS imaging of HD 377 shows an edge-on disc at a position angle of 47\degree\ with a radius of 2.2\arcsec\ ($\sim$86\,au) \citep{Choquet2016}. The disc has also 
been resolved using the SMA \citep{Steele2016} with a 850\,$\umu$m flux of 3.5\,$\pm$\,1\,mJy, just consistent with the SONS result, and a deconvolved disc radius of 47\,au 
at a PA of 30\degree. Although the mid-far IR region is reasonably well characterised by \emph{Spitzer} \citep{Chen2014}, there are few constraining points in the 
submillimetre/millimetre. Previous SED modelling suggested a two-component fit deriving ``warm'' and ``cold'' elements with dust temperatures of 240\,K and 57\,K \citep{Chen2014}. 
Using photometry at 1.2\,mm from the IRAM 30\,m telescope a dust mass of 0.058\,M$_{\earth}$ was derived \citep{Roccatagliata2009}. Modelling of the SED with the new 
850\,$\umu$m flux density measurement included gives a cold component dust temperature of 56\,K, but the poorly sampled long-wavelength slope means that $\beta$ is only 
constrained to be less than a value of 1.4. The estimated dust mass for the cold component of 0.036\,M$_{\earth}$ is lower than the IRAM result, but is just consistent within 
the measurement errors.

\subsection{HD 6798}
\label{sec:hd6798}

HD 6798 is a luminous A3 star in Cepheus lying at a distance of 83\,pc and an estimated age of around 320\,Myr but with an uncertainty spanning the range from 260 to 400\,Myr 
\citep{Moor2006}. An emission peak is seen at 850\,$\umu$m with a flux density of 7.2\,$\pm$\,1.0\,mJy (Fig.~\ref{fig:figureA1}b). As an unresolved disc the upper limit 
to the radius is 615\,au.

\vskip 1mm

The SED contains flux measurements from \emph{IRAS} and \emph{Herschel}/PACS (archival data) in the range 25 -- 160\,$\umu$m but no other long wavelength points. Modelling 
suggests, as in the case of HD 377, that there are both warm and cold disc components with the former having a dust temperature of 203\,K. The fit to the far-IR/submillimetre 
detected cold disc component gives a dust temperature of 48\,K and constrains $\beta$ to be between 0.2 and 0.9. The dust mass is estimated to be 0.25\,M$_{\earth}$.

\subsection{HD 8907}
\label{sec:hd8907}

\begin{figure}
\includegraphics[width=85mm]{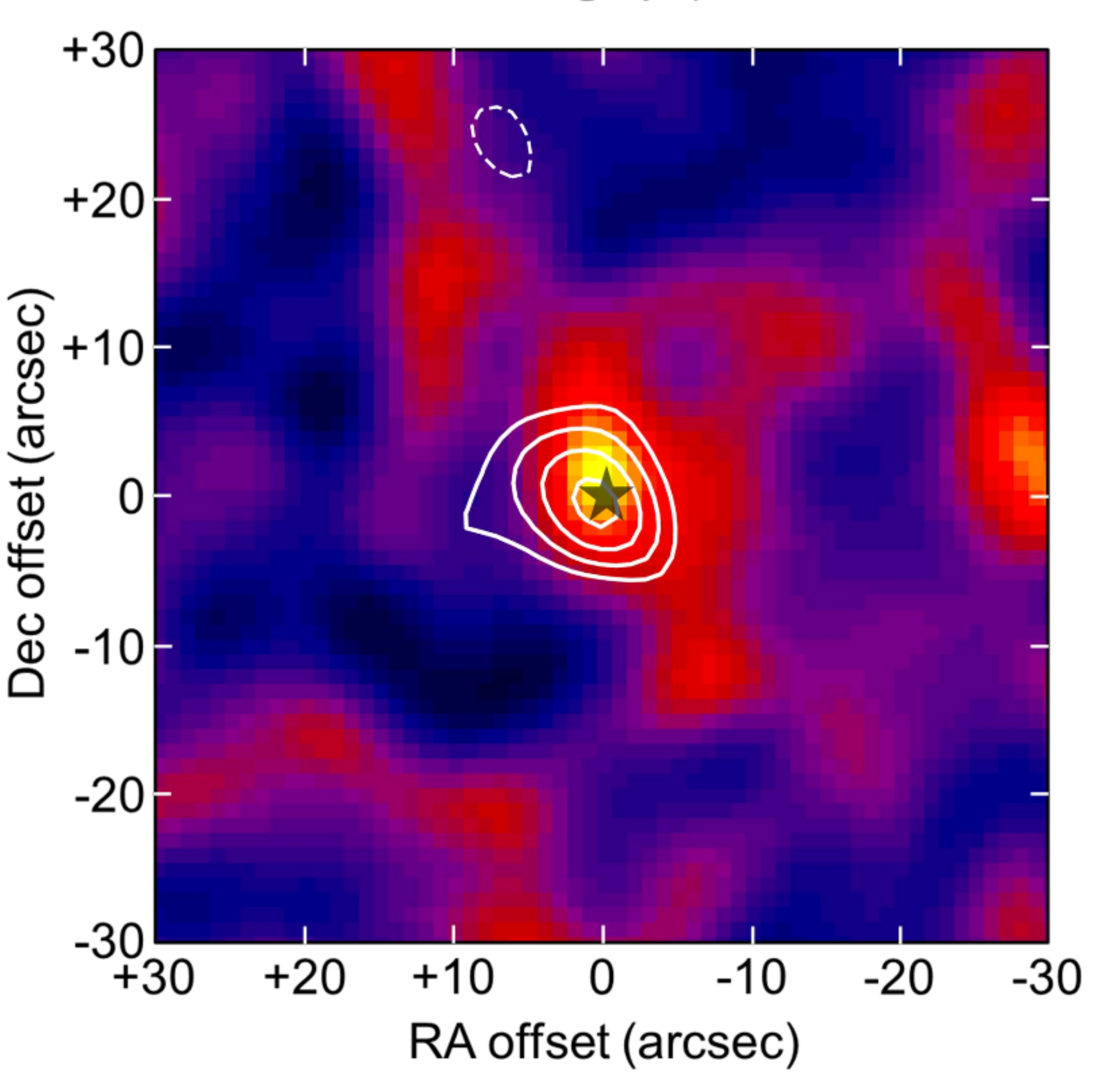}

\caption{The 450\,$\umu$m S/N image from observations of HD 8907 with contours from the 850\,$\umu$m image overlaid. The colours are scaled from $-$3\,$\sigma$ to the maximum S/N
in the image ($\sim$6\,$\sigma$). The contours start at $-$3\,$\sigma$ (dashed white) and then solid colours from 3\,$\sigma$ to the maximum in 1$\sigma$ steps. The star symbol
represents the position of the star with respect to the flux peak.}

\label{fig:hd8907}
\end{figure}

HD 8907 is an F8 star in Andromeda at a distance of 35\,pc and with an estimated age of around 320\,Myr \citep{Hillenbrand2008}, although there is significant uncertainty with 
lower and upper limits put at 110\,Myr and 870\,Myr, respectively \citep{Moor2006}. The peak in emission is coincident with the star position, and is detected at a flux density 
of 7.8\,$\pm$\,1.2\,mJy at 850\,$\umu$m (Fig.~\ref{fig:figureA1}c), somewhat higher than the previous SCUBA measurement of 4.8\,$\pm$\,1.2\,mJy \citep{Najita2005}. Emission was 
also detected (and unresolved) at 450\,$\umu$m with a flux density of 51\,$\pm$\,10\,mJy. The 450\,$\umu$m image with 850\,$\umu$m contours overlaid is shown in 
Fig.~\ref{fig:hd8907}, showing both peaks to be coincident. Based on the 450\,$\umu$m image, and interpreting the peak as a disc about the star, gives an upper limit to the 
radius of 175\,au.

\vskip 1mm

The SED is well characterised in the IR and submillimetre/millimetre, and observations at 1.2\,mm using the SMA \citep{Steele2016} resolve the disc with a measured radius of 
54\,au at a PA of 55\degree. The SONS measurements help to constrain the SED fit, with a dust temperature of 51\,K, and a $\beta$ value in the range 0.4 -- 2.1. The dust mass is 
estimated to be 0.044\,M$_{\earth}$.

\subsection{49 Ceti (HD 9672)}
\label{sec:49ceti}

\begin{figure}
\includegraphics[width=85mm]{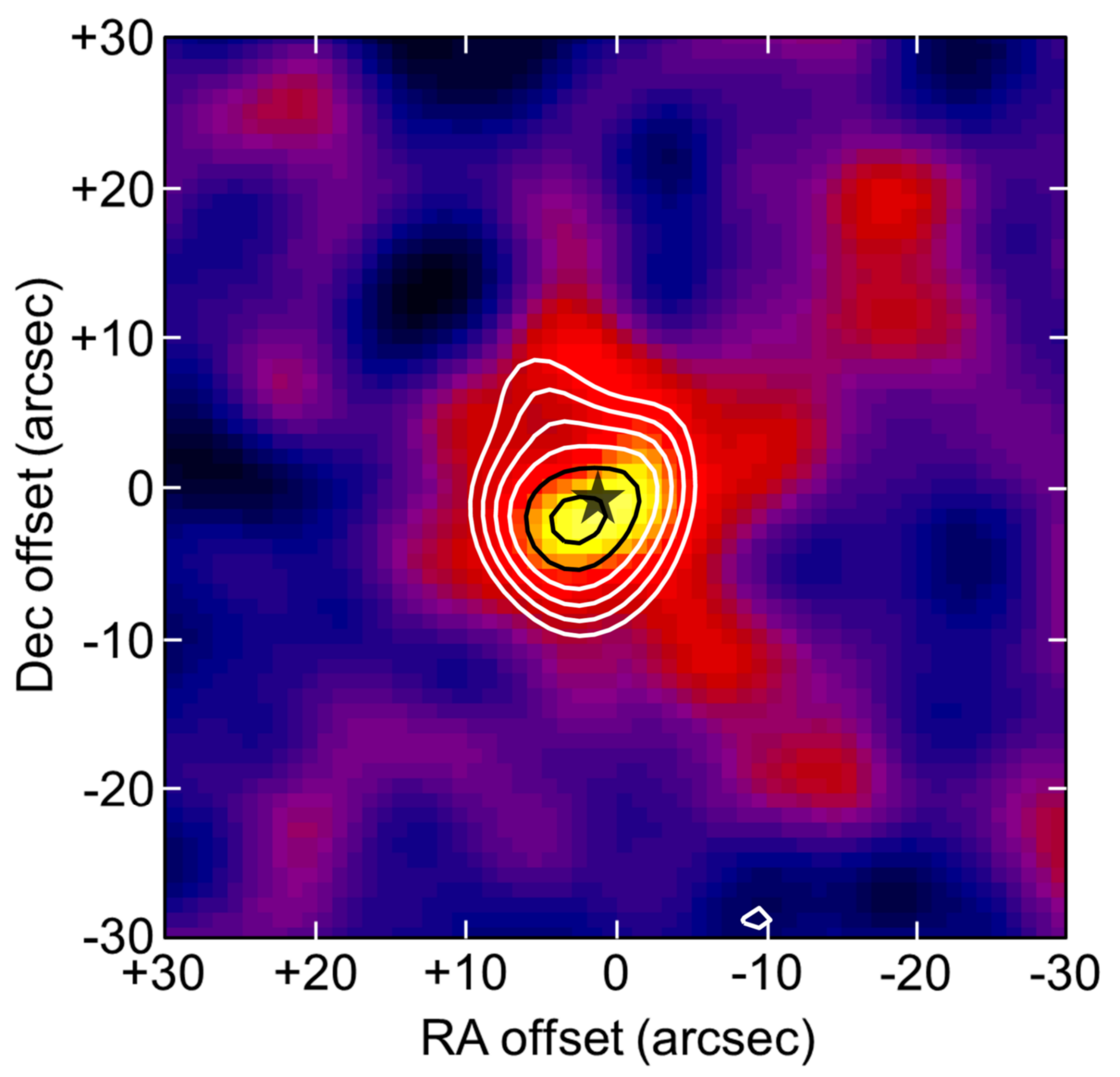}

\caption{The 450\,$\umu$m image from observations of 49 Ceti (HD 9672) with contours from the 850\,$\umu$m image overlaid. The contours and symbols are as described in 
Fig.~\ref{fig:hd8907}.}

\label{fig:49ceti}
\end{figure}

HD 9672 (49 Ceti) is a young A1 star in Cetus and a member of the Argus Association, at a distance of 59\,pc with an estimated age of 40\,Myr \citep{Torres2008} but could 
be as young as 20\,Myr \citep{Chen2014}. At 850\,$\umu$m the detected emission at the stellar position appears to be unresolved (Fig.~\ref{fig:figureA1}d) with a flux density 
of 13.5\,$\pm$\,1.5\,mJy. At 450\,$\umu$m, however, the emission morphology is elongated at a PA of 130\degree\, with a flux density of 125\,$\pm$\,18\,mJy, as shown in 
Fig.~\ref{fig:49ceti} \citep{Greaves2016}. The measured disc FWHM of 17.5\arcsec\ (deconvolved radius of 7.1\arcsec, corresponding to $\sim$420\,au) at 450\,$\umu$m, indicates 
that the structure is likely a disc about the star, that could also be marginally resolved at 850\,$\umu$m.

\vskip 1mm

\emph{Herschel}/PACS observations at 70\,$\umu$m reveal a resolved disc of radius 250\,au with a PA of 105\degree\ \citep{Roberge2013}. Recent ALMA observations also resolve the 
disc with a PA of 107\degree\ identifying dust that extends from just a few to around 300\,au from the star \citep{Hughes2017}. The fact that this is significantly less in extent 
than implied by the SONS result at 450\,$\umu$m remains an open issue. The observed structure has been modelled as having two components: an inner disc extending to a radius of 
60\,au (and depleted at less than 30\,au) \citep{Wahhaj2007}, and an outer disc of radius up to 400\,au \citep{Greaves2016}. Scattered light images, from \emph{HST}/NICMOS and 
coronograhic H-band images using VLT/SPHERE, show the outer disc extending from 1.1 to 4.6\arcsec\ ($\sim$65 -- 250\,au) with an inclination angle of 73\degree and a PA of 106 -- 
110\degree\ \citep{Choquet2017}. In addition, the system has a well-known molecular and atomic gas reservoir, which was originally purported to be consistent with the 
properties of a low-mass protoplanetary disc \citep{Zuckerman1995}. The age estimate for the system (likely to be 40\,Myr), however, suggested it was more likely the gas had a 
secondary origin, perhaps involving a high rate of comet destruction given the large observed dust mass \citep{Zuckerman&Song2012,Hughes2017}.

\vskip 1mm

The disc is well-characterised in the far-IR, with observations from \emph{Spitzer} and \emph{Herschel} contributing to points in the SED. In the millimetre the measured flux 
from IRAM at 1.3\,mm of 13.9\,$\pm$\,2.5\,mJy \citep{Walker+Butner1995} is significantly higher than expected based on the SED fit to the far-IR and submillimetre points. The 
reason for this is unknown. The SED, including 9\,mm photometry from the VLA \citep{MacGregor2016a} but not using the IRAM 1.3\,mm photometry, is modelled with a two-component 
fit. The ``warm'' (inner) element has a characteristic temperature of 165\,K. The dominant ``cold'' (outer) component has a dust temperature of 59\,K, a $\beta$ value of 
between 0.9 and 1.1, and a calculated dust mass of 0.20\,M$_{\earth}$.

\subsection{q$^1$ Eridani (HD 10647)}
\label{sec:q1eri}

q$^1$ Eridani is a nearby F9 star at a distance of 17\,pc and with an estimated age of 1600\,Myr \citep{Chen2014}. There is, however, considerable uncertainty in the age with 
estimates ranging from 300\,Myr to 7000\,Myr \citep{Moor2006}. Even though the southerly declination of HD\,10647 was quite a challenge for observing with the JCMT, a structure, 
peaking in emission at the star and extending eastwards, was detected at 850\,$\umu$m (Fig.~\ref{fig:figureA2}a). The measured integrated flux of 30.3 $\pm$ 3.9\,mJy, measured 
in a 50\arcsec\ diameter aperture centred on the star, is just consistent with the APEX/LABOCA result of 39.4 $\pm$ 4.1\,mJy at 850\,$\umu$m \citep{Liseau2008}, the data for 
which also show a roughly eastward extension to the disc. The structure is resolved within the 15\arcsec\ beam, and interpreted as a disc with a deconvolved radius of 
9.0\arcsec\ ($\sim$155\,au) at a PA of 72\degree, dominated by the eastward extension.

\vskip 1mm

The infrared excess was first detected by \emph{IRAS} \citep{Stencel1991}. The system has one known Jupiter-mass planet (HD 10647\,b; \citealt{Butler2006}) orbiting at a 
semimajor axis of 2\,au, so is unlikely to have any significant influence on a dust disc of radius $>$100\,au. \emph{HST}/ACS coronography revealed an 7.0 -- 8.2\arcsec\ 
($\sim$120 -- 140\,au) radius disc in scattered light with a PA of 56\degree\ \citep{Stapelfeldt2007}. \emph{Herschel}/PACS also resolved the disc with a PA of 54\degree\ and a 
beam-deconvolved radius of 6.7 $\times$ 3.8\arcsec\ at 160\,$\umu$m ($\sim$115 $\times$ 65\,au; \citealt{Liseau2010}). In the 850\,$\umu$m SONS image, the disc appears moderately 
extended compared to the beam in a roughly easterly direction with most of the flux concentrated at the position of the star. Within the 4$\sigma$ contour the PA of the disc in 
the 850\,$\umu$m SONS image is $\sim$65\degree, agreeing reasonably well with both the \emph{HST} and \emph{Herschel} observations, and on a similar scale ($\sim$15\arcsec) at 
least in the north-east direction. (There is also a hint of an extension to the south-west.)

\vskip 1mm

A more plausible explanation is that the eastward extension seen in the SCUBA-2 and LABOCA images is caused by a background object. This extra flux would explain why both 
850\,$\umu$m values are slightly high based on the SED model fit to the far-IR photometric points, and why there is not a symmetric disc seen about the star, as in the 
\emph{HST} and \emph{Herschel} images. In the \emph{Herschel}/PACS 70\,$\umu$m image there is evidence of a isolated source 20\arcsec\ to the east of the main flux peak 
\citep{Liseau2010}, which aligns well with the eastward extension seen at 850\,$\umu$m. A similar extension is also seen in the \emph{Herschel}/PACS 160\,$\umu$m. Hence, in 
this paper, the infrared excess flux surrounding q$^1$ Eri has been re-interpreted as an unresolved disc with a flux density of 20.1 $\pm$ 2.7\,mJy at 850\,$\umu$m. (This 
being the flux measured in a beam-sized aperture, centred on the star position.) The SED is well-constrained by \emph{Spitzer}, \emph{Herschel}, the long submillimetre points 
(including SCUBA-2 at 850\,$\umu$m) and 6.8\,mm photometry from ACTA \citep{Ricci2015b}, to give a dominant temperature component at 44\,K, a well-constrained $\beta$ between 
0.6 and 0.8, and an estimated dust mass of 0.033\,M$_{\earth}$. An upper limit to the disc radius from the 850\,$\umu$m image, assuming an unresolved source, is 125\,au.

\vskip 1mm

\subsection{$\tau$ Ceti (HD 10700)}
\label{sec:tauceti}

HD 10700 ($\tau$ Ceti) is a nearby G-type star, believed to be very similar to our Sun, at a distance of 3.7\,pc with an estimated age of 7650\,Myr, but with an uncertainty in 
the range 6130 to 8500\,Myr \citep{Pagano2015}. The SONS 850\,$\umu$m image shows a peak in emission at the stellar position with a flux density of 4.4 $\pm$ 0.6\,mJy 
(Fig.~\ref{fig:figureA2}b) of which an estimated 1\,mJy is emission from the photosphere. Interpreting the emission morphology as a disc gives an upper limit to the radius from 
the 850\,$\umu$m image of 27\,au.

\vskip 1mm

The infrared excess has been known since the early days of \emph{IRAS} and was later confirmed by \emph{ISO} \citep{Habing2001}. $\tau$ Ceti was first imaged by SCUBA 
\citep{Greaves2004} which suggested a disc extending some 55\,au from the star, and identified the structure as possibly being a massive Edgeworth-Kuiper Belt analogue. The disc was 
subsequently resolved by \emph{Herschel} at 70\,$\umu$m and 160\,$\umu$m (possibly also at 250\,$\umu$m) with the peak emission occurring at a radius of 5\arcsec\ ($\sim$20\,au) in 
the photosphere-subtracted image at 70\,$\umu$m (Figure 3 of \citealt{Lawler2014}), and modelled as a wide annulus with an inner edge between 1 and 10\,au and an outer edge at 
$\sim$55\,au inclined from a face-on orientation by 35 $\pm$ 10\degree\ \citep{Lawler2014}. More recently 1.3\,mm interferometric observations with ALMA have revealed a nearly 
face-on belt of cold dust at a radius of 44\,au with a PA of 90\degree\ surrounding an unresolved central source at the stellar position \citep{MacGregor2016a}. Modelling of the 
belt suggests an inner edge of 6\,au, consistent with \emph{Herschel} observations. There are also 5 candidate planets orbiting $\tau$ Ceti, all of which are believed to lie within 
a radius of 1.35\,au of the star \citep{Tuomi2013}. Whilst the outermost planet may have some influence on the disc inner edge, the majority of the disc extending from a few au to 
over 50\,au will be largely unaffected by the planetary system. The existence of the planetary system, however, remains unconfirmed, following recent work in which periodic 
signals were not detected from Doppler measurements obtained in the California Planet Survey \citep{Howard2016}.

\vskip 1mm 

\begin{figure}
\includegraphics[width=85mm]{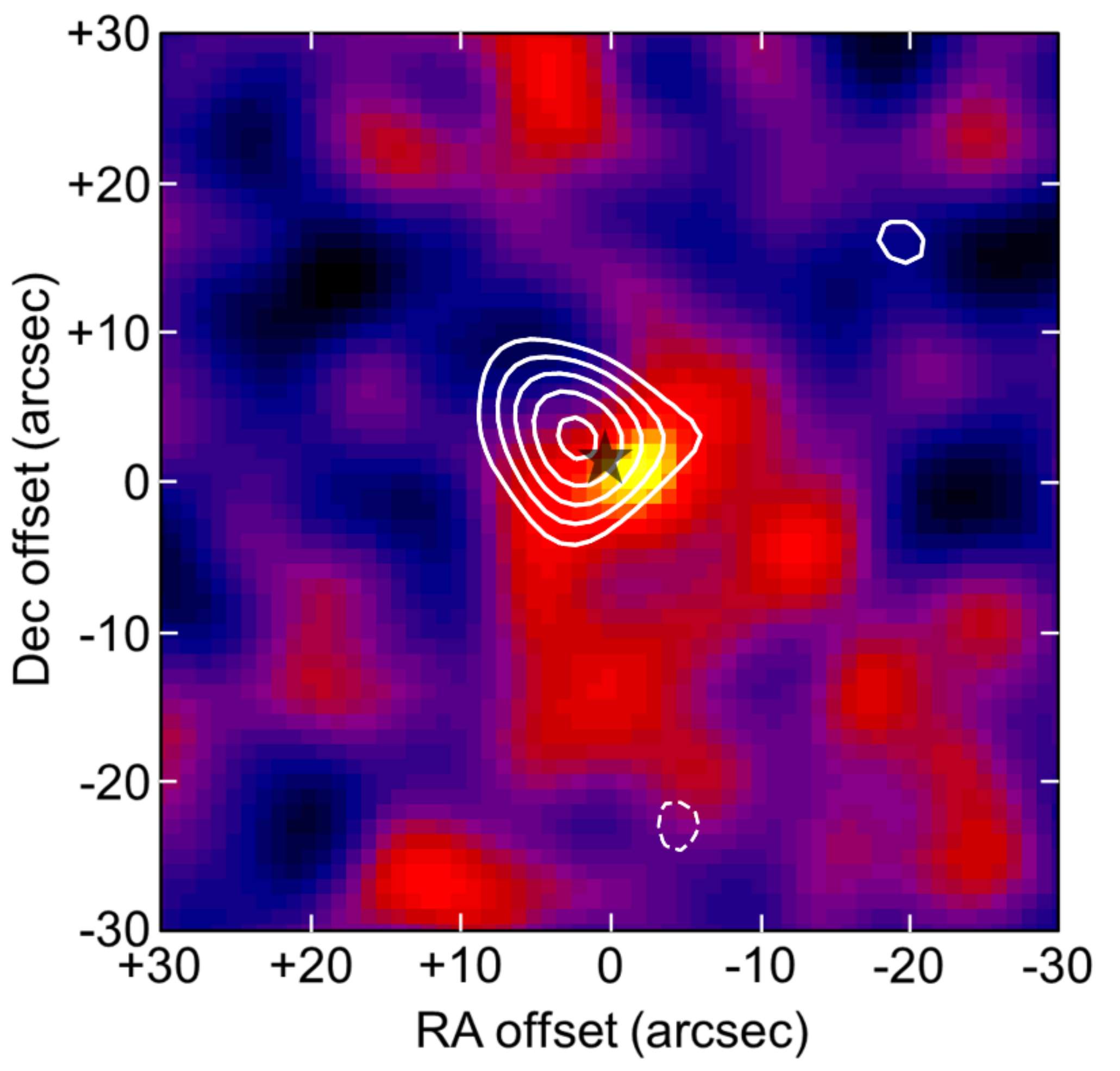}

\caption{The 450\,$\umu$m image from observations of $\tau$ Ceti (HD 10700) with contours from the 850\,$\umu$m image overlaid. The contours and symbols are as described in
Fig.~\ref{fig:hd8907}.}

\label{fig:tauceti}
\end{figure}

The disc is also detected at 450\,$\umu$m within the SONS survey with a flux density of 25.2 $\pm$ 4.5\,mJy (Fig.~\ref{fig:tauceti}) including an estimated 5\,mJy from the 
photosphere. The separation of the 450 and 850\,$\umu$m peaks is $\sim$5\arcsec, consistent within the expected statistical uncertainties (see Section~\ref{sec:offsets}), with 
the star positioned equidistant between the two peaks. The upper limit to the radius from the 450\,$\umu$m image of 18\,au. \citet{Lawler2014} suggested that since no SED 
model can adequately fit both the \emph{Herschel}/PACS and SCUBA/SCUBA-2 points, a separate, cooler disc component, may be required to explain the submillimetre observations. 
The SED shown in Fig.~\ref{fig:figureA2}b is the best single fit to all the points beyond 70\,$\umu$m, and results in a single-temperature component of 71\,K (but with large 
errors of $\pm$ 23\,K), a $\beta$ constrained between 0 and 1.1, and implying a dust mass of 2.1 $\times$ 10$^{-4}$ M$_{\earth}$ surrounding the star, only approximately 20 
times that of the Edgeworth-Kuiper belt \citep{Greaves2004}.

\vskip 1mm

\subsection{HD 10638}
\label{sec:hd10638}

HD 10638 is a luminous A3 star at a distance of 69\,pc with an estimated age of 50\,Myr, with lower and upper limits of 20 and 150\,Myr \cite{Moor2006}. Although the SONS survey 
image is somewhat uneven (i.e., there is a prominent gradient in the background towards the west), an emission peak is detected at 850\,$\umu$m with a flux density of 5.1 $\pm$ 
0.9\,mJy (Fig.~\ref{fig:figureA2}c) that is centred on the stellar position. As an unresolved disc about the star the upper limit to the radius is 515\,au.

\vskip 1mm

\emph{IRAS} detected excess far-IR emission from the star \citep{Silverstone2000}, but no other far-IR photometric measurements have been made. The SED is therefore loosely 
constrained by only the IRAS 60\,$\umu$m and SONS 850\,$\umu$m photometry, but the fit suggests a dominant cold component at 34\,K but with a large uncertainty, a $\beta$ 
value of $>$0.6, and an estimated dust mass of 0.18\,M$_{\earth}$. Modelling of the near-mid IR photometric points of the SED also hints at the presence of a warmer, inner 
component at 138\,K.

\subsection{$\beta$ Trianguli (HD 13161)}
\label{sec:hd13161}

HD 13161 ($\beta$ Tri) is a spectroscopic binary (primary of spectral type A5-IV) at a distance of 39\,pc, with a period of 31\,days. It has an estimated age of 730\,Myr with an 
uncertainty of $\pm$\,300\,Myr \citep{Vican2012}. The 850\,$\umu$m image shows a peak in emission at the stellar position with a flux density of 5.1 $\pm$ 0.9\,mJy 
(Fig.~\ref{fig:figureA2}d). Interpreting this unresolved peak to be a disc about the star, gives an upper limit to the radius of 290\,au.

\vskip 1mm

The disc is well-characterised in the far-IR, particularly with \emph{Herschel} contributing to points in the SED. \emph{Herschel}/PACS observations at 70\,$\umu$m reveal a 
deconvolved radius of 102\,au at a PA of 66\degree\ \citep{Booth2013}. Modelling of the \emph{Herschel} images concluded that the disc is consistent with being aligned with the 
binary orbital plane \citep{Kennedy2012}. The SED is modelled as a single modified blackbody with a characteristic dust temperature of 84\,K and a $\beta$ in the range 0.6 to 
1.7. The estimated dust mass is 0.022\,M$_{\earth}$.

\subsection{$\gamma$ Trianguli (HD 14055)}
\label{sec:hd14055}

HD 14055 ($\gamma$ Tri) has a spectral type A1 and is at a distance of 34\,pc. It has an estimated age of 230\,Myr with a lower and upper limits of 160\,Myr \citep{Thureau2014} 
and 300\,Myr \citep{Chen2014}, respectively. The 850\,$\umu$m image shows an emission morphology which peaks at an offset of 3\arcsec\ from the star, and appears to be slightly 
elongated in the north-south direction. The peak flux is 7.2 $\pm$ 1.0\,mJy/beam (Fig.~\ref{fig:figureA3}a), a value slightly higher than previously reported \citep{Panic2013}. 
The integrated flux density, determined over a 40\arcsec\ diameter aperture, however, reveals no significant evidence of extended emission compared to the beam. The flux is also 
consistent with that measured using SCUBA of 5.5 $\pm$ 1.8\,mJy \citep{Williams&Andrews2006}. There is no SCUBA-2 detection at 450\,$\umu$m with the 5$\sigma$ upper limit of 
70\,mJy being close to the predicted flux from the SED. Interpreting this structure as a disc about the star gives an upper limit to the disc radius from the 850\,$\umu$m image 
of 260\,au.

\vskip 1mm

The disc has been resolved by \emph{Herschel}/PACS observations at 70, 100 and 160\,$\umu$m and interpreted as a fairly broad disc extending from a radius of 3.7 to 5\arcsec\ 
($\sim$125 -- 170\,au), at a PA of 163\degree\ from the 160\,$\umu$m image \citep{Booth2013}. Based on the \emph{Herschel} results the north-south extension hinted at in the 
SONS 850\,$\umu$m image could be real, although, as previously mentioned, the integrated flux measurement indicates that it is not statistically significant. The 850\,$\umu$m 
SONS survey photometry helps to anchor the long-wavelength side of the SED, allowing a single blackbody fit with a dust temperature of 77\,K, a $\beta$ emissivity value of 
between 0.7 and 1.5, and an estimated dust mass of 0.028\,M$_{\earth}$.

\subsection{HD 15115}
\label{sec:hd15115}

HD 15115 is a young F2 star within the $\beta$ Pictoris Moving Group at a distance of 45\,pc, with an estimated age of 23 $\pm$ 3\,Myr \citep{Mamajek&Bell2014}. The 850\,$\umu$m 
image shows a structure with a slightly elongated morphology (PA of $\sim$150\degree), with a peak in emission 5\arcsec\ offset to the north from the star position 
(Fig.~\ref{fig:figureA3}b). The flux density of 8.2 $\pm$ 1.1\,mJy is marginally lower than that previously reported \citep{Panic2013}. The elongation is not significant 
compared to the beam size, and so interpreting the structure as an unresolved disc at the star gives an upper limit to the radius of 340\,au.

\vskip 1mm

HD\,15115 has been observed in optical scattered light with images showing a remarkable asymmetry between the eastern and western sides of the disc \citep{Kalas2007}, with the 
western side reaching a radius of 12\arcsec\ ($\sim$550\,au) from the star. The PA of the major axis of the disc from the scattered light image is 98.5\degree. Observations 
with the SMA at 1.3\,mm resolved the HD 15115 disc into an asymmetric structure extending to the west of the star with a measured radius of 6\arcsec\ \citep{MacGregor2015}. 
Modelling of the image suggests a well-defined emission belt at a radius of 110\,au from the star, and a $\sim$3\,$\sigma$ feature aligned with the asymmetric western 
extension of the scattered light disc \citep{MacGregor2015}. A fit to the \emph{Herschel}/PACS, 850\,$\umu$m and 6.8\,mm photometry from ACTA \citep{MacGregor2016a}  
suggests a dominant cold disc of characteristic temperature 57\,K, a $\beta$ in the range 0.5 to 0.9, and an estimated dust mass of 0.073\,M$_{\earth}$.

\subsection{12 Trianguli (HD 15257)}
\label{sec:hd15257}

\begin{figure}
\includegraphics[width=85mm]{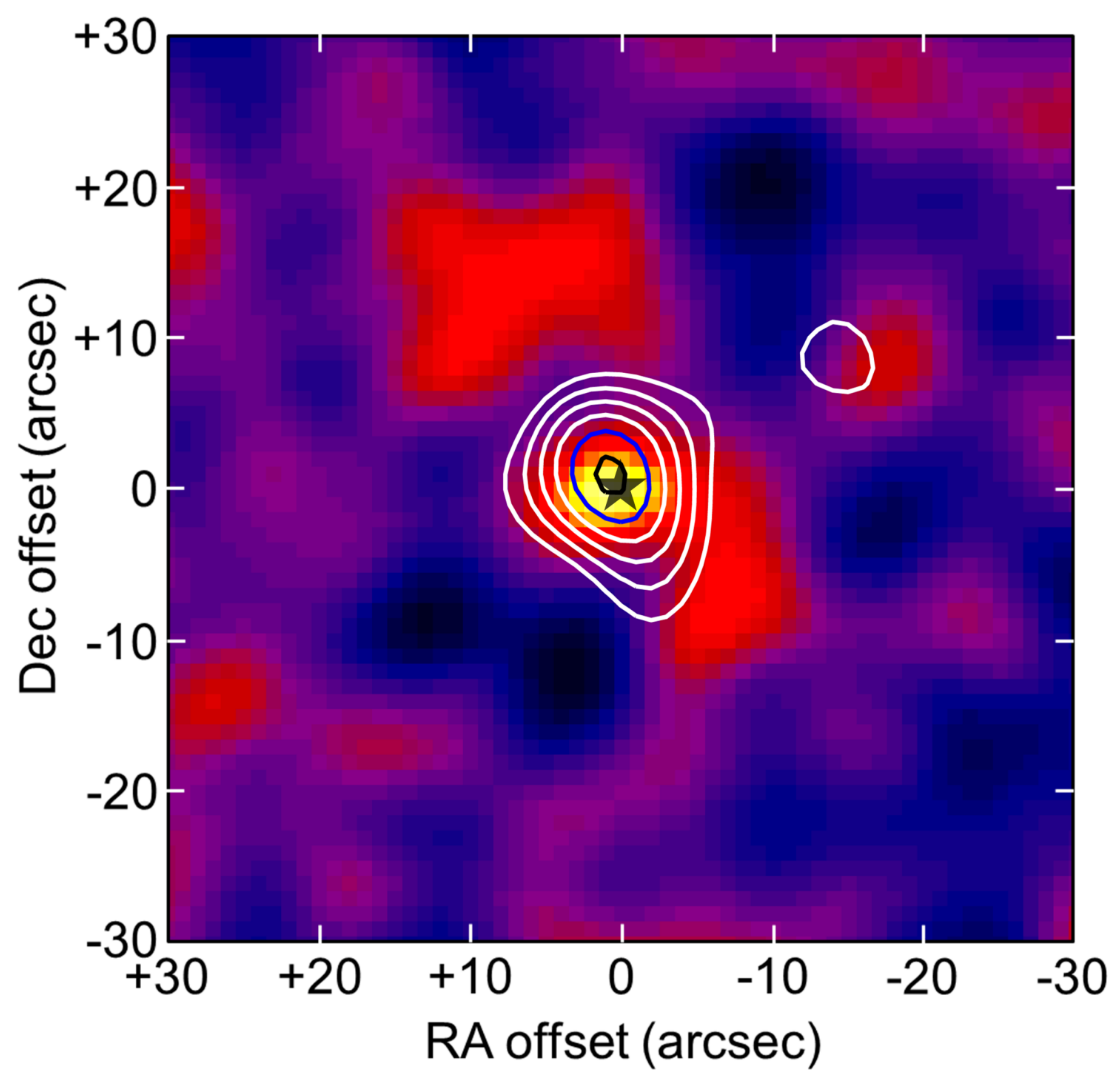}

\caption{The 450\,$\umu$m image from observations of HD 15257 with contours from the 850\,$\umu$m image overlaid. The contours and symbols are as described in 
Fig.~\ref{fig:hd8907}.}

\label{fig:hd15257}

\end{figure}

HD 15257 (12 Tri) is an F0 star at a distance of 50\,pc, with an estimated age of 1000\,Myr \citep{Chen2014}. The 850\,$\umu$m image reveals a well-detected emission peak with a 
flux density of 10.3 $\pm$ 1.2\,mJy (Fig.~\ref{fig:figureA3}c) well-centred on the star position. Emission is also detected at 450\,$\umu$m in the SONS survey with a flux of 56 
$\pm$ 11\,mJy ($\sim$ 5.1$\sigma$) as shown in Fig.~\ref{fig:hd15257}. The emission is unresolved at both wavelengths, and interpreting the results as being a disc about the 
star gives an upper limit to the disc radius from the 450\,$\umu$m image of 240\,au.

\vskip 1mm

\emph{IRAS} detected excess far-IR emission from the star at 60\,$\umu$m, but no other far-IR photometric measurements have been made. The SED is therefore sparsely sampled by 
only the \emph{IRAS} and SONS 450 and 850\,$\umu$m photometry leading to a poorly constrained value of $\beta$ in the range 0 to 1.9. The SED is modelled by a two-component fit, 
the warm element of which has a dust temperature 161\,K, based on a fit to the near/mid-IR photometric points from \emph{AKARI} \citep{Ishihara2010}, \emph{Spitzer} 
\citep{Chen2014}, and \emph{WISE} \citep{Wright2010}. The dominant cold component is at 53\,K, and the disc has an estimated dust mass of 0.11\,M$_{\earth}$.

\subsection{HD 15745}
\label{sec:hd15745}

HD 15745 is a young F2 star also in the $\beta$ Pictoris Moving Group, lying at a distance of 64\,pc with an estimated age of 23\,Myr \citep{Mamajek&Bell2014}. The 850\,$\umu$m 
image shows a well-detected, elongated structure with a peak flux of 8.4 $\pm$ 1.0\,mJy/beam (Fig.~\ref{fig:figureA3}d). The peak is offset to the south from the stellar 
position by $\sim$3\arcsec. Even though the star is at 64\,pc, the structure appears to be marginally resolved with an integrated flux, as measured in a 50\arcsec\ diameter 
aperture centred on the star, of 12.0 $\pm$ 1.4\,mJy. The fitted radial extent of the structure gives a deconvolved radius of 8.1\arcsec\ ($\sim$510\,au), suggesting the existence 
of a very large disc about the star at a PA of 164\degree.

\vskip 1mm

\begin{figure}
\includegraphics[width=85mm]{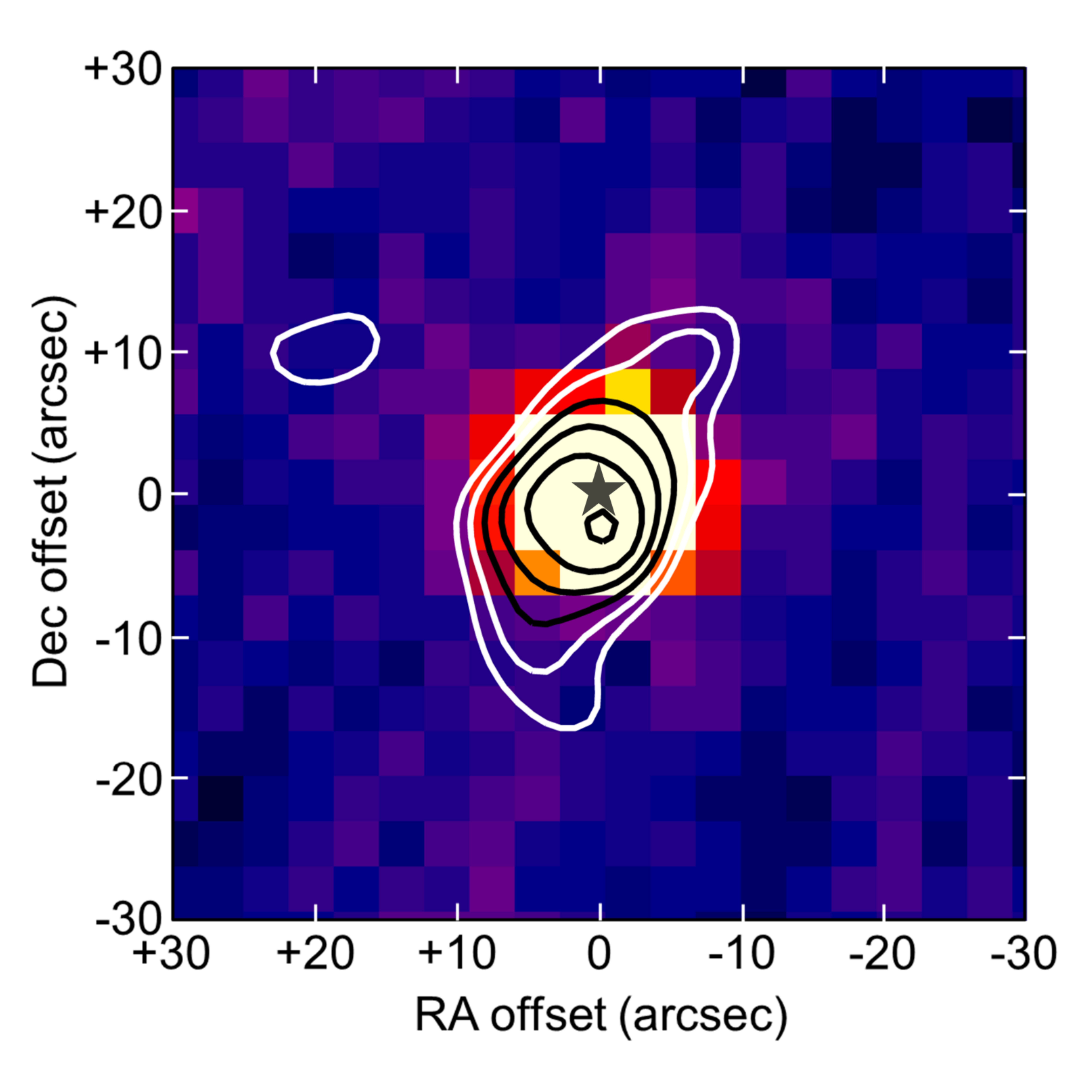}

\caption{\emph{Herschel}/PACS 160\,$\umu$m image of HD 15745 with contours from the SONS survey 850\,$\umu$m image overlaid. The contours are 3 and 4$\sigma$ (white) and then 5 to 
8$\sigma$ (black). The star symbol represents the position of the star with respect to the flux peak. The \emph{Herschel} image is taken from the \emph{Herschel} Science Archive.}

\label{fig:hd15745_herschel}
\end{figure}

In optical scattered light observations using the \emph{HST}/ACS coronograph, the disc appears asymmetric about the star at a PA of $\sim$22.5\degree, with fan-shaped emission 
extending to 480\,au in the region between position angles of 190 and 10\degree\ \citep{Kalas2007b,Schneider2014}. The scattered light emission seems to extend to a radius close 
to that seen in the SONS 850\,$\umu$m image ($\sim$8\arcsec; $\sim$510\,au) but the orientation of the disc in the submillimetre does not show the same asymmetry as in 
scattered light. \emph{Herschel}/PACS images at 70 and 160\,$\umu$m show that the disc appears to be well-centred on the stellar position but there is little evidence of the 
extension seen in the 850\,$\umu$m image (as shown in Fig.~\ref{fig:hd15745_herschel}).

\vskip 1mm

The SED is reasonably well-sampled in the mid-far IR from \emph{IRAS} \citep{Rhee2007}, \emph{Spitzer}/MIPS at 70\,$\umu$m \citep{Moor2011b}, and \emph{Herschel}/PACS (archival 
data) leading to a constraint on $\beta$ between 0 and 0.9. The model fit through the photometric points yields a dust temperature of 89\,K, leading to an estimated dust mass of 
0.13\,M$_{\earth}$. The relatively high derived dust temperature means that the modelled disc radius, assuming that the grains behave as blackbodies, lies close to the star at 
only 18\,au. Hence, there is a large discrepancy (factor of 28) in the modelled disc radius and that measured from the resolved image. This is further discussed in 
Section~\ref{sec:disc_morphology}.

\subsection{38 Arietis (HD 17093)}
\label{sec:hd17093}

HD 17093 (38 Ari) is a luminous A7 star at a distance of 36\,pc with an estimated age of 580\,Myr \citep{Vican2012}. The SONS survey 850\,$\umu$m image was first reported by 
\citet{Panic2013} and shows a well-detected peak of flux density 8.8 $\pm$ 0.9\,mJy, which is offset from the stellar position by 10\arcsec\ (Fig.~\ref{fig:figureA4}a). This offset 
is far in excess of what would be expected for a disc about the star, allowing for positional and statistical uncertainties in the measurement (see Section~\ref{sec:offsets}). It is 
therefore likely that the peak is due to a separate source, perhaps a background galaxy. This hypothesis is further supported by \emph{Herschel}/PACS archival images showing 
significant 100\,$\umu$m and 160\,$\umu$m emission at the same offset position to the SCUBA-2 850\,$\umu$m map \citep{Panic2013}. A fit to the far-IR photometric and submillimetre 
points (Fig.~\ref{fig:figureA4}a) suggests an object with a very cold ($\sim$14\,K) dust temperature. The conclusion is that there is no current evidence for a debris disc 
surrounding HD\,17093.

\subsection{Algol (HD 19356, $\beta$ Persei)}
\label{sec:algol}

HD 19356 (Algol) is a triple-star system with the primary of spectral type B8-V, lying at a distance of 28\,pc with an age estimated at 450\,Myr, although it could be as young 
as 300\,Myr \citep{Rhee2007} or as old as 570\,Myr \citep{Soderhjelm1980}. The bright primary (Algol/$\beta$ Per Aa1) is eclipsed by the fainter secondary (Aa2; separation of 
0.062\,au) on a period of 2.9\,days. The third star in the system (Ab) lies at a distance of 2.7\,au from the primary. The SONS survey 850\,$\umu$m image shows well-detected 
emission with a peak flux 6.4 $\pm$ 0.9\,mJy/beam, centred on the stellar position (Fig.~\ref{fig:figureA4}b). Although the morphology of the emission looks somewhat extended at 
850\,$\umu$m, the integrated flux, determined in a 30\arcsec\ diameter aperture, is in good agreement to the peak value, suggesting the structure is unresolved. Given that all 
three stars lie within an area subtended by the telescope beam, it is impossible to ascertain which star is responsible for the excess flux at submillimetre wavelengths.

\vskip 1mm

Although \emph{Spitzer}/MIPS detected emission from Algol at 24 and 70\,$\umu$m \citep{Su2006}, a modified blackbody fit to these points and the new SONS survey 850\,$\umu$m 
flux is not possible. Multiple discs may be present, with an ultra-cold component dominating the emission at 850\,$\umu$m. The SED model fit through the 850\,$\umu$m point 
(Fig.~\ref{fig:figureA4}b) is only indicative, as there are no other constraining points. Indeed, a cold disc component with a temperature of around 27\,K could exist (noting 
a large uncertainty of $\pm$ 17\,K in this temperature estimation). The estimated dust mass based on the 850\,$\umu$m flux density is 0.045\,M$_{\earth}$. Algol Aa1 is 
$\sim$100 times more luminous than the Sun. Combined with a relatively low derived dust temperature, this high luminosity suggests the radius of a disc (assuming the grains 
behave as blackbodies) could be as large as $\sim$1100\,au (again, with large uncertainties). This value compares to only 210\,au based on the upper limit to the disc radius 
from the 850\,$\umu$m image.

\vskip 1mm

The emission from the Algol system is known to be highly variable and of very high brightness temperature ($\sim$10$^9$\,K) \citep{Lestrade1988}. More specifically, it has 
been identified as gyro-synchrotron emission from mildly relativistic electrons accelerated by magnetic reconnections in its stellar chromosphere (as also known for the Sun 
and other stars). The spectrum of such emission peaks at a few centimetres, and may extend into the millimetre wavelength domain \citep{Dulk1985}. The flux observed at 
850\,$\umu$m therefore might be such a tail, rather than thermal emission from a debris disc. The recently measured variability of the millimetre flux density observed at the 
SMA strengthens this interpretation (Wyatt/Wilner, \emph{priv. comm.}).

\vskip 1mm

There are, however, also a number of possible scenarios to explain the presence of cool debris around the star. First, a dust ring may surround at least one of the stars but 
companion perturbations in the system may cause it to be disrupted, hampering the formation of planets but allowing for the formation of enough planetesimals to generate 
significant debris material \citep[e.g.][]{Thebault2016}. Second, a significant part of the emission may be related to mass loss from the secondary (and/or tertiary) to the 
primary, resulting in the presence of circumstellar material \citep{Miller2006}. Finally, there could be a mass outflow from the mass-gaining star, so the system loses angular 
momentum (the ``non-conservative'' problem) with the outflow, which then forms a shock as it encounters the interstellar medium \citep{Deschamps2015}. High angular resolution 
observations are needed to resolve the proposed structures and give further insight on the true nature of any ``debris'' surrounding Algol. Hence, although it may be possible 
for a debris disc to exist around Algol, evidence from longer wavelengths points to the emission at 850\,$\umu$m being due to radio variability rather than a debris disc.

\subsection{HD 21997}
\label{sec:hd21997}

HD 21997 is a young A3 star in the Columba Moving Group, at a distance of 72\,pc and an estimated age of 30\,Myr \citep{Torres2008} with an uncertainty for the group in the 
range 20 to 50\,Myr \citep{Marois2010}. At 850\,$\umu$m, the emission appears slightly elongated to the south at a PA of 25\degree\ with a peak flux of 7.9 $\pm$ 1.1\,mJy/beam 
at the stellar position (Fig.~\ref{fig:figureA4}c). The peak flux is also consistent with that measured using SCUBA photometry of 8.3 $\pm$ 2.3\,mJy 
\citep{Williams&Andrews2006}. The integrated flux density, measured over a 40\arcsec\ diameter aperture, is 10.7 $\pm$ 1.5\,mJy, suggesting that, even at a distance of 72\,pc, 
the structure is marginally resolved. This extent, however, depends largely on the significance of the feature to the south within the 3$\sigma$ contour. The fitted radial 
extent taking in this extension suggests that HD 21997 could be surrounded by an enormous disc of radius 810\,au at a PA of 27\degree.

\vskip 1mm

\emph{Herschel}/PACS and SPIRE observations resolved the disc, identifying a structure with a radius of at least 2.8\arcsec ($\sim$200\,au) at a PA of $\sim$25\degree\ 
\citep{Moor2015}. The disc has also been resolved by ALMA at 886\,$\umu$m revealing an inclined ring-like structure of radius $\sim$2.1\arcsec\ ($\sim$150\,au) at a PA of 
22.5\degree\ \citep{Moor2013}. Modelling of the morphology of the ring suggests inner and outer radii of $\sim$55 and $\sim$150\,au, respectively. The measured flux from the 
ALMA image is 2.7 $\pm$ 0.3\,mJy, considerably lower than previously measured at 850\,$\umu$m, suggesting that some of the emission has been resolved out by the interferometric 
observations, and leading to a significant underestimate of the amount of dust present in the system. Such a hypothesis would be consistent with detectable emission extending 
beyond the outer radius of $\sim$150\,au derived from ALMA, as indicated in the \emph{Herschel} and 850\,$\umu$m SONS observations (perhaps extending to $\sim$500\,au or more).

\vskip 1mm

HD\,21997 is also one of the few systems with an age greater than 10\,Myr that contains a detectable amount of cold CO gas \citep{Moor2011a}. The SED contains photometric points 
in the mid-far IR from \emph{Spitzer}/MIPS and \emph{Herschel}/PACS and SPIRE. The well-constrained fit through the photometric points, gives a dust temperature of 64\,K, and a 
$\beta$ of between 0.4 to 1.4. The estimated dust mass of 0.22\,M$_{\earth}$ is considerably higher than the value of 0.09\,M$_{\earth}$ reported based on the ALMA observations, 
due to the aforementioned larger flux.

\subsection{$\epsilon$ Eridani (HD 22049)}
\label{sec:epseri}

\begin{figure}
\includegraphics[width=85mm]{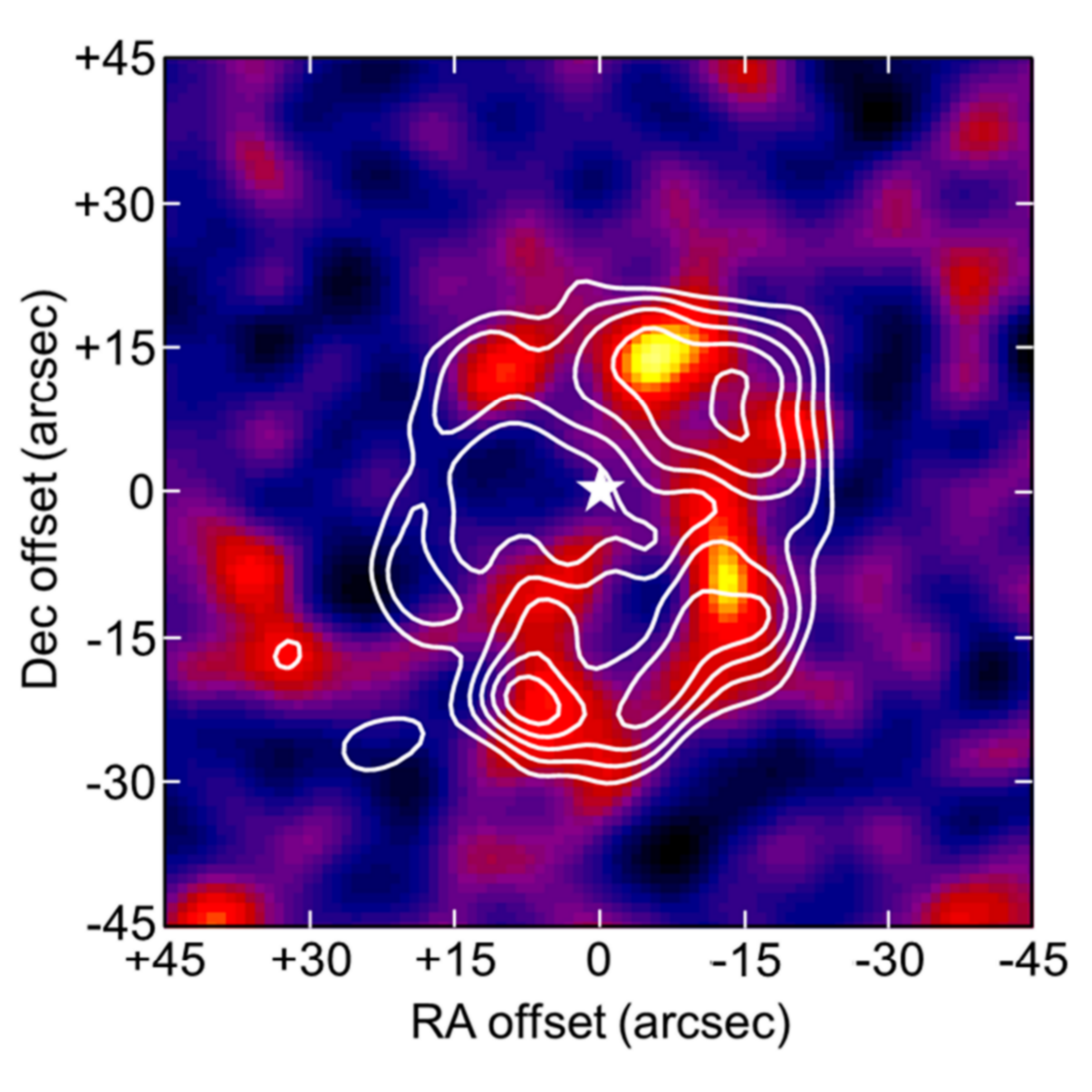}

\caption{The 450\,$\umu$m image from observations of $\epsilon$ Eridani (HD 22049) with contours from the 850\,$\umu$m image overlaid. The contours and symbols are as described in
Fig.~\ref{fig:hd8907}.}

\label{fig:epseri}

\end{figure}

$\epsilon$ Eridani (HD 22049) is a nearby K2 star at a distance of 3.2\,pc with an estimated age of 850\,Myr within a range spanning 800\,Myr \citep{DiFolco2004, 
Mamajek&Hillenbrand2008} and 1.4\,Gyr \citep{Bonfanti2015}. The SONS image at 850\,$\umu$m reveals a well-resolved, largely circularly-symmetric, ring structure with considerably 
less emission at the stellar position (Fig.~\ref{fig:figureA4}d). There appears to be significant sub-structure in the ring with the brightest clump to the south-east having a flux 
of 5.3 $\pm$ 0.7\,mJy/beam. The integrated flux, measured within a 70\arcsec\ diameter aperture centred on the star, is 31.3 $\pm$ 1.9\,mJy. Both the peak clump flux and the total 
integrated flux agree well with the previous SCUBA observations \citep{Greaves1998, Greaves2005}, although the latter is some 15\% lower, but within the measured uncertainties. 
Interpreting the structure as a dust ring about the star, gives a deconvolved radius, based on the Gaussian fit to the 850\,$\umu$m image, of 20.8\arcsec\ ($\sim$67\,au). The 
inclination angle to the plane of the sky is estimated to be 26\degree\, also consistent with the SCUBA results. The image at 450\,$\umu$m only shows clumps with low S/N 
($\sim$3--5$\sigma$), but nevertheless suggests a ring structure that is not inconsistent with the results at 850\,$\umu$m (Fig.~\ref{fig:epseri}). The integrated flux, within 
a 70\arcsec\ diameter aperture, is 181 $\pm$ 15\,mJy, which compares to 250 $\pm$ 20\,mJy determined by \citet{Greaves2005}. Significantly more integration time is ideally needed to 
improve upon the 450\,$\umu$m results.

\vskip 1mm

Given its age and close proximity to the Sun in both distance and spectral type, the $\epsilon$ Eridani dust ring has been widely studied at multiple wavelengths. The original 
850\,$\umu$m SCUBA image from 1998 first revealed the ring-like structure, peaking at a radius of 60\,au from the star and with an estimated dust mass of at least 
0.01\,M$_{\earth}$ \citep{Greaves1998}. It was concluded that the ring structure was probably akin to the Edgeworth-Kuiper belt in our Solar System with a central region 
partially cleared of dust by forming planetesimals. Further deep imaging with SCUBA over the period 1999 -- 2004 confirmed the dust ring morphology as well as producing the 
first image of the structure at 450\,$\umu$m \citep{Greaves2005}. The latter work also presented tentative evidence that two clumps in the ring were following the stellar 
motion (i.e. were not background objects) and furthermore showed evidence of counter-clockwise rotation. It was suggested that the structure in the ring could be caused by 
perturbations of a planet orbiting the star at a few tens of AU, and led to the speculation that the $\epsilon$ Eridani system could resemble an early version of our Solar 
System.

\vskip 1mm

Far-infrared observations with \emph{Spitzer} indicated the likely presence of two unresolved warm, inner dust rings, in addition to a wider, icy outer belt \citep{Backman2009}. The 
inner belt lies at a radius of around 3\,au, very similar to the distance of the Asteroid belt in the Solar System. There is also at least one candidate Jupiter-mass planet 
($\epsilon$ Eri b) in the system believed to lie just outside of this belt at a orbital radius of 3.5\,au, discovered using the radial-velocity (RV) technique \citep{Hatzes2000}. 
Astrometric observations also provide evidence for the existence of a planet and allow its inclination to be determined \citep{Benedict2006,Reffert2011}. Although the planet's orbit 
appears to be inclined at a similar angle ($\sim$30\degree) to the disc, the inclination is in a very different direction, meaning the disc and planetary orbit are not likely 
coplanar. The planet has also proven somewhat controversial as extensive followup RV observations have so far failed to confirm its existence \citep{Zechmeister2011}. 
\emph{Herschel} observations at 160\,$\umu$m to 350\,$\umu$m resolved two belts of debris emission. The inner belt has a radius of 12 -- 16\,au, whilst the outer cold belt orbited 
the star at a radius of 54 -- 68\,au \cite{Greaves2014}. There has also been considerable speculation about the existence of a planet orbiting at a distance close to the outer dust 
belt. The brightness asymmetry between the north and south ends of the belt from the \emph{Herschel} observations indicates a pericentre glow that could be attributable to such a 
planet.

\vskip 1mm

The clumpy structure of the observed disc was presented as evidence that dynamical interaction between an unseen planetary companion and the debris belt 
\citep{Quillen&Thorndike2002}. Millimetre-wave observations with MAMBO-2 on the IRAM 30\,m telescope at 1.2\,mm also confirmed the clumpy structure seen in the submillimetre images 
\citep{Lestrade&Thilliez2015}. However, other observations, most notably millimetre-wave observations using the bolometer array SIMBA on SEST \citep{Schutz2004} and the SMA 
interferometer \citep{MacGregor2015} did not confirm the presence of significant substructure within the ring. More recently, the AzTEC camera on the 50\,m Large Millimeter 
Telescope imaged the $\epsilon$ Eridani dust belt reaching an RMS noise level of 0.12\,mJy at 1.1\,mm \citep{Chavez2016}. The measured radius for the outer belt of 64\,au agrees 
well with other measurements, and although there are some inhomogeneities, the observed structure indicates a morphology that is essentially smooth. The deep LMT image also 
highlights the presence of numerous point-like sources, with the likelihood, based on the near 20 years of observations (and noting the large stellar proper motion of 1\arcsec 
yr$^{-1}$), that at least one of the original submillimetre clumps within the ring is likely to be a background source. The northern arc of the ring has also been imaged by 
ALMA at a wavelength of 1.3\,mm, revealing tentative evidence of the presence of a clumpy structure in the ring \citep{Booth2017}. The ring is very narrow at just 11 -- 13\,au, 
making the fractional disc width comparable with the Edgeworth-Kuiper belt in our Solar System.

\vskip 1mm

The SED is well-sampled in the far-IR through to the millimetre. There is somewhat of a discrepancy in the total disc flux between the recent millimetre-wave observations 
(IRAM/MAMBO-2 and the LMT/AzTec) and a fit to the \emph{Spitzer}, \emph{Herschel}, SCUBA, SCUBA-2 and the original IRAM photometry as shown in the SED in 
Fig.~\ref{fig:figureA4}d. The reason for this is unclear, but may be very dependent on the precision of the aperture photometry with respect to the ring and the presence of 
background sources. Modelling of the SED suggests a two component fit to the overall IR excess consistent with the results from \emph{Spitzer} and the resolved imaging from 
\emph{Herschel}. The fit to the warmer, inner component suggests the existence of an ``asteroid belt'' at an orbital radius of $\sim$14\,au with a characteristic temperature 
of 133\,K. The colder component is the more familiar outer ring from the submillimetre observations at a radius of $\sim$ 60\,au, modelled as a disc/belt with a dust 
temperature 44\,K, and a $\beta$ value in the range of 0.6 to 1.0, with an estimated dust mass of only 0.002\,M$_{\earth}$ calculated from the 850\,$\umu$m flux. The derived dust 
temperature for the cold disc is also in good agreement with previous estimates \citep{Greaves1998, Greaves2005}. The equivalent disc radius, assuming blackbody emission, is 
23\,au, this being significantly lower than the value measured from the 850\,$\umu$m image of 67\,au.

\subsection{HD 22179}
\label{sec:hd22179}

HD 22179 is a young G5 star, at a distance of 16\,pc and an estimated age of only 16\,Myr \citep{Carpenter2009}. At 850\,$\umu$m, there is a bright emission peak with a flux 
density of 7.0 $\pm$ 1.3\,mJy, located 18\arcsec\ from the stellar position to the south west (Fig.~\ref{fig:figureA5}a). Given the large offset it is unlikely that the peak is 
associated with a putative disc around the star, and is probably a background object. At far-IR wavelengths \emph{Spitzer} detected excess flux from the star at 70\,$\umu$m 
(just over 3$\sigma$), but no detection was made in the longer wavelength 160\,$\umu$m channel \citep{Hillenbrand2008}. Photometry shortward of 24\,$\umu$m indicated clear 
evidence of a flux excess, but it is also possible that \emph{Spitzer} detected the same offset peak seen in the SONS 850\,$\umu$m image. Hence, there is some uncertainty as 
to whether a debris disc does exist around HD\,22179.

\subsection{HD 25457}
\label{sec:hd25457}

HD 25457 is a F6 star in the AB Doradus Moving Group, at a distance of 19\,pc with a somewhat controversial age but generally accepted to be very similar to the Pleiades at 
130 +/- 20 Myr \citep{Barrado2004}. The SONS image at 850\,$\umu$m shows an unresolved peak of flux density 6.4 $\pm$ 1.4\,mJy, offset from the stellar position by 2\arcsec 
(Fig.~\ref{fig:figureA5}b). Interpreting the peak as an unresolved disc associated with the star, gives an upper limit for the disc radius of 140\,au. The 850\,$\umu$m image 
shows some unevenness, with ridges of excess flux towards the east and corresponding negative features in the west.

\vskip 1mm

\emph{Spitzer} detected excess flux from the star at wavelengths between 3.6\,$\umu$m and 160\,$\umu$m \citep{Hillenbrand2008}. Although there have been several observations 
of the star in the submillimetre and millimetre, the SONS 850\,$\umu$m image is the only clear detection of a disc at such long wavelengths. Modelling of the SED suggests a two 
component fit to the overall IR excess consistent with the results from \emph{Spitzer} and the imaging from SONS. The fit to the warmer component, dominated by the near-mid IR 
photometry, gives a characteristic temperature of 124\,K. The colder component has a dust temperature of 50\,K, a $\beta$ value only loosely defined as $>$0.1, and an 
estimated dust mass of 0.011\,M$_{\earth}$ derived from the 850\,$\umu$m flux.

\subsection{HD 35841}
\label{sec:hd35841}

HD 35841 is a young F3 star, also in the Columba Moving Group, at a distance of 96\,pc with an estimated age of 30\,Myr. The SONS survey image at 850\,$\umu$m reveals several 
peaks in the vicinity of the star, including a faint 4$\sigma$ peak offset from the star by $\sim$5\arcsec\ to the south, with a flux density of 3.5 $\pm$ 0.8\,mJy 
(Fig.~\ref{fig:figureA5}c). To the south-west, some 23\arcsec\ offset, is another peak with a flux of 3.1 $\pm$ 0.8\,mJy and a low-level ($\sim$2.5$\sigma$) ridge running 
between the two that extends to the north-east of the star. If the peak just to the south is an unresolved disc associated with the star, then the upper limit for the disc 
radius is 725\,au.

\vskip 1mm

The disc has been detected in scattered light via \emph{HST}/NICMOS observations that show a very compact, nearly edge-on disc with a radius of 1.5\arcsec\ ($\sim$144\,au) from 
the star \citep{Soummer2014}. The two lobes are not diametrically aligned with position angles of $\sim$180 and $\sim$335\degree. Only \emph{Spitzer}/MIPS photometry at 24 and 
70\,$\umu$m \citep{Chen2014} contributes to an otherwise poorly constrained SED. A fit to the SED, assuming the peak detected at 850\,$\umu$m is representative of a disc about 
the star, gives a dust temperature of 71\,K, a $\beta$ value very loosely constrained in the range 0 to 2.7, and an estimated dust mass of 0.11\,M$_{\earth}$.

\subsection{HD 38858}
\label{sec:hd38858}

HD 38858 is a nearby (15\,pc) Sun-like star of spectral type G4, hosting a super-Earth planet within 1\,AU, with a highly uncertain stellar age usually adopted as being 
4700\,Myr \citep{Beichman2006}, but likely to be in the range of 3200\,Myr \citep{Chen2014} to 6200\,Myr \citep{Vican2012}. The SONS observation at 850\,$\umu$m 
(Fig.~\ref{fig:figureA5}d) reveals a large extended structure with an integrated flux density, measured in a 60\arcsec\ diameter aperture, of 11.5 $\pm$ 1.3\,mJy 
\citep{Kennedy2015}. In terms of the interpretation of the image, the peak is clearly offset from the star position by 9\arcsec\ to the east, although there is still 
significant emission at the star position itself ($\sim$5\,mJy/beam). One possibility is that the extended structure to the south-east is a separate background source.

\vskip 1mm

\emph{Herschel}/PACS observations at 70\,$\umu$m reveal a disc of radius 7.5\arcsec\ ($\sim$110\,au) at a PA of 67\degree, which is well-centred at the stellar position 
\citep{Kennedy2015}, and similar to the results reported by \emph{Spitzer}/MIPS \citep{Krist2012}. Observations at 160\,$\umu$m, however, reveal a separate source to the south, 
coincident with the peak flux of the 850\,$\umu$m extension. Hence, there is a strong likelihood that this source is a background object (perhaps a high redshift galaxy given 
the absence of such a source at 70\,$\umu$m), and therefore the remainder of the discussion on this target assumes this to be the case. Finally, the image is also likely 
``contaminated'' at some level by the Orion Complex background emission to the east of the star \citep{Kennedy2015}.

\vskip 1mm

By constructing models of the HD 38858 disc it is possible to estimate the likely contribution to the true disc flux at 7.5 $\pm$ 1.5\,mJy, taking into account the background 
contamination and assuming the southern peak is not associated with a disc \citep{Kennedy2015}. This interpretation still assumes that the flux peak is at a distance of 
9\arcsec\ from the star. With this single long wavelength point, the disc spectrum is now close to a pure blackbody with $\beta$ around zero. A fit to the SED gives a 
characteristic dust temperature of 50\,K and an estimated dust mass of 0.0086\,M$_{\earth}$.

\vskip 1mm

The radius of the disc, derived from the SED fit and assuming blackbody grain properties, is only 28\,au. This value is significantly smaller than the measured deconvolved 
radius of the disc of 12.7\arcsec\ ($\sim$190\,au) at a PA of 75\degree (determined after subtracting a point-source at the location of the southern peak). Such a result is 
surprising given the requirement of a nearly pure blackbody spectrum fit to the SED (as described above), which would imply that the blackbody and resolved disc sizes should be 
in agreement, i.e., the disc material should lie at a single radius from the star. Blackbody spectra have been seen for other discs at larger radii than their blackbody 
temperatures would suggest. An example is AU Mic \citep{Matthews2015} where it is likely that the submillimetre observations trace the parent-body distribution (which may have 
blackbody properties), whilst the far-IR data sample a halo of smaller grains from highly eccentric orbits due to radiation pressure. In summary, the SONS 850\,$\umu$m image of 
HD 38858 is interpreted as a resolved disc, with a peak flux offset from the star position by 9\arcsec\, and with the presence of a background object contaminating the observed 
structure to the south.

\subsection{56 Aurigae (HD 48682)}
\label{sec:hd48682}

HD 48682 (56 Aur) is an old F9 star, at a distance of 17\,pc with an estimated age of around 6000\,Myr \citep{Barry1988}, though could be as young as 3200\,Myr 
\citep{Holmberg2009} or as old as 8900\,Myr \citep{Beichman2006}. The field at 850\,$\umu$m shows faint multiple peaks, of which the nearest to the star has a flux of 2.7 $\pm$ 
0.6\,mJy/beam (Fig.~\ref{fig:figureA6}a). The peak is offset by 6.5\arcsec\ from the star, significantly greater than the maximum offset of 3.7\arcsec\ expected due to 
statistical and positional uncertainties (Section~\ref{sec:offsets}). The structure near the star could also be marginally extended compared to the beam, mainly in the east-west 
direction, and has an integrated flux density (in a 30\arcsec\ diameter aperture) of 3.9 $\pm$ 0.8\,mJy. Interpreting the structure nearest the star to be a disc, the radial 
fitting gives a deconvolved disc radius of 11\arcsec\ ($\sim$184\,au) at a PA of 94\degree, which is larger than the estimated radius of 57\,au based on the assumption of pure 
blackbody emission from grains (see discussion in Section~\ref{sec:spectral_slopes}).

\vskip 1mm

Observations with \emph{Herschel}/PACS at 100\,$\umu$m reveal an elongated disc with a measured deconvolved radius 16.4\arcsec\ ($\sim$280\,au) at a PA of 107\degree\ 
\citep{Eiroa2013, Pawellek2014} approximately 50\% larger than the estimate from the SONS 850\,$\umu$m image. The peak to the SW is not considered to be associated with this 
main disc. A fit to the SED, mainly sampled by \emph{Herschel} photometry, and including the 850\,$\umu$m SONS flux, gives a dust temperature of 43\,K and a $\beta$ in the 
range 0.7 to 1.8, leading to an estimated dust mass of 0.0062\,M$_{\earth}$.

\subsection{HD 61005 (``The Moth'')}
\label{sec:hd61005}

HD 61005 is a G8 star and a possible member of the Argus Association, at a distance of 35\,pc with an estimated age of 40\,Myr \citep{Desidera2011}, although other estimates 
include 100\,Myr \citep{Hillenbrand2008} and in the range of range of 72\,Myr to 186\,Myr \citep{Desidera2011}. The 850\,$\umu$m image shows emission peaking just offset from 
the stellar position and extending somewhat to the south-east, with a flux density of 13.5 $\pm$ 2.0\,mJy (Fig.~\ref{fig:figureA6}b). The elongation is not significant 
compared to the beam size, however, and interpreting the structure as a disc about the star gives an upper limit to the disc radius of 265\,au.

\vskip 1mm

The disc has been studied in detail at many wavelengths and results include the fan-like scattered light observation that led to the nickname of ``The Moth'' 
\citep{Hines2007}. Observations with the SMA first resolved the disc at millimetre wavelengths, revealing a double-peaked structure, viewed close to edge-on, with a measured 
major axis radius of 2.2\arcsec\ ($\sim$80\,au) at a PA of 71\degree\ \citep{Ricarte2013}. ALMA observations at 1.3\,mm also resolve the disc into a highly-inclined 
($i$\,$\sim\,$85\degree) structure with the emission peaking at a radius of 1.5\arcsec\ ($\sim$55\,au) from the star, with the disc having a PA of 71\degree\ 
\citep{Olofsson2016}. Although the disc was not detected at 450\,$\umu$m in the SONS survey, the SED is further well-constrained in the mid-far IR with photometry from both 
\emph{Spitzer}/MIPS \citep{Chen2014} and \emph{Herschel}/PACS \citep{Morales2016}, and in the millimetre by photometry at 9\,mm from the VLA \citep{MacGregor2016a}. The fit to 
the SED gives a characteristic dust temperature of 61\,K, and a $\beta$ value tightly constrained within the range 0.4 -- 0.7. The calculated dust mass from the 850\,$\umu$m 
flux is 0.069\,M$_{\earth}$.

\subsection{GJ 322}
\label{sec:gj322}

\begin{figure}
\includegraphics[width=85mm]{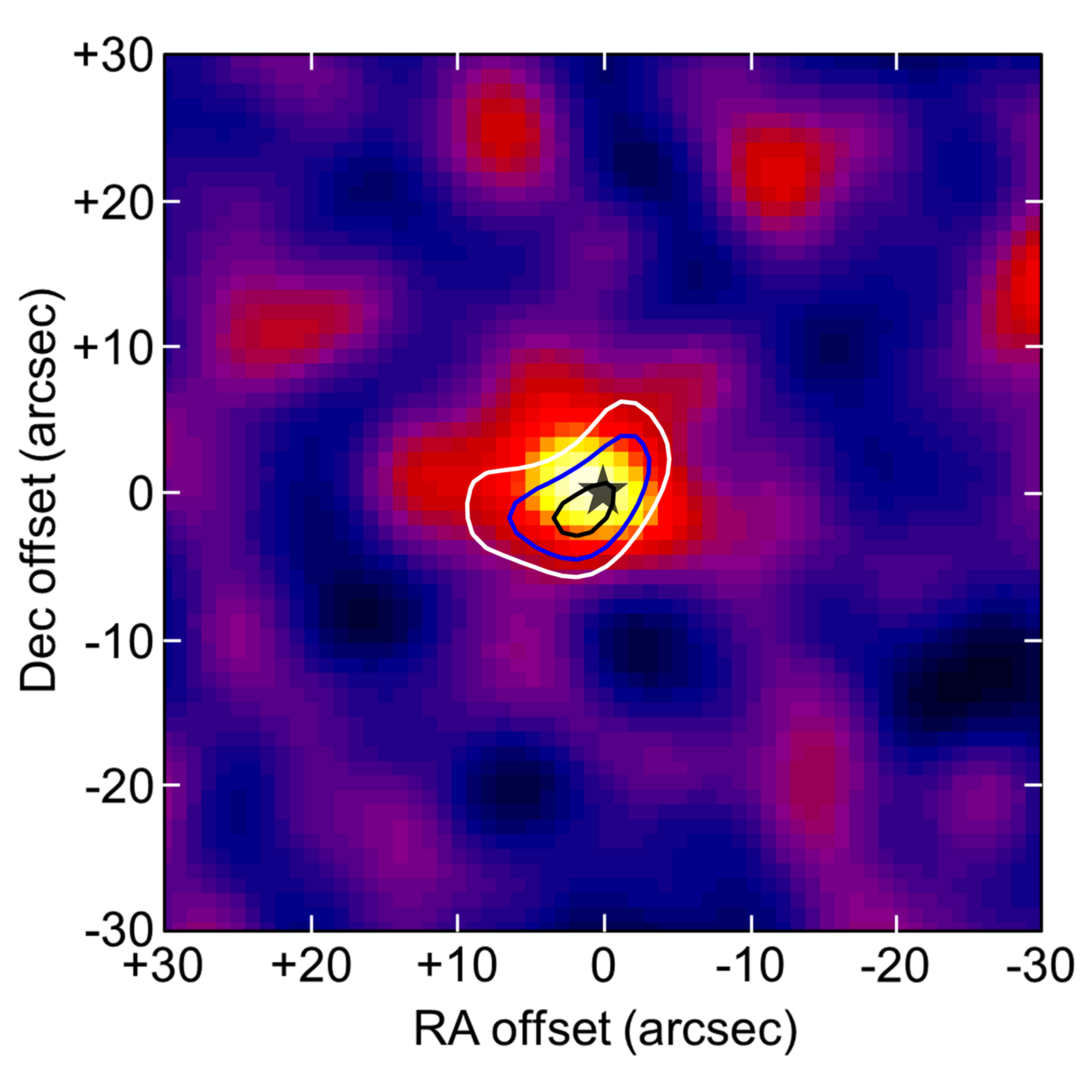}

\caption{The 450\,$\umu$m image from observations of GJ 322 with contours from the 850\,$\umu$m image overlaid. The contours and symbols are as described in
Fig.~\ref{fig:hd8907}.}

\label{fig:gj322}
\end{figure}

GJ 322 is a K5 star in the Ursa Major Moving Group, at a distance of 16.5\,pc with an estimated age of 490 $\pm$ 100\,Myr \citep{Jones2015}. The SONS survey results show an 
unresolved peak, well-centred on the stellar position, at both 450\,$\umu$m and 850\,$\umu$m with flux densities of 57 $\pm$ 11\,mJy and 7.3 $\pm$ 1.4\,mJy, respectively 
(Fig.~\ref{fig:figureA6}c and Fig.~\ref{fig:gj322}). Interpreting the peak to be a disc about the star gives an upper limit to disc radius of 85\,au based on the 450\,$\umu$m 
image. The SED, however, is poorly constrained in the mid-far IR with only \emph{Spitzer}/MIPS photometry at 70\,$\umu$m \citep{Chen2014}, which together with the steep spectral 
slope between 450 and 850\,$\umu$m leads to unconstrained values of both $\lambda_0$ and $\beta$. A fit to the SED gives a characteristic dust temperature of 24\,K (but with a 
large error of $\pm$\,10\,K), and a derived dust mass of 0.021\,M$_{\earth}$. The disc radius from the SED fit is also poorly estimated at 43 $\pm$ 24\,au, assuming emission 
from pure blackbody grains.

\subsection{63 Cancri (HD 76582)}
\label{sec:hd76582}

\begin{figure}
\includegraphics[width=85mm]{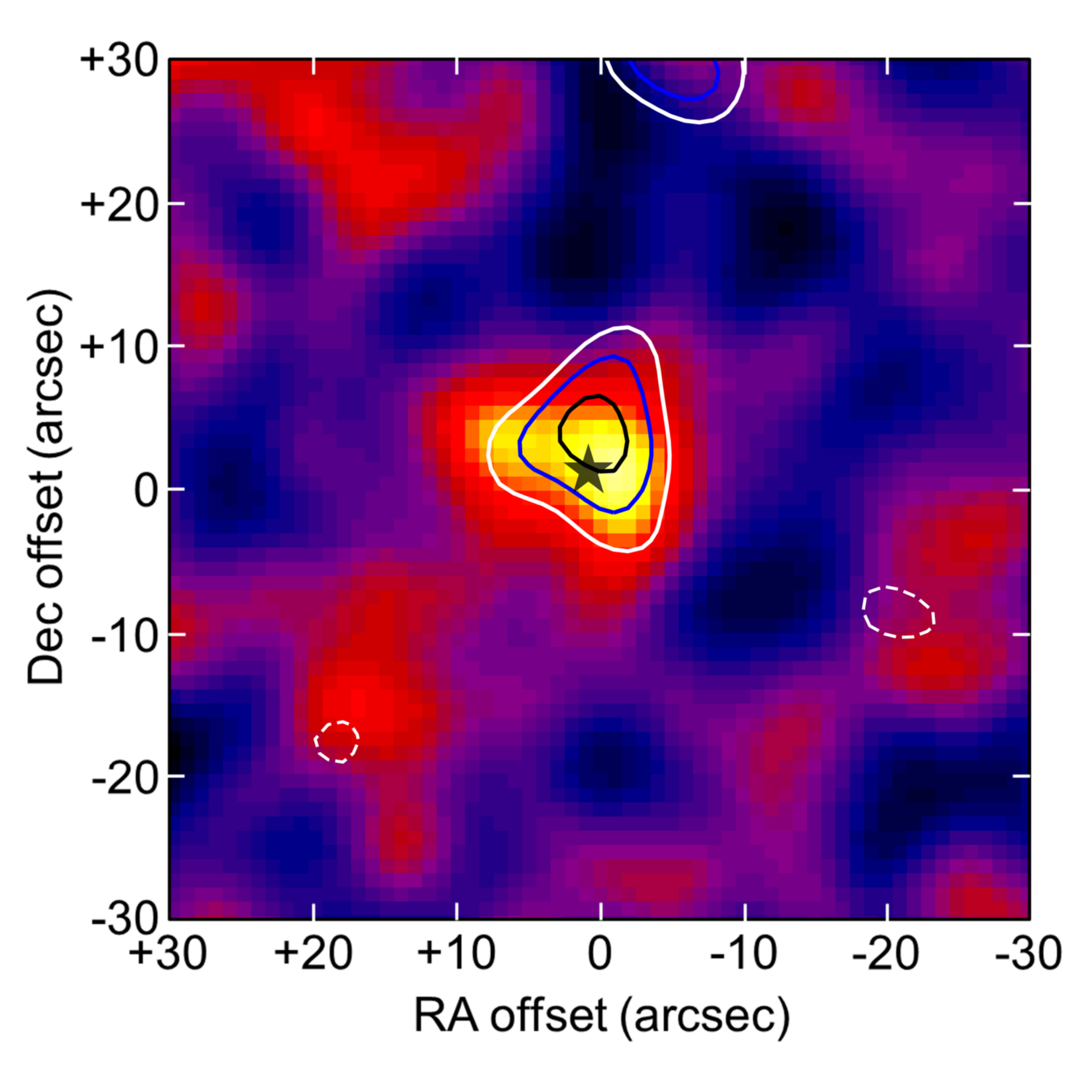}

\caption{The 450\,$\umu$m image from observations of HD 76582 (63 Cnc) with contours from the 850\,$\umu$m image overlaid. The contours and symbols are as described in
Fig.~\ref{fig:hd8907}.}

\label{fig:hd76582}
\end{figure}

HD 76582 (63 Cnc) is a moderately old F0 star, at a distance of 46\,pc with an estimated age of 540\,Myr \citep{Moor2006} and a likely age in the range 300\,Myr to 2130\,Myr 
\citep{Zorec+Royer2012}. The SONS survey observations show unresolved emission at both 450 and 850\,$\umu$m at the stellar position, with peak fluxes of 89 $\pm$ 17 and 5.7 
$\pm$ 1.0\,mJy/beam, respectively (Fig.~\ref{fig:figureA6}d and Fig.~\ref{fig:hd76582}). The 450\,$\umu$m flux estimate is based on a re-analyis of the data, and is 
slightly lower than previously published \citep{Marshall2016}. The 450\,$\umu$m image looks slightly extended in a roughly east-west direction, but aperture photometry reveals 
an integrated flux density that is not significant higher than the peak value. Based on the 450\,$\umu$m image, and interpreting the structure as a disc about the star, gives an 
upper limit to the radius of the disc of 230\,au.

\vskip 1mm

The disc was originally detected by \emph{IRAS} and has been well-characterised in the mid-far infrared \citep{Zuckerman&Song2004,Moor2006}. The disc has also been resolved by 
\emph{Herschel}/PACS, giving a radius of 5.9\arcsec\ ($\sim$ 271\,au) at a PA of 103\degree\ for the 100\,$\umu$m image, and 8.5\arcsec\ ($\sim$ 390\,au) at a PA of 115\degree 
at 160\,$\umu$m \citep{Marshall2016}. The disc is inclined at 64\degree\ to the plane of the sky \citep{Marshall2016}. Based on these measurements, it is possible that the 
disc might be marginally resolved at 450\,$\umu$m (even though the integrated flux suggests otherwise). At least a two-component, modified, blackbody model is needed to fit 
the photometric points from the near-IR to the submillimetre in the SED. The inner component, based on a fit to the near-mid IR photometry, has an estimated dust temperature 
of 157 $\pm$ 31\,K. The relative sparse coverage at far-IR and submillimetre wavelengths, coupled with the steep spectral slope between 450 and 850\,$\umu$m, means that both 
$\lambda_0$ and $\beta$ are poorly constrained by the modelling. A fit to the long wavelength photometric points gives a characteristic dust temperature of 52\,K, and a lower 
limit on $\beta$ of 1.1. The estimated dust mass from the 850\,$\umu$m flux is 0.057\,M$_{\earth}$. \citealt{Marshall2016} also suggest that there could actually be three 
distinct components to the disc, a warm inner element that fits the near-mid IR photometry (radius of $\sim$20\,au) and two annuli for the far-IR and submillimetre points 
(with radii of $\sim$80\,au and $\sim$270\,au).

\subsection{HD 84870}
\label{sec:hd84870}

HD 84870 is a luminous A3 star at a distance of 88\,pc with an estimated age of 100\,Myr \citep{Rhee2007}. At 850\,$\umu$m, the emission peaks on the stellar
position with a flux density of 6.2 $\pm$ 1.0\,mJy (Fig.~\ref{fig:figureA7}a). There are a couple of lower S/N peaks also in the vicinity which are most likely background
sources. Interpreting the emission as a disc about the star gives an upper limit to the disc radius from the 850\,$\umu$m image of 660\,au. 

\vskip 1mm

The infrared excess was originally detected by \emph{IRAS} \citep{Rhee2007}. The disc has recently been resolved by \emph{Herschel}/PACS revealing a ring-like structure in the 
star-subtracted 100\,$\umu$m image \citep{Vican2016}. Modelling of the disc indicates that the dust extends to a radius of 252\,au from the star at a PA of 147\degree. The disc 
was not detected at 450\,$\umu$m but a fit to the other SED points gives a dust temperature of 50\,K, and constrains $\beta$ in the range of 0.3 to 1.2. The estimated dust mass 
is 0.24\,M$_{\earth}$. There is also evidence for a warmer disc component at 146\,K, based on a fit to the near-mid IR data from \emph{AKARI} \citep{Ishihara2010}, \emph{Spitzer} 
\citep{Chen2014}, and \emph{WISE} \citep{Wright2010}.

\subsection{CE Antliae (TWA 7, 2MASS J10423011--3340162)}
\label{sec:ceant}

CE Ant (TWA 7) is an M2 dwarf star in the TW Hydrae Association, with an estimated age of 9\,Myr in a range spanning 3\,Myr to 20\,Myr \citep{Barrado2006}. The distance of the 
star is uncertain but is likely to be similar to other members of the association ($\sim$50\,pc). The 850\,$\umu$m image shows a clear, unresolved dust emission peak that is 
offset by approximately 6\arcsec\ to the east of the star (Fig.~\ref{fig:figureA7}b). The offset is greater than the 3.3\arcsec\ expected due to statistical and positional 
uncertainties (see Section~\ref{sec:offsets}). The peak has a flux density of 7.2 $\pm$ 1.3\,mJy, consistent with the measurement made by SCUBA at 850\,$\umu$m, of 9.7 $\pm$ 
1.6\,mJy \citep{Matthews2007}. The SCUBA measurement was carried out using photometry mode \citep{Holland1999}, and hence it is not possible to ascertain whether there was 
also an offset associated with that particular observation. Interpreting the structure as a disc about the star gives an upper limit to the radius from the 850\,$\umu$m image 
of 380\,au.

\vskip 1mm

\begin{figure}
\includegraphics[width=85mm]{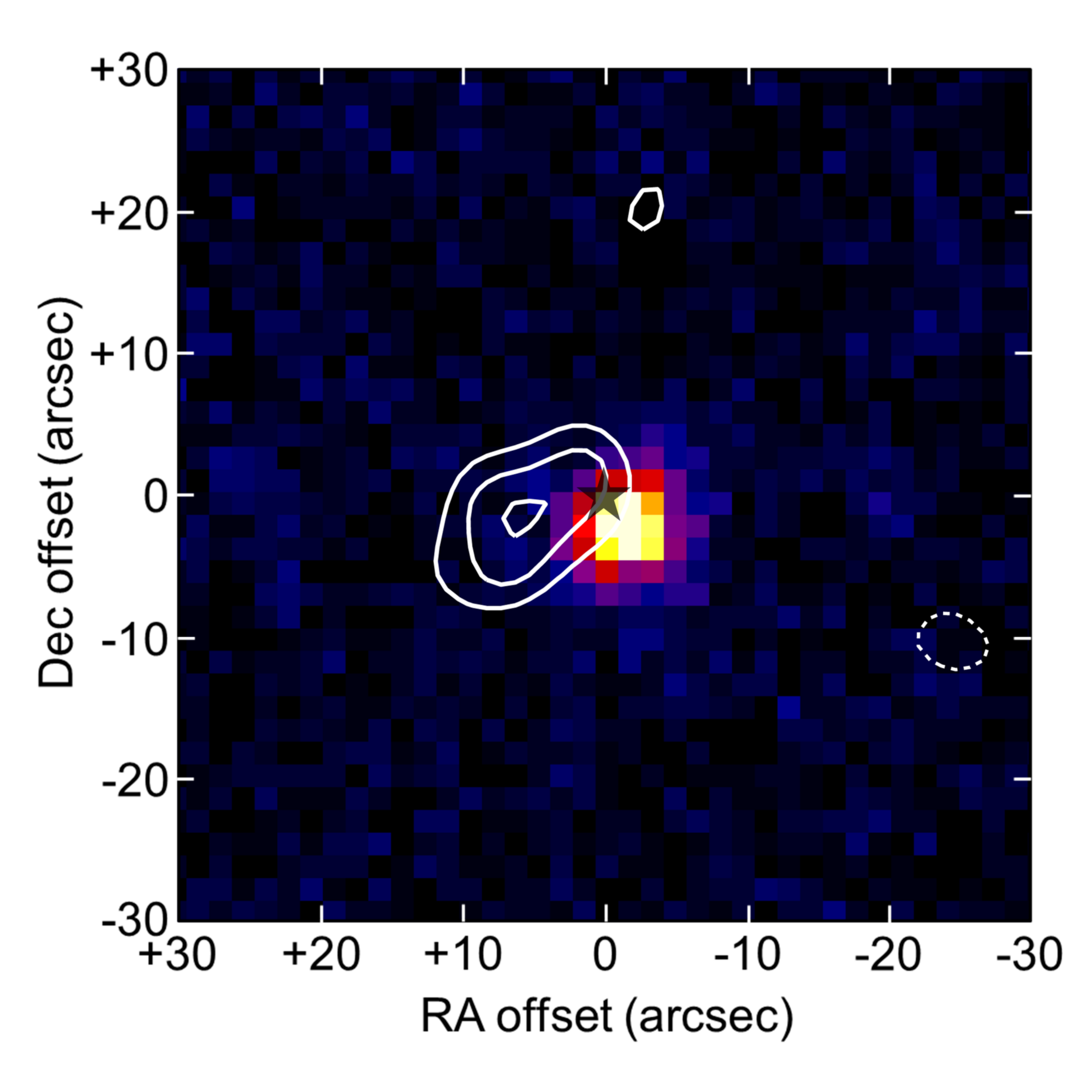}

\caption{\emph{Herschel}/PACS 70\,$\umu$m image of CE Ant (TWA 7) with the SONS survey 850\,$\umu$m contours overlaid. The contours start at 3$\sigma$ and increase in 1$\sigma$ 
steps. The \emph{Herschel} image is taken from the \emph{Herschel} Science Archive.}

\label{fig:twa7_herschel}
\end{figure}

\emph{HST}/NICMOS imaging of the scattered light reveals an inclined disc extending to a radius of 1\arcsec\ ($\sim$35\,au) at a PA of 53\degree\ \citep{Choquet2016}. 
Modelling suggests the observed structure is more likely to be a ring rather than a continuous disc. \emph{Herschel}/PACS observations at 70\,$\umu$m resolve the disc with a 
deconvolved radius of 6.2\arcsec\ ($\sim$210\,au) \citep{Cieza2013}. The \emph{Herschel}/PACS image shows that the disc is reasonably well-aligned to the the stellar position, 
unlike the case for the SONS 850\,$\umu$m result (see Fig.~\ref{fig:twa7_herschel}). It is therefore possible that the peak seen at 850\,$\umu$m could be a background object, 
rather than a debris disc associated with CE Ant, although there is no evidence from \emph{Herschel} observations that such an object exists at this position.

\vskip 1mm

The SED contains both \emph{Spitzer}/MIPS \citep{Low2005} and \emph{Herschel} photometry \citep{Cieza2013} and is best fitted with two modified blackbodies, consistent with a 
previous analysis \citep{Riviere-Marichalar2013}. The first blackbody fits the photometric points shortward of $\sim$100\,$\umu$m with a dust temperature of 81\,K and suggests 
an inner disc at a radius of 2.5\,au. Assuming the observed structure at 850\,$\umu$m is a real disc associated with CE Ant, a second blackbody fit to the long-wavelength 
submillimetre photometry indicates the presence of a very cold structure, with a dust temperature of 19\,K, orbiting at a mean radius of 49\,au, assuming the dust grains have 
pure blackbody emission (both of which have large uncertainties, as indicated in Table~\ref{tab:table3}). The value of $\beta$ is unconstrained due to the sparse photometric 
coverage, and lies in the range of 0 to 2.5, whilst the estimated dust mass is 0.24\,M$_{\earth}$.

\vskip 1mm

\vskip 1mm

\subsection{HD 92945 (V419 Hya)}
\label{sec:hd92945}

HD 92945 is a K1 dwarf star at a distance of 21\,pc with an estimated age of 200\,Myr, lying between the lower and upper estimates of 100\,Myr and 300\,Myr \citep{Song2004, 
Chen2014}, respectively. The 850\,$\umu$m SONS survey image shows a structure where the emission peaks at the stellar position, but with an apparent asymmetric extension to the 
south (Fig.~\ref{fig:figureA7}c). The peak flux is 8.6 $\pm$ 1.1\,mJy/beam, that being consistent with the previously published result \citep{Panic2013}. The integrated flux, 
measured over an aperture of 40\arcsec\ diameter, is 12.6 $\pm$ 1.5\,mJy confirming that the structure is resolved, if the extension to the south is really associated with a disc 
about the star. From the 850\,$\umu$m image, the measured deconvolved disc radius from the 2D Gaussian fitting is 11\arcsec\ ($\sim$210\,au) in the major axis at a PA of 
178\degree, i.e. dominated by the north-south extension.

\vskip 1mm

\begin{figure}
\includegraphics[width=85mm]{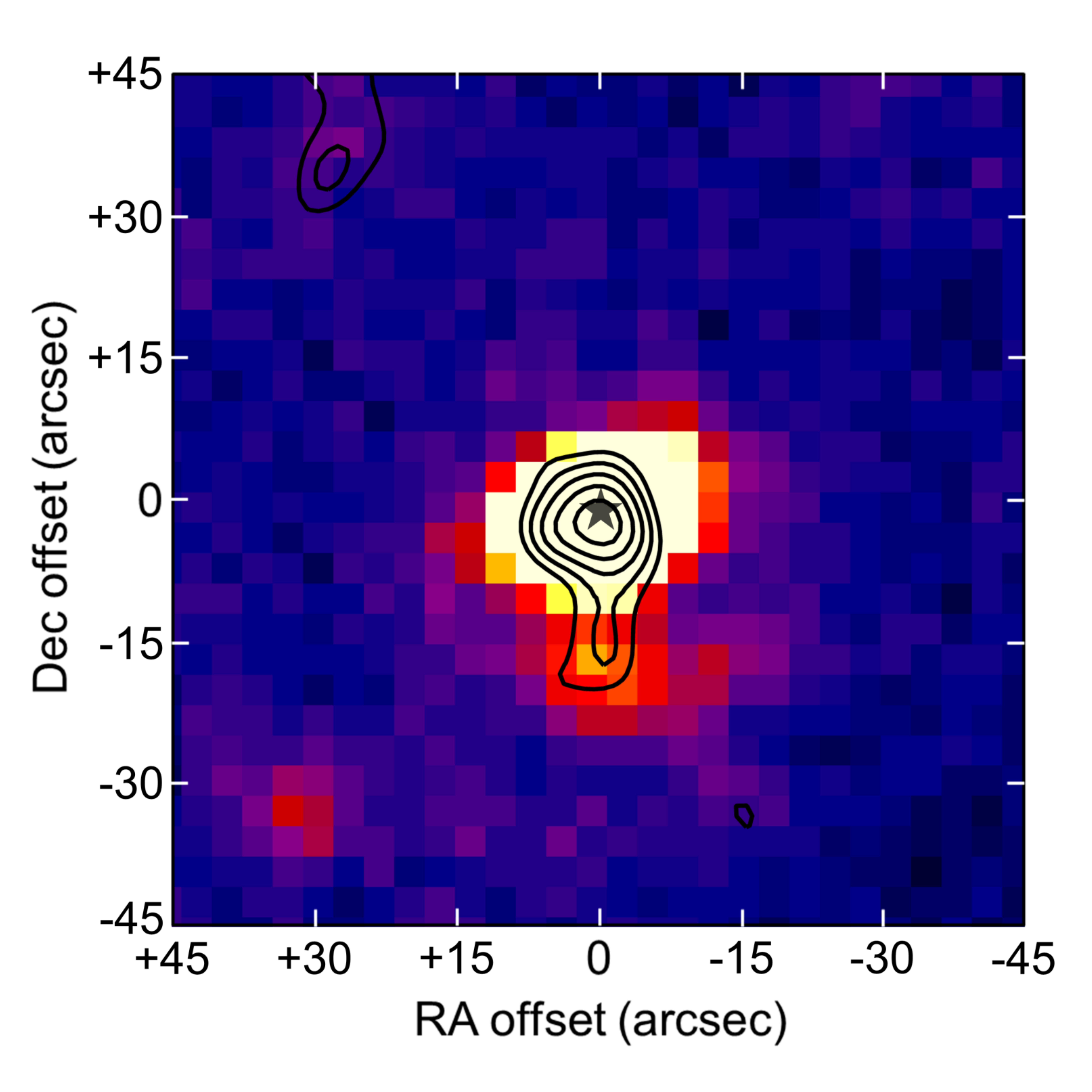}

\caption{\emph{Herschel}/PACS 160\,$\umu$m image of HD 92945 with the SONS survey 850\,$\umu$m contours overlaid. The contours start at 3$\sigma$ and increase in 1$\sigma$ 
steps. The \emph{Herschel} image is taken from the \emph{Herschel} Science Archive.}

\label{fig:hd92945_herschel}
\end{figure}

The disc has been imaged in scattered light using \emph{HST}/ACS, revealing an inclined, axisymmetric structure with an inner ring of radius 2 $- $3\arcsec\ ($\sim$43 $-$ 
65\,au) from the star and an outer disc declining in brightness to 5\arcsec\ ($\sim$110\,au) roughly oriented east-west \citep{Golimowski2011}. \emph{Herschel}/PACS imaging at 
70 and 160\,$\umu$m reveal an extension (or possible second source) to the south, coincident with that seen in the 850\,$\umu$m image (Fig.~\ref{fig:hd92945_herschel}). Given 
the asymmetric nature of the extension, it is more likely that this feature is caused by a background object, and is not part of a disc (see 
Section~\ref{sec:background_galaxies}). In this scenario, the peak flux of 8.6 $\pm$ 1.1\,mJy/beam, and an upper limit to the disc radius of 160\,au, are better estimates to be 
adopted for a disc about the star. The SED contains both \emph{Spitzer}/MIPS \citep{Golimowski2011} and \emph{Herschel}/PACS photometry (archival data), and using the 
850\,$\umu$m peak flux, is well-fitted by a single temperature modified blackbody with a dust temperature of 42\,K, a $\beta$ in the range 0.4 to 1.1, and a dust mass of 
0.024\,M$_{\earth}$.

\vskip 1mm

\vskip 1mm

\subsection{HD 98800}
\label{sec:hd98800}

HD 98800 is a pre-main-sequence K5 dwarf star in the TW Hydrae association, at a distance of 50\,pc with an estimated age of 9\,Myr \citep{Barrado2006}. HD 98800 is actually a 
quadruple star system, containing a pair of spectroscopic binary stars (HD 98800A and B, separated by $\sim$36\,au) and in which only HD 98800B harbours a circumstellar disc 
\citep{Torres1995, Koerner2000}. Together with Fomalhaut Section~\ref{sec:fomalhaut}, the 850\,$\umu$m image represents by far the brightest disc in the SONS survey with a flux 
density of 94 $\pm$ 1.5\,mJy (Fig.~\ref{fig:figureA7}d). The disc is also well-detected at 450\,$\umu$m, revealing an unresolved disc with a flux density of 242 $\pm$ 14\,mJy. 
The upper limit to the disc radius from the 450\,$\umu$m image is 255\,au. 

\vskip 1mm

The HD 98800B disc has been resolved by the SMA revealing that it is closely aligned with the orbit of the spectroscopic binary (HD 98800A), and extends to a radius of only 
10--15\,au from HD 98800B at a PA of 158\degree\ \citep{Andrews2010}. The disc has a fractional luminosity of $\sim$0.1 and so is outside the generally accepted range for a 
debris disc (see Section~\ref{sec:fractional_luminosities}). Indeed, it is commonly referred to as a ``transitional disc'', i.e., a disc evolving between the younger 
protoplanetary and the older debris phases. The SED in the infrared is well-characterised by \emph{WISE} \citep{Wright2010}, \emph{AKARI} \citep{Ishihara2010} and 
\emph{Spitzer}/MIPS photometry \citep{Chen2014}. In the submillimetre, the photometry has considerable more scatter, with seemingly low flux estimates at 870\,$\umu$m from 
APEX/LABOCA (a 4$\sigma$ detection; \citealt{Nilsson2010}) and 1.3\,mm from JCMT/UKT14 (also a 4$\sigma$ detection; \citealt{Sylvester1996}), compared to the SED fit. The near 
blackbody fit suggests a dust temperature of 156\,K, a well-constrained value of $\beta$ in the range 0.0 to 0.1, and an estimated dust mass of 0.37\,M$_{\earth}$.

\subsection{HD 104860}
\label{sec:hd104860}

HD 104860 is a F8 star located at a distance of 46\,pc with an estimated age of 200\,Myr \citep{Morales2013}. The 850\,$\umu$m image (Fig.~\ref{fig:figureA8}a) shows a 
structure in which the peak emission is well-centred on the stellar position. The peak flux density is 6.5 $\pm$ 1.0\,mJy/beam, in good agreement with previous SCUBA flux 
estimates \citep{Najita2005}. The integrated flux, measured in a 50\arcsec\ diameter aperture centred on the star, is 8.7 $\pm$ 1.4\,mJy, indicating the overall structure 
could be marginally resolved. Another possibility, however, given the asymmetry of the structure, is that the 4$\sigma$ peak to the east may be a background object. 
Interpreting the emission as that from a disc gives an upper limit to the radius from the 850\,$\umu$m image of 340\,au (assuming an unresolved structure).

\vskip 1mm

The disc has been resolved by \emph{Herschel}/PACS, with the modelled image at 160\,$\umu$m suggesting a slightly elongated disc of radius 3.5\arcsec\ ($\sim$166\,au) from the 
star at a PA of 12\degree\ \citep{Morales2013}. There is also a hint of a slight eastward extension to the image at 160\,$\umu$m, which is coincident with the second peak seen 
in the 850\,$\umu$m image, perhaps supporting the hypothesis that this is a background object. The disc also been resolved by the SMA at 1.3\,mm, revealing a disc with a peak 
in emission at a radius of 2.5\arcsec\ ($\sim$110\,au) and elongated at a PA roughly in a north-south direction \citep{Steele2016}. The SMA images, however, show no signs of a 
second source to the east. The SED in the infrared is well characterised by \emph{Spitzer}/MIPS \citep{Hillenbrand2008}, in the far-IR by \emph{Herschel}/PACS 
\citep{Morales2013}, CSO and IRAM \citep{Roccatagliata2009} in the submillimetre/near-millimetre, and the VLA at 9\,mm \citep{MacGregor2016a}. In fitting the SED it has been 
assumed that the flux at 850\,$\umu$m is best represented by the peak value, giving a dust temperature of 47\,K, a $\beta$ index in the range of 0.4 to 1.2, and an estimated 
dust mass of 0.071\,M$_{\earth}$.

\subsection{HD 107146}
\label{sec:hd107146}

HD 107146 is a young solar analogue star (G2V) located at a distance of 28.5\,pc with an estimated age of 100\,Myr \citep{Chen2014}, although it could be as young as 30\,Myr 
\citep{Williams2004}. The 850\,$\umu$m image (Fig.~\ref{fig:figureA8}b) reveals a bright structure, with a peak flux of 20.6 $\pm$ 2.1\,mJy/beam slightly offset by 
$\sim$4\arcsec\ from the star. The structure appears slightly extended to the south-west, although the integrated flux, measured over a 40\arcsec\ diameter 
aperture, indicates that this feature is not significant compared to the peak flux value. Interpreted as a disc the upper limit to the radius from the 850\,$\umu$m image is 
210\,au. The measurement is in excellent agreement with the previous SCUBA measurement of 20 $\pm$ 4\,mJy, and the disc was also resolved by SCUBA at 450\,$\umu$m with a 
deconvolved radius of 150\,au from the star at a PA of 155\degree\ \citep{Williams2004}.

\vskip 1mm

The disc was first detected in scattered light using \emph{HST}/ACS, showing a circularly-symmetric structure with maximum opacity occurring at 130\,au, with a PA of 
$\sim$160\degree\ and inclined at 25\degree\ to the plane of the sky \citep{Ardila2004}. The scattered light images also reveal an object $\sim$7\arcsec\ south-west of the star, 
which was identified as a spiral galaxy. In the nine years between the \emph{HST} and SCUBA-2 observations, the proper motion of the star means that the separation to the galaxy 
is now 5\arcsec. Hence, it is possible that the SONS survey image at 850\,$\umu$m is contaminated by the presence of a background galaxy causing the apparent extension to the 
disc.

\vskip 1mm

SMA observations at 880\,$\umu$m show that the emission is distributed in a ring of radius 3.5\arcsec\ from the star, and this emission is modelled with inner and outer radii of 
50 and 170\,au with a disc PA of 148\degree\ \citep{Hughes2011}. The total flux measured by the SMA was 36 $\pm$ 1\,mJy, significantly higher than the SONS result at 
850\,$\umu$m. Given the 5\arcsec\ separation between the star and the galaxy, it is possible that the total SMA emission also contains flux from the background source. Similarly, 
with ALMA at 1.3\,mm the disc is seen extending from 30 to 150\,au at a PA of 144\degree, whilst models suggest a clear decrease in the dust emission at a radius of $\sim$80\,au 
extending over a width of 9\,au \citep{Ricci2015a}. The total flux density for the disc of 12.5 $\pm$ 1.3\,mJy from the ALMA observations may also be contaminated by 
flux from the nearby galaxy. Given that the star is a young Solar analogue, the disc could be a larger version of the Edgeworth-Kuiper belt in our Solar System 
\citep{Ricci2015a}.

\vskip 1mm

The emission from the disc is well-sampled in the infrared through millimetre region, including measurements from \emph{Spitzer}/MIPS \citep{Chen2014}, \emph{Herschel}/PACS 
\citep{Morales2016}, CSO \citep{Corder2009}, OVRO \citep{Carpenter2005} and ACTA \citep{Ricci2015b}. The fit to the SED, using the peak flux at 850\,$\umu$m, tightly 
constrains $\beta$ to a value of between 0.8 and 1.0 and gives a dust temperature of 41\,K, leading to an estimated dust mass of 0.093\,M$_{\earth}$. The fit to the SED at 
near-mid IR wavelengths also suggests an inner, ``warm'' component radiating at 81\,K.

\subsection{$\eta$ Corvi (HD 109085)}
\label{sec:etacrv}

\begin{figure}
\includegraphics[width=85mm]{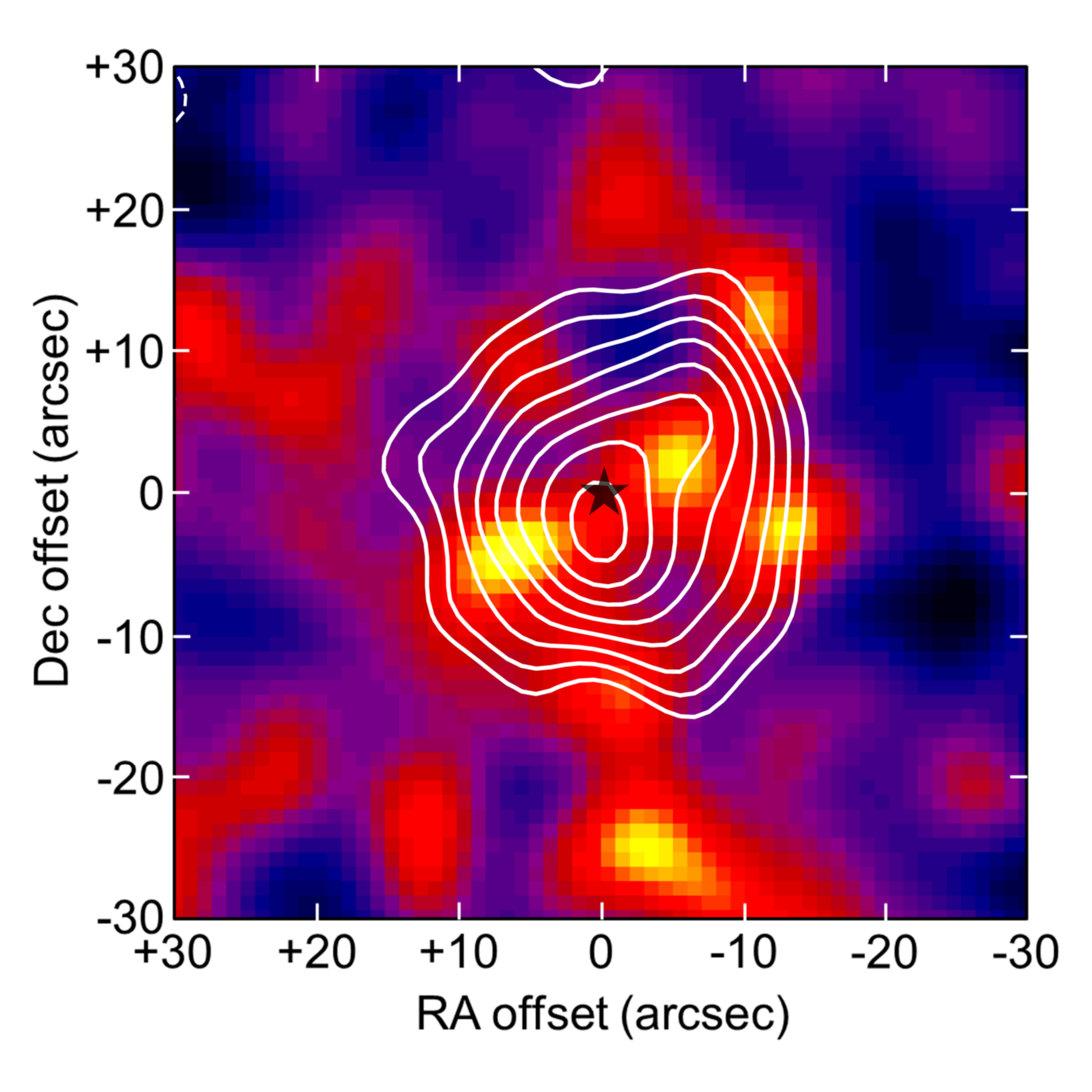}

\caption{The 450\,$\umu$m image from observations of HD 109085 ($\eta$ Crv) with contours from the 850\,$\umu$m image overlaid. The contours and symbols are as described in
Fig.~\ref{fig:hd8907}.}

\label{fig:etacorvi}
\end{figure}

$\eta$ Crv (HD 109085) is a nearby F2 star located at a distance of 18\,pc with an estimated age of 1380\,Myr \citep{Chen2014} in a range of 1000\,Myr to 2000\,Myr 
\citep{Vican2012}. The images and fluxes presented in this paper are slightly updated versions of those first published by \citet{Duchene2014} following a re-analysis of the 
data. The 850\,$\umu$m image (Fig.~\ref{fig:figureA8}c) shows emission peaking with a flux of 7.2 $\pm$ 0.7\,mJy/beam. The integrated flux density of 15.4 $\pm$ 1.1\,mJy, 
measured in a 40\arcsec\ diameter aperture centred on the star, in good agreement with previous SCUBA flux estimates \citep{Wyatt2005}. Interpreting the emission as a disc about 
the star gives deconvolved major and minor axis radii of 10.4 and 7.5\arcsec\ (190 and 135\,au), respectively, at a PA of 132\degree\ and inclined to the plane of the sky by 
44\degree. The SONS survey image at 450\,$\umu$m is low S/N due to the relatively poor weather conditions at the time of the observations, but shows the disc is resolved into 
the two prominent clumps seen previously \citep{Wyatt2005} albeit only at the $\sim$4$\sigma$ significance level (Fig.~\ref{fig:etacorvi}).

\vskip 1mm

The first infrared excess flux measurements of $\eta$ Crv were made by \emph{IRAS} \citep{Stencel1991}, and the disc has been observed at many wavelengths over the past two 
decades. SCUBA observations at 450\,$\umu$m resolved the disc for the first time revealing a ring-like structure oriented at PA of 130\degree\, with model fits indicating a disc 
inclination of 45\degree\ and radius of 150\,au \citep{Wyatt2005}. \emph{Herschel}/PACS images at 70\,$\umu$m show a central flux peak on the star surrounded by an inclined ring 
at $\sim$47\degree\ to the plane of the sky \citep{Duchene2014}, the latter being the same emission seen in the submillimetre. Modelling of the disc suggests that there are both 
warm and cold dust belts in the system, the cold component peaking in emission at 164\,au with a estimated width of $\sim$9\,au, at a PA of 116\degree\ \citep{Duchene2014}.

\vskip 1mm

More recently, ALMA observations at 880\,$\umu$m reveal an asymmetric belt of emission of mean radius of 8.4\arcsec\ ($\sim$152\,au) with a width of 2.6\arcsec\ ($\sim$46\,au), 
at a PA of 117\degree\ and an inclination of 35\degree\ \citep{Marino2017}. The total flux emission measured from the ALMA image of 10.1 $\pm$ 0.4\,mJy is 2--3$\sigma$ lower 
than would be expected compared to the SCUBA-2 850\,$\umu$m flux if extrapolated to 880\,$\umu$m, assuming a spectral index of 3. As pointed out by \citet{Marino2017} the 
difference could be due either to extended emission being missed by ALMA as a result of having an insufficient number of short baselines, or to the image reconstruction method 
adopted for the ALMA data. 

\vskip 1mm

The photometric data from the SED is fitted by a two-component model. The fit to the near-mid IR data suggests an inner belt at a radius of $\sim$3\,au composed of warm dust 
radiating at 254\,K, whilst the fit to far-IR/submillimetre data indicates the presence of a cold belt with a characteristic dust temperature of 45\,K, and a $\beta$ in the 
range of 0 to 0.7, leading to an estimated dust mass of 0.028\,M$_{\earth}$.

\vskip 1mm

\subsection{HR 4796 (TWA11, HD 109573)}
\label{sec:hr4796}

\begin{figure}
\includegraphics[width=85mm]{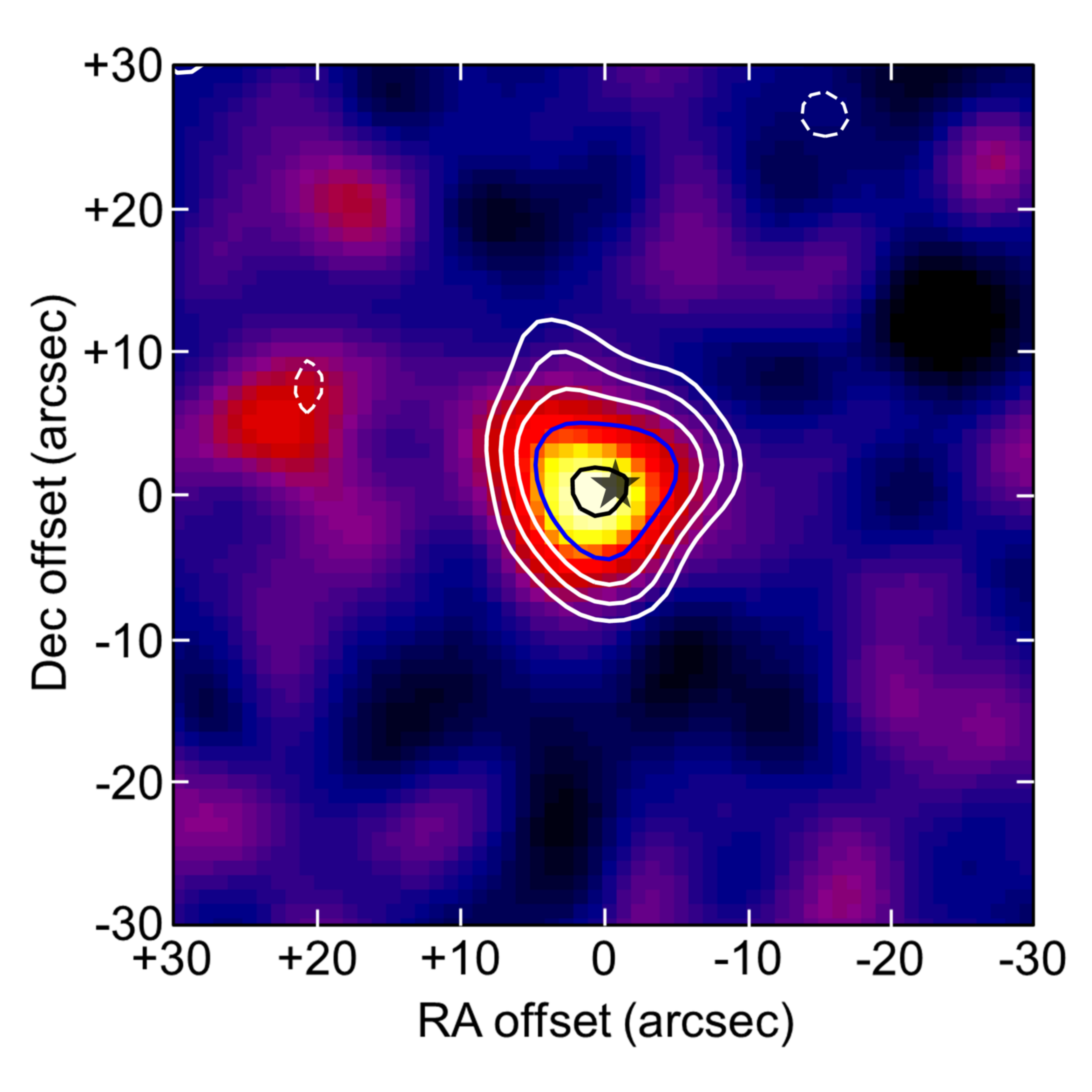}

\caption{The 450\,$\umu$m image from observations of HR 4796 (TWA 11; HD 109573) with contours from the 850\,$\umu$m image overlaid. The contours and symbols are as described in
Fig.~\ref{fig:hd8907}.}

\label{fig:hr4796}
\end{figure}

HR 4796 (HD 109573) is a binary star system in the constellation of Centaurus, at a distance of 73\,pc and is part of the TW Hydrae Moving Group with an age of 9\,Myr 
\citep{Barrado2006}. The two stars of the system are separated by 7.7\arcsec, with the primary having a spectral type A0, whilst the smaller companion is a red dwarf of type 
M2.5. Emission is detected at both 850 and 450\,$\umu$m with flux densities of 14.4 $\pm$ 1.9 and 117 $\pm$ 21\,mJy, respectively (Fig.~\ref{fig:figureA8}d and 
Fig.~\ref{fig:hr4796}). Interpreting the emission as a disc structure about the star gives an upper limit to the radius from the 450\,$\umu$m image is 370\,au.

\vskip 1mm

The disc around HR 4796 was first resolved in the mid-IR \citep{Koerner1998, Jayawardhana1998} at 12.5 and 20\,$\umu$m revealing an elongated structure with the dust peaking 
at a radius of 70\,au surrounding the central primary star. The PA of the disc was measured at 30\degree, consistent with the binary companion, and suggesting that the 
disk-binary system is being seen nearly along the orbital plane. The disc was interpreted as a ring-like structure, having two lobes similar to HD 216956 (Fomalhaut), with an 
inner hole devoid of dust extending to a radius of 55\,au. The disc has also been imaged in scattered light using \emph{HST}/NICMOS and STIS which show a distinctive ring-like 
symmetrical structure, consistent with the mid-IR observations \citep{Schneider1999, Schneider2009}. The images revealed that the ring has a peak intensity at 70\,au, with a 
resolved width of 17\,au, and a major axis lying at a PA of 27\degree. The observed brightness variation in across the disc in thermal emission has been attributed to 
pericentre glow, in which the asymmetry is caused by the gravitational perturbations from an unseen planetary system \citep{Wyatt2000}.

\vskip 1mm

The SED is well-sampled from the near-IR to the millimetre and includes photometric points from both \emph{Spitzer} \citep{Chen2014} and \emph{WISE} \citep{Wright2010}. 
Fitting the SED suggests a very well-defined, single temperature disc of 99\,K, and a $\beta$ value of 0.7 to 1.0, and an estimated dust mass of 0.19\,M$_{\earth}$ from the 
850\,$\umu$m flux.

\subsection{61 Virginis (HD 115617)}
\label{sec:hd115617}

HD 115617 (61 Vir) is a nearby main sequence star (G7-V) hosting a system of at least two known planets, both of which are sub-Saturn mass and have orbits of less than 
0.5\,AU. The star lies at a distance 8.6\,pc with an estimated age of 6300\,Myr, within a range of 6100 to 6600\,Myr \citep{Mamajek&Bell2014}. The SONS survey image at 
850\,$\umu$m reveals a main emission peak a few arcseconds offset from the star together with a second isolated 3.5$\sigma$ peak 12\arcsec\ to the north 
(Fig.~\ref{fig:figureA9}a). The peak flux for the main structure is 3.9 $\pm$ 0.8\,mJy and is possibly marginally resolved with an integrated flux of 5.8 $\pm$ 1.0\,mJy 
measured in a 40\arcsec\ diameter aperture centred on the star, but not including the peak to the north. The image reported here benefits from further integration time than 
the image published by \citet{Panic2013}. The structure has a PA of $\sim$66\degree\, and interpreting as a disc about the star gives a deconvolved radius of 4.7\arcsec\ 
($\sim$40\,au). The updated SONS results are reported in \citet{Marino2017b}.

\vskip 1mm

61 Vir was also included in a survey of FGK stars by \emph{Spitzer} that detected the disc at 24 and 70\,$\umu$m \citep{Tanner2009}. \emph{Herschel} has provided the most 
definitive study of the HD 115617 debris disc so far with images from 70 to 500\,$\umu$m as part of the DEBRIS survey. The images show a nearly edge-on configuration extending 
out to at least 100\,au from the star at a PA of $\sim$65\degree, with the dust population peaking in density in a belt between 30 and 100\,au \citep{Wyatt2012}. Thus, there 
is expected to be very little interaction between the disc and the known planets, although theory suggests that planetesimal belts such as this one might hold the key to 
understanding super-Earth systems such as HD 115617. The northern source, seen in the SONS 850\,$\umu$m image, is most likely a background object based on the interpretation 
of the \emph{Herschel}/PACS 160\,$\umu$m image \citep{Wyatt2012}. Observations with ALMA at 870\,$\umu$m resolve 3 compact sources \citet{Marino2017b}, one of which seems 
coincident with the northern source seen in both SONS 850\,$\umu$m and \emph{Herschel}/PACS 160\,$\umu$m images. By combining SCUBA-2 and ALMA observations it is inferred that 
the disc is likely very extended covering radius range of 30\,au to at least 150\,au, implying the existence of a broad parent planetesimal belt.

\vskip 1mm

The SCUBA-2 850\,$\umu$m result seems to indicate a rapid fall off in flux towards the millimetre region ($\lambda_0$\,$>$\,650\,$\umu$m), although this sharp decrease is not 
totally consistent with the \emph{Herschel}/SPIRE photometry. Hence the fit to the SED is not convincing with a characteristic dust temperature of 65\,K, and only constrains 
$\beta$ in the range between 0 and 1.5. The dust mass is estimated to be 1.6 $\times$ 10$^{-3}$ M$_{\earth}$. The radius of the disc, derived from the SED fit and assuming 
blackbody grain properties, is only 17\,au, significantly smaller than the value of 40\,au obtained from the marginally resolved 850\,$\umu$m image.

\subsection{$\lambda$ Bo\"{o}tis (HD 125162)}
\label{sec:hd125162}

HD 125162 ($\lambda$ Boo) is a luminous A0 star at a distance 30\,pc with an estimated age of 2800\,Myr, but with a large uncertainly stretching from 2000\,Myr to 3900\,Myr 
\citep{Montesinos2009}. At 850\,$\umu$m, unresolved emission is detected, well-centred on the stellar position, with a flux density of 3.9 $\pm$ 0.8 mJy 
(Fig.~\ref{fig:figureA9}b). If interpreted as a disc about the star, the upper limit to the radius from the 850\,$\umu$m image is 230\,au.

\vskip 1mm

The HD 125612 debris disc has been resolved by \emph{Herschel}/PACS at 70\,$\umu$m revealing a slightly elongated structure of radius 40\,au at a PA of 42\degree\ 
\citep{Booth2013}. Photometry from \emph{Spitzer} \citep{Chen2014} also helps to define the mid-IR section of the SED. The 850\,$\umu$m photometric point helps to anchor the SED 
fit to give a characteristic dust temperature of 87\,K, but still a poorly constrained $\beta$ value in the range of 0.3 to 1.8. The dust mass is estimated to be 
0.010\,M$_{\earth}$.

\subsection{HD 127821}
\label{sec:hd127821}

HD 127821 is an F4 star at a distance of 32\,pc with an estimated age of 1020\,Myr \citep{Chen2014}, although it could be as old as 3400\,Myr \citep{Moor2006}. The SONS image at 
850\,$\umu$m appears to show a predominantly double-peaked structure with one peak reasonably-well centred on the stellar position, and the other offset by $\sim$13\arcsec\ to 
the south-east (Fig.~\ref{fig:figureA9}c). The peak nearest to the star has a flux of 5.8 $\pm$ 0.7 mJy/beam. The integrated flux in a 40\arcsec\ diameter aperture, centred 
between the two peaks, is 10.5 $\pm$ 1.4 mJy. This flux is just consistent with previous SCUBA photometry of 13.2 $\pm$ 3.7 mJy/beam \citep{Williams&Andrews2006}, noting also 
the slightly larger FWHM beam for SCUBA. In addition, given that the two peaks are approximately the same brightness, and that the integrated flux is equal to their combined 
levels, it is most likely these are two separate sources. There is also other structure seen to the north, offset from the star by $\sim$18\arcsec.

\vskip 1mm

\begin{figure}
\includegraphics[width=85mm]{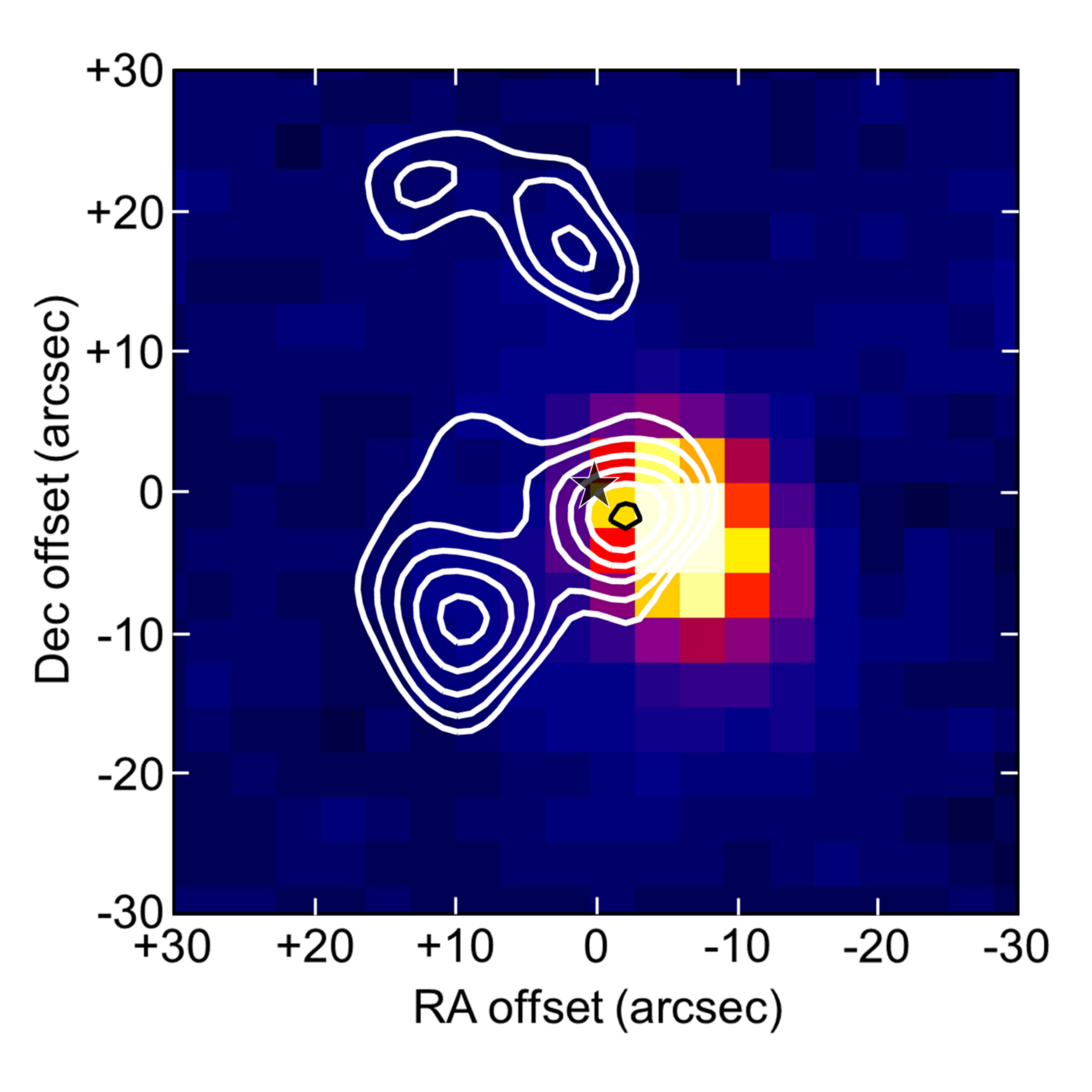}

\caption{\emph{Herschel}/PACS 160\,$\umu$m image of HD 127821 with the SONS survey 850\,$\umu$m contours overlaid. The \emph{Herschel} image is taken from the
\emph{Herschel} Science Archive.}

\label{fig:hd127821_herschel}
\end{figure}

\emph{Herschel} imaging with PACS and SPIRE at 160\,$\umu$m and 250\,$\umu$m reveals a single source, well-aligned with the central peak in the SONS 850\,$\umu$m image. 
Fig.~\ref{fig:hd127821_herschel} shows the \emph{Herschel}/PACS 160\,$\umu$m image with the SONS survey 850\,$\umu$m contours overlaid. Based on these observations, it is 
therefore possible that the peak to the south-east (as well as the one to the north) are indeed background objects, although these are not seen in the \emph{Herschel} images. It 
is therefore likely that only the central peak is representative of a disc, and under this assumption (i.e., only using the flux of the central peak at 850\,$\umu$m) the SED can 
be fitted with a single-temperature 47\,K modified blackbody and a $\beta$ value in the range 0.9 -- 1.9. The estimated dust mass is 0.031\,M$_{\earth}$, whilst the upper limit 
to the disc radius from the 850\,$\umu$m image is 235\,au.

\subsection{$\sigma$ Bo\"{o}tis (HD 128167)}
\label{sec:hd128167}

HD 128167 ($\sigma$ Boo) is an F2 star at a distance of 16\,pc with an estimated age of 1000\,Myr but it could be as old as 4780\,Myr \citep{Rhee2007}. The SONS survey image 
at 850\,$\umu$m shows emission, offset from the star by $\sim$11\arcsec\ to the north-west, with a flux density of 4.1 $\pm$ 0.9\,mJy (Fig.~\ref{fig:figureA9}d). Given the 
large offset it is unlikely that this peak is associated with the star and is probably a separate source. Observations of HD 128167 with SCUBA at 850\,$\umu$m reported a flux 
of 6.2 $\pm$ 1.7\,mJy \citep{Sheret2004}. These observations used the technique of ``extended photometry'', in which a 9--point map was made around the source with a spacing 
of 5\arcsec. Such a map would have detected flux from the offset position shown in the SCUBA-2 image. Unpublished imaging observations with SCUBA also seem to indicate an 
offset peak in the same direction (Wyatt, \emph{priv comm.}).

\vskip 1mm

\begin{figure}
\includegraphics[width=85mm]{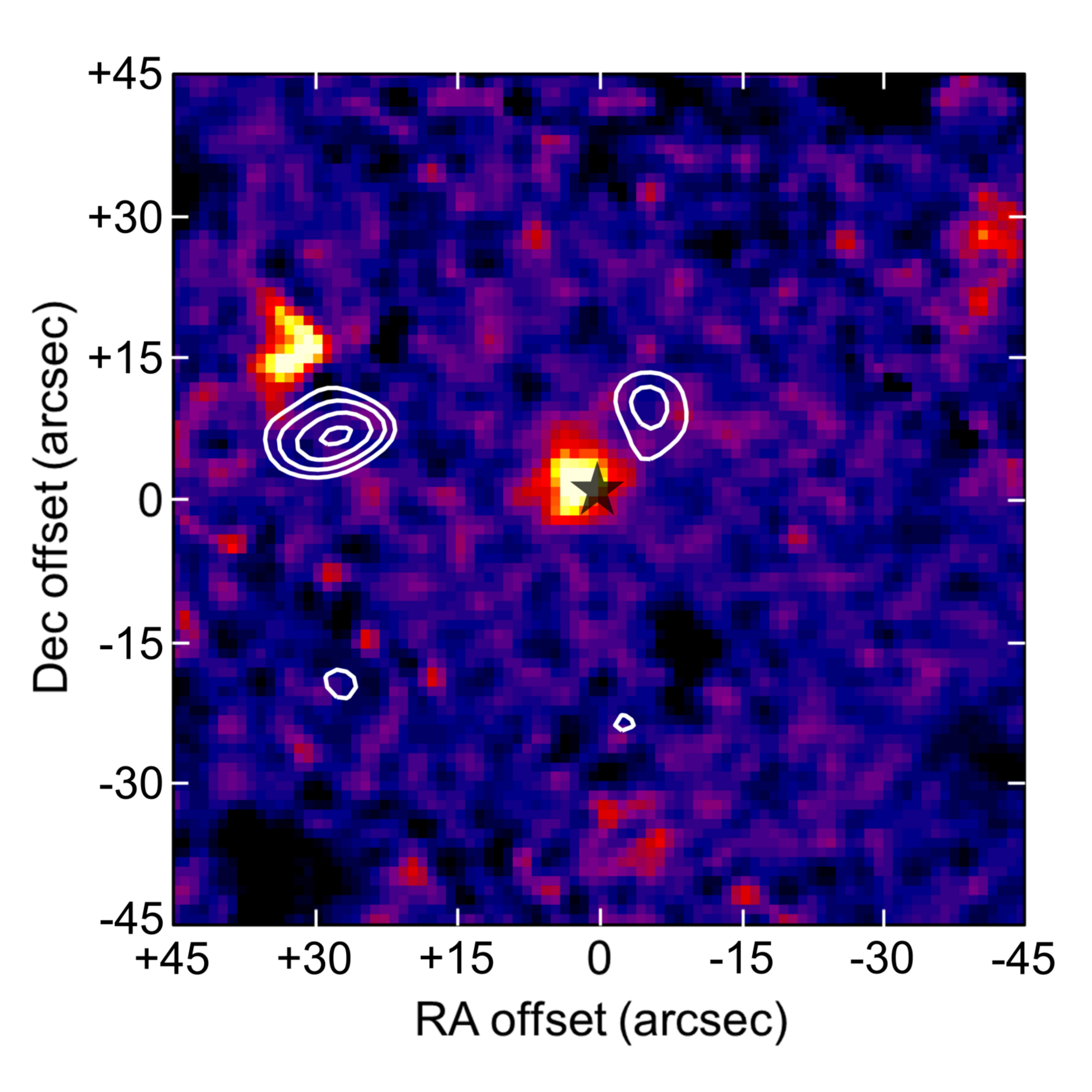}

\caption{\emph{Herschel}/PACS 160\,$\umu$m image of HD 128167 ($\sigma$ Boo) with the SONS survey 850\,$\umu$m contours overlaid. The \emph{Herschel} image is taken from the
\emph{Herschel} Science Archive.}

\label{fig:sigboo}
\end{figure}

\emph{Herschel}/PACS observations at 160\,$\umu$m reveal a peak at 17.5 $\pm$ 3.5\,mJy from a position coincident with the star (Fig.~\ref{fig:sigboo}). A second peak detected 
in the PACS image to the north-east is $\sim$8.5\arcsec\ offset from the brighter peak in the 850\,$\umu$m SONS image. If a rotational shift of $\sim$25\degree\ was applied in 
either image then there would be close alignment between the peak near the centre of the field and that to the east. Given the accuracy of pointing and image reconstruction for 
both telescope facilities, such problems would be unlikely explanations for the discrepancies reported here. It is therefore concluded that the $\sigma$ Boo disc was not 
detected by SCUBA-2 and the peak seen to the north-west is a likely background object. The SED (Fig.~\ref{fig:figureA9}a) only includes photometric points from the position at 
the star (i.e., the SCUBA-2 850\,$\umu$m point is presented as an upper flux limit). If the fit is truly representative of the SED at long wavelengths, then the disc would 
have a flux density of significantly less than 1\,mJy at 850\,$\umu$m (i.e., below the sensitivity threshold of this observation).

\vskip 1mm

\subsection{HD 141378}
\label{sec:hd141378}

HD 141378 is a luminous A5 star at a distance of 54\,pc with an estimated age of 150\,Myr \citep{Rieke2005}. The SONS survey 850\,$\umu$m image reveals a number of features, 
all of which are offset from the star to the south (Fig.~\ref{fig:figureA10}a). The nearest peak is 13\arcsec\ offset and has a flux density 8.5 $\pm$ 1.8\,mJy. This offset is 
far in excess of what would be expected for a disc about the star, allowing for positional and statistical uncertainties in the measurement (see Section~\ref{sec:offsets}). It 
is therefore likely that this peak is due to a separate source, perhaps a background galaxy.

\vskip 1mm

\begin{figure}
\includegraphics[width=85mm]{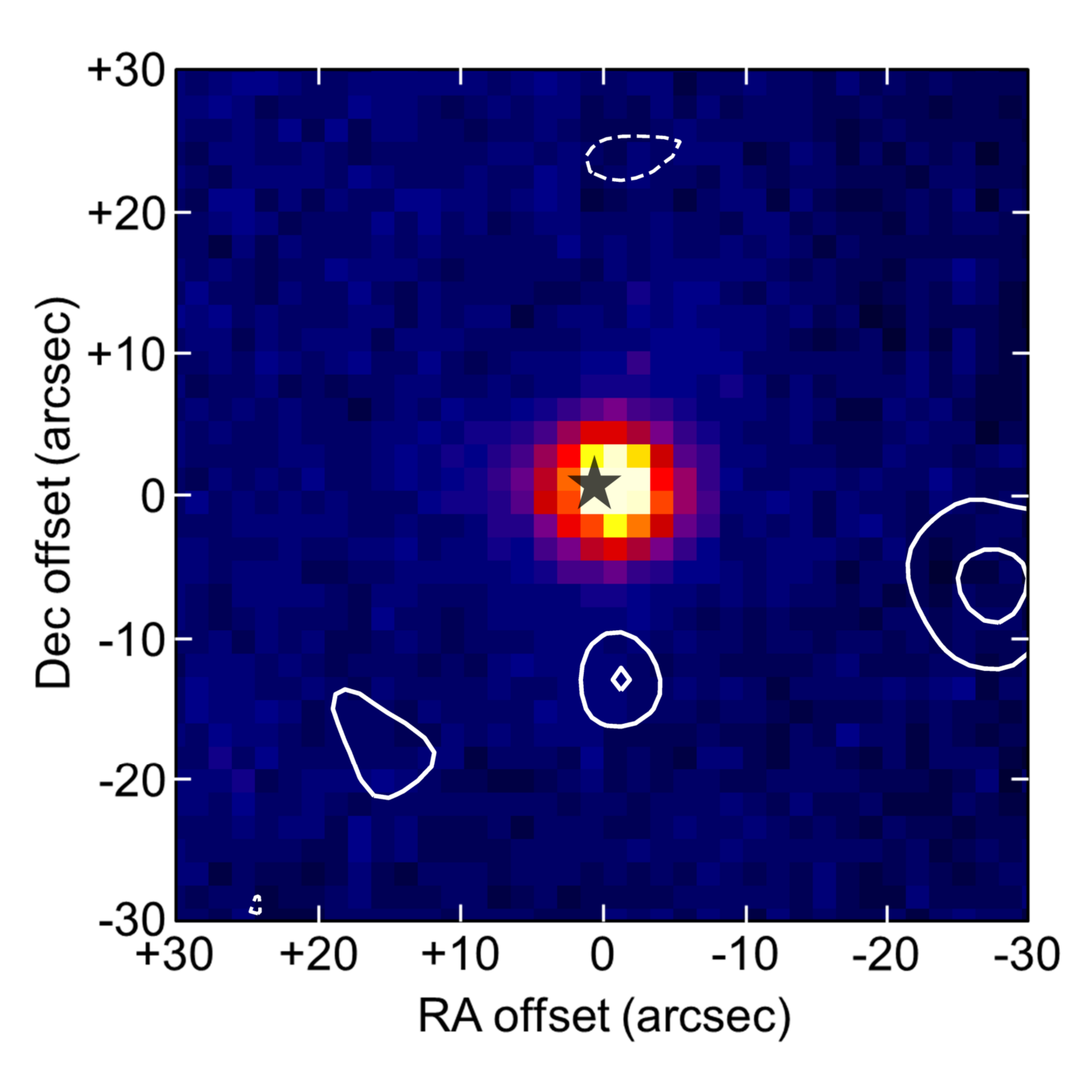}

\caption{\emph{Herschel}/PACS 100\,$\umu$m image of HD 141378 with the SONS survey 850\,$\umu$m contours overlaid. The contours start at 3$\sigma$ and increase in 1$\sigma$ 
steps. The \emph{Herschel} image is taken from the \emph{Herschel} Science Archive.}

\label{fig:hd141378_herschel}
\end{figure}

\emph{Herschel}/PACS imaging at 100 and 160\,$\umu$m reveals an emission peak that is well-centred on the star position, and does not show any evidence of the multiple peaks 
seen at 850\,$\umu$m to the south (see Fig.~\ref{fig:hd141378_herschel}). Furthermore, the SED fit to the \emph{Herschel} photometry suggests that the disc would not have been 
detected by SCUBA-2 at the sensitivity levels reached in the observation. Hence, it is probable that a debris disc does exist around HD 141378, but would be very faint at 
850\,$\umu$m ($\sim$2\,mJy).

\subsection{44 Serpentis (HD 143894)}
\label{sec:hd143894}

HD 143894 (44 Ser) is a luminous A3 star at a distance of 55\,pc and has an estimated age of 300\,Myr \citep{Rhee2007} although it could be as old as 530\,Myr \citep{Chen2014}. 
The SONS 850\,$\umu$m image clearly shows emission with a peak flux density of 5.6 $\pm$ 0.9\,mJy/beam (Fig.~\ref{fig:figureA10}b), but offset from the stellar position by 
$\sim$4\arcsec\ (slightly above the level of 3.1\arcsec\ expected due to positional and statistical uncertainties; Section~\ref{sec:offsets}). The integrated flux in a 
40\arcsec\ diameter aperture, centred on the stellar position, is 10.1 $\pm$ 1.2\,mJy, indicating that, even at a distance of 55\,pc, the emission is resolved. Fitting the 
radial extent of the emission gives a deconvolved radius of 10.2\arcsec\ ($\sim$560\,au) at a PA of 70\degree, almost four times the radius of 150\,au derived assuming the dust 
grains have pure blackbody emission. If the emission is indeed indicative of a disc, it represents a very large structure about the star (i.e., 10$\times$ the size of 
the Edgeworth-Kuiper belt in our Solar System). Given the offset between the star and flux peak, one possibility is that the morphology could be somewhat distorted by a 
background object, resulting in an artificially enlarged structure. Further observations are needed to resolve this issue.

\vskip 1mm

This target was placed in the ``hard to quantify'' flux category, having a very poorly constrained dust temperature due to lack of far-IR photometry. Hence, the SED is 
sparsely defined with only photometry from \emph{Spitzer}/MIPS at 24\,$\umu$m \citep{Chen2014} and \emph{IRAS} at 60\,$\umu$m contributing to the disc model SED fit. The 
850\,$\umu$m flux helps to constrain $\beta$ to a value in the range 0 to 1.1. A temperature of 53\,K and a dust mass of 0.14\,M$_{\earth}$ are derived from the fit and 
850\,$\umu$m flux, respectively.

\subsection{37 Herculis (HD 150378)}
\label{sec:hd150378}

HD 150378 (37 Her) is a luminous A1 star at a distance of 90\,pc with an estimated age of 200\,Myr \citep{Chen2014}. The SONS survey image at 850\,$\umu$m shows a bright peak, 
offset from the stellar position by 17\arcsec\ with a flux density of 10.2 $\pm$ 1.1 mJy (Fig.~\ref{fig:figureA10}c). This emission is highly unlikely to be associated with the star 
and is probably another object, although no identification has been possible from standard catalogues. There is also an \emph{IRAS} detection of excess far-IR radiation at 
60\,$\umu$m, but the large \emph{IRAS} beamsize means it is impossible to ascertain whether this is the same object as detected by SCUBA-2 (the most likely scenario) or a true 
debris disc associated with the star. The SED contains \emph{Spitzer}/IRS photometry that suggests the possibility of a warm, inner disc component. The fit to the two long 
wavelength photometric points in the SED (Fig.~\ref{fig:figureA10}c) suggests an object with a dust temperature of 32\,K. Given the large offset in the 850\,$\umu$m image, it is 
concluded that there is no strong evidence for a cold debris disc surrounding HD\,150378 based on the far-IR and submillimetre data.

\subsection{39 Herculis (HD 150682)} 
\label{sec:hd150682} 

HD 150682 (39 Her) is an F3 star at a distance of 44\,pc with an estimated age of 1700\,Myr \citep{Chen2014}. The SONS survey image at 850\,$\umu$m shows an unresolved peak, 
well-centred on the star with a flux density of 5.5 $\pm$ 0.9 mJy (Fig.~\ref{fig:figureA10}d). The upper limit to the disc radius from the 850\,$\umu$m image is 330\,au. The 
only other far-IR photometry detection is from \emph{Spitzer}/MIPS at 70\,$\umu$m \citep{Trilling2007}. The fit to the poorly-sampled SED suggests a cold component disc with a 
dust temperature of 32\,K (but with large error of $\pm$\,14\,K), $\beta$ in the range from 0.0 to 2.2, and an estimated dust mass of 0.081\,M$_{\earth}$.

\subsection{HD 158352}
\label{sec:hd158352}

HD 158352 is a luminous A8 star at a distance of 60\,pc with an estimated age of 750\,Myr, with an uncertainty of around $\pm$ 150\,Myr \citep{Moor2006}. The SONS image at 
850\,$\umu$m shows an unresolved peak, well-centred on the star, with a flux density of 5.3 $\pm$ 1.0 mJy (Fig.~\ref{fig:figureA11}a). Interpreting the emission as being 
representative of a disc about the star gives an upper limit to the disc radius from the 850\,$\umu$m image of 450\,au. There is another peak some 25\arcsec\ to the north-west 
that is slightly brighter. The only other far-IR photometry detection is from \emph{Spitzer}/MIPS at 24 and 70\,$\umu$m \citep{Chen2014} and so $\lambda_0$ and $\beta$ are not 
well constrained. The fit to the SED suggests a single component disc with a dust temperature of 62\,K, and an estimated dust mass of 0.075\,M$_{\earth}$.

\subsection{$\gamma$ Ophiuchi (HD 161868)}
\label{sec:hd161868}

HD 161868 ($\gamma$ Oph) is a luminous A0 star at a distance of 32\,pc with an estimated age of 185\,Myr with lower and upper limits of 50\,Myr and 277\,Myr, respectively 
\citep{Song2001}. There are, however, other studies that suggest the star could be as old as 450\,Myr \citep{Moor2015}. The 850\,$\umu$m observations reported in this paper are 
deeper than those previously published \citep{Panic2013}, and show a marginally extended structure, with the peak emission having a flux density of 4.8 $\pm$ 0.8 mJy/beam 
(Fig.~\ref{fig:figureA11}b). The integrated flux, measured in an aperture of 50\arcsec\ in diameter and centred on the peak of the emission is 7.1 $\pm$ 1.0 mJy, indicating that 
the structure is resolved. Interpreted as a disc about the star, the radial extent gives a deconvolved radius of 7.8\arcsec\ ($\sim$250\,au) at a PA of 75\degree, compared to 
$\sim$85\,au assuming pure blackbody emission from the dust grains (see Table~\ref{tab:table3}).

\vskip 1mm

\emph{Spitzer}/MIPS observations first resolved the disc showing dust extending out to a large radius of 18\arcsec\ ($\sim$520\,au) at 70\,$\umu$m, with an inclination angle to 
the plane of the sky of $\sim$50\degree\ and a PA of 55\degree\ \citep{Su2008}. These data suggest a disc almost twice as large as the SCUBA-2 observations indicate. 
\emph{Herschel}/PACS observations showed a disc of radius 4.3 to 5.5\arcsec\ ($\sim$135 -- 172\,au) over the observed wavelength range of 70 to 160\,$\umu$m \citep{Moor2015}, 
inclined at 58\degree\ to the plane of the sky and at a PA of 61\degree\ \citep{Moor2015}. Subsequent modelling of the \emph{Herschel}/PACS results suggested inner and outer 
radii for the disc of 49 and 278\,au, respectively \citep{Moor2015}. A background object has also been identified at a position 21\arcsec\ to the east of the star 
\citep{Moor2015}, and hence it is possible that the low-level emission (which is less than 3$\sigma$ significance) seen in the 850\,$\umu$m image at this position is due to 
this.

\vskip 1mm

The far-IR SED is well constrained by measurements from \emph{Spitzer}/MIPS \citep{Su2008, Chen2014} and \emph{Herschel} observations from PACS and SPIRE 
\citep{Pawellek2014,Moor2015}. The fit to the SED suggests a dual-component disc with the cold component having a characteristic dust temperature of 68\,K, a $\beta$ value in 
the range of 0.8 to 1.4, and an estimated dust mass of 0.025\,M$_{\earth}$. The warmer component, fitting to the near-mid IR photometric points, has a dust temperature of 174\,K 
at an estimated radius from the star of 13\,au, assuming blackbody emission from the grains.

\subsection{HD 170773}
\label{sec:hd170773}

HD 170773 is an F5 star at a distance of 37\,pc with an estimated age of 200\,Myr \citep{Rhee2007} but which could be as young as 150\,Myr and as old as 300\,Myr 
\citep{Zuckerman&Song2004}. The SONS survey 850\,$\umu$m image shows a resolved structure, well-centred on the stellar position but extending to the south-east 
(Fig.~\ref{fig:figureA11}c). The peak flux is 16.5 $\pm$ 1.8\,mJy/beam, whilst the integrated flux, measured in an aperture of 40\arcsec\ in diameter, is 26.0 $\pm$ 2.3\,mJy. 
Assuming the structure represents a disc about the star the deconvolved radius is 6.8\arcsec\ ($\sim$250\,au) at a PA of 140\degree\ (roughly aligned in a south-east 
direction; see Table~\ref{tab:table4}), compared to a radius of 73\,au obtained assuming pure blackbody emission from the dust grains. The 850\,$\umu$m peak flux also agrees 
well with a previous photometry using LABOCA on APEX at 870\,$\umu$m of 18 $\pm$ 5.4\,mJy/beam \citep{Liseau2010}.

\vskip 1mm

\emph{Herschel}/PACS observations reveal a disc of radius 5.9\arcsec\ ($\sim$217\,au) over the observed wavelength range of 70 to 160\,$\umu$m \citep{Moor2015}, inclined at
31\degree\ to the plane of the sky and at a PA of 118\degree\ \citep{Moor2015}. These are in reasonably good agreement with the SCUBA-2 results. Modelling of the
\emph{Herschel}/PACS results suggested inner and outer radii for the disc of 81 and 265\,au, respectively \citep{Moor2015}. The far-IR SED is well-characterised by
measurements from \emph{Spitzer}/MIPS \citep{Chen2014} and \emph{Herschel} PACS and SPIRE \citep{Pawellek2014, Moor2015}. The fit to the SED gives a single component disc
with a dust temperature of 46\,K, a $\beta$ value of 0.5 to 1.4, and an estimated dust mass of 0.19\,M$_{\earth}$.

\subsection{Vega (HD 172167, $\alpha$ Lyr)}
\label{sec:vega}

\begin{figure}
\includegraphics[width=85mm]{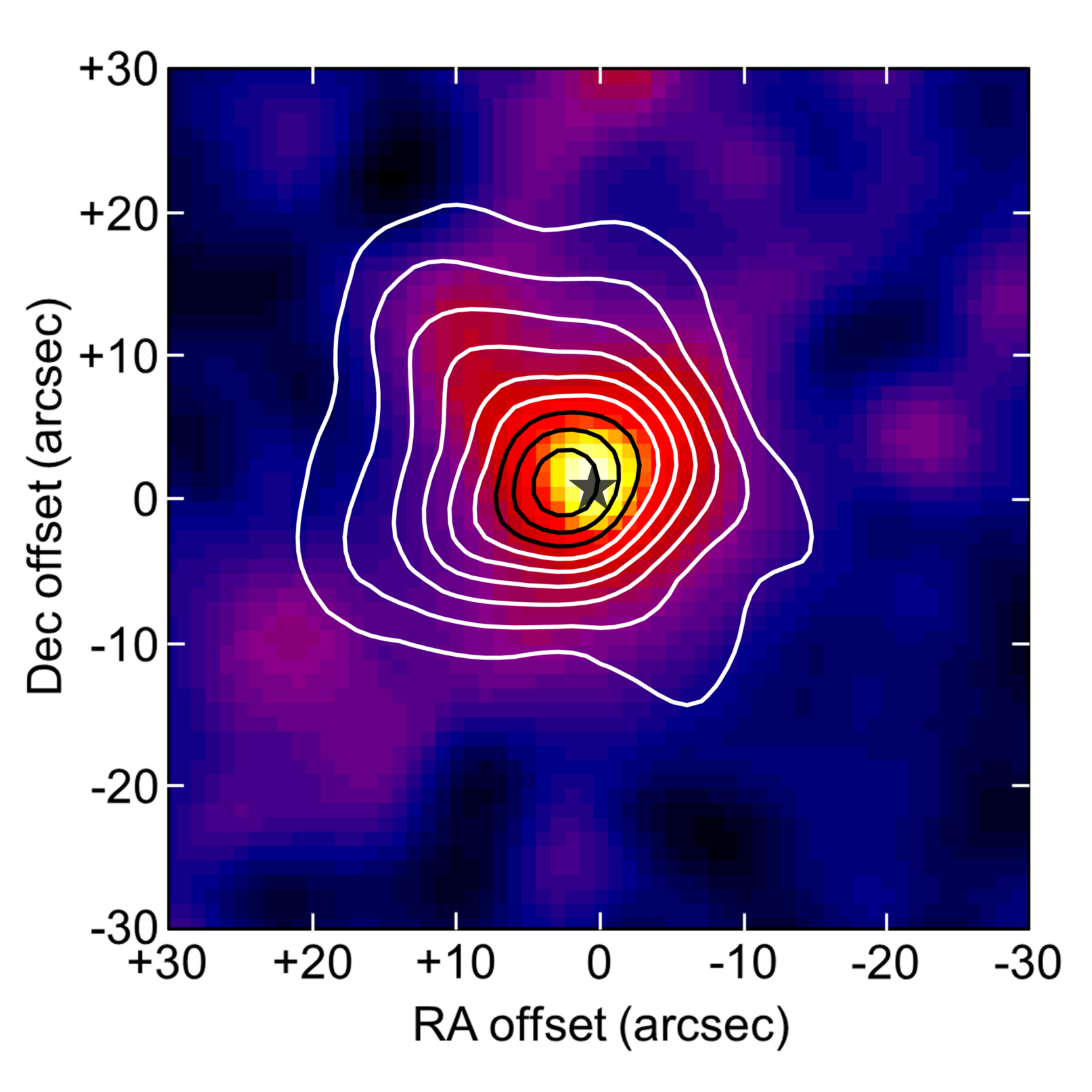}

\caption{The 450\,$\umu$m S/N image from observations of Vega ($\alpha$ Lyr; HD 172167) with contours from the 850\,$\umu$m image overlaid. The contours are scaled from
$-$3\,$\sigma$ to the maximum S/N in the image. The contours start at $-$3\,$\sigma$ (dashed white) and then solid colours from 4\,$\sigma$ to the maximum in 1.5$\sigma$ steps.
The star symbol represents the position of the star with respect to the disc.}

\label{fig:vega}
\end{figure}

Vega (HD 172167) is a luminous A0 star at a distance of 7.7\,pc with an estimated age of 700\,Myr with a range extending from 625\,Myr to 850\,Myr \citep{Monnier2012}. The SONS 
image at 850\,$\umu$m reveals a well-resolved, largely circularly-symmetric, smooth structure, with the peak in emission offset from the proper motion-corrected stellar position 
by only 2\arcsec\ (Fig.~\ref{fig:figureA11}d). The peak flux is 13.7 $\pm$ 0.8 mJy/beam, whilst the integrated flux, measured within a 50\arcsec\ diameter aperture, is 34.4 
$\pm$ 1.4\,mJy. Both of these flux measurements are approximately 25\% lower than the original SCUBA measurements \citep{Holland1998}, and just outside the threshold of the 
estimated measurement errors. Emission was also detected with high S/N at 450\,$\umu$m with an integrated flux density of 229 $\pm$ 14 mJy (Fig.~\ref{fig:vega}). Interpreting 
the structure as a disc about the star, gives deconvolved fitted radii, based on the measured 450\,$\umu$m and 850\,$\umu$m images, of 9.5 and 17.5\arcsec\ ($\sim$73 and 135\,au), 
respectively. 

\vskip 1mm

The original SCUBA 850\,$\umu$m result suggested a more clumpy structure than seen in the new SONS images, with the main peak being offset from the star by $\sim$6\arcsec\ 
\citep{Holland1998}. With approximately 13 years separating the observations it is unlikely that there would be any significant change to the physical structure of the disc that 
might affect the offset. Both the SONS 450\,$\umu$m and 850\,$\umu$m images appear to be slightly elongated in the north-east direction, and it is possible that this extension 
is caused by another source. Given the proper motion of the star over the period between the SCUBA and SCUBA-2 observations (+2, +3.6\arcsec\ in RA and dec, respectively) a 
possible scenario is that the north-east ``clump'', seen in the original SCUBA 850\,$\umu$m image, is a background object (see Section~\ref{sec:background_galaxies}). There are, 
however, issues with this explanation in terms of the relative brightness of the peaks, and it is perhaps more plausible that the difference is due to a combination of 
statistical errors and pointing shifts from the two datasets (see Section~\ref{sec:offsets} for a further discussion of offsets).

\vskip 1mm

The Vega debris disc has been extremely well-studied in the near-IR to millimetre region since the original discovery of its large infrared excess by \emph{IRAS} 
\citep{Aumann1984}. Interferometric observations at 1.3\,mm detected dust peaks offset from the star by 8--14\arcsec\ that were speculated to be part of a ring of emission at a 
radius of 60--95\,au \citep{Koerner2001, Wilner2002}. Multi-band imaging using \emph{Spitzer}/MIPS in the mid-far IR revealed a resolved, smooth disc, devoid of any clumpiness 
\citep{Su2005}. The disc also appeared to be much larger than originally thought, with measurements at 160\,$\umu$m suggesting a radius as large as 800\,au. Further observations 
with the CSO at 350/450\,$\umu$m appeared to show a ring-like morphology with inhomogeneous azimuthal structure, having inner and outer radii of 6.9 and 13.9\arcsec\ ($\sim$53 
and $\sim$107\,au), respectively \citep{Marsh2006}. These observations tended to support a hypothesis that the disc changes from a smooth, axisymmetric structure in the infrared 
to a more clumpy structure in the millimetre, perhaps with an intermediate structural state governed by a grain population of around 1\,mm in size at short submillimetre 
wavelengths. Theories to explain such a structure included a collision between two massive planetesimals (statistically unlikely) or that the clumps (dominated by grains of 
1\,mm and larger) are trapped in resonance with a Neptune-mass planet, whilst smaller grains are perturbed by radiation pressure and have a more uniform distribution in the disc 
\citep{Wilner2002, Wyatt2006, Marsh2006}.

\vskip 1mm

\emph{Herschel} observations with PACS and SPIRE brought into question the clumpy nature of the disc at submillimetre wavelengths. All images between 70 and 500\,$\umu$m show 
a well-resolved, but largely smooth disc, leading to the conclusion that the disc is steady-state in nature \citep{Sibthorpe2010}. The star-subtracted \emph{Herschel} images at 
70 and 160\,$\umu$m show that the peak surface brightness of the disc occurs at a radius of $\sim$11\arcsec\ ($\sim$85\,au) from the star. Furthermore, observations from SMA, 
CARMA and GBT at wavelengths of 870\,$\umu$m, 1.3\,mm and 3\,mm, respectively, do not reveal any clumpy structure on the size-scale expected from previous observations 
\citep{Hughes2012}. Analysis of these more recent interferometric results, in particular, demonstrate that the observations are consistent with a smooth disc having an inner 
radius of between 20 and 100\,au and a broad width estimated at $>$50\,au.

\vskip 1mm

Modelling of the well-sampled SED of Vega suggests a two component fit to the overall IR excess. The fit to the warmer, inner component is consistent with the analysis of 
\emph{Spitzer}/IRS and 24\,$\umu$m, and MSX mid-infrared photometry, carried out by \citet{Su2013}, in which the existence of a warm, unresolved ``asteroid belt'' at a radius of 
$\sim$14\,au was postulated. The colder component is modelled as a disc/belt with a dust temperature 46\,K, and a $\beta$ value in the range of 1.2 to 1.6, with an estimated dust 
mass of 0.011\,M$_{\earth}$ derived from the 850\,$\umu$m flux. The derived dust temperature for the cold disc is also in good agreement with previous estimates 
\citep{Su2005}. The relatively high $\beta$ range means that the equivalent disc radius, assuming blackbody emission, is 260\,au, this being significantly lower than the values 
measured from the images. Since the measured disc radius at 450\,$\umu$m is also less than that at 850\,$\umu$m by almost a factor of two, one possibility to explain this 
discrepancy is a lack of sensitivity in the SONS observations to low-level, extended emission, i.e., the overall emission extends well beyond the lowest contours of the current 
maps, possibly to several hundred au. This is further discussed in Section~\ref{sec:disc_morphology}.

\vskip 1mm

The overall conclusion, supported by the new SONS images, is that the Vega debris disc is largely smooth and circularly symmetric at infrared to millimetre wavelengths. Vega is, 
however, certainly a complex system, most likely containing an inner ``asteroid belt'' analogue at a radius of $<$14\,au, in addition to a smooth, wide outer disc, 
perhaps containing another belt at 30 to 100\,au, as well as a diffuse halo extending to many hundreds of au.

\subsection{HD 181327}
\label{sec:hd181327}

HD 181327 is an F5 star and a member of the $\beta$ Pictoris moving group, at a distance of 52\,pc with an estimated age of 23\,Myr $\pm$ 3\,Myr \citep{Mamajek&Bell2014}. At 
850\,$\umu$m the SONS image shows a bright, unresolved peak, well-centred on the stellar position with a flux density of 23.6 $\pm$ 3.4\,mJy (Fig.~\ref{fig:figureA12}a). The 
relatively high noise level is due to the short integration time (1\,hr) and the low declination of the source. Interpreted as a disc about the star, the upper limit to the 
radius from the 850\,$\umu$m image is 390\,au. The LABOCA camera on APEX measured a peak flux of 24.2\,mJy/beam at 870\,$\umu$m \citep{Nilsson2009}, consistent with the 
850\,$\umu$m SONS result. The integrated flux from the LABOCA measurement, however, is far higher at 51.7 $\pm$ 6.2\,mJy, which suggests the disc is well resolved. This is 
contrary to the result from the SONS 850\,$\umu$m image.

\vskip 1mm

The debris disc around HD 181327 has been well-studied in the infrared through to the millimetre. The disc was first resolved by \emph{HST}/NICMOS and modelling revealed a 
face-on ring peaking in intensity at a radius of 1.7\arcsec\ ($\sim$86\,au) with a width of 0.7\arcsec\ ($\sim$36\,au), inclined at 32\degree\ to the plane of the sky and at a 
PA of 107\degree\ \citep{Schneider2006, Stark2014}. More recently, the disc has been resolved using ALMA at 1.3\,mm revealing a circular ring with the emission radially confined 
between 1.0 and 2.4\arcsec\ ($\sim$50 -- 125\,au) with a peak intensity occurring at a radius of 1.7\arcsec\ ($\sim$86\,au) from the star at a PA of 99\degree\ 
\citep{Marino2016}. The ALMA observations also detected $^{12}$CO(2--1) emission around an F star for the first time.

\vskip 1mm

Both the \emph{HST}/NICMOS and ALMA results suggest the existence of one or more compact, background objects that could have contaminated the LABOCA photometry, leading to an 
artificially high flux estimate for the disc. There are also some low-level ($<$\,3$\sigma$) features in the vicinity of the disc in the SONS 850\,$\umu$m image, which could 
provide extra evidence to support this hypothesis. For the modelling of the SED it is therefore assumed that the peak flux value from LABOCA of 24.2\,mJy/beam at 870\,$\umu$m is 
more representative of a true disc flux (although the SED in Fig.~\ref{fig:figureA11}a shows the integrated flux measurement this is not used in the fit). The SED is modelled 
as a two component disc with the warmer element having a dust temperature of 95\,K and located at a radius of 16\,au from the star (assuming the grains behave like perfect 
blackbodies). Photometry from \emph{Spitzer}/MIPS at 24, 70 and 160\,$\umu$m \citep{Chen2008,Chen2014}, and \emph{Herschel}/PACS observations (archival data) help constrain 
the mid-far IR part of the SED, in addition to a 7\,mm photometric point from the VLA \cite{MacGregor2016a}. Fitting the SED submillimetre points using the SONS result at 
850\,$\umu$m suggests a cold disc component with a dust temperature of 63\,K, a $\beta$ value tightly constrained between 0.4 and 0.6, and an estimated dust mass of 
0.25\,M$_{\earth}$. The total flux measured of 7.9 $\pm$ 0.2\,mJy, measured by ALMA at 1.3\,mm, is also consistent with the fit to the SED \citep{Marino2016}.

\vskip 1mm

\subsection{HD 182681}
\label{sec:hd182681}

HD 182681 is a luminous B8 star at a distance of 70\,pc with an estimated age of 145\,Myr \citep{Chen2014}, but is sometimes referred to as a pre-main-sequence star with an
age as young as 19\,Myr, but also as old as 208\,Myr \citep{Moor2015}. The 850\,$\umu$m image shows a unresolved disc, just offset from the stellar position, with a flux density of
6.8 $\pm$\,1.2 mJy (Fig.~\ref{fig:figureA12}b). There is a second source offset by 19\arcsec\ to the south-east, which is likely to be a background object, although there is a
low-level ($<$3$\sigma$ significance) ridge running approximately north-west to south-east. Assuming only the central peak is part of a disc structure, the upper limit to the
disc radius from the 850\,$\umu$m image is 525\,au.

\vskip 1mm

The disc was resolved by \emph{Herschel}/PACS with the 160\,$\umu$m image indicating a disc of radius 3.3\arcsec\ ($\sim$230\,au), inclined at 67\degree\ to the line-of-sight
and at a PA of 56\degree\ \citep{Moor2015}. Modelling of the PACS images suggests a structure where the flux density peaks at 159\,au, with inner and outer radii of 52\,au and
263\,au, respectively. The south-east peak is not visible in any of the \emph{Herschel} images, including longer wavelength results from SPIRE at 350\,$\umu$m and 450\,$\umu$m.
Photometry from \emph{Herschel} also helps to constrain the mid-far IR part of the SED \citep{Pawellek2014,Moor2015}. Fitting the SED suggests a single-component disc with a
dust temperature of 80\,K, a $\beta$ value in the range of 0.2 -- 1.0, and an estimated dust mass of 0.104\,M$_{\earth}$.

\subsection{HD 191089}
\label{sec:hd191089}

HD 191089 is an F5 star and a member of the $\beta$ Pictoris moving group. It lies at a distance of 52\,pc with an estimated age of 23\,Myr \citep{Mamajek&Bell2014}. The SONS 
survey image at 850\,$\umu$m reveals a compact, unresolved structure, with a flux density of 4.9 $\pm$ 0.9\,mJy (Fig.~\ref{fig:figureA12}c). The peak emission is offset from 
the star position by $\sim$6\arcsec, i.e., greater than the maximum offset of 3.5\arcsec, caused by statistical uncertainties and expected pointing errors (see 
Section~\ref{sec:offsets} for a discussion of offsets). Based on other observations (see below) it is believed that the observed structure is indeed a disc associated with the 
star, but there exists the possibility that there could be some contamination from a background source to account for the offset (see Section~\ref{sec:background_galaxies}). 
The upper limit to the disc radius from the observed 850\,$\umu$m image is 395\,au.

\vskip 1mm

The disc has been resolved from observations made with the T-ReCS instrument on the Gemini South telescope at 18.3\,$\umu$m, which show the structure extending to a maximum radius 
of 0.85\arcsec\ ($\sim$90\,au) at position angle of 80\degree\ \citep{Churcher2011}. Modelling indicates that the emission arises from a dust belt with radius 28\,au to 90\,au from 
the star, inclined at 35\degree\ to edge on, and with a central cavity largely devoid of emission. The disc has also been imaged in scattered light using the \emph{HST}/NICMOS, 
revealing a ring-like structure that peaks in scattered light at $\sim$73\,au, consistent with the mid-IR image \citep{Soummer2014}. The far-IR part of the SED is well characterised 
by measurements from \emph{Spitzer}/MIPS \citep{Chen2014}, the CSO SHARC-II instrument \citep{Roccatagliata2009}, and \emph{Herschel} PACS and SPIRE (archival data). The fit to the 
SED gives a single component disc with a dust temperature of 89\,K, a $\beta$ value in the range 0.3 to 2.3, and an estimated dust mass of 0.037\,M$_{\earth}$.

\subsection{AU Microscopii (HD 197481)}
\label{sec:aumic}

\begin{figure}
\includegraphics[width=85mm]{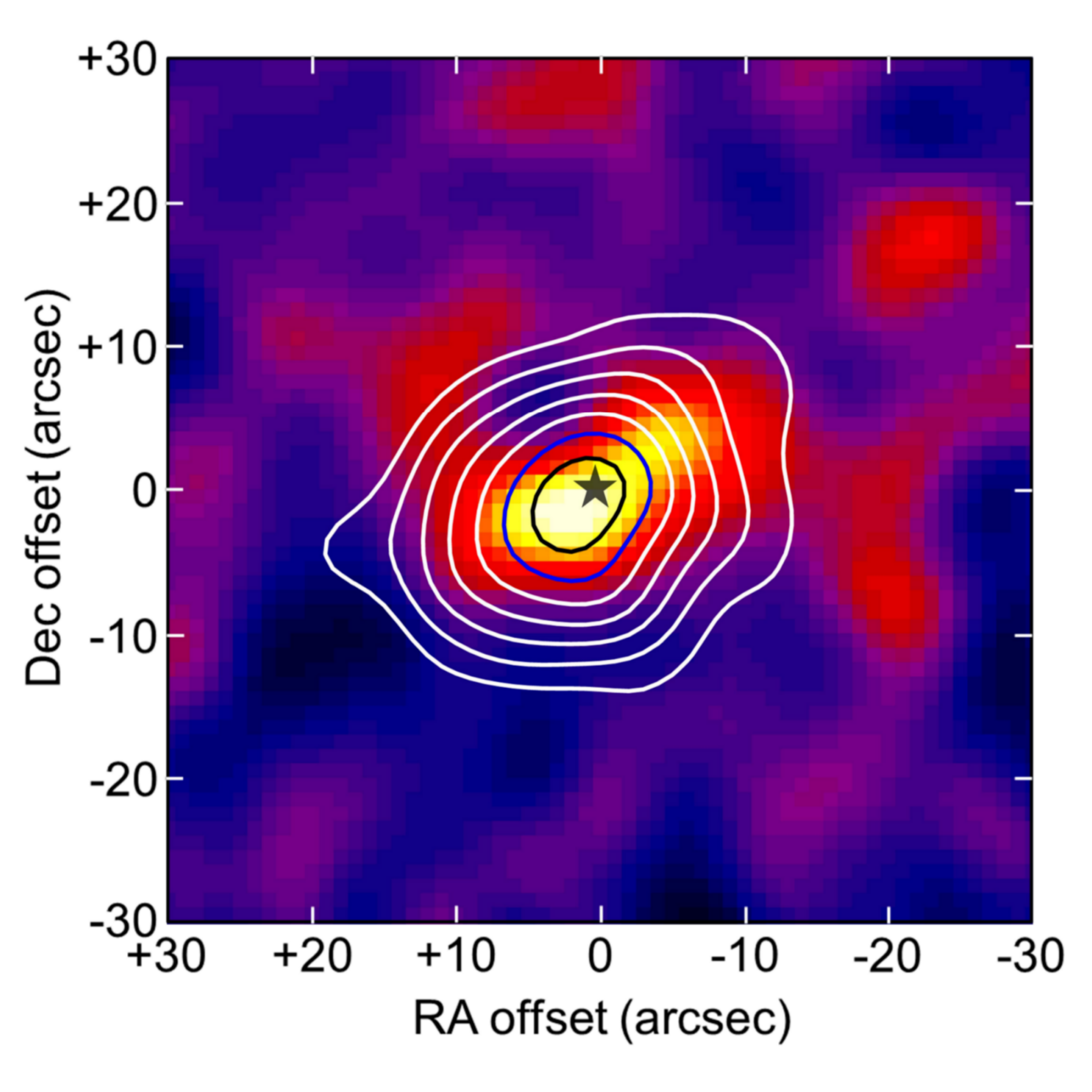}

\caption{The 450\,$\umu$m S/N image from observations of HD 197481 (AU Mic) with contours from the 850\,$\umu$m image overlaid. The colours are scaled from $-$3$\sigma$ to the
maximum S/N in the image. The contours start at $-$3$\sigma$ (dashed white) and then solid colours from 4$\sigma$ to the maximum in 2$\sigma$ steps. The star symbol represents
the position of the star with respect to the disc.}

\label{fig:aumic}
\end{figure}

AU Mic (HD 197481) is a young M1 star and a member of the $\beta$ Pictoris moving group, at a distance of 10\,pc with an estimated age of 23\,Myr \citep{Mamajek&Bell2014}. The 
disc appears to be unresolved at 850\,$\umu$m with a flux density of 12.9 $\pm$ 0.8\,mJy (Fig.~\ref{fig:figureA12}d), despite apparently showing a slight extension roughly in 
the north-west to south-east direction \citep{Matthews2015}. Indeed, this extension is confirmed with 450\,$\umu$m observations that reveal a disc of deconvolved radius 
7.1\arcsec\ ($\sim$70\,au) at a PA of $\sim$127\degree\ (as shown in Fig.~\ref{fig:aumic}, with details of the disc size given in Table~\ref{tab:table4}). The 450\,$\umu$m 
integrated flux, measured within a 40\arcsec\ diameter aperture centred on the star, is 57 $\pm$ 9\,mJy, consistent with previous photometry using JCMT/SCUBA at 450\,$\umu$m 
and 850\,$\umu$m \citep{Liu2004} and CSO/SHARC-II at 350\,$\umu$m \citep{Chen2005}.

\vskip 1mm

The far-IR excess from AU Mic was first detected by \emph{IRAS} \citep{Moshir1990}, and since then it has been studied across the spectrum from the optical to millimetre, with 
much of the interest focusing on the fact that it was the first M star to show evidence of a debris disc. Scattered light images of the disc from \emph{HST}/ACS, the second of 
which to be resolved at optical wavelengths, revealed an edge-on disc extending to a radius of 10\arcsec\ ($\sim$100\,au) at a PA of 128\degree\ \citep{Krist2005}. The disc was 
also resolved from \emph{Herschel}/PACS observations at 70\,$\umu$m and 160\,$\umu$m, confirming the wide edge-on belt extending roughly from 1\arcsec\ to 4\arcsec\ ($\sim$9\,au 
$-$ 40\,au) at a PA of 135\degree\ \citep{Matthews2015}, somewhat less in size than the SONS 450\,$\umu$m image shows. The \emph{HST} results, and those from \emph{Herschel} at 
70\,$\umu$m, suggest the presence of an extra component of small grains (sometimes referred to as a ``halo''), extending beyond the radius of the main AU Mic belt, possibly to a 
radius of 140\,au \citep{Matthews2015}.

\vskip 1mm

The disc was first resolved at millimetre wavelengths by the SMA, revealing the broad belt with the emission peaking at a radius of 3.5\arcsec\ ($\sim$35\,au) at a PA of 
130\degree\ \citep{Wilner2012}. Subsequent imaging with ALMA at 1.3\,mm shows a 10:1 aspect ratio for the edge-on belt, extending from a radius of 0.9\arcsec ($\sim$9\,au) 
outwards to 4\arcsec\ ($\sim$40\,au), at a PA of 128\degree\ \citep{MacGregor2013} in agreement with the \emph{Herschel} results. The ALMA image was modelled as a narrow 
``birth-ring'' or ``parent belt'' of planetesimals at 40\,au \citep{MacGregor2013}. The SED is well-characterised in the far-IR with additional photometry from 
\emph{Herschel}/SPIRE \citep{Matthews2015}, and is fit by an almost pure black body spectrum ($\beta$ = 0 -- 0.3; see Section~\ref{sec:spectral_slopes} for further discussion) 
at a dust temperature of 50\,K, and a estimated dust mass of 6$\times$10$^{-3}$\,M$_{\earth}$. The radius of the disc determined from the fit to the SED, and assuming perfect 
blackbody emission from the dust grains, is 9\,au, consistent with the inner edge estimates from ALMA imaging.

\subsection{HD 205674}
\label{sec:hd205674}

HD 205674 is an F0 star and possible member of the AB Doradus moving group lying at a distance of 52\,pc with an estimated age of 130\,Myr in a range of 110\,Myr to 150\,Myr 
\citep{Barrado2004}. The image at 850\,$\umu$m shows that the the emission is largely concentrated in two peaks running in a line from north to south 
(Fig.~\ref{fig:figureA13}a). The image presented in this paper benefits from a considerable increase in integration time over that published in \citet{Panic2013}, which 
clearly showed a peak to the south offset by 10\arcsec\ from the star. The main dust peak has a flux of 4.0 $\pm$ 0.7\,mJy/beam, and is offset by $\sim$6\arcsec\ from the 
stellar position, greater than the maximum offset of 3.2\arcsec\ expected due to statistical uncertainties and pointing errors (see Section~\ref{sec:offsets}). The total flux, 
including the second peak to the south, and measured in a 40\arcsec\ diameter aperture centred between the two peaks, is 7.5 $\pm$ 1.1\,mJy. Both the offset of the main peak 
from the star and the elongated structure are difficult to explain. The observed structure certainly hints at two distinct peaks, perhaps suggesting the peak to the south 
could be a background source.

\vskip 1mm

\begin{figure}
\includegraphics[width=85mm]{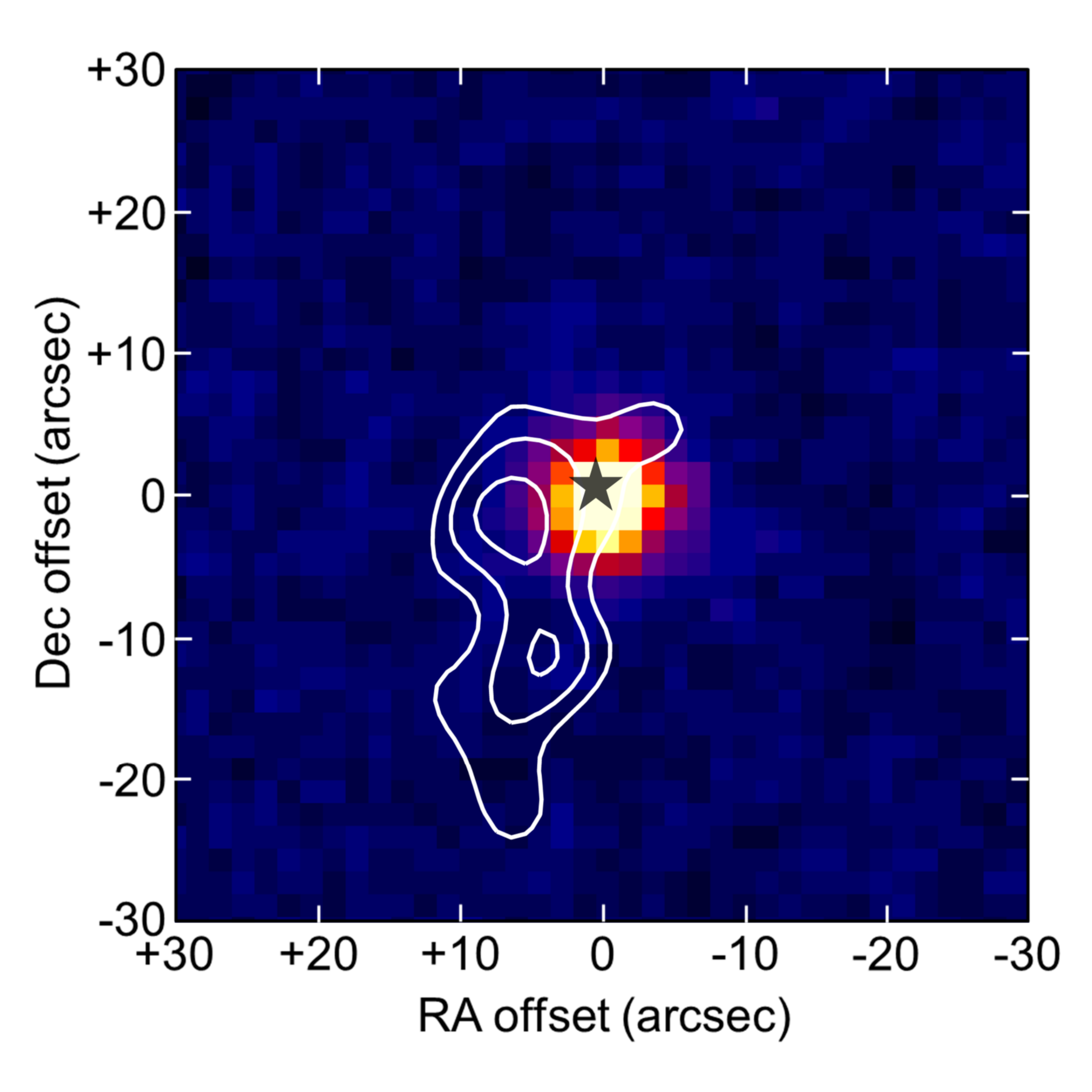}

\caption{\emph{Herschel}/PACS 100\,$\umu$m image of HD 205674 with the SONS survey 850\,$\umu$m contours overlaid. The contours start at 3$\sigma$ and increase in 1$\sigma$ 
steps. The \emph{Herschel} image is taken from the \emph{Herschel} Science Archive.}

\label{fig:hd205674_herschel}
\end{figure}

\emph{Herschel}/PACS imaging at 100 and 160\,$\umu$m reveals an emission peak that is well-centred on the star position, and does not show any evidence of an extension or second 
source to the south (see Fig.~\ref{fig:hd205674_herschel}). This strengthens the argument that the the peak seen to the south in the SONS 850\,$\umu$m image is most likely a 
background source, although no such source has been identified from catalogues. Hence, in this paper the emission surrounding HD 205674 is interpreted as an offset disc, 
assuming the southern peak is a background source. The far-IR SED is well constrained by measurements from \emph{Spitzer}/MIPS \citep{Moor2011b} and archival data from 
\emph{Herschel} PACS and SPIRE but the 850\,$\umu$m SONS flux is the only other long wavelength photometric data. Fitting the SED suggests a disc with a dust temperature of 
60\,K, $\beta$ in the range 0.1 to 1.1, and an estimated dust mass of 0.044\,M$_{\earth}$.

\vskip 1mm

\subsection{HD 206893}
\label{sec:hd206893}

HD 206893 is an F5 star, lying at a distance of 38\,pc with an age estimated to be 860\,Myr \citep{Pace2013}, but could be as young as 200\,Myr \citep{Zuckerman&Song2004} or 
as old as 2.1\,Gyr \citep{David&Hillenbrand2015}. The SONS image at 850\,$\umu$m shows a faint, unresolved peak of flux density 4.5 $\pm$ 1.1\,mJy, offset from the stellar 
position by 4\arcsec\ (Fig.~\ref{fig:figureA13}b). Interpreting the peak as an unresolved disc associated with the star, gives an upper limit for the disc radius of 290\,au. 

\vskip 1mm

\emph{Spitzer} detected excess flux from the star at wavelengths between 3.6\,$\umu$m and 160\,$\umu$m \citep{Moor2006}. The disc has also been marginally resolved by 
\emph{Herschel}/PACS observations at 70\,$\umu$m, revealing a dust ring extending from 1.3 to 5.2\arcsec\ ($\sim$50 -- 200\,au) at a PA of 60\degree\ \citep{Milli2017}. The 
disc has also been imaged in scattered light extending to a radius of 4 -- 5\arcsec\ ($\sim$150 -- 190\,au) at a very similar PA \citep{Milli2017}. Recently, a low-mass 
companion brown-dwarf (mass in the range 24 to 73\,M$_{Jup}$) was also detected using VLT/SPHERE at H-band, lying in an orbit of radius 10\,au from HD\,206893 
\citep{Milli2017}. The detection of a low-mass companion inside a massive debris disc makes this system analogous to other young planetary systems, such as $\beta$ Pictoris 
and HR\,8799.

\vskip 1mm

The SED is reasonably well-sampled in the near-far IR, but only has the SONS 850\,$\umu$m photometric point at long wavelengths. The SED is modelled with a single-temperature 
modified blackbody deriving a dust temperature of 54\,K, a $\beta$ value only loosely constrained between values of 0.5 and 2.7, and an estimated dust mass of 
0.030\,M$_{\earth}$ derived from the 850\,$\umu$m flux.

\subsection{HD 207129}
\label{sec:hd207129}

HD 207129 is a Solar analogue (G2 star) at a distance of 16\,pc with an estimated age of 3800\,Myr but with large uncertainties spanning the range of 1300\,Myr to 7400\,Myr 
\citep{Holmberg2007}. The image at 850\,$\umu$m shows emission in two 4 -- 5$\sigma$ peaks with the star equidistant between the two. The peak to the west has a flux of 7.0 
$\pm$ 1.2\,mJy/beam (Fig.~\ref{fig:figureA13}c), whilst the integrated flux within a 40\arcsec\ diameter aperture, covering both peaks, is 10.8 $\pm$ 1.8\,mJy, indicating the 
disc is marginally resolved by the JCMT beam. The symmetric nature of the peaks about the star suggests a possible inclined toroidal geometry for the structure (perhaps similar 
to Fomalhaut, as described in Section~\ref{sec:fomalhaut}). Interpreting the structure as a disc, extended in a roughly east-west direction, gives a deconvolved radius of 10\arcsec\ 
($\sim$160\,au) at a PA angle of $\sim$115\degree.

\vskip 1mm

Being a relatively close solar analogue, HD\,207129 has been well-studied at most wavelengths between the optical and millimetre. The debris disc was first discovered by 
\emph{IRAS} \citep{Walker&Wolstencroft1988} with subsequent far-IR photometry from \emph{ISO} \citep{Jourdain1999} and \emph{Spitzer} \citep{Krist2010}. The disc is also well 
resolved in scattered light observations revealing a narrow ring of radius 10\arcsec\ ($\sim$160\,au) at a PA of 127\degree\ \citep{Krist2010}. The HD\,207129 disc has been 
resolved by \emph{Herschel}/PACS as part of the DUNES key programme \citep{Marshall2011, Lohne2012}. Modelling of the 100\,$\umu$m PACS image suggests a ring of radius 
8.7\arcsec\ ($\sim$140\,au), consistent with an interpretation of collisional dust being produced in an icy ``exo-Kuiper'' belt. The disc appears inclined at 53\degree\ from 
pole-on with a PA of 122\degree. The SCUBA-2 image at 850\,$\umu$m is in good agreement with both the scattered light and the \emph{Herschel} results. The SED is well-sampled by 
\emph{Spitzer}/MIPS \citep{Trilling2008, Tanner2009, Krist2010} and \emph{Herschel} PACS and SPIRE \citep{Marshall2011, Eiroa2013} in the far-IR. Fitting the SED suggests a disc 
with a single dust temperature of 46\,K, a $\beta$ value of 0.4 -- 1.1, and an estimated dust mass of 0.015\,M$_{\earth}$.

\subsection{HD 212695}
\label{sec:hd212695}

HD 212695 is an old F5 star at a distance of 47\,pc with an estimated age of 2300\,Myr \citep{Trilling2007}. The 850\,$\umu$m image shows unresolved emission, just offset from 
the star, with a flux density of 5.7 $\pm$ 1.1\,mJy (Fig.~\ref{fig:figureA13}d). Interpreting this structure as a disc about the star gives an upper limit to the disc radius from 
the 850\,$\umu$m image of 350\,au. The SED is not well-constrained in the far-IR, only having photometry from \emph{Spitzer}/MIPS at 70\,$\umu$m \citep{Chen2014}. A fit to this 
limited SED suggests the single-component disc is cold with a dust temperature of 35\,K (but with errors of $\pm$ 19\,K) and an estimated dust mass of 0.086\,M$_{\earth}$.

\subsection{39 Pegasi (HD 213617)}
\label{sec:hd213617}

39 Peg is a F1 star at a distance of 50\,pc with an estimated age of 1200 +/- 300 Myr \citep{Moor2011b}. The SONS image at 850\,$\umu$m shows a marginally-detected, unresolved 
peak of flux density 4.6 $\pm$ 1.3\,mJy, offset from the stellar position by 4\arcsec\ (Fig.~\ref{fig:figureA14}a). Interpreting the peak as an unresolved disc associated with 
the star, gives an upper limit for the disc radius of 380\,au.

\vskip 1mm

\emph{Spitzer} detected excess flux from the star at wavelengths up to 70\,$\umu$m \citep{Moor2011b}, but the SONS 850\,$\umu$m photometry is the only data at longer 
wavelengths. Modelling of the SED suggests a single component fit having a dust temperature of 59\,K, a $\beta$ value only loosely constrained between values of 0 and 2.1, and 
an estimated dust mass of 0.049\,M$_{\earth}$ derived from the 850\,$\umu$m flux.

\subsection{Fomalhaut (HD 216956, $\alpha$ PsA)}
\label{sec:fomalhaut_obs}

Fomalhaut (HD 216956) is a luminous A4 star at a distance of 7.7\,pc with an estimated age of 440\,Myr of ($\pm$40\,Myr) \citep{Mamajek&Bell2014}. The SCUBA-2 850\,$\umu$m 
image shows the familiar double-lobed disc structure at a position angle of 160\degree, with the star positioned equidistant between the lobes (Fig.~\ref{fig:figureA14}b). The 
two peaks have fluxes of 26.3 $\pm$ 1.0\,mJy/beam (north-west) and 25.2 $\pm$ 1.0\,mJy/beam (south-east), in good agreement with previous submillimetre results 
\citep{Holland1998, Holland2003}. The integrated flux within a 60\arcsec\ diameter aperture is 91.2 $\pm$ 2.5\,mJy\footnote{The calibration uncertainty at 850\,$\umu$m adds an 
additional 7\,per cent to this error estimate.}, also in good agreement with the previous estimates (Section~\ref{sec:fomalhaut}). As previously discussed in 
Section~\ref{sec:disc_radius} the emission morphology is not well-represented by a Gaussian profile, and the resultant model-subtracted map shows a residual peak at the 
position. Nevertheless, for this paper such an approximation gives a reasonable estimation for the overall size of the structure. Interpreted as a disc/ring about the star, 
the deconvolved major and minor radii from the 2D Gaussian fitting are 19.7\arcsec\ and 7.5\arcsec\ ($\sim$151\,au and $\sim$57\,au), respectively.

\vskip 1mm

The IR excess from Fomalhaut was first discovered and resolved by \emph{IRAS} \citep{Gillet1986,Backman+Paresce1993}. Further indication that the excess could be an extended 
disc structure came from point-by-point photometry using the JCMT \citep{Zuckerman&Becklin1993}. In the late 1990's, the SCUBA camera \citep{Holland1998} provided the first 
true image of the thermal emission from the Fomalhaut debris disc, resolving the double-lobed structure. The SCUBA image suggested a massive torus-like Kuiper Belt surrounds 
the star, with the possibility of one or more planets acting as shepherds to the disc structure. The first optical observations of the disc in scattered light revealed a 
narrow elliptical belt with semi-major and minor axes of 18.5\arcsec\ and 7.6\arcsec\ ($\sim$141\,au and $\sim$58\,au), respectively \citep{Kalas2005}, at a PA of 156\degree\ 
and with an estimated thickness of 15\,au. Modelling of the geometry of the dust belt concluded that the observed features, such as the sharp inner edge and variation in the 
azimuthal brightness, were likely attributable to a planetary system. 

\vskip 1mm

The discovery of the first planet in the system, Fomalhaut b, was also made via \emph{HST} observations with a location just inside the inner radius of the ring 
\citep{Kalas2008}, and subsequent observations have provided further insights on both the main belt structure and the possible planetary system \citep{Kalas2013}. However, the 
orbit of Fomalhaut b is not apsidally aligned with the dust ring, and so it may be the case that additional planets are responsible for shaping the dust morphology. It has 
also been suggested that Fomalhaut b is not actually a planet, but either a dust cloud (perhaps orbiting a planet) \citep{Kennedy&Wyatt2011} or the dusty aftermath of a 
collision between two Kuiper belt-like objects \citep{Galicher2013}.

\vskip 1mm

\emph{Herschel} observations provided high dynamic range images of the dust belt in the far-IR, as well as evidence that the system is remarkably active with dust grains being
produced at a very high rate by a collisional cascade of planetesimals \citep{Acke2012}. Modelling of the 70\,$\umu$m PACS image showed a ring with a mean radius of 18\arcsec\ 
($\sim$138\,au) at a PA of 157\degree, consistent with the submillimetre and optical images. Further constraints on the Fomalhaut planetary system have been obtained by high
angular resolution observations with ALMA of the north-west side of the belt. Observations at 850\,$\umu$m with $\sim$1.5\arcsec\ resolution show a ring that peaks in surface
brightness at a radius of 18.4\arcsec\ ($\sim$142\,au), with a PA of 168\degree. The ring is inclined to the plane of the sky by 66\degree\ and has a width of $\sim$17\,au
\citep{Boley2012}. Moreover, the image showed that almost all of the emission is confined to the ring, with the flux estimates also agreeing with previous submillimetre
measurements at 850\,$\umu$m (for the half of the ring that was observed). More recent observations with ALMA on the entire ring at 1.3\,mm, tightly constrain the radius to
$\sim$136\,au with a width of 13.5\,au \citep{White2016,MacGregor2017}.

\vskip 1mm

\begin{figure}
\includegraphics[width=85mm]{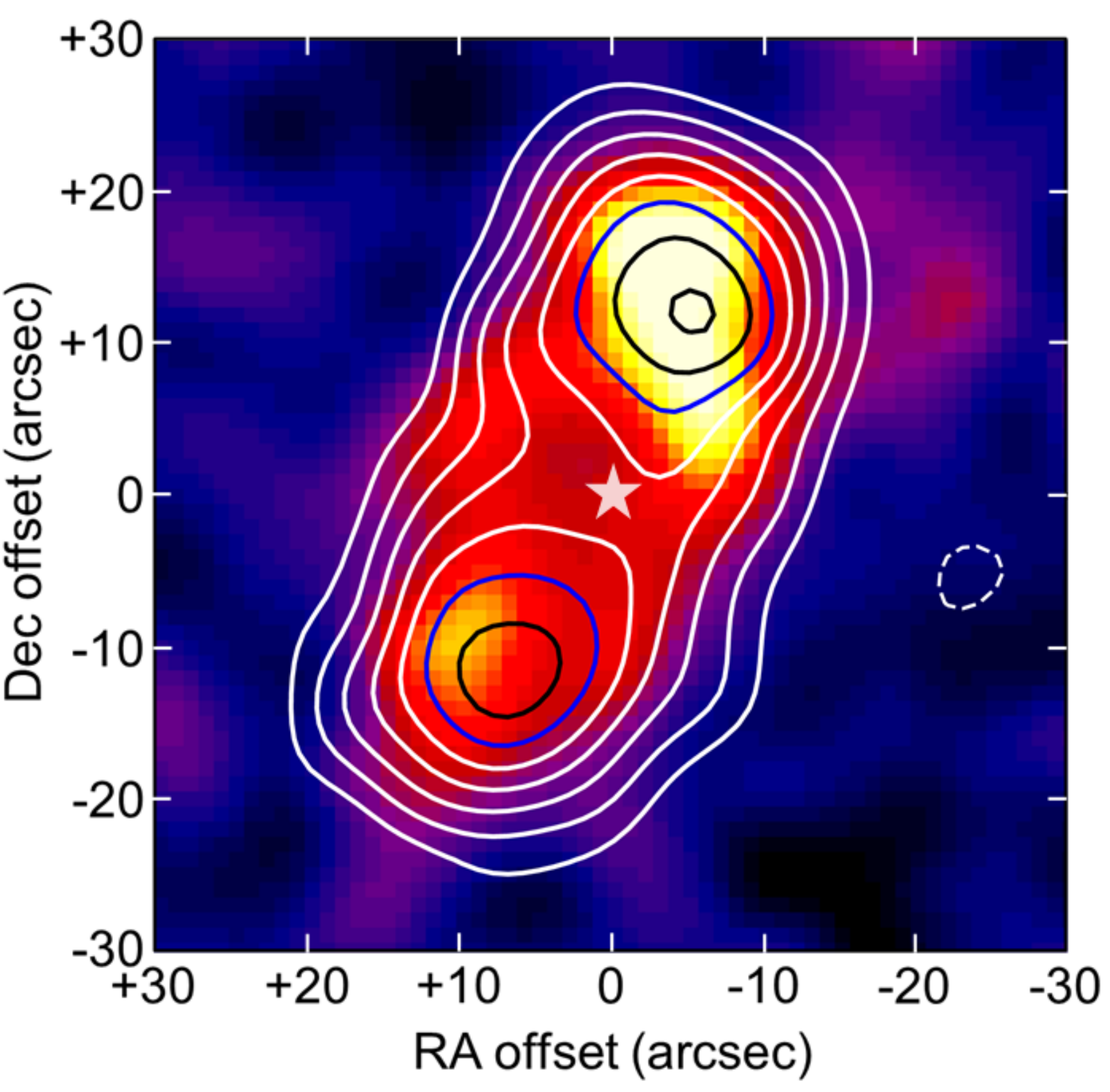}

\caption{The 450\,$\umu$m S/N image from observations of Fomalhaut (HD 216956) with contours from the 850\,$\umu$m image overlaid. The colours are scaled from $-$3$\sigma$ to
the maximum S/N in the image. The contours start at $-$5\,$\sigma$ (dashed white) and then solid colours from 5\,$\sigma$ to the maximum in 5\,$\sigma$ steps. The star symbol
represents the position of the star with respect to the disc.}

\label{fig:fomalhaut_450}
\end{figure}

The Fomalhaut disc was also detected with high S/N at 450\,$\umu$m, and Fig.~\ref{fig:fomalhaut_450} shows the 450\,$\umu$m S/N image with the 850\,$\umu$m contours overlaid. The 
general morphology is consistent with previous SCUBA observations, and there is even evidence of a slight flux excess to the east (e.g., a flux difference of 10\,mJy compared to 
that at the equivalent radius from the star on the west side) as reported in the previous asymmetric description of the disc about the star \citep{Holland2003}. The striking 
difference here, however, is that the new image appears to show a brightness asymmetry between the north-west and south-east lobes at approximately the 3$\sigma$ level. This 
asymmetry is also supported by CSO observations of the Fomalhaut disc \citep{Marsh2005}. The theory of ``apocentre glow'' predicts that a steady-state disc will show an 
over-density of dust at apocentre, due to the Keplerian orbital velocity in an eccentric disc being slower at apocentre rather than at pericentre \citep{Pan2016}. The most recent 
results from high resolution observations with ALMA provide conclusive evidence that the apocentre (north-west peak) has excess submillimetre/millimetre dust 
\citep{MacGregor2017,Matra2017}. The SED in the far-IR and submillimetre is extremely well-sampled with a fit suggesting a dominant cold component disc with a dust temperature of 
41\,K, a $\beta$ value well-constrained in the range of 1.0 to 1.3, and an estimated dust mass of 0.032\,M$_{\earth}$.

\subsection{HR 8799 (HD 218396)}
\label{sec:hr8799}

HR 8799 (HD 218395) is a young luminous A5 star in the Columba Moving Group \citep{Doyon2010, Zuckerman2011}, lying at a distance of 39\,pc with an estimated age of 30\,Myr 
\citep{Torres2008, Marois2010}. The SONS survey image at 850\,$\umu$m reveals a significantly extended structure, with a peak flux of 10.9 $\pm$ 1.0\,mJy/beam, well-centred on 
the stellar position (Fig.~\ref{fig:figureA14}c). The integrated flux, measured in a 60\arcsec\ diameter aperture, is 23.0 $\pm$ 2.0\,mJy which takes in all the structure including 
the extension (possibly a separate background source) to the north-west (see discussion below). The peak flux is also consistent with the previous SCUBA flux estimate of 10.3 $\pm$ 
1.8\,mJy \citep{Williams&Andrews2006}, which was measured in photometry mode and so not sensitive to any extended structure. Fitting the radial extent, but without including the 
north-west peak, gives a deconvolved disc radius of 10\arcsec\ ($\sim$395\,au) at a PA of 71\degree\ (see Table~\ref{tab:table4}).

\vskip 1mm

The HR 8799 system also contains at least four known giant planets, all of which have orbital radii within $\sim$70\,au of the star \citep{Marois2008, Marois2010}. The thermal 
emission from the debris disc was first detected by \emph{IRAS} \citep{Sadakane&Nishida1986}. The disc was first resolved by \emph{Spitzer} at 24\,$\umu$m and 70\,$\umu$m and 
the results of SED modelling suggested there were three distinct components to the debris system \citep{Su2009}. The first is a warm ($\sim$150\,K) dust cloud orbiting within 
planets ``d'' and ``e'' at a radius of $<$0.6\arcsec\ ($\sim$24\,au) from the star. The second is a broad zone of cold ($\sim$45\,K) dust with a sharp inner edge and orbiting 
just outside the outermost planet (``planet b'') at a radius of 2.3\arcsec\ ($\sim$90\,au) but extending out to 7.7\arcsec\ ($\sim$300\,au). The final component is a halo of 
small grains, perhaps originating in the cold belt, but extending from 7.7\arcsec\ ($\sim$300\,au) to at least a radius of 25\arcsec\ ($\sim$1000\,au), and suggesting dust 
grains are prevalent at very large radii from the host star \citep{Su2009}. Subsequent observations by \emph{Herschel} also resolved the disc between wavelengths of 70\,$\umu$m 
and 250\,$\umu$m \citep{Matthews2014} with deconvolved radii of 7.2\arcsec\ and 7.7\arcsec\ ($\sim$284\,au and 300\,au) at a PA of 62\degree, based on the 70 to 160\,$\umu$m 
PACS results. Modelling the images also supported the \emph{Spitzer} findings that the disc consists of three distinct components: a warm inner asteroid belt analogue, a 
planetesimal belt extending from 100\,au to 310\,au and an outer halo reaching a radius of 1500\,au from the star. Both the \emph{Spitzer} and \emph{Herschel} observations also 
show a compact source some 15\arcsec\ to the north-west, coincident with the extension seen in the 850\,$\umu$m image. It is therefore likely that this source is a separate 
background object.

\vskip 1mm

The HR 8799 disc has also recently been observed with ALMA at 1.3\,mm, resolving the planetesimal belt. The emission was modelled as a broad ring at a radius of between 
3.7\arcsec\ and 10.7\arcsec\ ($\sim$145\,au and 429\,au), with a PA of 51\degree\ \citep{Booth2016}. The outermost planet in the system (``planet b'') was expected to be 
responsible for shaping the inner edge of the belt, but the size of the belt seems inconsistent with the planet's orbit. This inconsistency suggests that either the orbit has 
varied over time or there is another (smaller) planet further out from the star. The measured flux density of 2.8\,mJy at 1.3\,mm also seems inconsistent with the well-sampled 
SED, suggesting that the ALMA data is missing significant flux on a scale larger than its primary beam.

\vskip 1mm

\begin{figure}
\includegraphics[width=85mm]{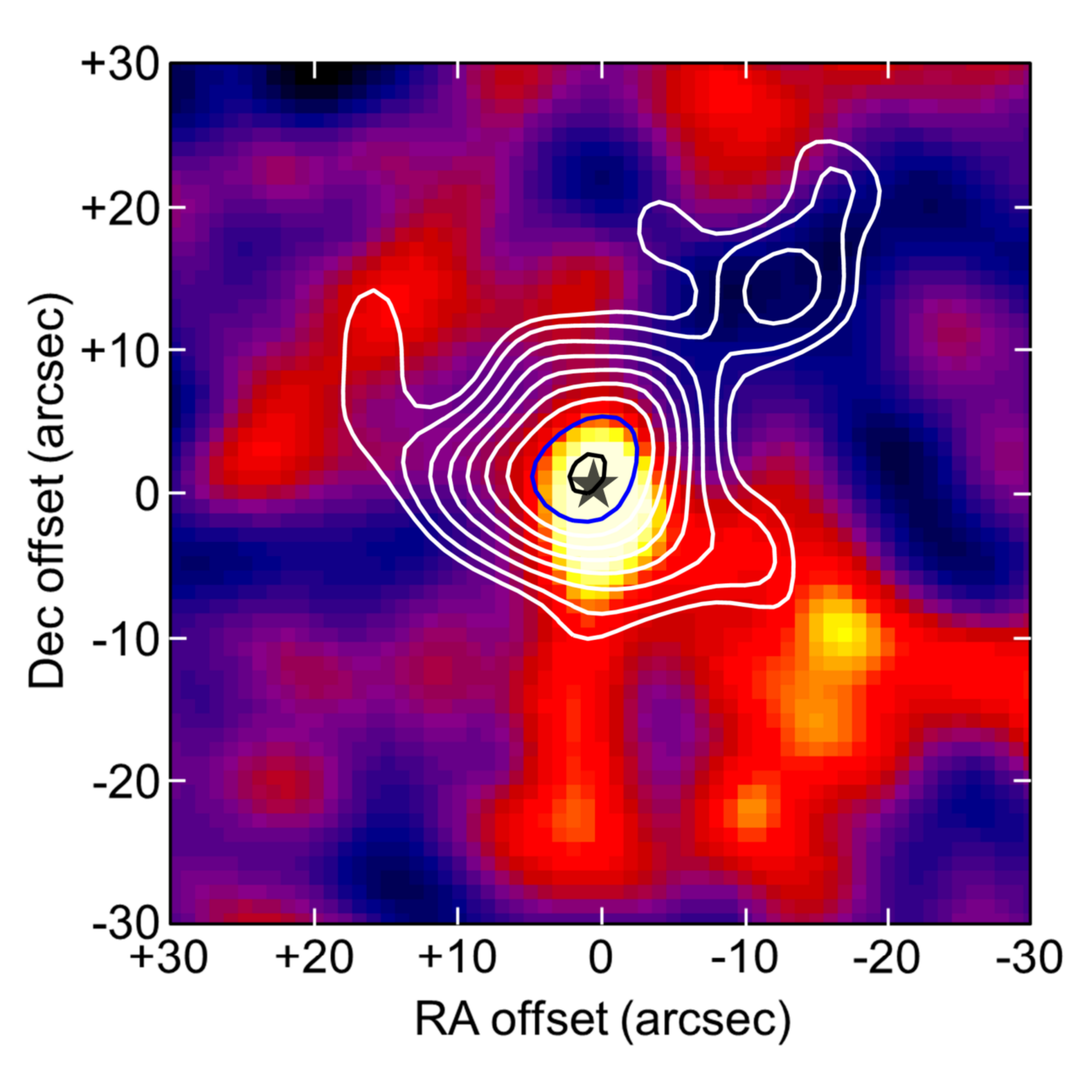}

\caption{The 450\,$\umu$m S/N image from observations of HR 8799 (HD 218396) with contours from the 850\,$\umu$m image overlaid. The colours are scaled from $-$3$\sigma$ to the
maximum S/N in the image. The contours start at $-$3$\sigma$ (dashed white) and then solid colours from 4$\sigma$ to the maximum in 2$\sigma$ steps. The star symbol represents
the position of the star with respect to the disc.}

\label{fig:hr8799}
\end{figure}

The SED in the far-IR is well-sampled, including the aforementioned photometry from \emph{Spitzer} and \emph{Herschel}. The disc is also detected with SCUBA-2 at 450\,$\umu$m 
with a peak flux of 122 $\pm$ 22\,mJy/beam ($\sim$5\,$\sigma$) and Fig.~\ref{fig:hr8799} shows the 450\,$\umu$m image with the 850\,$\umu$m contours overlaid. In common to the 
other (mainly shorter wavelength) observations from \emph{Herschel}, there appears to be significant low-level extended emission seen to the south and south-west at 
450\,$\umu$m, outside of the reasonably compact central source. This wide extent is reflected in an integrated flux of 346 $\pm$ 34\,mJy, measured in a 60\arcsec\ diameter 
aperture centred on the star. This emission is likely due to a known dust cloud in the vicinity of HR 8799 \citep{Matthews2014}. Hence, taking this cloud into account, the 
peak flux value has been adopted as the photometric point at 450\,$\umu$m. At 850\,$\umu$m, the flux of the background source to the north-west (5.6\,mJy) has been removed 
from the total flux estimate (revised value is 17.4 $\pm$ 1.5\,mJy), and it has been assumed there is no contribution from the local dust cloud at this wavelength. Based on 
the modelling of the SED, there appears to be evidence of two components, and fitting to the near to mid-IR photometry suggests an inner disc of 190\,K at a radius of 5\,au 
from the star. The colder component, representative of the planetesimal disc, has a dust temperature of 43\,K, a $\beta$ value in the range 0.6 to 1.9, and an estimated dust 
mass of 0.156\,M$_{\earth}$. Assuming the dust grains to be pure blackbodies implies that the disc would be at a radius of 80\,au from the star.

\vskip 1mm

\section{Discussion}
\label{sec:discussion}

\subsection{Detection rate}

\begin{figure}
\includegraphics[width=85mm]{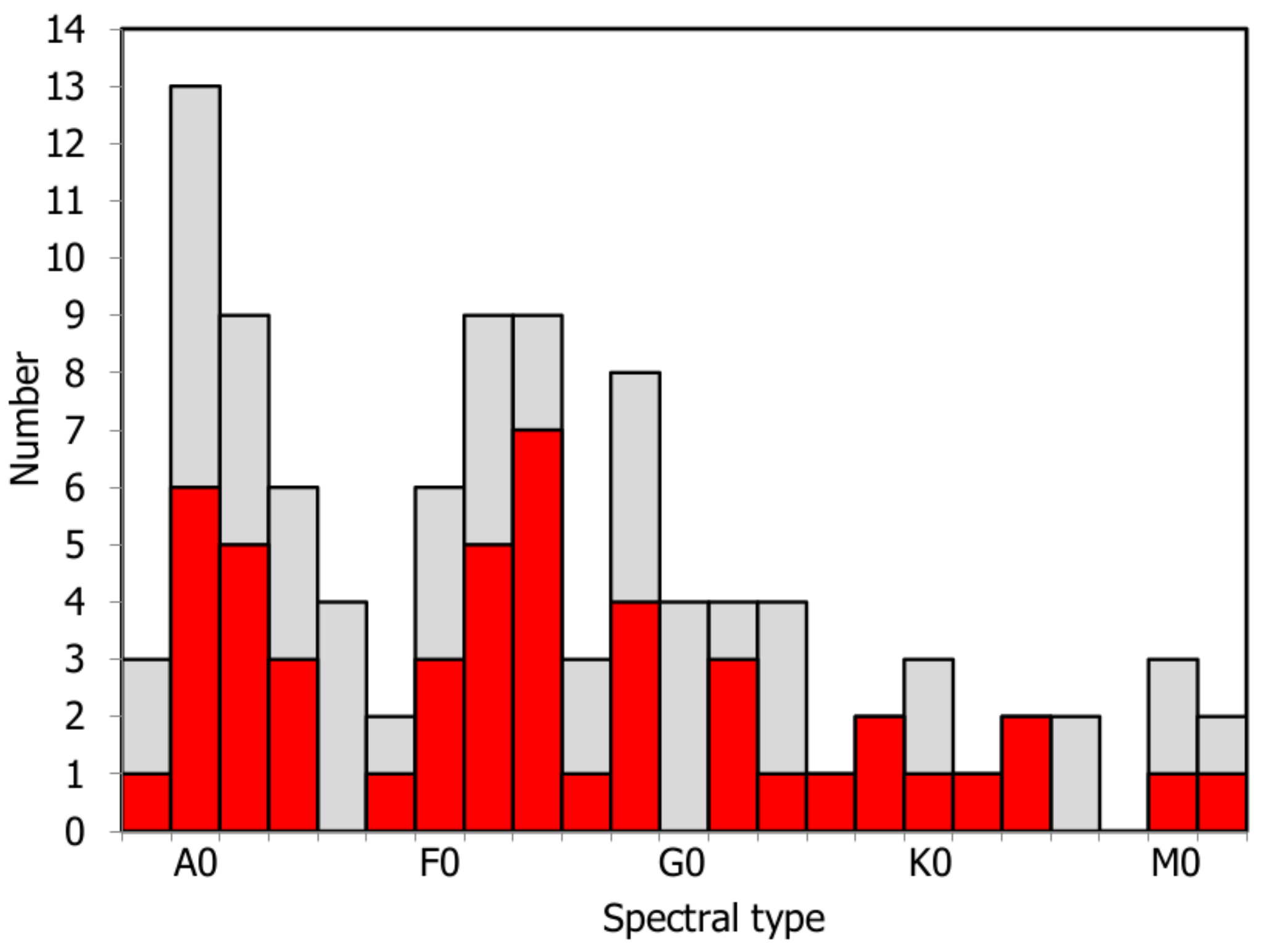}

\caption{The number of discs detected at 850\,$\umu$m (red bars) as a function of host star spectral type. The grey bars represent the non-detections within the survey.}

\label{fig:spectral_type}
\end{figure}

The detection rate for discs at 850\,$\umu$m (3$\sigma$+ peaks) within a half-beam diameter of the stellar position is 49 per cent (i.e., not including the five cases where 
the peak is significantly offset from the star that they are considered to be background objects, namely HD 17093, HD 22179, HD 128167, HD 141378 and HD 150378, and also 
Algol, where the emission is most likely related to the radio variability). Given that all the targets in the sample are known disc hosts, and assuming the fits to the 
photometric points in the SEDs are accurate, it is clear that the SONS observations do not go deep enough for the non-detected disc cases. Fig.~\ref{fig:spectral_type} shows 
that the number of discs detected is largely uniform as a function of spectral type (stellar luminosity) which is to be expected from such a targeted survey. 
Table~\ref{tab:table5} summarises the sample numbers and detection rate by spectral type. Approximately 30 per cent of the targets that were detected at 850\,$\umu$m also 
showed flux excesses at 450\,$\umu$m, i.e., 14 out of the original list of 100 stars showed excess flux at the 5$\sigma$+ level within a 7\arcsec\ radius of the stellar 
position. As discussed for individual targets (Section~\ref{sec:target_discussion}), the 450\,$\umu$m photometry has been used to constrain further the fitting of the SED and 
in some cases offers improved angular resolution ($\sim$7.5\arcsec, or $\sim$10\arcsec\ with smoothing) for extended structures.

\begin{table}

  \centering

  \caption{Summary of the sample numbers and detections (at 850\,$\umu$m) by spectral type. The detections do not include the extreme offset cases (HD 17093, HD 22179, HD 
128167, HD 141378 and HD150378), as well as Algol (HD 19356) where the emission is likely linked to radio variability i.e. not originating from a circumstellar disc. The 
errors in the detection fraction are derived assuming a binomial distribution according to $df = \sqrt{f(1-f)/N}$, where $f$ is the detection fraction and $N$ the number in the 
sample.}

  \label{tab:table5}

  \begin{tabular}{lccc}
\hline

Spectral           &   Number          &     Number           &   Detection                  \\
type               &   in sample       &     detected         &   fraction (per cent)        \\

\hline

B                  &   3               &     1                &   33 $\pm$ 4.7               \\
A                  &   34              &     15               &   44 $\pm$ 5.0               \\
F                  &   35              &     20               &   57 $\pm$ 5.0               \\
G                  &   15              &     7                &   47 $\pm$ 5.0               \\
K                  &   8               &     4                &   50 $\pm$ 5.0               \\
M                  &   5               &     2                &   40 $\pm$ 4.9               \\

\hline
  \end{tabular}

\end{table}

\subsection{Disc morphology and sizes}
\label{sec:disc_morphology}

\begin{figure}
\includegraphics[width=84mm]{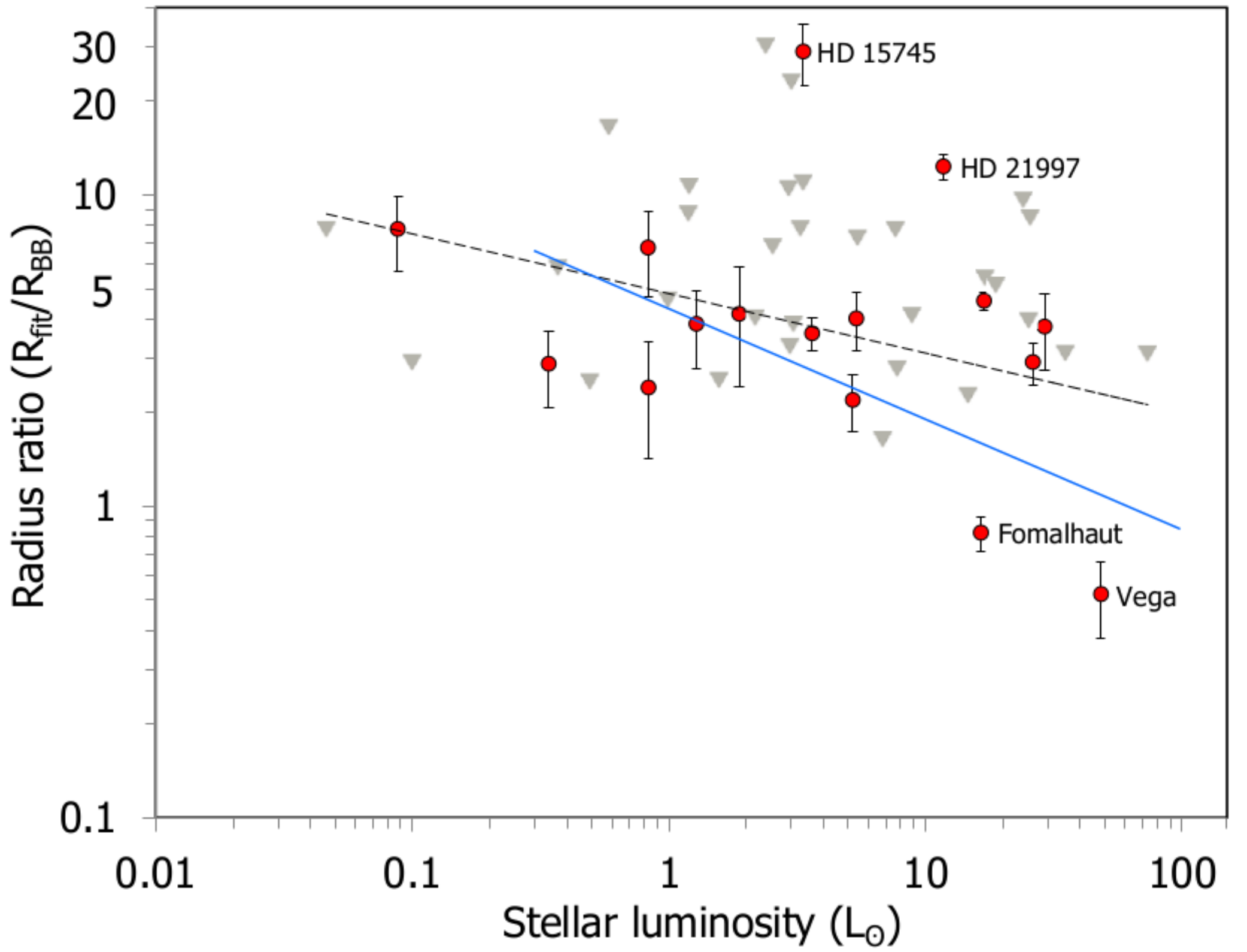}

\caption{The ratio of the measured disc radius ($R_{\rm{fit}}$) to that derived from the SED (assuming the dust grains behave as perfect blackbodies; ($R_{\rm{BB}}$) plotted against 
stellar luminosity. The inverted triangles represent upper limits for the cases in which the disc is unresolved. The dashed line is a fit to the data points for which the disc is 
resolved (i.e. the upper limits are not included in the fit). The solid blue line is based on fitting the observed SED from \emph{Herschel} observations assuming a dust composition 
dominated by ice and silicates \citep{Pawellek+Krivov2015}.}

\label{fig:radius_compare}
\end{figure}

Within the sample, 16 discs have been spatially resolved, with deconvolved measured radii ranging from 40\,au to 800\,au (as shown in Table~\ref{tab:table3} and 
Table~\ref{tab:table4}). If the disc is composed of small grains these will be hotter than their blackbody equivalent, and hence the disc size will be larger than that 
determined from the SED fit, according to equation~\ref{eq:radius} (see Section~\ref{sec:disc_radius}). Larger grains will behave more like blackbodies with the ratio of the 
measured (observed) radius ($R_{\rm{fit}}$) to that derived from the SED fit ($R_{\rm{BB}}$) approaching unity. Fig.~\ref{fig:radius_compare} shows the ratio of 
$R_{\rm{fit}}$/$R_{\rm{BB}}$ plotted against stellar luminosity. It can be seen that the ratio is greater than unity in the vast majority of cases and typically less than 10. 
This behaviour implies the existence of planetesimals at larger radii than those derived by assuming a simple blackbody fit to the SED.

\vskip 1mm

\vskip 1mm

The general trend is for the discs around lower luminosity stars to be larger than expected based on their blackbody dust grain properties. The ratio is seen to decline as 
$\propto L_*^{-0.2}$, as shown in Fig.~\ref{fig:radius_compare}. A number of papers have compared the disc sizes resolved by, for example, \emph{Herschel}, relative to the 
expectation from blackbody emission, and found that, although the ratio was typically less than 4, there was clear evidence of a decline with increasing stellar luminosity 
\citep{Booth2013, Morales2013, Pawellek2014, Pawellek+Krivov2015}. This decline was interpreted as being consistent with the expectation from blowout, where higher luminosity 
stars have removed the smallest grains via radiation pressure, leaving relatively large, blackbody-like dust grains, whilst stars of lower luminosity should have super-heated 
grains. For comparison, Fig.~\ref{fig:radius_compare} also shows the ratio based on fitting the observed SED from \emph{Herschel} observations, assuming a model in which the 
dust composition mixture is dominated by ice and silicates \citep{Pawellek+Krivov2015}. The SONS results support a similar behaviour, although with a larger scatter than seen 
in the \emph{Herschel} data and a lower number of resolved sources. Some of the difference between the SCUBA-2 and \emph{Herschel} results is to be expected due to the different 
methods used to measure the resolved disc sizes. The \emph{Herschel} papers listed above all fit narrow rings to the data, whilst in this paper 2D Gaussians are adopted. In 
cases where the disc is wide, a Gaussian fit will tend to give a radius closer to the outer radius, whereas a narrow ring fit will give a radius close to that dominating the 
emission and hence closer to the equivalent blackbody radius.

\vskip 1mm

There are two prominent (high) outliers in the plot with ratios above a value of 10 (namely HD 15745 and HD 21997). As discussed in Section~\ref{sec:hd15745}, the 
disc around HD 15745 appears to be extended at 850\,$\umu$m, but shows no such morphology at far-IR wavelengths (appearing unresolved in \emph{Herschel}/PACS 100\,$\umu$m and 
160\,$\umu$m observations). The scattered light image \citep{Kalas2007b,Schneider2014} seems to suggest that the disc does extend to a radius close to that seen in the SCUBA-2 
image ($\sim$500\,au), although the scattered light disc has a PA of $\sim$22.5\degree, compared to the $\sim$164\degree\ from the SONS 850\,$\umu$m image. It is unclear what 
could cause such a discrepancy, although it has been hypothesised that it may be related to interstellar winds \citep{Schneider2016}. HD 21997 is only marginally resolved based 
on an extended 3$\sigma$ contour to the south (Section~\ref{sec:hd21997}). Nonetheless, a previous multi-wavelength analysis does suggest the possibility of a disc extending 
to at least 490\,au \citep{Moor2013}, and it is likely that the radius ratio is so high because the disc is very wide and the SED is dominated by warmer grains at the inner edge 
($\sim$50\,au). 

\vskip 1mm

Vega, on the other hand, has a low ratio of 0.52 based on the fitted 850\,$\umu$m image (or 0.28 based on the 450\,$\umu$m image), meaning that the observed disc is 
significantly smaller than that implied assuming the dust grains are radiating as pure blackbodies. As discussed in Section~\ref{sec:vega}, one possibility is that there is 
undetected, low-level emission in the SONS images, extending well beyond the lowest contours of the map, resulting in an underestimate for the measured radius of the disc. It is 
already known that the disc radius could extend to many hundreds of au \citep{Su2005}, and so it is possible that the far-IR emission (which sets the temperature) is dominated 
by halo grains at a radius significantly larger than the main belt (peaking around 200 $-$ 300\,au). This would, however, mean that the halo properties for Vega are different 
than those around other stars (e.g., AU Mic) in that such grains would have to be cooler than their blackbody equivalents. This remains an unexplained issue. 

\vskip 1mm

In summary, the median $R_{\rm{fit}}$ to $R_{\rm{BB}}$ ratio for all the targets is 3.8 (or 3.6 if the high and low outliers are excluded). The greater than unity ratio in the 
vast majority of cases implies the existence of planetesimals at larger radii than those derived from a blackbody fit. There are several anomalies that still need to be 
explained, including both high ratio outliers such as HD 15745 and targets, such as Vega, that have a low ratio.

\vskip 1mm

\begin{figure}
\includegraphics[width=84mm]{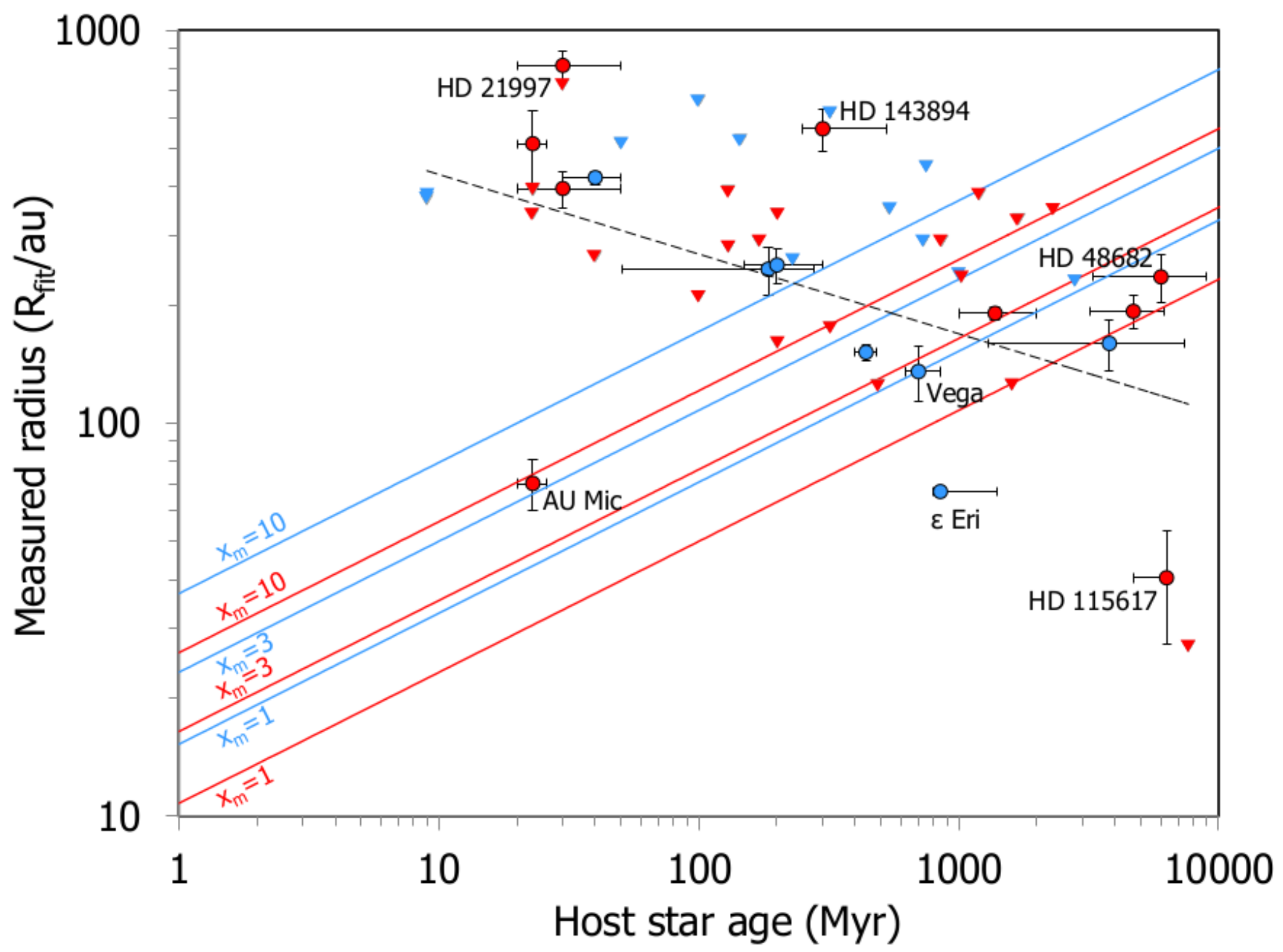}

\caption{The measured radius as a function of host star age for the 16 targets in the SONS survey for which the disc radius was resolved, as represented by the red and blue circles 
(with error bars). The inverted triangles are upper limits to the radius for the cases in which the disc is unresolved. For both the circles and triangles, red symbols represent 
stars in the mass range of 0.3 $-$ 1.5\,M$_{\odot}$, whilst blue covers the 1.5 $-$ 3.3\,M$_{\odot}$ range. The red and blue lines represent self-stirring models computed for the 
host stars with masses of 1\,M$_{\odot}$ (red) and 2\,M$_{\odot}$ (blue) and for three different scaling factors, $x_m$, adapted from \citet{Moor2015}. The dashed line is a fit to 
the resolved disc data points, and tentatively indicates a decline in disc radius as a function of stellar age $\propto$\,$t^{-0.2}$, where $t$ is the host star age.}

\label{fig:radius_age}
\end{figure}

After the gas dissipates from a circumstellar disc, planetesimals need to be stirred to initiate a collisional cascade. \citet{Moor2015} investigate stirring in a number of 
massive discs (many of which are also in this survey). Fig.~\ref{fig:radius_age} shows a tentative decline in the measured radius of the resolved discs in the SONS sample as a 
function of the host star age. The red and blue lines represent ``self-stirring'' models\footnote{To achieve ``self-stirring'' requires the formation of Pluto-sized 
planetesimals.} computed for the host stars with masses of 1\,M$_{\odot}$ (red) and 2\,M$_{\odot}$ (blue) and for three different scaling factors, $x_m$, adapted from 
\citet{Moor2015}, and where $x_m$ is the ratio of the assumed mass of the initial protostellar disc to the minimum mass Solar nebula. The results are also in agreement with 
\citet{Moor2015} in that that self-stirring can explain large discs but only around old systems. For young stars with discs extending out to radii of greater than 
$\sim$100\,au, planet stirring is currently the only explanation for the detection of dust at such large distances from the star. In this survey such an explanation would 
apply to the discs detected around 49 Cet, HD 15745, HD 21997, 44 Ser, $\gamma$ Oph, HD 170773 and HR 8799.

\vskip 1mm

\subsection{Spectral slopes and grain properties}
\label{sec:spectral_slopes}

\begin{figure}
\includegraphics[width=84mm]{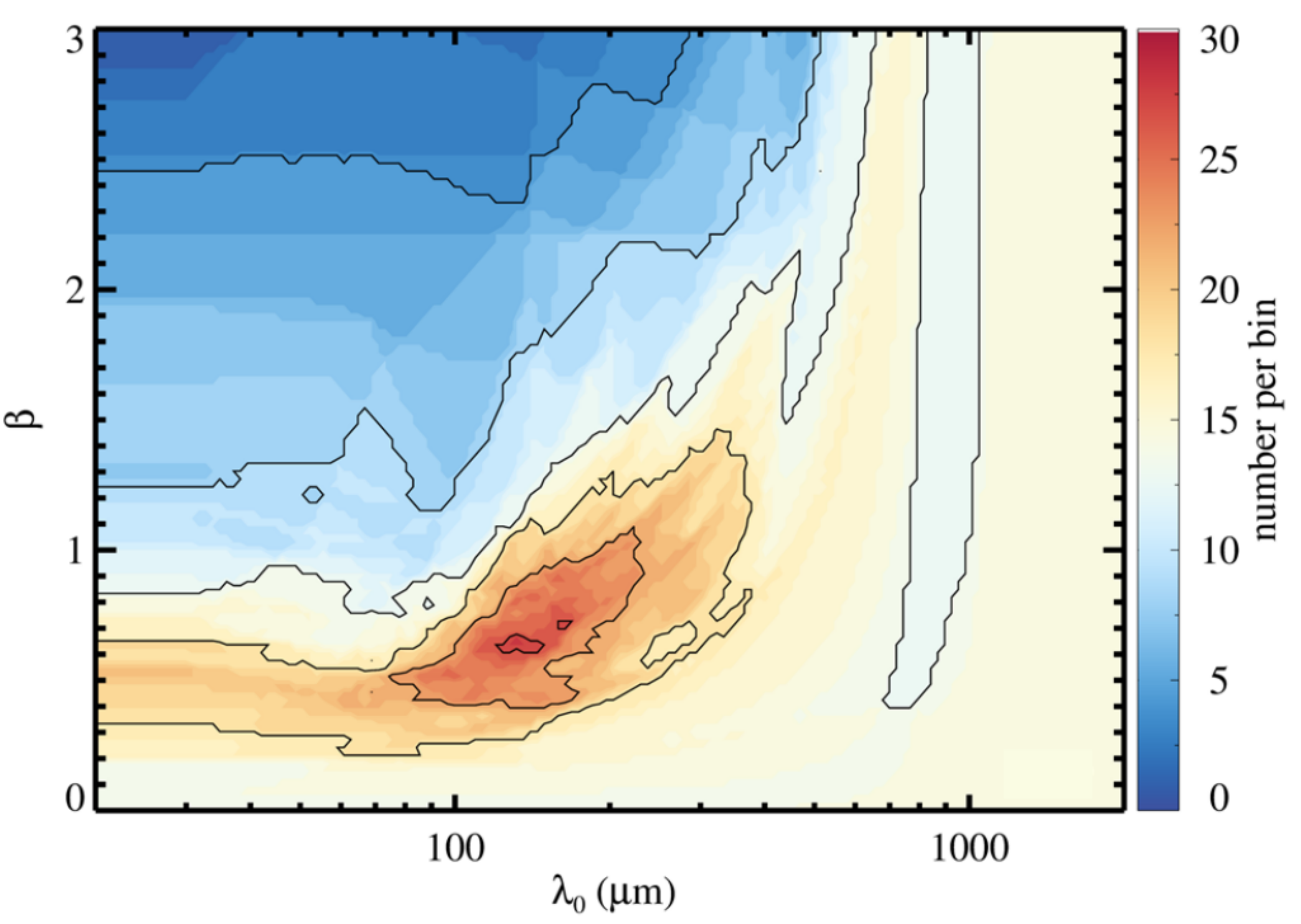}

\caption{The 3$\sigma$ constraints of dust emissivity ($\beta$) plotted against critical wavelength ($\lambda_0$) for all the detections in the SONS survey. The contours are at 
levels of 5, 10, 15, 20, 25 and 30.}

\label{fig:beta_plot}
\end{figure}

\begin{figure}
\includegraphics[width=83mm]{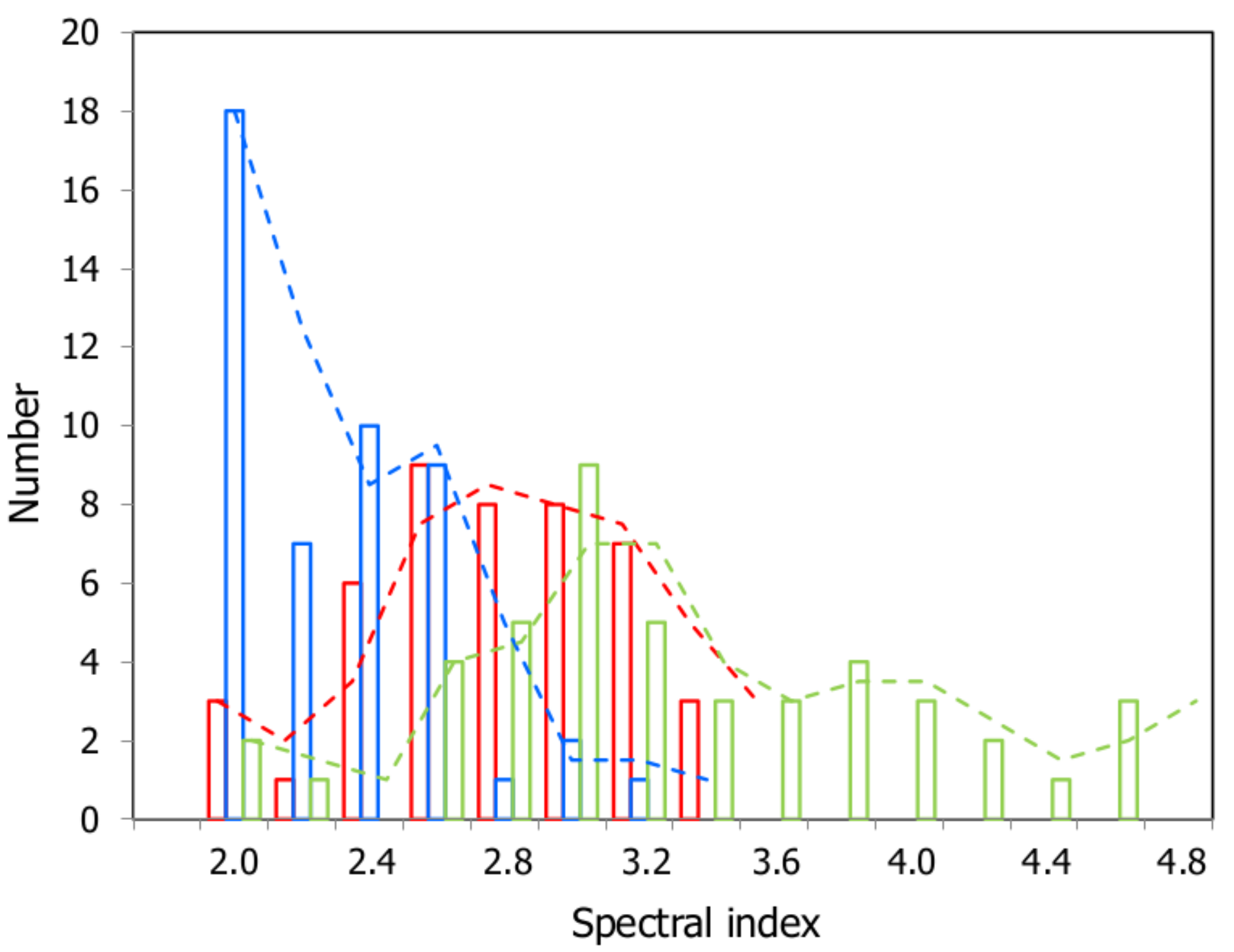}

\caption{Histogram of the spectral indices derived from the SED modelling. The three coloured bars represent the range of $\alpha$ ($\alpha$ = 2 + $\beta$) based on the derived 
$\beta$ values (blue the minimum value, red the median and green the maximum of the range).}

\label{fig:spectral_index}
\end{figure}

There are many instances where the absence of photometric data beyond 160\,$\umu$m means that the critical wavelength ($\lambda_0$) and the dust emissivity index ($\beta$) are 
degenerate and independent constraints are not possible. The 850\,$\umu$m (and 450\,$\umu$m, when available) data greatly improve this situation for the majority of the 
targets in the SONS survey. Fig.~\ref{fig:beta_plot} shows a combined contour plot of the 3$\sigma$ constraints on $\beta$ against $\lambda_0$ for all the detections in the 
survey, using the SED modelling described in Section~\ref{sec:dust_temperature}. The plot highlights that whilst there are plenty of systems in which $\lambda_0$ could be 
quite long (and hence $\beta$ quite high), where there are constraints they tend to be around $\lambda_0$$\sim$\,100 -- 300\,$\umu$m and $\beta$$\sim$\,0.5 -- 1.0. Since it is 
unlikely that $\lambda_0$ will be much greater than 300\,$\umu$m, this assumption means that there are reasonable constraints on $\beta$ being less than 1.5. The spectral 
index (defined as $\alpha$, where $\alpha$ = 2 + $\beta$) of grains in the interstellar medium typically has a value of $\alpha$ = 4 (i.e., $\beta$ = 2) whilst debris discs 
tend to have shallower spectral indices (smaller $\beta$), as shown by the majority of discs having $\alpha$ $<$ 4 ($\beta$ $<$ 2) in Fig.~\ref{fig:spectral_index}. There are 
a few cases for which the spectral index could approach 4 or more, based on the range of $\beta$ values derived from the SED modelling. Such a steep spectrum, if confirmed, 
could signify a disc dominated by small grains, most likely not greater than 10\,$\umu$m in size \citep{Ertel2012, Booth2013}.

\vskip 1mm

It is clear that there is a paucity of data between 160\,$\umu$m and 850\,$\umu$m for all but the brightest discs. For example, the \emph{Herschel}/SPIRE data at 250\,$\umu$m, 
350\,$\umu$m and 500\,$\umu$m reach the background confusion limit too quickly for fainter disc candidates to be confirmed, and so the SCUBA-2 450\,$\umu$m data potentially become 
critical in sampling this important region of the SED for debris discs. In addition, the fact that the 450\,$\umu$m and 850\,$\umu$m data are obtained simultaneously with SCUBA-2 
means that systematic biases in flux calibration \emph{should} impact both wavelengths in the same way. This inherent behaviour would suggest that any high spectral indices derived 
directly from the ratio of 450\,$\umu$m to 850\,$\umu$m fluxes would appear to be genuine, but defy interpretation in terms of our understanding of the dust size distribution for 
discs. From the SONS survey, there is only one example in which the spectral index, derived directly from the 450:850 flux ratio, is 4 or higher, namely HD 76582. This source has a 
spectral index of 4.3 $\pm$ 1.5, with the estimated error representing a significant uncertainty in the value derived. There are also examples from \emph{Herschel} observations 
where a steep spectrum is observed beyond 70\,$\umu$m \citep{Ertel2012}. Further data are needed to confirm disc signatures and to improve the S/N of (at least) some of the 
450\,$\umu$m detections to make any definitive conclusions.

\vskip 1mm

The use of the modified blackbody is a simplified way of taking into account the absorption and emission efficiencies of the grains in the disc \citep{Backman+Paresce1993}. The 
radiative efficiency of individual grains depends on a variety of different properties including the molecular composition, molecular structure, size and porosity 
\citep[e.g.][]{Draine2006}. Even if it is assumed that the composition and structure of particles is the same throughout the disc, there will be a distribution of particle 
sizes, and, potentially, grains at a wide variety of distances from the star. Hence, the shape of the SED for the disc as a whole will depend on the summation of the emission 
from particles at a range of temperatures \citep{Backman+Paresce1993, Gaspar2012}. Resolved images can help break this degeneracy. For example, the low $\beta$ value for AU Mic 
is largely due to the main belt being wide, extending from $\sim$8\,au to 40\,au, with a halo of small grains beyond this radius \citep{Matthews2015, Schuppler2015}. This means 
there are grains at a wide range of temperatures contributing to the SED fit, resulting in a shallow spectral slope that masks any size distribution effects. Conversely, if it 
is known that the ring is narrow then the spectral slope is likely to be largely due to the size distribution \citep{Draine2006}. For a small sample of discs, 
\citet{MacGregor2016a} found values of $\beta$ between roughly 0.4 and 1.1, and showed how these suggest size distributions with power law indices around 3.4, slightly less than 
the canonical \citet{Dohnanyi1969} value of 3.5, and considerably less than the values predicted by \citet{Gaspar2012} and \citet{Pan+Schlichting2012}, taking into account more 
complicated physical processes. The authors propose that a wavy size distribution is a possible cause of these lower $\beta$ values. The SONS survey larger sample backs up the 
results of \citet{MacGregor2016a} as most discs have $\beta$ values in the 0.4 $-$ 1.1 range.

\subsection{Flux peak offsets}
\label{sec:offsets}

As discussed in Section~\ref{sec:target_discussion} a significant number of detections show emission offset from the star, at a distance not explainable by positional errors. 
The statistical uncertainty in a position is commonly expressed by 0.6\,$\theta\,(\rm{S/N})^{-1}$ \citep{Ivison2007}, where $\theta$ is the FWHM of the telescope beam, and S/N 
is the peak signal-to-noise ratio, under the assumption of Gaussian noise properties in the maps. In addition, there is also a potential error associated with the telescope 
pointing accuracy, which being uncorrelated, is added in quadrature with the positional error. In the case where a target is observed over multiple nights, leading to several 
hours of integration time, it is possible that cumulative pointing errors may conspire to create an artificial offset between the star and the disc. No correlation, however, 
has been found between the magnitude of the offset and the total number of observations.

\begin{figure}
\includegraphics[width=82mm]{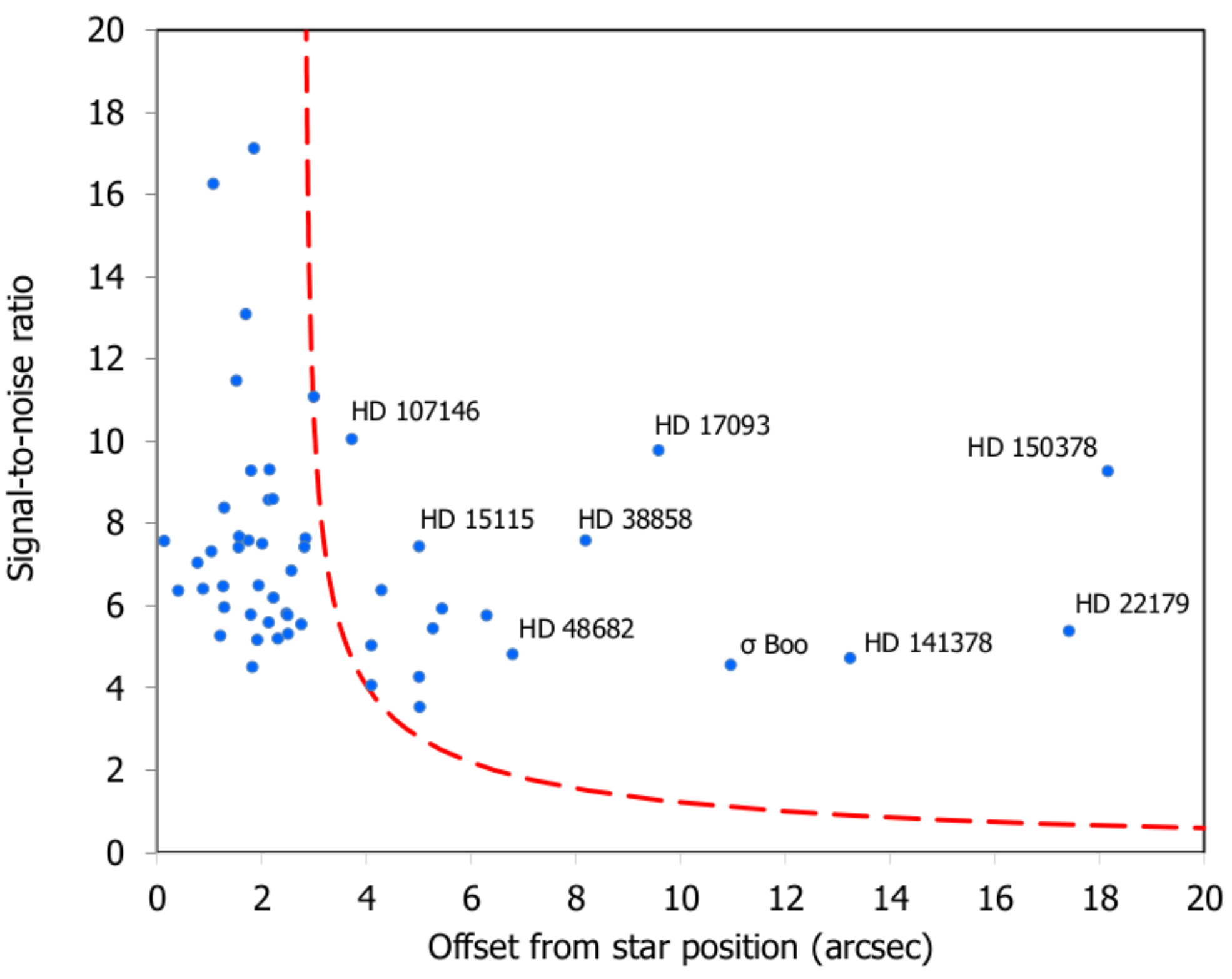}

\caption{The offset of the flux peak from the star position as a function of S/N at 850\,$\umu$m measured from the SONS survey images. The red curve represents the expected 
positional offset based on the statistical uncertainties and RMS telescope pointing errors of 2\arcsec\ in both azimuth and elevation. The peaks detected in the vicinity of HD 
22179, HD 17093, $\sigma$ Boo, HD 141378 and HD 150378, are all believed to be background objects.}

\label{fig:offsets}
\end{figure}

\vskip 1mm

As an example, taking the RMS pointing accuracy of the telescope as 2\arcsec\ in azimuth and elevation\footnote{See: http://www.eaobservatory.org/jcmt/about-jcmt/}, then at 
850\,$\umu$m the overall positional uncertainty for a peak detected at 4$\sigma$ should be no greater than 3.5\arcsec\ for the effective FWHM beam size of 15\arcsec\ (after 
smoothing). Fig.~\ref{fig:offsets} shows the measured offset of the flux peak from the stellar position as a function of the peak S/N of the detection\footnote{For HD 207129 and 
Fomalhaut the offset is measured from the geometric centre between the two peaks, i.e., it is assumed that the emission is distributed in a ring about the star. Likewise, for 
$\epsilon$ Eri the offset is measured from the geometric centre of the dust ring.}. Also plotted (red line) is the expected positional offset based purely on the statistical 
uncertainties and a RMS telescope pointing error of 2\arcsec\ in both azimuth and elevation. As can be seen the majority of the points lie to the left of the curve, suggesting that 
the pointing accuracy of the telescope is routinely better than 2\arcsec\ in both axes (which based on the pointings in this survey, taken prior to target observations, seems to be 
the case). In some of the more extreme cases (i.e., HD 17093, HD 22179, $\sigma$ Boo, and HD 150378), where the offset is $\sim$10\arcsec\ or more, it is very unlikely that the 
detected peak is associated with the star (as discussed for individual cases in Section~\ref{sec:target_discussion}). In total, 12 out of 49 peaks fall strictly outside the expected 
positional error (not including the five extreme cases) based on the statistical and nominal telescope pointing uncertainties.

\vskip 1mm

\begin{figure}
\includegraphics[width=82mm]{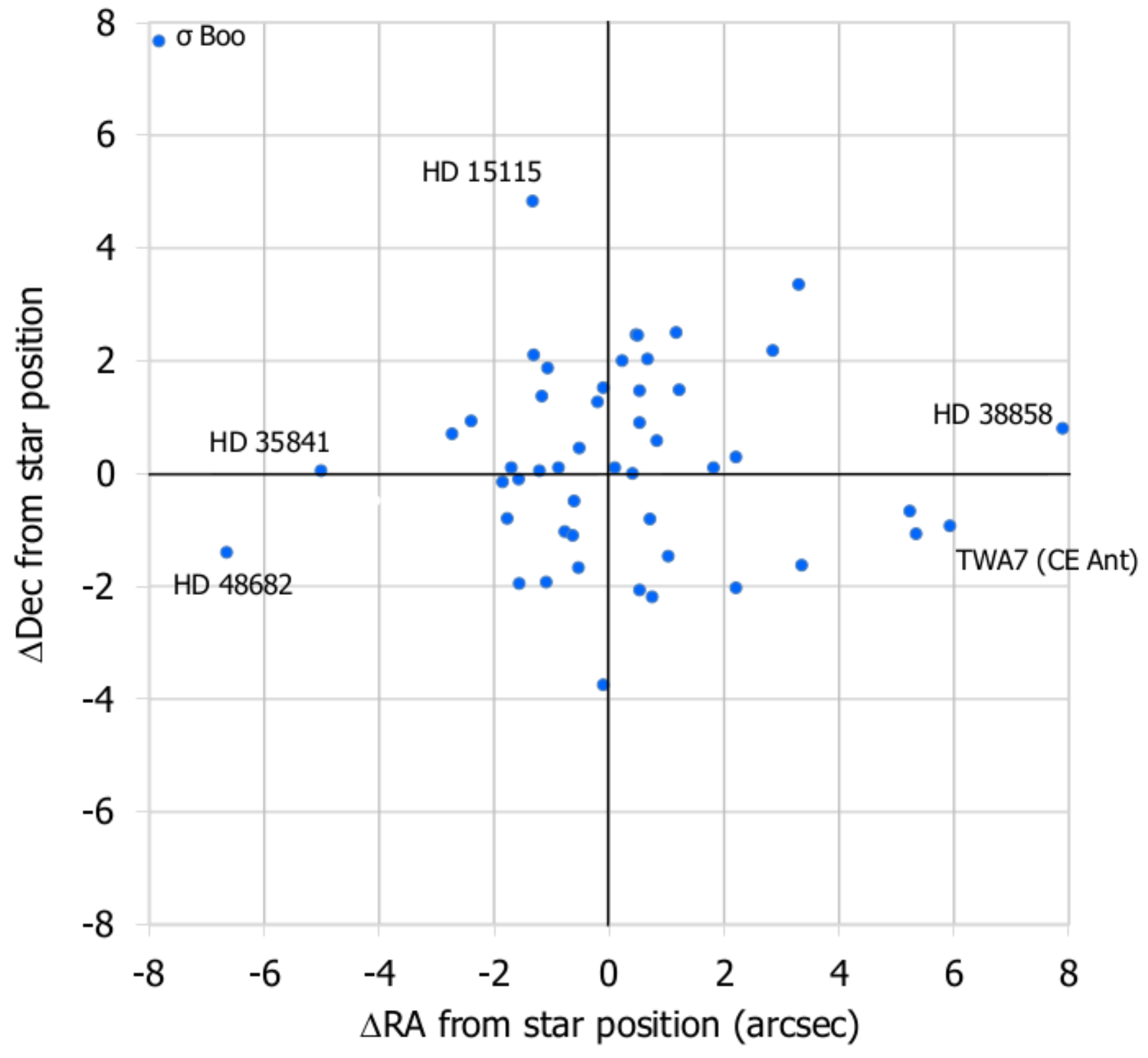}

\caption{Correlation between the measured difference in RA and dec of the flux peak from the stellar position for the SONS survey sample. The names of some of the stars that 
show a significant offset in RA or dec are also given. Three of the extreme offset cases (as indicated in Fig.~\ref{fig:offsets}, namely HD 17093, HD 22179 and HD 150378) are 
not shown.}

\label{fig:offsets_radial}
\end{figure}

Why there are so many ``offset'' disc cases remains a mystery. Fig.~\ref{fig:offsets_radial} shows the measured difference in RA and dec of the flux peak from the stellar position 
(ensuring corrections are made for the proper motion of the star). It is interesting to note that there is a slight tendency for the offsets to occur predominantly in RA, i.e., in 
an east-west direction in the images. Periodic tracking errors from the telescope are ruled out as the the JCMT is an Alt-Az mounted telescope, and indeed, rebinning a selection of 
the offset case datasets in an Azimuth-Elevation coordinate frame showed no indication of any preferential positional shift. Some of these cases have speculative (i.e., unproven) 
explanations, usually citing the possible presence of a background source as in the case of several of the more extreme offset cases shown (see 
Section~\ref{sec:background_galaxies}). The association of an offset peak with a putative disc about the star can also have a physical explanation. For instance, such emission might 
indicate a perturbed disc that could have detectable volatiles. Within the SONS survey sample there are a few examples of systems containing molecular CO gas, including the young 
($<$50\,Myr) A-stars HD 21997 and 49 Ceti \citep{Zuckerman1995, Moor2011a}. No CO, however, has been detected around HD 38858, the most extreme offset disc in the sample that is not 
believed to be a background object based on current understanding \citep{Kennedy2015}.

\subsection{Background contamination}
\label{sec:background_galaxies}

At the sensitivity levels achieved within the SONS survey (e.g., an average 1$\sigma$ noise level of 1.2\,mJy/beam at 850\,$\umu$m in the central 3\,arcmin$^2$ region), images 
will have some level of ``contamination'' by high redshift galaxies (submillimetre galaxies; SMGs). Indeed, SMG occurrence is a growing issue in terms of the interpretation of 
debris discs as surveys go deeper. There are a number of examples in the SONS survey where asymmetric features or close-by peaks could be due to background galaxies. An example 
is the disc around q$^1$ Eri (Fig.~\ref{fig:figureA2}a) as already discussed in Section~\ref{sec:q1eri}. In this case, 870\,$\umu$m observations with APEX/LABOCA 
\citep{Liseau2010} showed evidence of the east-west extension, which was very clearly seen by SCUBA-2, and identified as an isolated point-source at shorter wavelengths by 
\emph{Herschel} \citep{Liseau2010}.

\vskip 1mm

It is possible to estimate the chances of a background galaxy lying, say, within a half-beam distance from the star of interest. Assuming an average detection threshold of 
4\,mJy (3$\sigma$) at 850\,$\umu$m, the source count number is $\sim$1400 per sq. degree \citep[e.g.,][]{Blain1999}. This surface density corresponds to $\sim$2.1 sources in 
the 3\,arcmin$^2$ central area of a SONS field. Alternatively, simply counting the number of 3.5$\sigma$ peaks in the same area (avoiding the central region of radius 
15\arcsec, i.e., a distance of $\sim$2 beam radii from the star), gives an average of 2.3 peaks per field, in good agreement with the galaxy count models. Therefore, the 
probability of a background galaxy lying within a half-beam distance of any star is crudely estimated to be around 5 per cent. This expectation means that 2--3 fields within 
the SONS sample could have a background galaxy within a half-beam distance of the star. Based on the discussion outlined for individual targets in 
Section~\ref{sec:target_discussion}, there are potentially as many as nine out of the 49 cases (ignoring the five extreme offsets and Algol) for which the interpretation of a 
disc structure could be influenced by a background object. The stars for which the observed structure (or part thereof) is believed to be a background object, supported by 
multi-wavelength imaging, are q$^1$ Eri, $\epsilon$ Eri, HD 107146 and $\sigma$ Boo, whilst those which require further investigation are HD 38858, HD 92945, HD 104860, HD 
127821, and HD 205674.

\vskip 1mm

Finally, the SONS survey amounts to 100 fields of $\sim$10\,arcmin$^2$, equating to $\sim$3\,deg$^2$ of total area. Even though the use of the DAISY mode means that the noise
increases almost linearly with radius outside the central 3\,arcmin$^2$ region (estimated maximum 1\,$\sigma$ noise level of 2\,mJy/beam at the edge of the field;
\citealt{Holland2013}), this performance still represents a significant depth and area to carry out an unbiased survey count of SMGs. Such a study is planned in the future.

\subsection{Disc mass evolution}

In describing how a disc evolves the key parameters to investigate are how the fractional luminosity ($f$) and the dust mass ($M_d$) vary with stellar age. Understanding how
stellar ages are derived and the accuracy (and consistency) of the results is fundamental in this regard \citep{Zuckerman&Song2004}. Different diagnostics can give conflicting
results \citep{Moor2006} and so conclusions about disc evolution depend on stellar ages being determined in a consistent manner. The values adopted in this paper, and referenced
for individual stars in Section~\ref{sec:target_discussion}, have been deliberately conservative in terms of the quoted uncertainties in stellar age. For the cases in which no
age uncertainty is available, a value of $\pm$25 per cent has been adopted.

\vskip 1mm

\begin{figure}
\includegraphics[width=83mm]{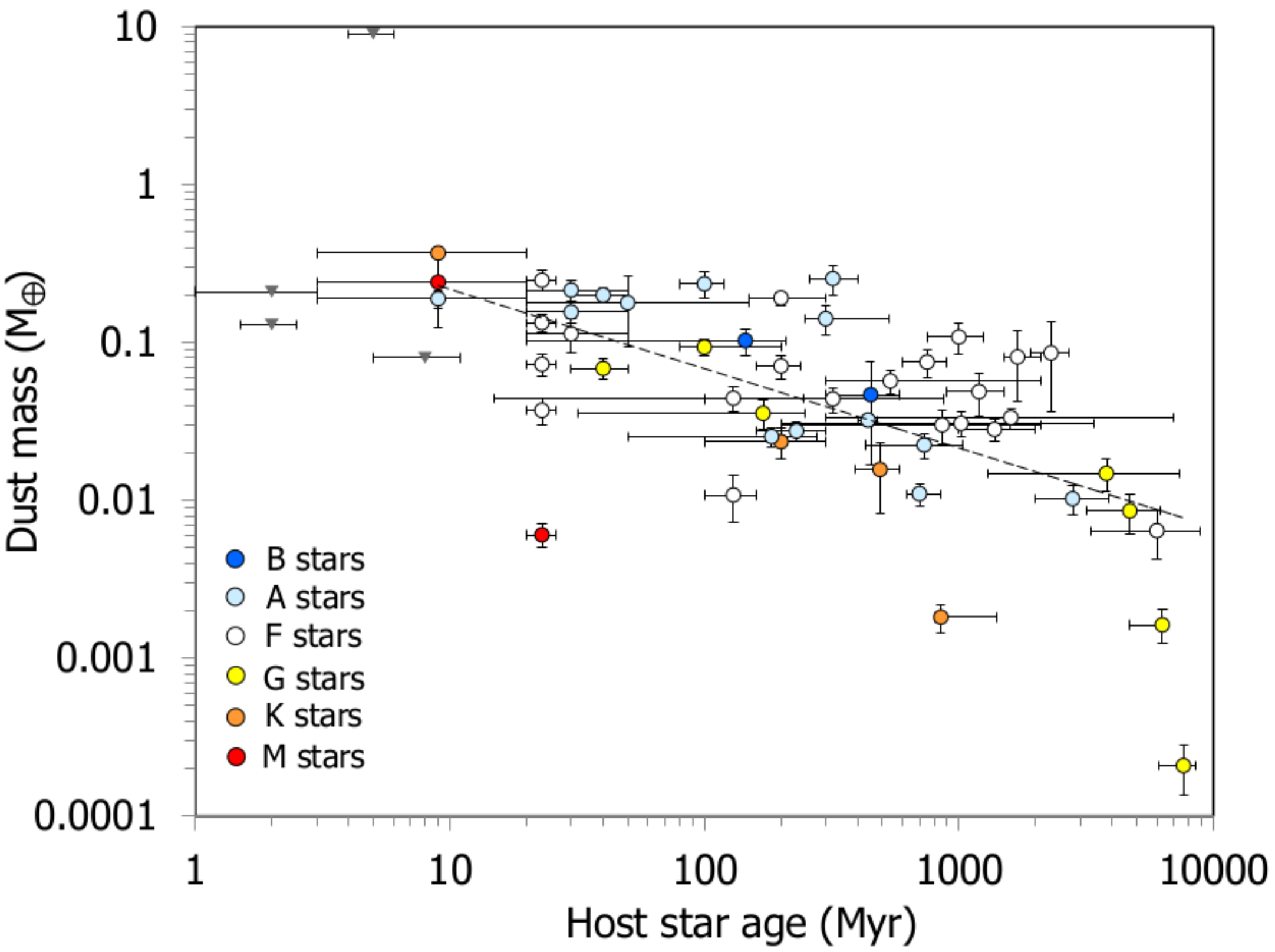}

\caption{The derived dust mass variation with host star age for the SONS survey sample. The circular symbols represent the dust mass estimates for the debris discs in the SONS 
sample (all the stars have estimated ages of $>$ 9\,Myr). A fit to the SONS results (shown by the dashed line) suggests a decline in dust mass according to $0.8t^{-0.5}$, 
where $t$ is the host star age. The inverted grey triangles represent the approximate 3\,$\sigma$ mass sensitivity levels from recent published surveys of circumstellar discs 
within clusters/associations: Upper Scorpius around 8\,$\pm$\,3\,Myr \citep{Barenfeld2016}, $\lambda$ Orionis around 5\,$\pm$\,2\,Myr \citep{Ansdell2015}, Lupus 
2\,$\pm$\,1\,Myr \citep{Ansdell2016} and Chamaeleon I around 2\,$\pm$\,0.5\,Myr \citep{Pascucci2016}. Although these targets are mainly classified as pre-main sequence 
protoplanetary discs, it should be noted that some debris disc candidates are also contained in the Upper Scorpius sample.}

\label{fig:dust_evolution}
\end{figure}

\begin{figure}
\includegraphics[width=83mm]{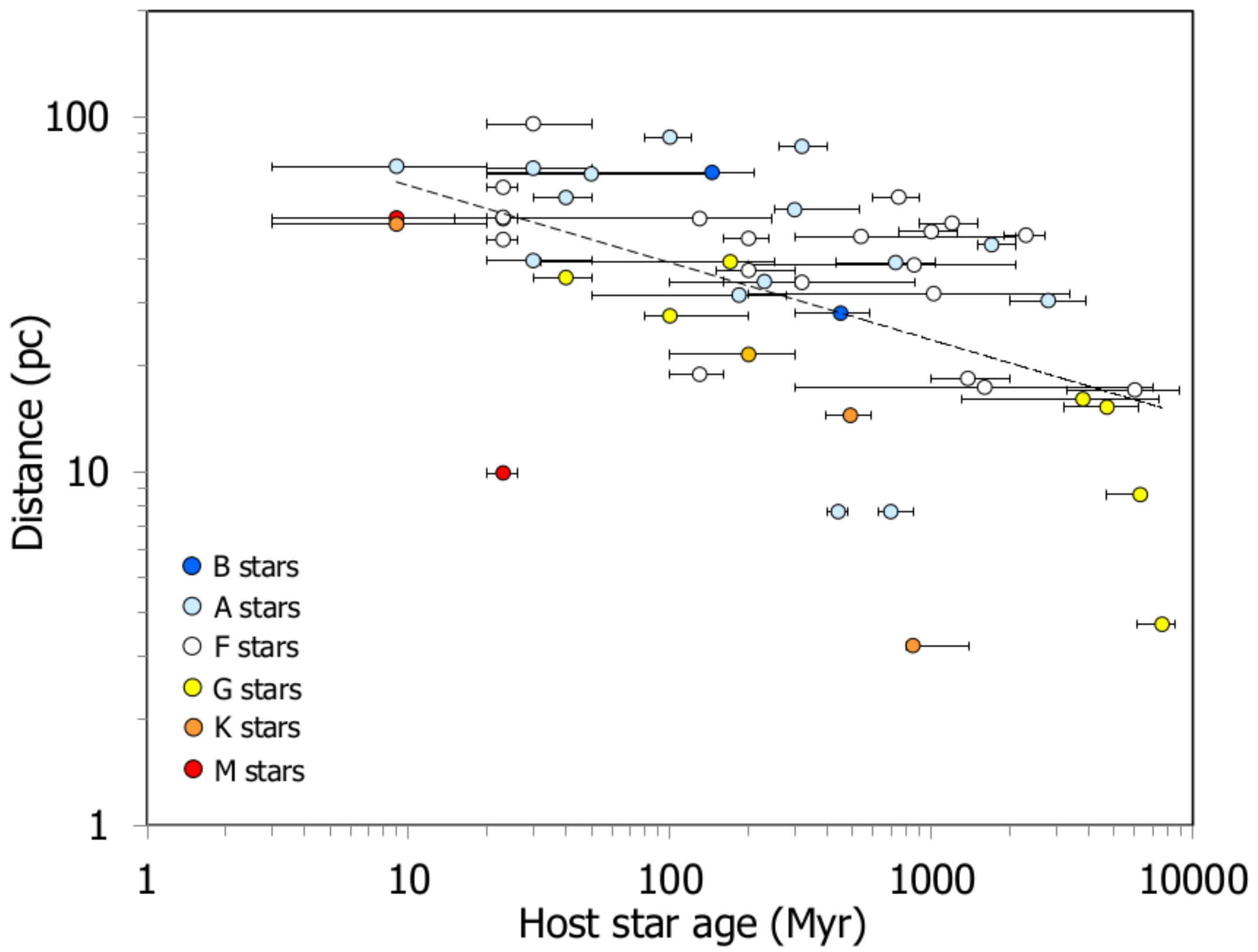}

\caption{Distance of the star plotted against host star age for the SONS survey sample. The dashed trend line ($\propto t^{-0.2}$) illustrates the bias of younger stars being 
more distant within the survey sample.}

\label{fig:distance_age}
\end{figure}

Emission from debris discs is expected to diminish over time as the reservoir of large planetesimal bodies is depleted and the collisional timescales increase \citep{Wyatt2008}. 
Fig.~\ref{fig:dust_evolution} shows how the dust mass, calculated from the 850\,$\umu$m flux and the dust temperature derived from the SED fit, varies with stellar age for the SONS 
survey sample (circular symbols). The largest measured dust masses within the sample over the range 10 to 10000\,Myr are around 0.4\,M$_{\earth}$, whilst the smallest are around the 
two old G-stars, namely HD\,115617 and $\tau$ Cet, with a minimum value of 2 $\times$10$^{-4}$\,M$_{\earth}$) for the latter. Although there is considerable scatter (as shown by the 
errors), the general trend is for a decline in dust mass as the star gets older according to $\sim$$t^{-0.5}$, where $t$ is the host star age. This trend is less steep than expected 
from a condition of steady-state collisional evolution in which $M_d \propto t^{-1}$ \citep{Wyatt2008}. As discussed in \citet{Panic2013}, there is, however, a degree of uncertainty 
in this interpretation due to observational bias. As shown in Fig.~\ref{fig:distance_age}, the younger stars in the sample tend to be more distant, meaning that the mass sensitivity 
for these targets is lower than for older, but closer stars (around which low-mass discs are easier to detect). This effect can certainly be seen in Fig.~\ref{fig:dust_evolution} 
for a few very old stars with derived masses lower than 0.01\,M$_{\earth}$, but not around younger stars since such stars are more distant. Hence, it is likely that at least some of 
the variation seen in the dust mass with stellar age plot is due to an observational bias, and not a true indication of disc mass evolution.

\vskip 1mm

In more general terms, the mass in millimetre-sized grains within circumstellar discs has previously been shown to apparently decline sharply in the period around 10\,Myr 
(generally considered to be the ``transitional phase'' between protoplanetary and debris discs, \citealt{Wyatt2008}), experiencing a drop of at least two orders of magnitude to 
the level shown for the youngest stars \citep[e.g.][]{Panic2013}. It is suggested that this age may represent a period of rapid accretion of material onto planetesimals, 
potentially one of the final phases in the evolution of a transition disk to a debris disc \citep{Wyatt2015}. 

\vskip 1mm

Until the recent advent of interferometers the mass sensitivity of single-dish telescopes in the submillimetre was rather limited for protoplanetary discs since they tend to be more 
distant (e.g., 2\,kpc for Taurus) than the debris discs chosen in most surveys (e.g., for SONS, the criterion was a distance of $<$100\,pc). The new surveys with SMA and ALMA have 
shown mass sensitivities to levels as low as, for example, $\sim$0.1\,M$_{\earth}$ for the Upper Scorpius OB association \citep{Barenfeld2016}. Fig.~\ref{fig:dust_evolution} also 
shows the 3$\sigma$ mass sensitivity from a number of recent such surveys of early circumstellar disc candidates (denoted by the inverted grey triangles in the plot). The improved mass 
sensitivity now suggests the decline in mass is not as steep as previously suggested i.e., the concept that protoplanetary discs suddenly dissipate at $\sim$10\,Myr is no longer 
strongly supported by the data. Given that there also debris discs in the recent interferometer surveys of young associations, it is likely that some such discs form at a much 
earlier stage than previously believed ($<$10\,Myr), and therefore co-exist with protoplantary discs within star-forming regions. Finally, the significant fraction of the mass is 
expected to be contained in larger (km-sized) planetesimals, particularly for debris discs, and so the total mass of the disc may be considerably higher than the estimates presented 
in this paper.

\vskip 1mm

\vskip 1mm

Future surveys might target younger stars, allowing us to quantify how disc properties, such as the size distribution and mass, vary with age, at stages when the evolution of 
these properties is most rapid. At such ages, the discs may not yet have reached equilibrium following the dispersal of the protoplanetary disc, a point that would be evidenced 
in a difference in the properties derived for the youngest discs in this sample. For example, ALMA observations of the young disc HD 21997 have shown that whilst the dust 
emission is consistent with a debris disc (i.e., the fractional luminosity is too low for a protoplanetary disc), it also hosts a CO disc \citep{Greaves2016}, with well-defined 
Keplerian rotation that may be primordial \citep{Kospal2013}. It is not unequivocal that systems do take on ``debris-like'' dust properties before the primordial gas is 
dispersed, and it is clear from the most recent surveys that there is a less significant change in the bulk of the millimetre-sized grains over a short period of time around the 
10\,Myr mark than previously hypothesised \citep{Panic2013}.

\subsection{Planet hosts}

\begin{figure}
\includegraphics[width=83mm]{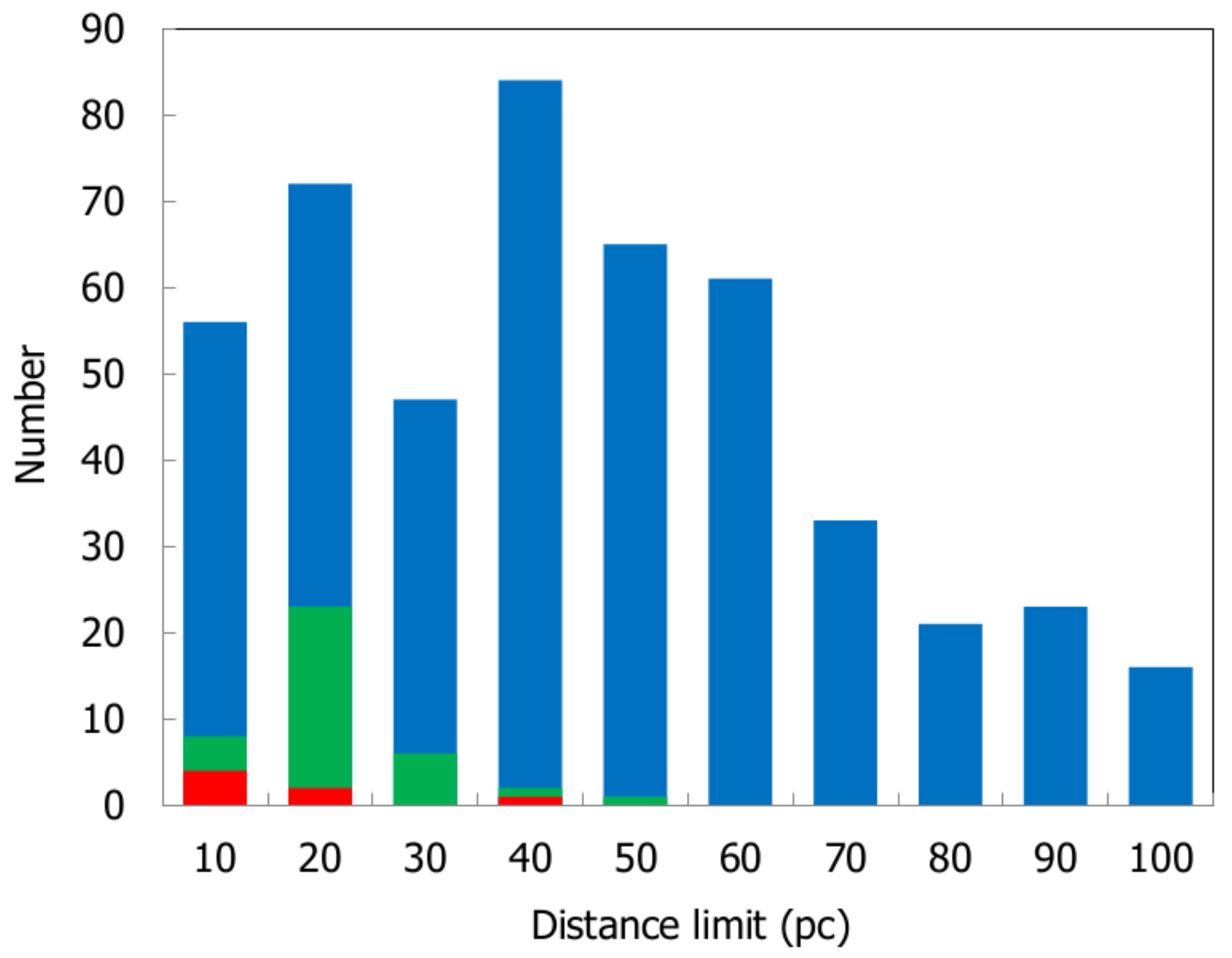}

\caption{Histogram of the known planet hosts (blue bars) in distance bins of 10\,pc from the Sun \citep{Han2014} (i.e. the distance bin labelled ``10'' contains stars up to 
10\,pc distance from the Sun). The green and red bars represent the stars that have both known planets and detected debris discs based on recent \emph{Herschel} surveys 
\citep{Marshall2014a,Moro-Martin2015} and the SONS survey, respectively.}

\label{fig:planet_hosts}
\end{figure}

As outlined in Section~\ref{sec:history} the targets in the SONS survey are selected on detectability at 850\,$\umu$m and distance, and hence do not represent a comprehensive 
exoplanet sample (i.e. it is a reasonably unbiased sample with respect to known disc-planet occurrences). There are, however, ten known planet hosts in the sample (as summarised in 
Table~\ref{tab:table6} and including the $\tau$ Ceti planetary system and planet ``b'' around $\epsilon$ Eridani, which remain controversial) with seven having a detected debris 
disc based on the observations carried out in this survey. A Fisher exact probability test \citep{Fisher1922}, carried out on the planet/disc data within the survey, suggests the 
likelihood of the planet-hosting stars having a higher disc frequency than the rest of the sample is only 7.7 per cent. Three other planet hosts were not detected within the SONS 
survey, HD 82943 \citep{Kennedy2013}, GJ 581 \citep{Lestrade2012}, and HD 113337 \citep{Su2013}, all of which have known debris discs from \emph{Herschel} observations, but were too 
faint to be detected at 850\,$\umu$m based on the current observations. The majority of stars with known planets and a debris disc are naturally chosen to be close to the Sun, as 
shown in Fig.~\ref{fig:planet_hosts} for both SONS and recent \emph{Herschel} surveys \citep{Marshall2014a,Moro-Martin2015}. It is also the case that the few directly-imaged 
exoplanets (i.e. HD 216956 and HR 8799) are also found in debris systems. Indeed, with the exception of these systems, the planets discovered tend to be on orbits of just a few 
astronomical units and so will have little influence on discs with size-scales of 50\,au or more (see Table~\ref{tab:table6}).

\begin{table*}


  \caption{A summary of the information on the planet hosts in the SONS survey sample.}

  \label{tab:table6}

  \begin{tabular}{lllccc}
\hline

HD      &   Other                  & Planet name          & Orbit semi-major             & Eccentricity         & $Msini$                                \\
number  &   names                  &                      & axis (au)                    &                      & (M$_{Jup}$)                             \\

\hline

10647   & q$^1$ Eri                &  HD 10647 b          & 2.022 $\pm$ 0.082            & 0.16 $\pm$ 0.20      & 0.93 $\pm$ 0.24                        \\

10700   & $\tau$ Cet               &  $\tau$ Cet b        & 0.105 $\pm$ 0.005            & 0.16 $\pm$ 0.22      & 0.006 $\pm$ 0.002                      \\

        &                          &  $\tau$ Cet c        & 0.195 $\pm$ 0.011            & 0.03 $\pm$ 0.28      & 0.010 $\pm$ 0.005                      \\

        &                          &  $\tau$ Cet d        & 0.374 $\pm$ 0.02             & 0.08 $\pm$ 0.26      & 0.011 $\pm$ 0.005                      \\

        &                          &  $\tau$ Cet e        & 0.552 $\pm$ 0.03             & 0.05 $\pm$ 0.22      & 0.014 $\pm$ 0.007                      \\

        &                          &  $\tau$ Cet f        & 1.35 $\pm$ 0.09              & 0.03 $\pm$ 0.26      & 0.021 $\pm$ 0.011                      \\

22049   & $\epsilon$ Eri           &  $\epsilon$ Eri b    & 3.38 $\pm$ 0.32              & 0.25 $\pm$ 0.23      & 1.05 $\pm$ 0.19                        \\

38858   &                          &  HD 38858 b          & 1.038 $\pm$ 0.019            & 0.27 $\pm$ 0.17      & 0.096 $\pm$ 0.012                      \\

82943   &                          &  HD 82943 c          & 0.742 $\pm$ 0.013            & 0.43 $\pm$ 0.03      & 1.590 $\pm$ 0.103                      \\

        &                          &  HD 82943 b          & 1.185 $\pm$ 0.022            & 0.20 $\pm$ 0.07      & 1.589 $\pm$ 0.097                      \\

113337  &                          &  HD 113337 b         & 1.033 $\pm$ 0.035            & 0.46 $\pm$ 0.04      & 2.83 $\pm$ 0.24                        \\

115617  & 61 Vir                   &  61 Vir b            & 0.050 $\pm$ 0.001            & 0.12 $\pm$ 0.11      & 0.016 $\pm$ 0.002                      \\

        &                          &  61 Vir c            & 0.217 $\pm$ 0.004            & 0.14 $\pm$ 0.06      & 0.033 $\pm$ 0.004                      \\

        &                          &  61 Vir d            & 0.475 $\pm$ 0.008            & 0.35 $\pm$ 0.09      & 0.072 $\pm$ 0.009                      \\

        & GJ 581                   &  GJ581 e             & 0.028 $\pm$ 0.001            & 0.03 $\pm$ 0.01      & 0.006 $\pm$ 0.001                      \\

        &                          &  GJ581 b             & 0.041 $\pm$ 0.001            & 0.03 $\pm$ 0.01      & 0.050 $\pm$ 0.002                      \\

        &                          &  GJ581 c             & 0.073 $\pm$ 0.002            & 0.07 $\pm$ 0.06      & 0.017 $\pm$ 0.001                      \\

216956  & Fomalhaut                &  Fomalhaut b         & 115                          & 0.12 $\pm$ 0.01      & $<$0.5$^1$                             \\

218396  & HR 8799                  & HR 8799 e            & 14.5 $\pm$ 0.5               & 0                    & 7 $\pm$ 3$^1$                          \\

        &                          & HR 8799 d            & 24 $\pm$ 0                   & 0.1                  & 7 $\pm$ 3$^1$                          \\

        &                          & HR 8799 c            & 38 $\pm$ 0                   & 0                    & 7 $\pm$ 3$^1$                          \\

        &                          & HR 8799 b            & 68 $\pm$ 0                   & 0                    & 5 $\pm$ 2$^1$                          \\

\hline
  \end{tabular}

\begin{flushleft}
$^1$The mass is not $M sini$ in this case, but model estimates based on their photometry.
\end{flushleft}

\end{table*}

\begin{figure}
\includegraphics[width=83mm]{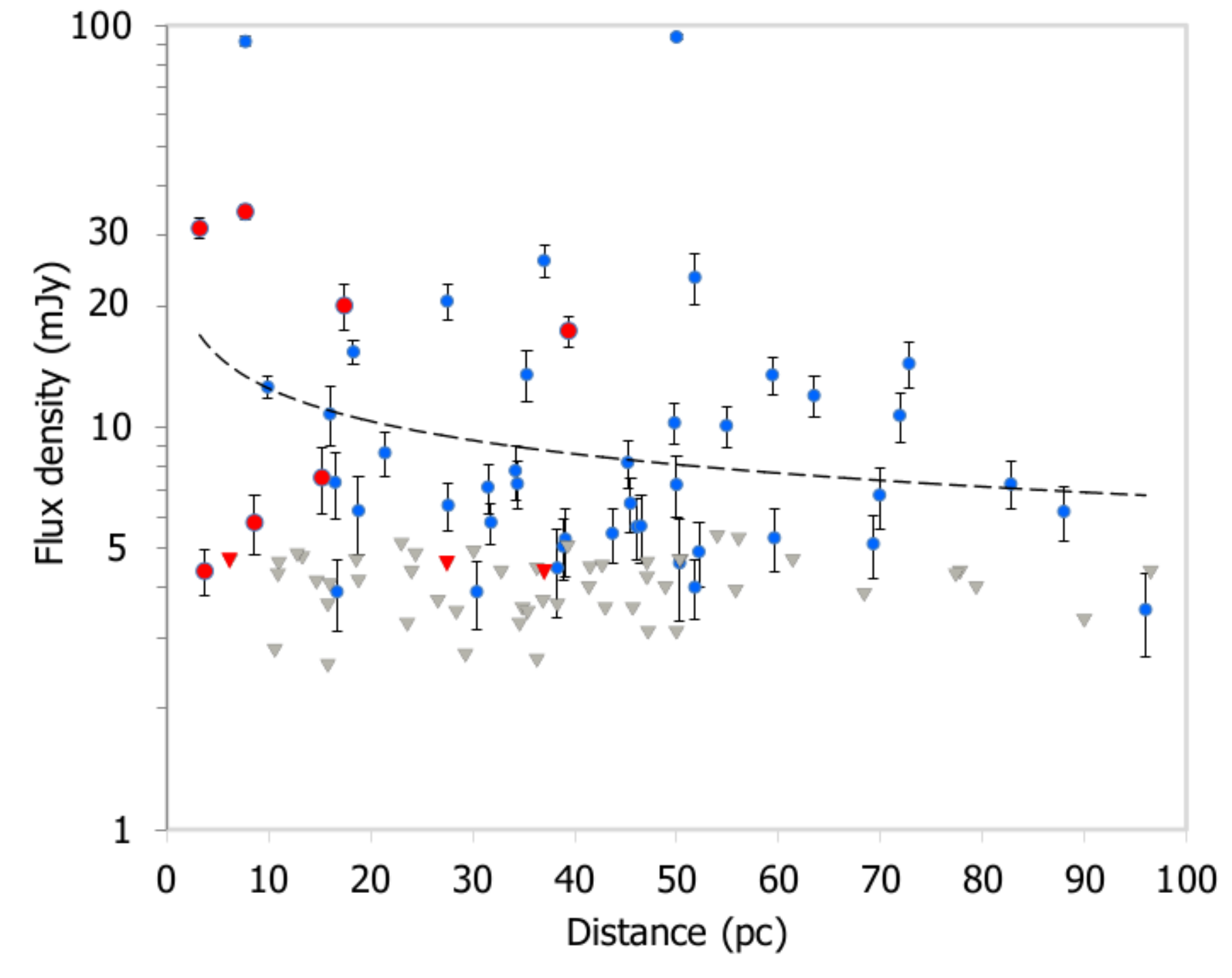}

\caption{The measured 850\,$\umu$m flux density against distance for the detected discs in the SONS survey. The blue circles represent disc detections, whilst the inverted triangles 
are the 3$\sigma$ upper flux limits. The red filled circles and triangles indicate the known planet hosts in the sample. The dashed curve is a fit to the detected discs only as a 
function of distance. }

\label{fig:flux_planets}
\end{figure}

\vskip 1mm

Other work has derived detection rates for disc systems around planet-hosting stars of 20 $-$ 30 per cent, based on more targeted surveys of planet hosting stars using
\emph{Spitzer} and \emph{Herschel} surveys \citep{Trilling2008, Eiroa2013, Marshall2014a}. Further studies identified possible trends between low-mass planets and the presence of
cool dust \citep{Wyatt2012}. Since debris discs and planets form and evolve together, the amount of solids present largely determines the outcome of each, implying that discs
around planet hosting stars will be brighter than those around stars with no planets \citep{Wyatt2007b}. Fig.~\ref{fig:flux_planets} shows the measured 850\,$\umu$m
flux density for all the detected SONS targets as a function of stellar distance. Based on the small number of planet hosts in this sample, there is no obvious evidence to
support this hypothesis. A more recent study of 204 FGK stars also did not find any compelling evidence that debris discs are more common (or, indeed, more dusty) around stars
harbouring planets \citep{Moro-Martin2015}. In the current era of planet hunting, dynamical studies of debris discs can provide a unique avenue for the detection of sub-Jupiter
mass planets at large orbital radii (e.g., see Fig. 11 in \citealt{Wyatt2008}).

\section{Conclusions}

The SONS survey has detected 850\,$\umu$m emission from the vicinity of 55 stars in a sample of 100 targets. At least five of these detections are believed to be background 
sources, and one is thought to be due to radio emission variability rather than emission from a disc. The remainder trace dust grains up to a few millimetres in size and, by 
inference, the distribution of planetesimals around the host stars. The legacy of the survey is a catalogue of fluxes and images, and better characterisation of the detected 
discs via modelling of their SEDs to derive dust temperatures and masses. Key results from the SONS survey include:

\begin{enumerate}[1.]
\item The number of detected discs at 850\,$\umu$m from single-dish telescope observations has doubled from 24 known pre-SONS (mainly from JCMT, CSO and APEX) to 49;

\vskip 1mm

\noindent \item Approximately one-third (16 out of 49 detected discs) have been spatially resolved by the 850\,$\umu$m beam of the JCMT, allowing a measurement of the disc size;

\vskip 1mm

\item From the observed radial profiles, the discs are around 1 -- 10 $\times$ the size of the Edgeworth-Kuiper belt in our own Solar System (assumed diameter of
100\,au);

\vskip 1mm

\item The majority of the measured 850\,$\umu$m disc radii are substantially larger than expected based on the blackbody fit to the SED, implying the existence of 
planetesimals at large radii from the host stars;

\vskip 1mm

\item The ratio of the observed (measured) disc radius to that derived from the blackbody fit also declines with stellar luminosity according to $L_*^{-0.2}$. This trend is 
consistent with the higher luminosity stars having removed the smallest grains, leaving behind relatively large (blackbody-like) dust grains;

\vskip 1mm

\item The dust spectral index ($\alpha$) is constrained to be between values of 2.5 and 3.5 for the vast majority of the discs in the SONS survey sample, less than the typical 
value of 4 found in the interstellar medium;

\vskip 1mm

\item As surveys go deeper in terms of sensitivity, interpretation of disc structures becomes more difficult due to the inevitable contamination by background (notably high
redshift) galaxies. There are a number of examples within SONS for which the presence of a background galaxy could help to explain the observed disc morphology;

\vskip 1mm

\item The fractional dust luminosity covers the wide range of 10$^{-1}$ to 10$^{-5}$ compared to a value of $\sim$10$^{-7}$ for the Edgeworth-Kuiper belt. All of the discs in 
the survey have f-values of less than the 10$^{-2}$ threshold usually adopted for debris discs, with the exception of HD 98800 (f $\sim$ 0.1) and is usually classified 
as a protoplanetary rather than a debris disc;

\vskip 1mm

\item The most massive discs are around 0.5\,M$_{\earth}$ whilst the least massive are in the 10$^{-2}$M$_{\earth}$ range (with one exception, a tiny 2 $\times$ 
10$^{-4}$\,M$_{\earth}$). The least massive range of 10$^{-2}$M$_{\earth}$ is a similar level to the Edgeworth-Kuiper belt;

\vskip 1mm

\item The amount of dust appears to decline roughly as $t^{-0.5}$, where $t$ is the host star age, consistent with theories of steady state collisional evolution models (with 
the caveat that at least some of this variation may be due to an observational bias);

\vskip 1mm

\item Although the SONS survey does not represent a planet-host sample, a disc detection around many of the nearest stars means that such systems will become key planet-search
targets for the future.

\end{enumerate}

Just as diverse outcomes of planetary system architectures are now recognised (e.g., via Kepler; \citealt{Batalha2014}), so have recent debris disc surveys (including SONS) 
revealed a large diversity in disc geometries. The observed diversity in exo-planetary system architectures implies that the debris material should be distributed in different 
ways. Establishing how varied debris discs are is thus a key piece of information that will help link the formation and evolution of planetary systems with the evolution of 
planetary building blocks (planetesimals). Building such empirical bridges between discs and planets brings us closer to unravelling the physical processes that led to the 
formation of planetary systems. Surveys such as SONS, via the study of the location, mass, morphology, and dust properties of these discs, highlight the submillimetre as an 
important waveband to address these questions. One of the key legacy goals of the SONS survey was indeed to lay the foundation for future, high angular resolution studies 
with existing and new facilities such as SMA, ALMA, and \emph{JWST}, with a view, for example, to studying belts of planetesimals in fine detail.

\section*{Acknowledgements}

During the period of these observations the James Clerk Maxwell Telescope was operated by the Joint Astronomy Centre on behalf of the Science and Technology Facilities Council of 
the United Kingdom, the National Research Council of Canada and the Netherlands Organisation for Pure Research. Additional funds for the construction of SCUBA-2 were provided by the 
Canada Foundation for Innovation. MCW acknowledges the support of the European Union through ERC grant number 279973. GMK is supported by the Royal Society as a Royal Society 
University Research Fellow. MB acknowledges support from a FONDECYT Postdoctoral Fellowship, project no. 3140479 and the Deutsche Forschungsgemeinschaft through project Kr 
2164/15-1. JPM is supported by a UNSW Vice Chancellor's postdoctoral research fellowship. The work of OP is supported by the Royal Society through a Royal Society Dorothy Hodgkin 
fellowship. GJW gratefully acknowledges support from the Leverhulme Trust. This research has made use of the Exoplanet Orbit Database, the Exoplanet Data Explorer at exoplanets.org 
and the SIMBAD database operated at CDS, Strasbourg, France. The Starlink software \citep{Currie2014} is currently supported by the East Asian Observatory. This research used the 
facilities of the Canadian Astronomy Data Centre operated by the National Research Council of Canada with the support of the Canadian Space Agency.






\vskip 10mm

\noindent List of affiliations

\vskip 5mm

\noindent $^{1}$UK Astronomy Technology Centre, Royal Observatory, Blackford Hill, Edinburgh, EH9 3HJ, UK\\
$^{2}$Institute for Astronomy, University of Edinburgh, Royal Observatory, Blackford Hill, Edinburgh, EH9 3HJ, UK\\
$^{3}$National Research Council of Canada Herzberg Astronomy \& Astrophysics Programs, 5071 West Saanich Road, Victoria, BC, V9E 2E7, Canada\\
$^{4}$Department of Physics \& Astronomy, University of Victoria, 3800 Finnerty Road, Victoria, BC, V8P 5C2, Canada\\
$^{5}$Institute of Astronomy, University of Cambridge, Madingley Road, Cambridge, CB3 0HA, UK\\
$^{6}$School of Physics and Astronomy, University of St. Andrews, North Haugh, St. Andrews, Fife, KY16 9SS, UK\\
$^{7}$Instituto de Astrofisica, Pontificia Universidad Catolica de Chile, Vicua Mackenna 4860, 7820436, Macul, Santiago, Chile\\
$^{8}$Astrophysikalisches Institut and Universit\"{a}t-Sternwarte, Friedrich-Schiller-Universit\"{a}t Jena, Schillerg\"{a}sschen 2-3, 07745 Jena, Germany\\
$^{9}$Centre de recherche en astrophysique du Qu{\'e}bec and Department de Physique, Universit{\'e} de Montreal, Montreal, QC, H3C 3J7, Canada\\
$^{10}$Jet Propulsion Laboratory, California Institute of Technology, 4800 Oak Grove Drive, Pasadena, CA 91109, USA\\
$^{11}$Department of Physics and Astronomy, James Madison University, MSC 4502-901 Carrier Drive, Harrisonburg, VA 22807, USA\\
$^{12}$Space Telescope Science Institute, 3700 San Martin Drive, Baltimore, MD 21218, USA\\
$^{13}$Centre for Astrophysics Research, Science and Technology Research Institute, University of Hertfordshire, College Lane, Hatfield, Herts, AL10 9AB, UK\\
$^{14}$Joint ALMA Observatory, Alonso de Cordova 3107, Vitacura 763-0355, Santiago, Chile\\
$^{15}$Astronomy Department, University of California, Berkely, CA, 94720-3411, USA\\
$^{16}$Univ. Grenoble Alpes/CNRS, IPAG, F-38000 Grenoble, France\\
$^{17}$Department of Physics \& Astronomy, University of British Columbia, 6224 Agricultural Road, Vancouver BC V6T 1Z1, Canada\\
$^{18}$JointAstronomy Centre, 660 N. A`oh\={o}k\={u} Place, University Park, Hilo, HI 96720, USA\\
$^{19}$European Southern Observatory, Karl-Schwarzschild-Str. 2, D-85738 Garching, Germany\\
$^{20}$Observatoire de Paris, PSL Research University, CNRS, Sorbonne Universit{\'e}s, UPMC, 61 Av. de l'Observatoire, F-75014 Paris, France\\
$^{21}$School of Physics, UNSW Australia, High Street, Kensington, NSW 2052, Australia\\
$^{22}$Australian Centre for Astrobiology, UNSW Australia, High Street, Kensington, NSW 2052, Australia\\
$^{23}$Computational Engineering and Science Research Centre, Uuniversity of Southern Queensland, Toowoomba, QLD 4350, Australia\\
$^{24}$Center for Astrophysical Sciences, John Hopkins University, Baltimore, MD 21218, USA\\
$^{25}$School of Physical Sciences, The Open University, Milton Keynes, MK7 6AA, UK\\
$^{26}$SRON Netherlands Institute for Space Research, NL-9747 AD Groningen, The Netherlands \\
$^{27}$Department of Physics \& Astronomy, University of California, Los Angeles, 90095, USA\\
$^{28}$Jeremiah Horrocks Institute, University of Central Lancashire, Preston, Lancashire, PR1 2HE, UK\\
$^{29}$Leiden Observatory, Leiden University, PO Box 9512, NL-2300 RA Leiden, The Netherlands\\
$^{30}$RAL Space, The Rutherford Appleton Laboratory, Chilton, Didcot, OX11 0NL, UK\\
$^{31}$Harvard-Smithsonian Center for Astrophysics, 60 Garden Street, Cambridge, MA, 02138, USA



\appendix

\section{The SONS survey images and spectral energy distributions}
\label{sec:appendixA}

This Appendix presents the 850\,$\umu$m S/N images from the SONS survey and spectral energy distributions together with model fits to the stellar photosphere
and infrared to millimetre thermal (disc) excess. Unless otherwise stated the colours are scaled from $-$3$\sigma$ to the maximum S/N in the images. Similarly
the contours start at $-$3$\sigma$ (dashed white) and then solid colours from 3$\sigma$ to the maximum in 1$\sigma$ steps. The white bar represents the spatial
scale corresponding to the FWHM beam in astronomical units at the star. The ``star symbol'' represents the position of the star with respect to the disc,
taking into account any proper motion corrections if required.

\vskip 1mm

For the spectral energy distributions, the black symbols represent measured fluxes whilst the brown symbols are stellar photosphere-subtracted values (i.e.
disc fluxes from the infrared to millimetre excess). Small green dots are the star-subtracted \emph{Spitzer}/IRS spectrum (if it exists). The grey and black
inverted triangles represent the 3$\sigma$ upper flux limits (again, star-subtracted). Photometry from the SONS survey are highlighted by the blue circles.

\begin{figure*}
\includegraphics[width=160mm]{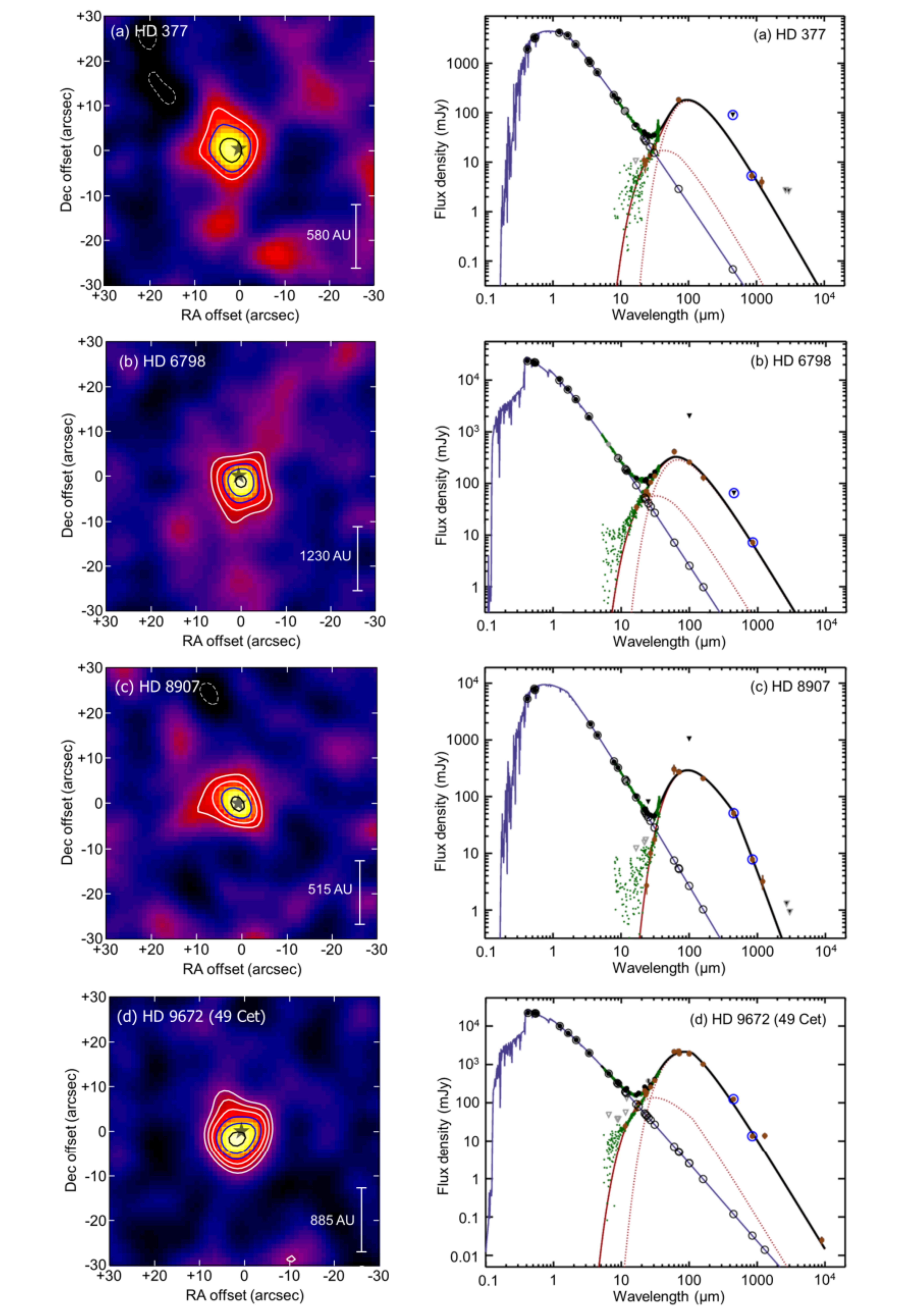}

  \caption{850\,$\umu$m S/N images and spectral energy distributions for the targets (a) HD 377, (b) HD 6798, (c) HD 8907 and (d) HD 9672 (49 Cet).}

\label{fig:figureA1}
\end{figure*}

\begin{figure*}
\includegraphics[width=160mm]{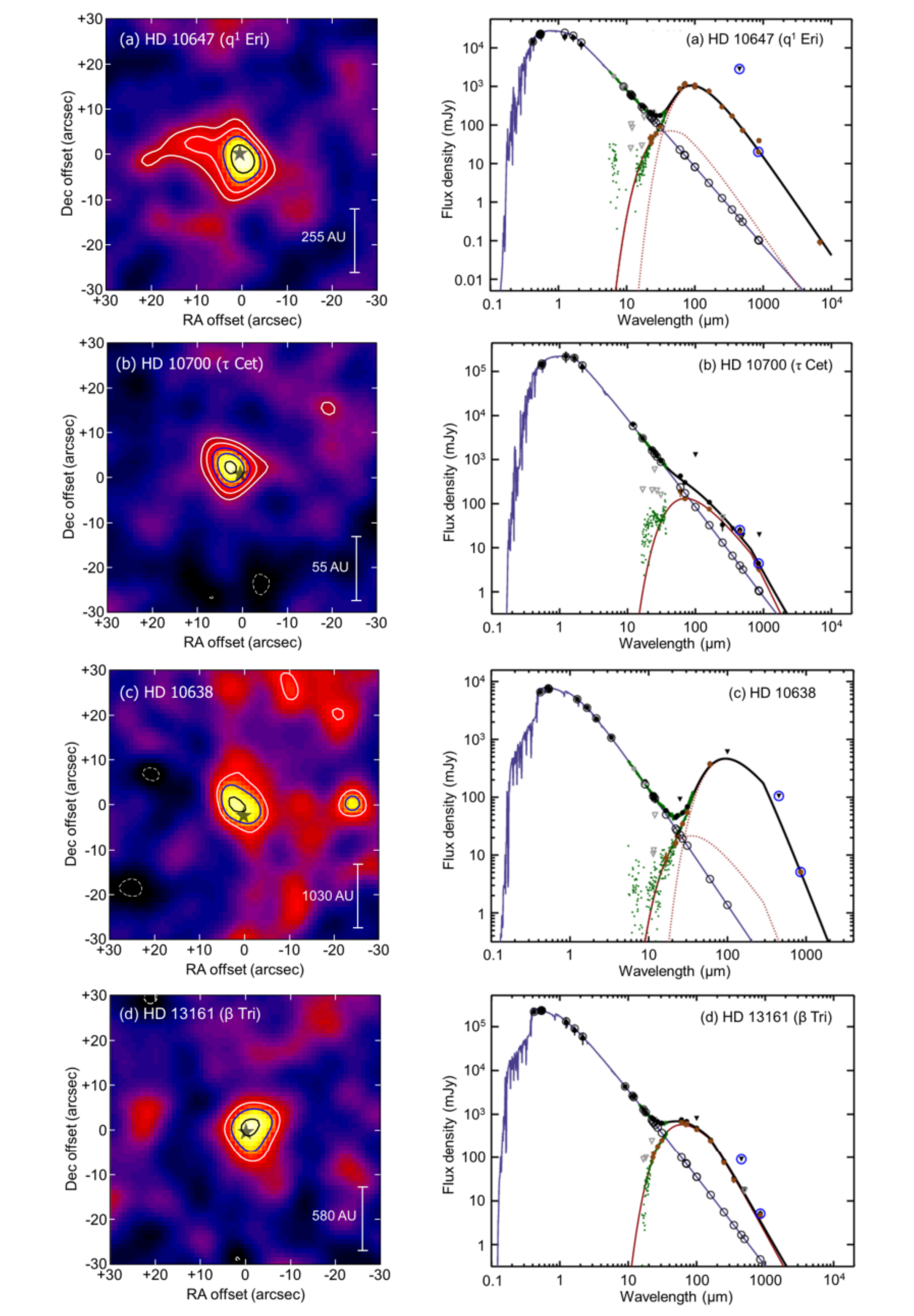}

  \caption{As for Figure A1 for the targets (a) HD 10647 (q$^1$ Eri), (b) HD 10700 ($\tau$ Cet), (c) HD 10638 and (d) HD 13161 ($\beta$ Tri).}

\label{fig:figureA2}
\end{figure*}

\begin{figure*}
\includegraphics[width=160mm]{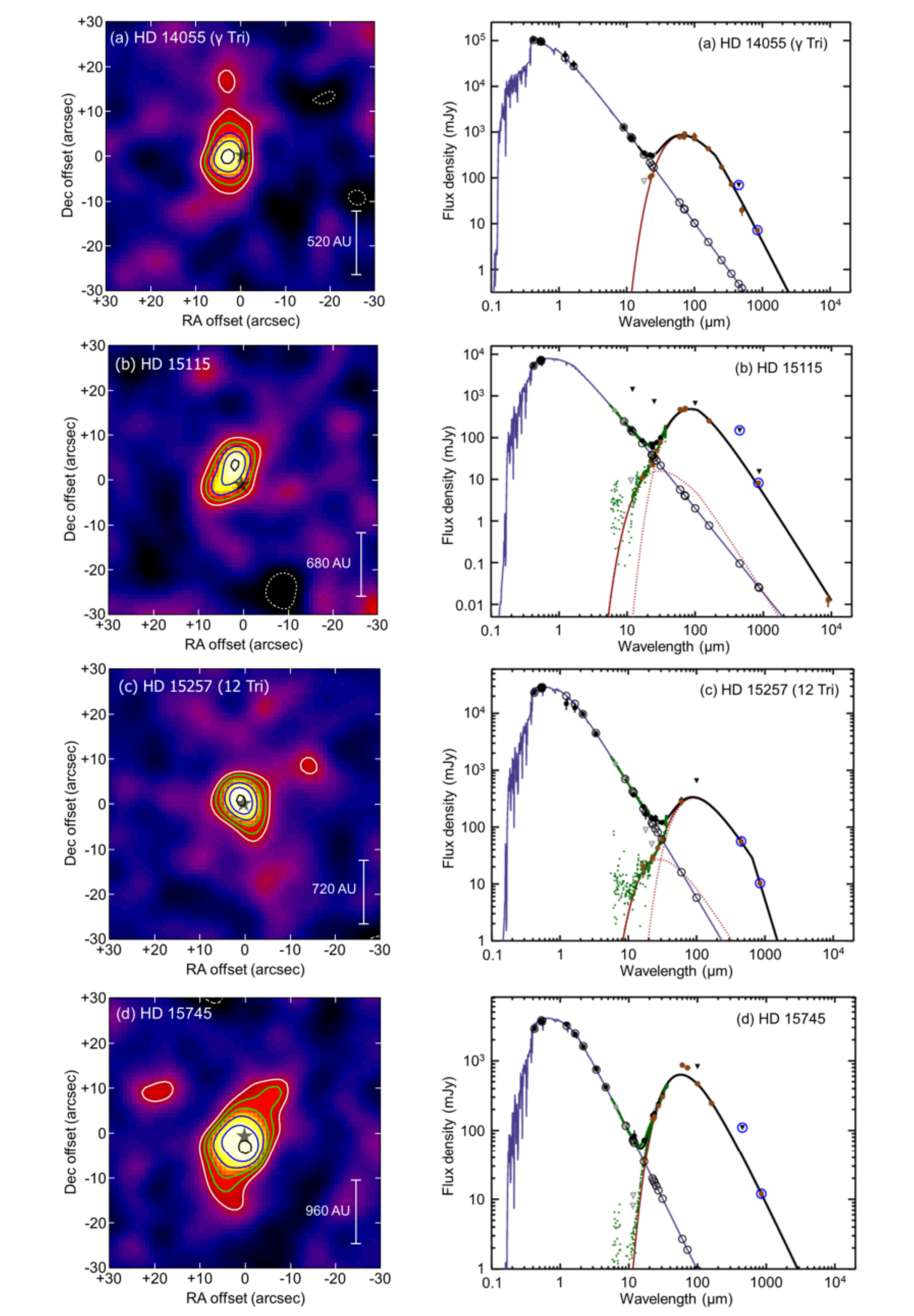}

  \caption{As for Figure A1 for the targets (a) HD 14055 ($\gamma$ Tri), (b) HD 15115, (c) HD 15257 (12 Tri) and (d) HD 15745.}

\label{fig:figureA3}
\end{figure*}

\begin{figure*}
\includegraphics[width=160mm]{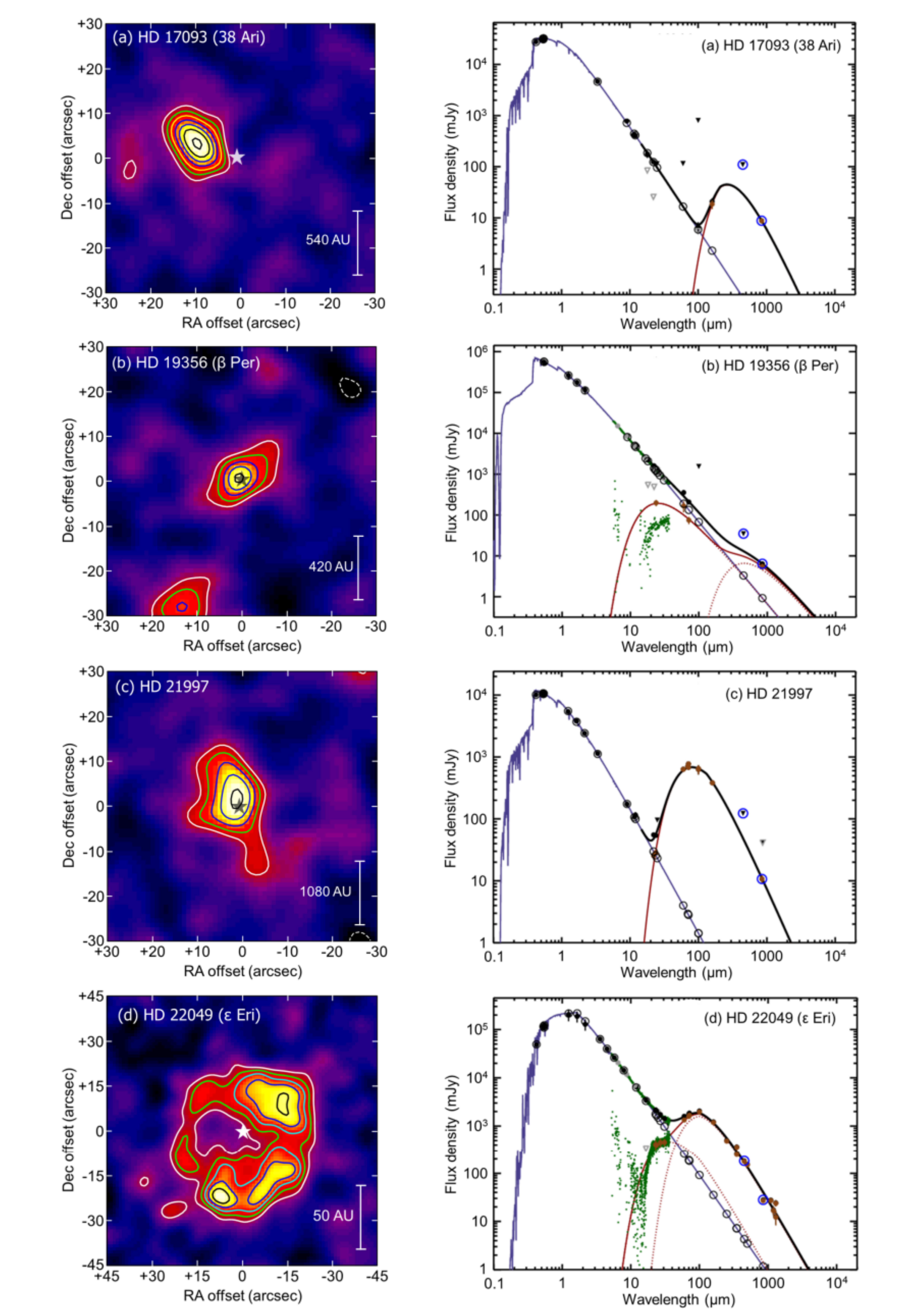}

  \caption{As for Figure A1 for the targets (a) HD 17093 (38 Ari), (b) HD 19536 ($\beta$ Per), (c) HD 21997 and (d) HD 22049 ($\epsilon$ Eri).}

\label{fig:figureA4}
\end{figure*}

\begin{figure*}
\includegraphics[width=160mm]{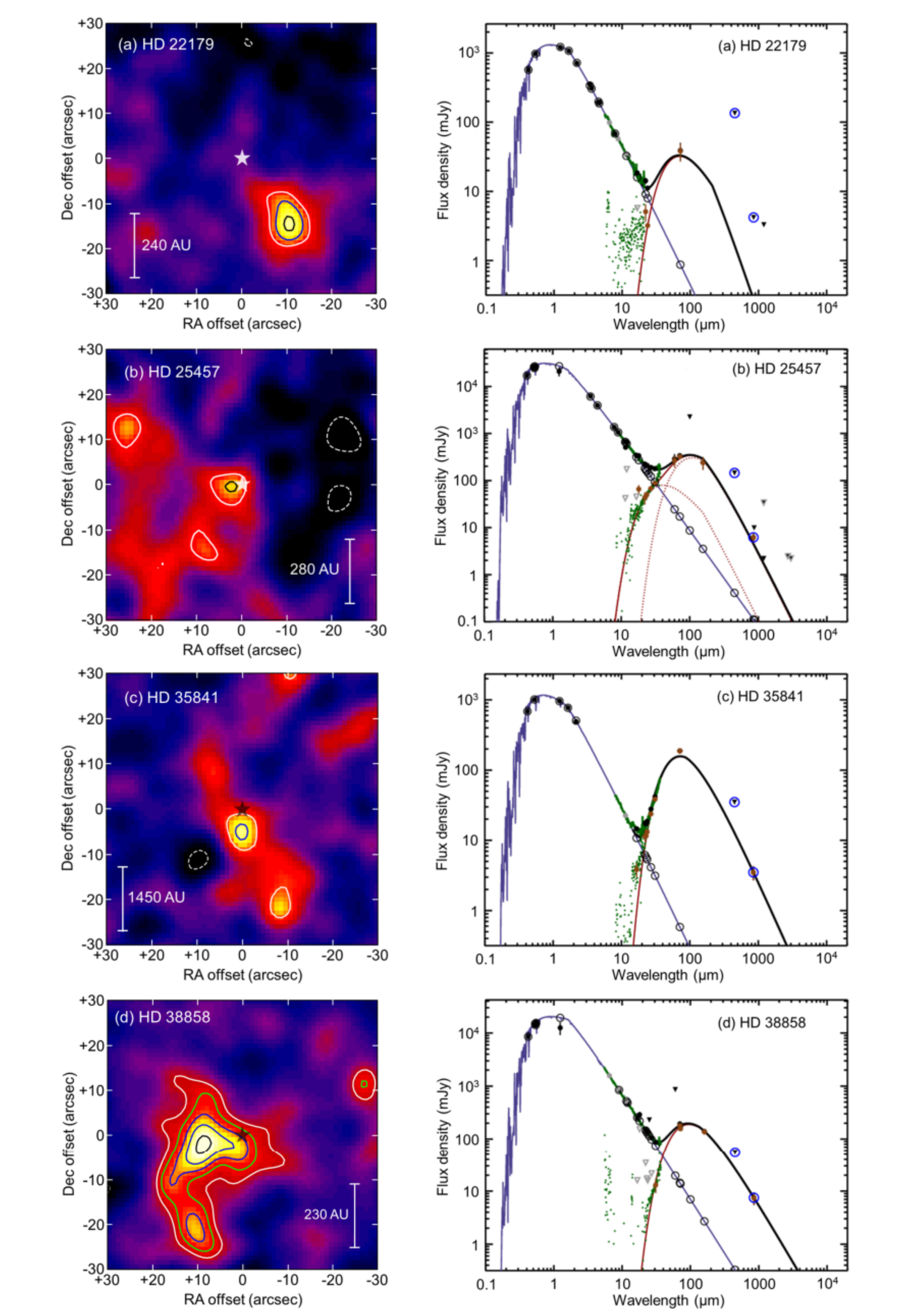}

  \caption{As for Figure A1 for the targets (a) HD 22179, (b) HD 25457, (c) HD 35841 and (d) HD 38858.}

\label{fig:figureA5}
\end{figure*}

\begin{figure*}
\includegraphics[width=160mm]{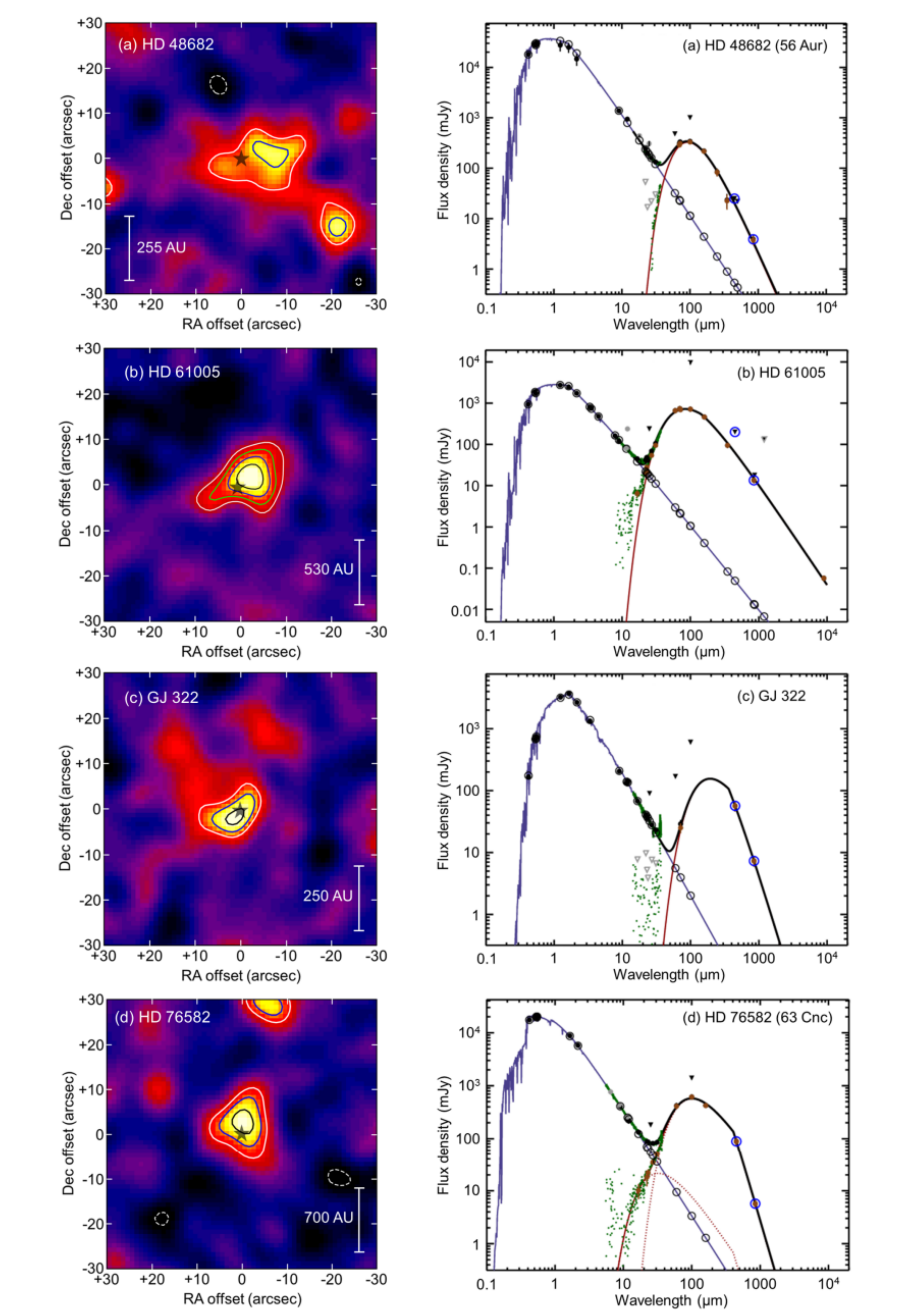}

  \caption{As for Figure A1 for the targets (a) HD 48682, (b) HD 61005, (c) GJ 322 and (d) HD 76582 (63 Cnc).}

\label{fig:figureA6}
\end{figure*}

\begin{figure*}
\includegraphics[width=160mm]{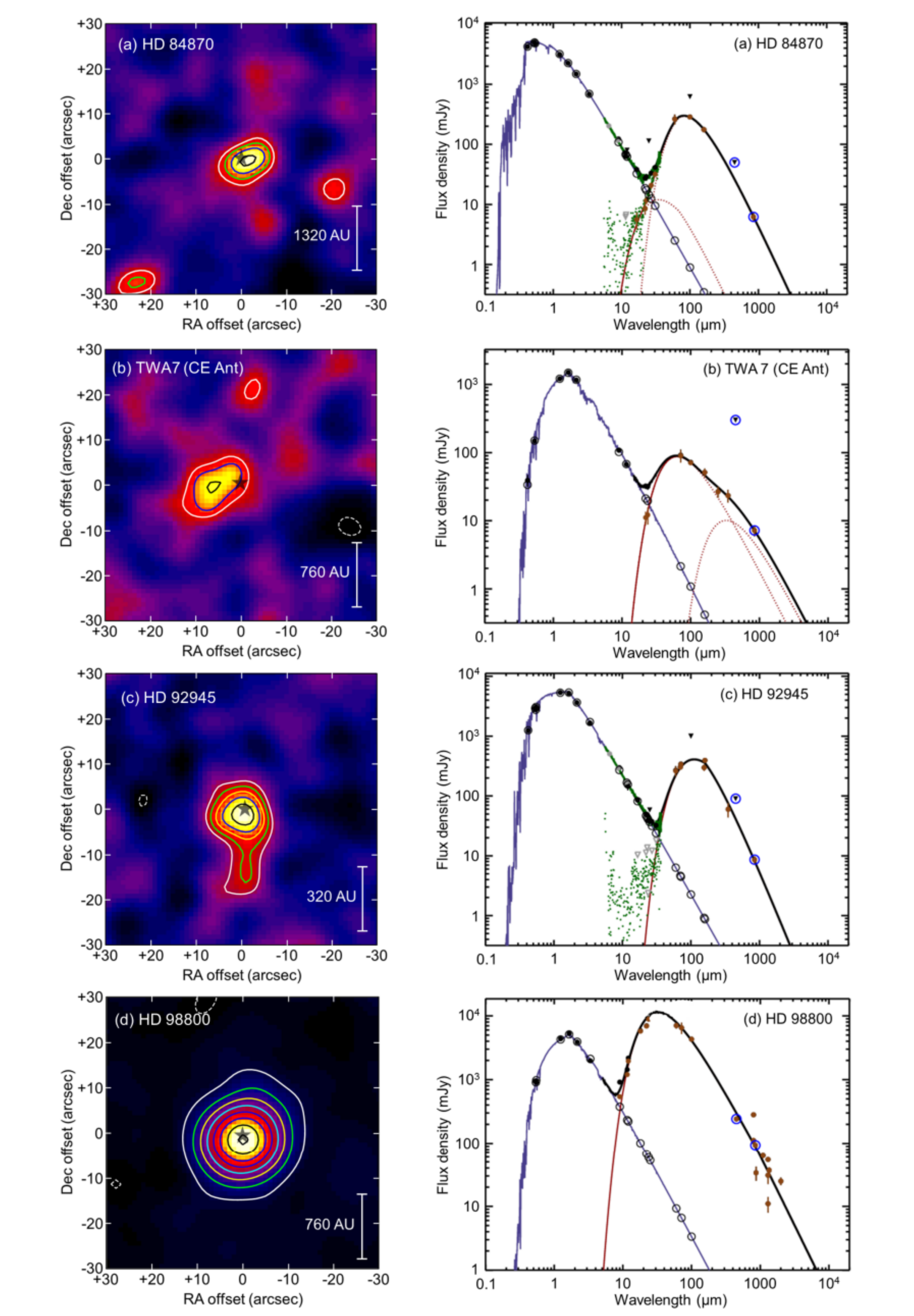}

  \caption{As for Figure A1 for the targets (a) HD 84870, (b) TWA7 (CE Ant), (c) 92945 and (d) HD 98800. The contours for HD 98800 are -4$\sigma$ (dashed) and
then solid contours starting at 4$\sigma$ and increasing in 8$\sigma$ steps.}

\label{fig:figureA7}
\end{figure*}

\begin{figure*}
\includegraphics[width=160mm]{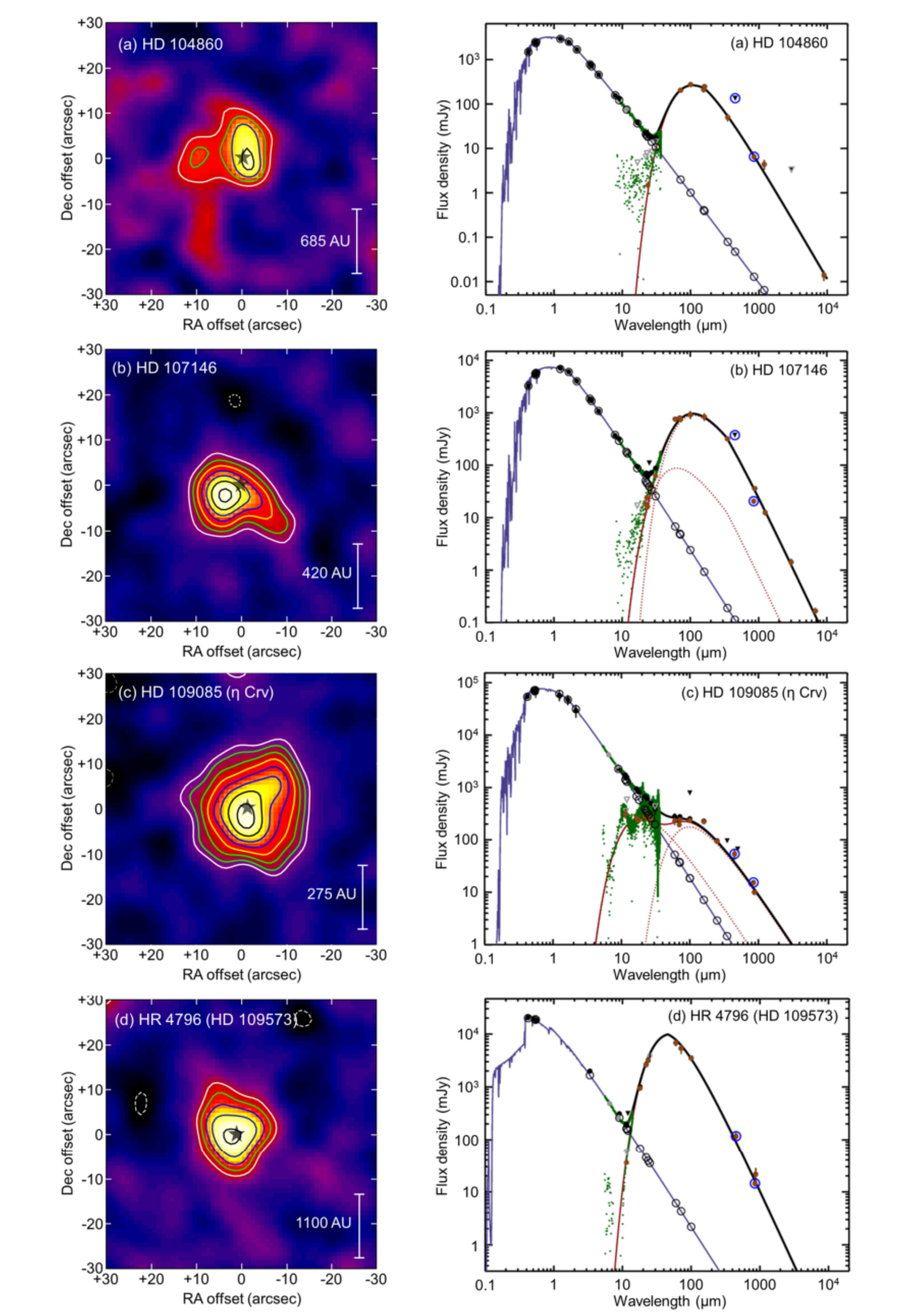}

  \caption{As for Figure A1 for the targets (a) HD 104860, (b) HD 107146, (c) HD 109085 ($\eta$ Crv) and (d) HR 4796 (HD 109573).}

\label{fig:figureA8}
\end{figure*}

\begin{figure*}
\includegraphics[width=160mm]{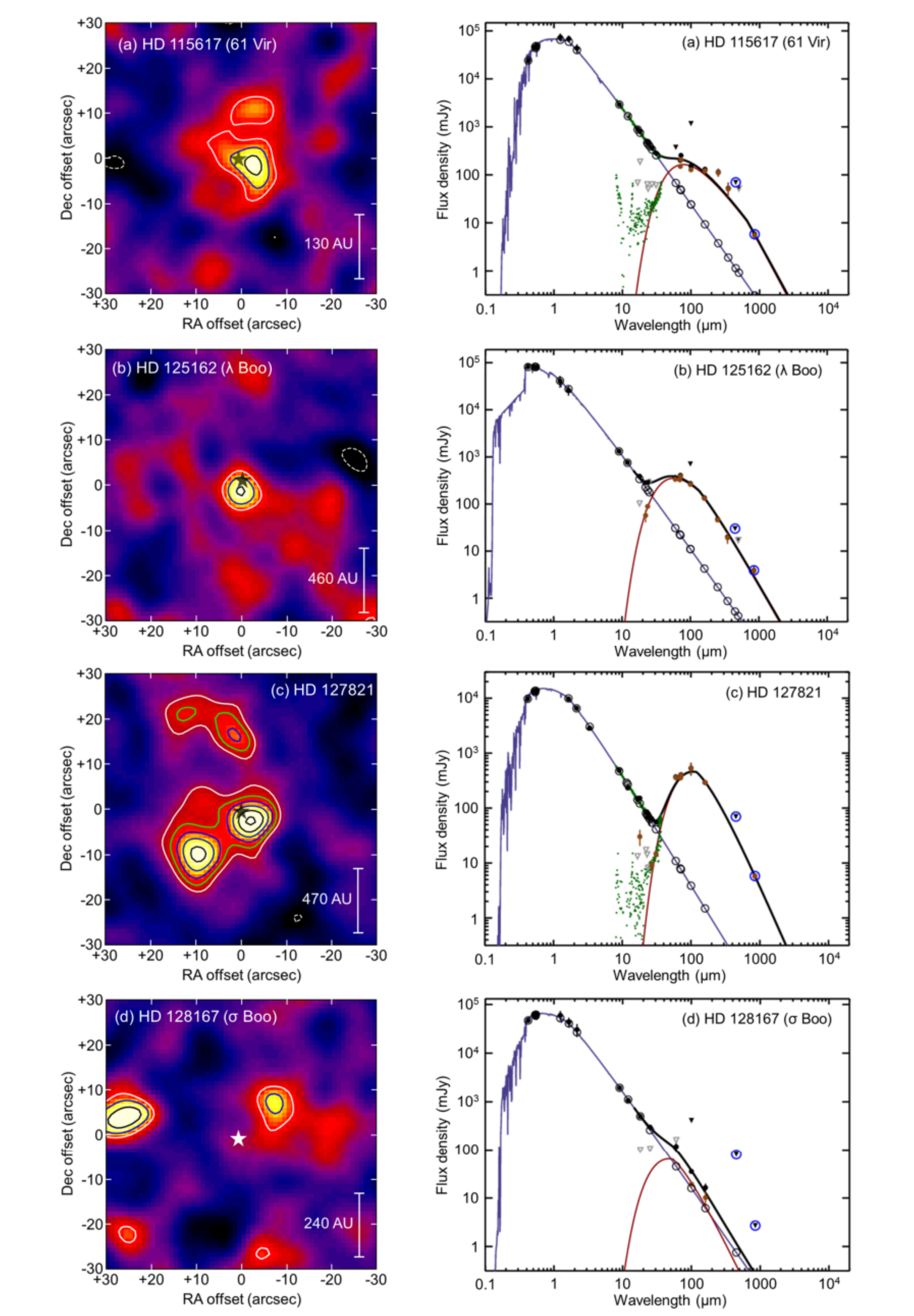}

  \caption{As for Figure A1 for the targets (a) HD 115617 (61 Vir), (b) HD 125162 ($\lambda$ Boo), (c) HD 127821 and (d) HD 128167 ($\sigma$ Boo).}

\label{fig:figureA9}
\end{figure*}

\begin{figure*}
\includegraphics[width=160mm]{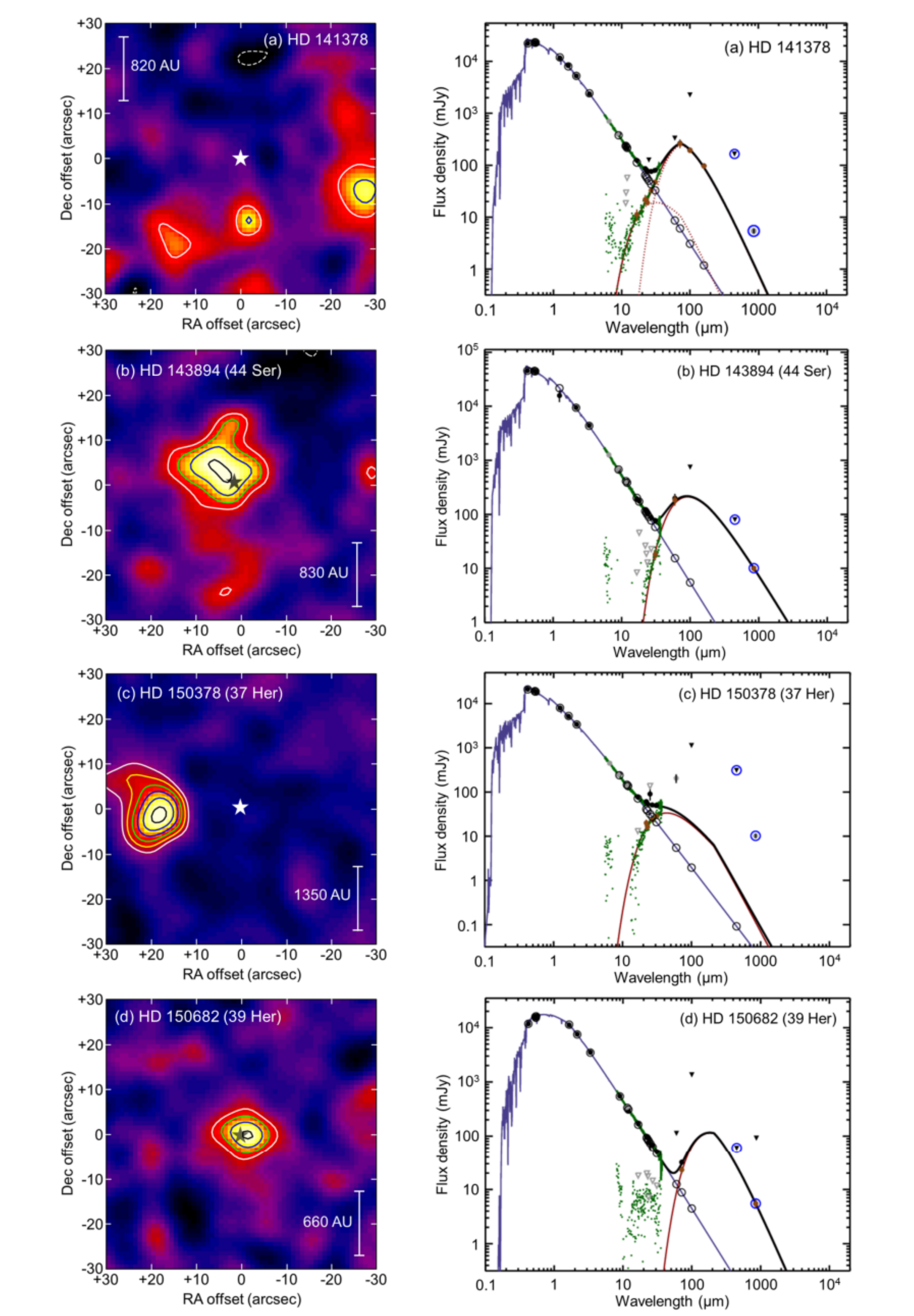}

  \caption{As for Figure A1 for the targets (a) HD 141378, (b) HD 143894 (44 Ser), (c) HD 150378 (37 Her) and (d) HD 150682 (39 Her).}

\label{fig:figureA10}
\end{figure*}

\begin{figure*}
\includegraphics[width=160mm]{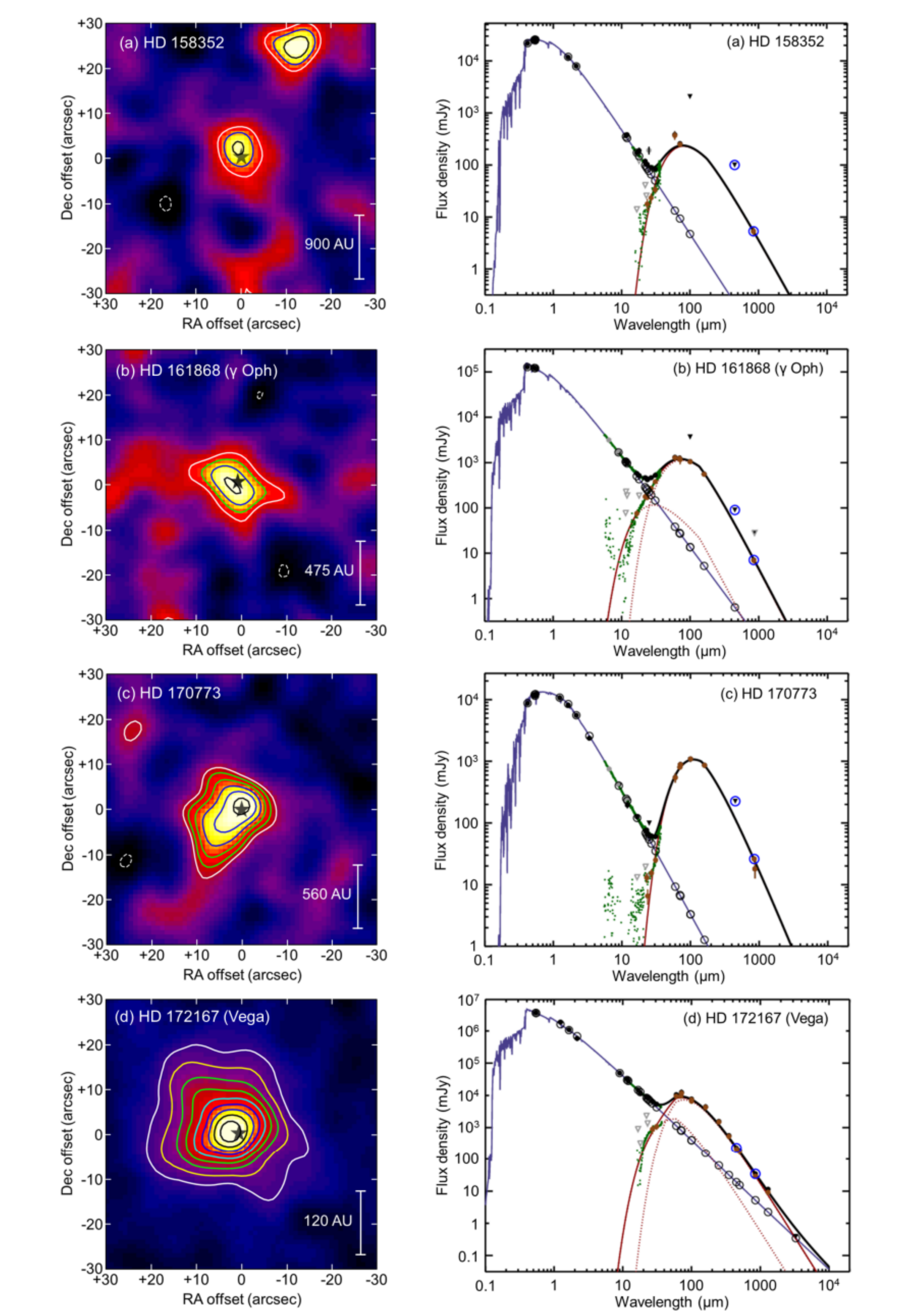}

  \caption{As for Figure A1 for the targets (a) HD 158352, (b) HD 161868 ($\gamma$ Oph), (c) HD 170773 and (d) HD 192167 (Vega).}

\label{fig:figureA11}
\end{figure*}

\begin{figure*}
\includegraphics[width=160mm]{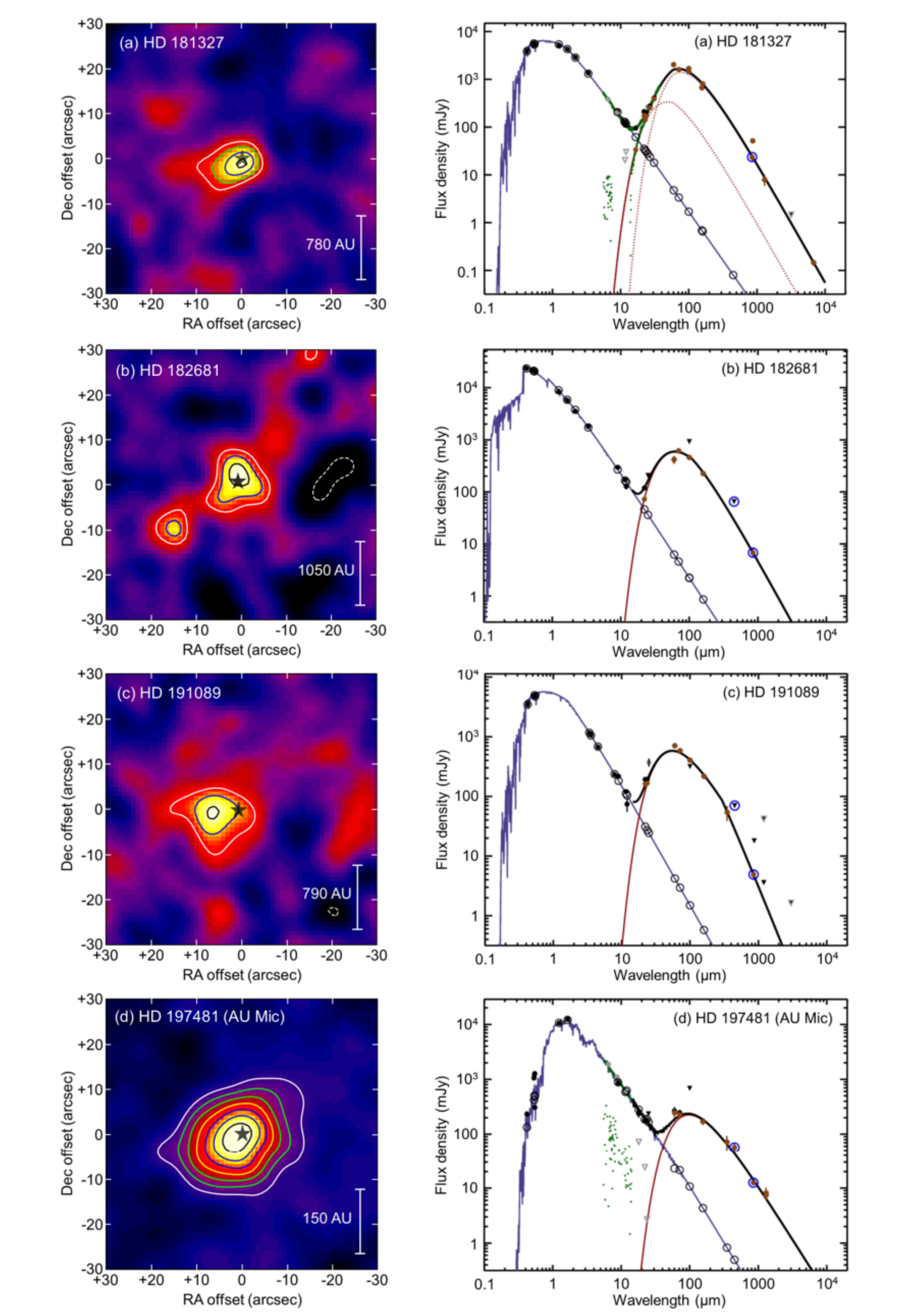}

  \caption{As for Figure A1 for the targets (a) HD 181327, (b) HD 182681, (c) HD 191089 and (d) HD 197481 (AU Mic). The contours for the HD
197481 are $-$3$\sigma$ (dashed) and then from +4$\sigma$ in 2$\sigma$ steps.}

\label{fig:figureA12}
\end{figure*}

\begin{figure*}
\includegraphics[width=160mm]{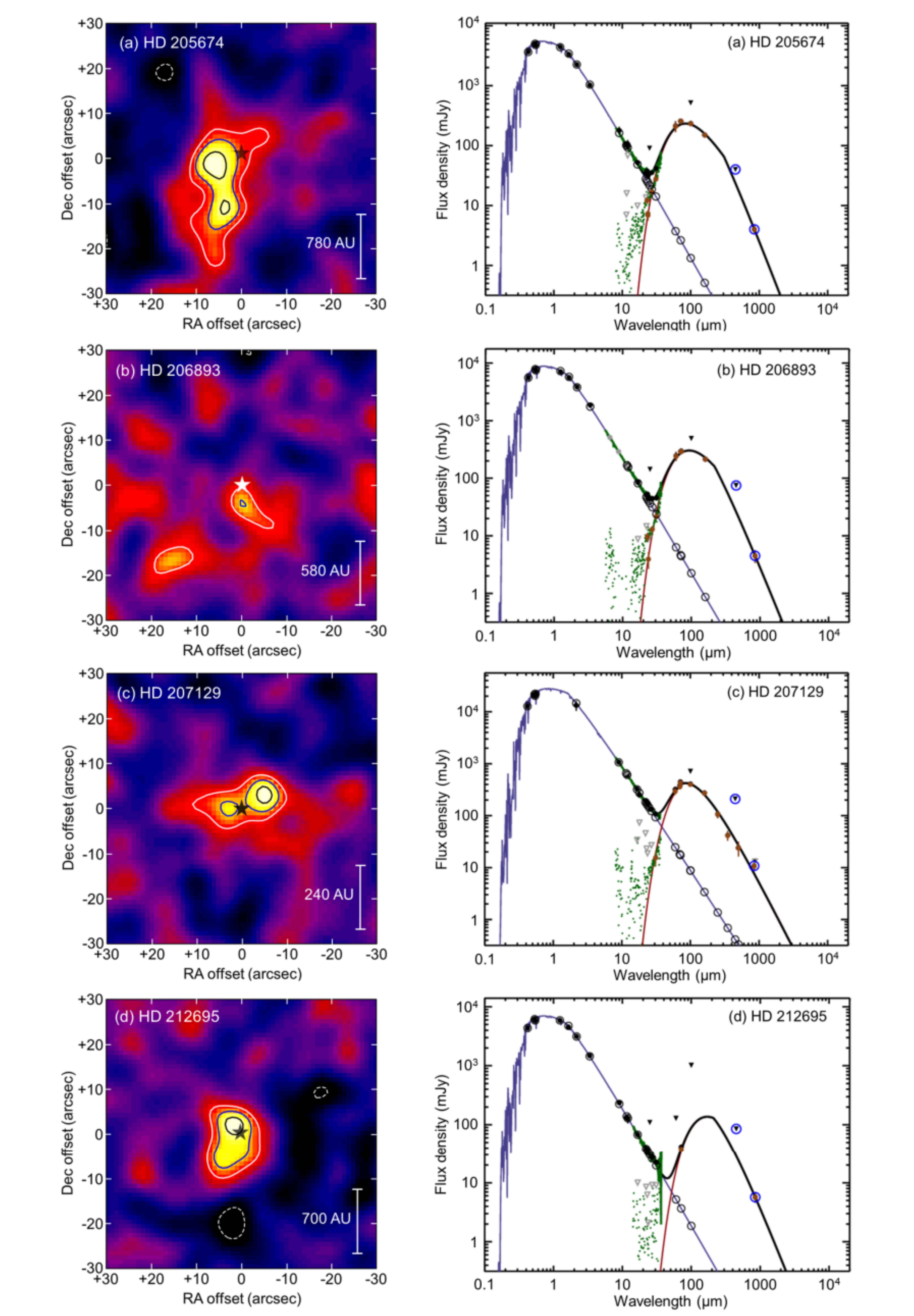}

  \caption{As for Figure A1 for the targets (a) HD 205674, (b) HD 206893, (c) HD 207129 and (d) HD 212695.}

\label{fig:figureA13}
\end{figure*}

\begin{figure*}
\includegraphics[width=160mm]{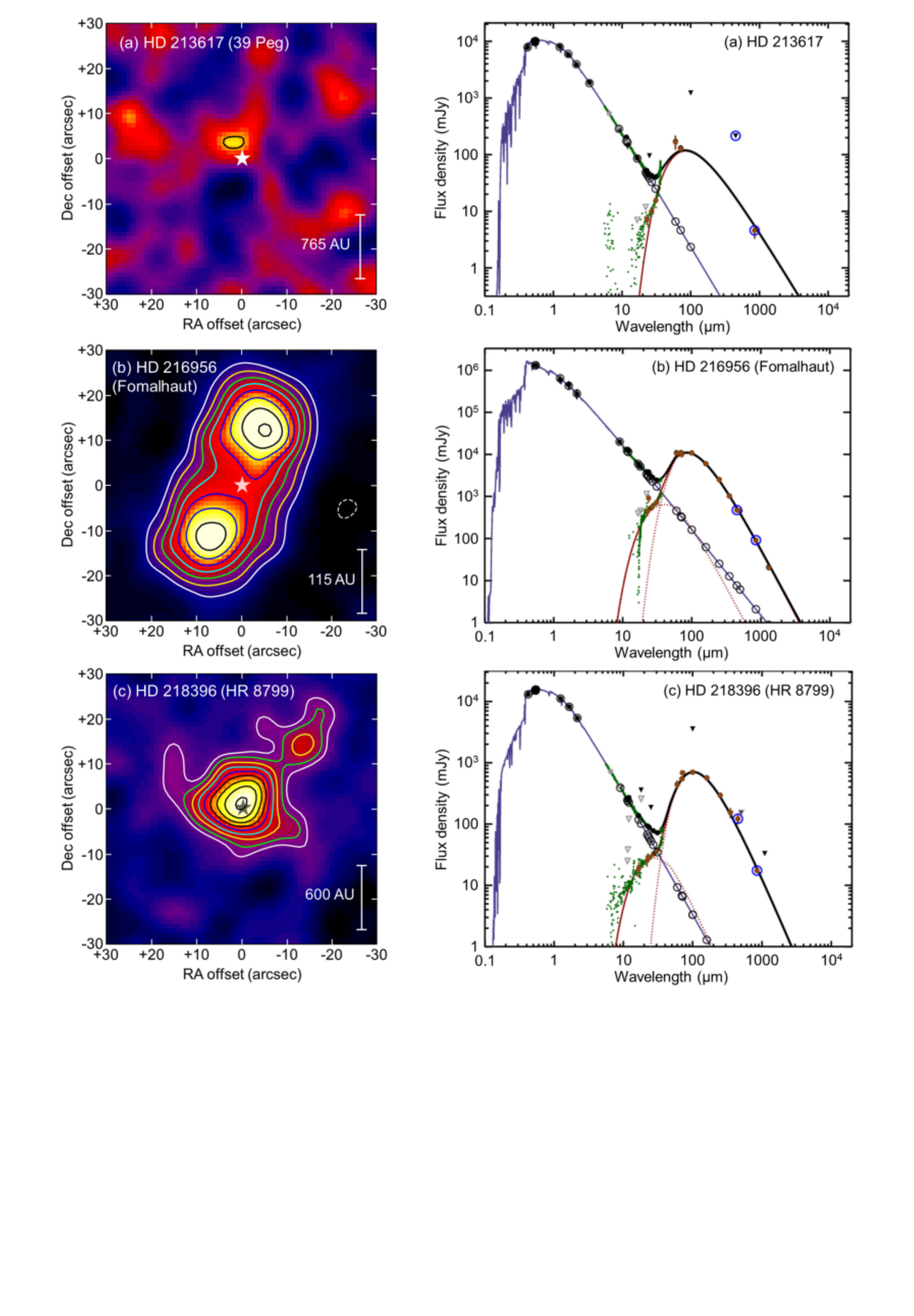}

  \caption{As for Figure A1 for the targets (a) HD 213617, (b) HD 216956 (Fomalhaut), and (c) HD218396 (HR 8799). The contours for the HD
216396 are $-$5$\sigma$ (dashed) and then from +5$\sigma$ in 3$\sigma$ steps.}

\label{fig:figureA14}
\end{figure*}


\vskip 5mm

\bsp    
\label{lastpage}
\end{document}